\documentclass[
aps,
prd,
reprint,
10pt,
preprintnumbers,
nofootinbib,
superscriptaddress,
floatfix
]{revtex4-2}

\usepackage{amsmath}
\usepackage{amssymb}
\usepackage{graphicx}
\usepackage{mathrsfs}
\usepackage{slashed}
\usepackage{array}
\usepackage[dvipsnames]{xcolor}
\usepackage{indentfirst}
\usepackage{multirow}
\usepackage[caption=false]{subfig}
\usepackage{placeins}

\usepackage[colorlinks=true, allcolors=blue]{hyperref}
\usepackage{cleveref}

\newcommand{\rme}{\mathrm{e}}
\newcommand{\rmi}{\mathrm{i}}
\newcommand{\rmd}{\mathrm{d}}

\begin{document}


\title{Baryon Light-Cone Distribution Amplitudes from Lattice QCD: Formalism, Renormalization, Extrapolation, and Matching}

\collaboration{\bf{Lattice Parton Collaboration ($\rm {\bf LPC}$)}}

\author{\includegraphics[scale=0.10]{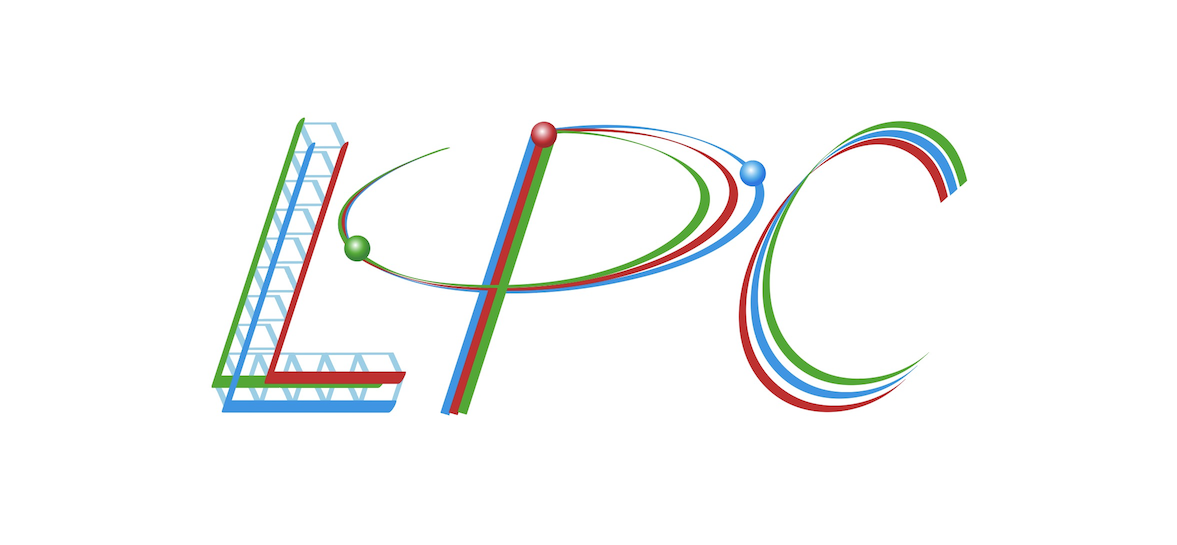}\\Mu-Hua~Zhang}
\affiliation{State Key Laboratory of Dark Matter Physics, Key Laboratory for Particle Astrophysics and Cosmology (MOE), Shanghai Key Laboratory for Particle Physics and Cosmology, Tsung-Dao Lee Institute and School of Physics and Astronomy, Shanghai Jiao Tong University, Shanghai 200240, China}

\author{Haoyang~Bai}
\affiliation{Institute of High Energy Physics, CAS, Beijing 100049, China}
\affiliation{School of Physics, University of Chinese Academy of Sciences, Beijing 100049, China}

\author{Min-Huan~Chu}
\affiliation{Faculty of Physics and Astronomy, Adam Mickiewicz University, ul.\ Uniwersytetu Pozna\'nskiego 2, 61-614 Pozna\'{n}, Poland}

\author{Jun~Hua}
\thanks{Corresponding author.}
\email{junhua@scnu.edu.cn}
\affiliation{State Key Laboratory of Nuclear Physics and Technology, Institute of Quantum Matter, South China Normal University, Guangzhou 510006, China}
\affiliation{Guangdong Basic Research Center of Excellence for Structure and Fundamental Interactions of Matter, Guangdong Provincial Key Laboratory of Nuclear Science, Guangzhou 510006, China}

\author{Xiangdong~Ji}
\affiliation{Tsung-Dao Lee Institute and School of Physics and Astronomy,
Shanghai Jiao Tong University, Shanghai 201210, China}

\author{Xiangyu~Jiang}
\affiliation{CAS Key Laboratory of Theoretical Physics, Institute of Theoretical Physics, Chinese Academy of Sciences, Beijing 100190, China}

\author{Jian~Liang}
\affiliation{State Key Laboratory of Nuclear Physics and Technology, Institute of Quantum Matter, South China Normal University, Guangzhou 510006, China}
\affiliation{Guangdong Basic Research Center of Excellence for Structure and Fundamental Interactions of Matter, Guangdong Provincial Key Laboratory of Nuclear Science, Guangzhou 510006, China}

\author{Cai-Dian~L\"u}
\affiliation{Institute of High Energy Physics, CAS, Beijing 100049, China}
\affiliation{School of Physics, University of Chinese Academy of Sciences, Beijing 100049, China}

\author{Andreas~Sch\"afer}
\affiliation{Institut f\"ur Theoretische Physik, Universit\"at Regensburg, D-93040 Regensburg, Germany}

\author{Wei~Wang}
\thanks{Corresponding author.}
\email{wei.wang@sjtu.edu.cn}
\affiliation{State Key Laboratory of Dark Matter Physics, Key Laboratory for Particle Astrophysics and Cosmology (MOE), Shanghai Key Laboratory for Particle Physics and Cosmology, School of Physics and Astronomy, Shanghai Jiao Tong University, Shanghai 200240, China}
\affiliation{Southern Center for Nuclear-Science Theory (SCNT), Institute of Modern Physics, Chinese Academy of Sciences, Huizhou 516000, Guangdong Province, China}

\author{Yi-Bo~Yang}
\affiliation{CAS Key Laboratory of Theoretical Physics, Institute of Theoretical Physics, Chinese Academy of Sciences, Beijing 100190, China}
\affiliation{School of Fundamental Physics and Mathematical Sciences, Hangzhou Institute for Advanced Study, UCAS, Hangzhou 310024, China}
\affiliation{International Centre for Theoretical Physics Asia-Pacific, Beijing/Hangzhou, China}
\affiliation{School of Physical Sciences, University of Chinese Academy of Sciences,
Beijing 100049, China}

\author{Jian-Hui~Zhang}
\affiliation{School of Science and Engineering, The Chinese University of Hong Kong, Shenzhen 518172, China}

\author{JiaLu~Zhang}
\affiliation{State Key Laboratory of Dark Matter Physics, Key Laboratory for Particle Astrophysics and Cosmology (MOE), Shanghai Key Laboratory for Particle Physics and Cosmology, Tsung-Dao Lee Institute and School of Physics and Astronomy, Shanghai Jiao Tong University, Shanghai 200240, China}

\author{Qi-An~Zhang}
\affiliation{School of Physics, Beihang University, Beijing 102206, China}

\begin{abstract}
Baryon light-cone distribution amplitudes (LCDAs) are inherently multi-dimensional objects parametrized by two independent longitudinal momentum fractions, making their first-principles determination substantially more challenging than that of meson LCDAs.  
We present a systematic large-momentum effective theory (LaMET) framework for determining baryon leading-twist  LCDAs from lattice QCD. 
The framework covers the complete path from equal-time three-quark quasi-distribution amplitudes to physical baryon LCDAs. 
We formulate the leading-twist $V$, $A$, and $T$ quasi-DAs and analyze their spin-flavor and coordinate-space symmetries, including antisymmetric amplitudes with vanishing local limits. 
We develop a hybrid renormalization prescription on the $(z_1,z_2)$ plane, introduce a newly developed large-$\lambda$ extrapolation strategy based on the asymptotic large-distance behavior of Euclidean correlators, and derive the corresponding one-loop LaMET matching relation in the hybrid renormalization scheme.
As a demonstration, we apply the complete analysis pipeline to the $\Lambda$-baryon $A$-structure quasi-DAs using seven $N_f=2+1$ lattice ensembles, and use this amplitude to examine the impact of large-distance extrapolation, perturbative matching, and extrapolation to the continuum, physical-pion-mass, and infinite-momentum limits, together with the associated systematic uncertainties.
This work provides the formalism, renormalization, extrapolation, and matching infrastructure for first-principles determinations of $x$-dependent baryon LCDAs.
\end{abstract}

\maketitle


\tableofcontents

\section{Introduction}

Understanding the internal structure of hadrons in terms of quark and gluon degrees of freedom is one of the central goals of quantum chromodynamics (QCD). In hard exclusive processes, this information is encoded in the light-cone distribution amplitudes (LCDAs), which describe the longitudinal momentum distribution of partons in the leading Fock state and enter QCD factorization formulas at large momentum transfer~\cite{Lepage:1980fj,Efremov:1978rn}. LCDAs therefore provide essential non-perturbative inputs connecting hadron structure to experimentally measurable observables.

Baryon LCDAs enter a broad class of exclusive reactions, including baryon electromagnetic and transition form factors and weak decays of heavy baryons. Their phenomenological relevance has become more pronounced with recent progress in baryonic flavor and CP physics. Following earlier evidence for CP violation in $\Lambda_b^0\to\Lambda K^+K^-$ decays~\cite{LHCb:2024yzj}, the first observation of CP violation in baryon decays by the LHCb Collaboration~\cite{LHCb:2025ray} has highlighted the need for improved theoretical control of hadronic amplitudes in heavy-baryon decays. In perturbative-QCD analyses of selected two-body $\Lambda_b$ decays, baryon LCDAs provide essential non-perturbative inputs to both decay amplitudes and CP asymmetries~\cite{Han:2024kgz,Han:2025tvc}. A complementary and more direct connection to light-hyperon structure has recently emerged from electric-dipole-moment (EDM) studies. The substantially improved experimental constraint on the $\Lambda$ EDM from BESIII~\cite{BESIII:2025vxm}, together with a perturbative-QCD analysis relating hyperon EDMs to convolutions involving their LCDAs~\cite{Chen:2025rab}, provides a new application sensitive to electric and chromoelectric dipole interactions of the underlying quarks. At present, however, many phenomenological applications still rely on model inputs or QCD sum-rule estimates~\cite{Chernyak:1987nu,Ball:2008fw}, leaving the limited knowledge of baryon LCDAs as a significant source of theoretical uncertainty.


Compared with meson LCDAs, baryon LCDAs are intrinsically more involved. The leading-twist baryon LCDAs depend on three valence-quark momentum fractions subject to $x_1+x_2+x_3=1$, or equivalently on two independent variables. It is therefore a genuinely two-dimensional non-perturbative distribution function. In addition, baryon LCDAs contain several independent Dirac-$\gamma$ structures already at leading-twist level and obey nontrivial spin-flavor and permutation symmetries. These features make baryon LCDAs richer than their mesonic counterparts, but also substantially more difficult to determine from first principles. 

The difficulty is amplified in lattice QCD. In coordinate space, baryon LCDAs are defined by matrix elements of nonlocal three-quark light-cone operators with two independent separations. A lattice calculation must therefore deal with three-quark operators depending on two independent coordinates, a two-dimensional Fourier transform, more severe signal degradation at large separations, and a renormalization structure involving multiple Wilson-line distances. These features prevent a direct extension of the standard meson-LCDA analysis and require a baryon-specific framework.

Historically, quantitative information on baryon LCDAs has mainly come from QCD sum rules and lattice calculations of local moments. Early QCD sum-rules studies related the lowest moments of baryon distribution amplitudes to matrix elements of local operators~\cite{Chernyak:1984bm,Chernyak:1987nu,King:1986wi}. These works provided important phenomenological guidance, but the resulting picture remained model-dependent. In particular, different QCD sum-rules analyses led to quantitatively different estimates for the shape of the nucleon distribution amplitudes, ranging from strongly asymmetric models to more moderate asymmetries~\cite{Chernyak:1984bm,King:1986wi}. The broader interpretation of exclusive baryonic processes has also been affected by the long-standing debate over the relative importance of perturbative hard-scattering contributions and soft end-point mechanisms~\cite{Isgur:1988iw}. Later developments, including treatments based on nonlocal condensates and light-cone sum rules, further refined this picture and generally supported less extreme deviations from the asymptotic form~\cite{Mikhailov:1986be,Braun:2006hz}.

On the lattice, the traditional approach to light-cone observables is based on the operator product expansion (OPE), where low order moments of distribution amplitudes can be obtained from local operators. This program has been successfully applied to baryons. In particular, the RQCD collaboration determined normalization constants and first moments of octet-baryon distribution amplitudes and studied SU(3)-flavor breaking effects~\cite{Bali:2015ykx,RQCD:2019hps,Bali:2024oxg}. These results constitute important first-principles constraints on baryon LCDAs.

Nevertheless, the information currently available from moment-based lattice calculations remains intrinsically limited. Existing studies have determined normalization constants and the lowest nontrivial moments, the first moments, which constrain only the lowest conformal coefficients of baryon LCDAs. For leading-twist baryon LCDAs, however, the relevant objects are two-dimensional functions of the valence-quark momentum fractions. Their detailed shape, correlations among the three quarks, and endpoint behavior cannot be reconstructed from such a small set of moments. This is insufficient for applications in which the LCDAs enter factorization formulas as functional inputs, and motivates a direct determination of the $x$-dependent baryon LCDAs.

Large-momentum effective theory (LaMET) provides a conceptually different route to light-cone physics from Euclidean lattice QCD~\cite{Ji:2013dva,Ji:2014gla,Ji:2020ect}. In LaMET, one computes equal-time spatial correlators in a boosted hadron state and relates the resulting quasi-distributions to the corresponding light-cone distributions through perturbative matching. Since the quasi-distributions and light-cone distributions share the same infrared structure, LaMET allows one to go beyond the moment expansion and access the full $x$-dependence. Over the past decade, this strategy has been developed and applied to a wide range of hadronic observables, including meson LCDAs, nucleon parton distribution functions (PDFs), transverse-momentum-dependent distributions (TMDs), and generalized parton distributions (GPDs)~\cite{Zhang:2017bzy,Zhang:2020gaj,Holligan:2023rex,Hua:2020gnw,LatticeParton:2022zqc,Baker:2024zcd,Cloet:2024vbv,Xiong:2013bka,Alexandrou:2016eyt,Chen:2017mie,Zhang:2017zfe,Xu:2018mpf,Liu:2018hxv,Wang:2019msf,Zhang:2019qiq,Chen:2020ody,LatticeParton:2020uhz,Lin:2020rxa,Bhattacharya:2021moj,Gao:2021hxl,Li:2021wvl,Deng:2022gzi,Gao:2022iex,Gao:2022uhg,LatticeParton:2022xsd,LatticePartonCollaborationLPC:2022myp,LatticePartonLPC:2022eev,Zhang:2022xuw,Zhu:2022bja,Ji:2023pba,LatticeParton:2023xdl,Zhao:2023ptv,Liu:2023onm,Avkhadiev:2024mgd,Good:2024iur,Han:2024cht,Han:2024min,Holligan:2024umc,Holligan:2024wpv,Ji:2024hit,Wang:2024wwa,LatticeParton:2024zko,Zhang:2024omt,Bollweg:2025iol,Wang:2025uap,Ji:2025mvk,Chen:2025cxr,Tan:2025ofx,Chu:2025kew,Chu:2025jsi,LPC:2026vyv,LPC:2026ffe,Gao:2025inf,Gao:2026hix,Grebe:2026qmt,Francis:2026czb}.
In parallel, closely related approaches based on pseudo-distributions~\cite{Radyushkin:2017lvu,Zhang:2018ggy,Karpie:2018zaz,Joo:2019jct,Joo:2019bzr,HadStruc:2021qdf,Bhat:2022zrw,Kovner:2024pwl,Bhattacharya:2024qpp,HadStruc:2024rix,NieMiera:2025inn,Dutrieux:2026grg} and current-current correlators~\cite{Bali:2017gfr,Sufian:2019bol,Bali:2018spj,Sufian:2020vzb,Zimmermann:2024zde} have provided complementary routes to light-cone parton physics from Euclidean lattice QCD.

Extending LaMET to baryon LCDAs introduces several new theoretical and practical issues. The baryon quasi-DAs depend on two independent spatial separations $z_1$ and $z_2$, thus the Fourier transform and matching are genuinely two-dimensional, and the leading-twist sector contains several Dirac-$\gamma$ structures with different $z_1\leftrightarrow z_2$ exchange symmetries. For the $\Lambda$-baryon, the antisymmetric $V$ and $T$ structures vanish in the local limit, so their quasi-DAs cannot be normalized through their own local matrix elements. This affects both the reduction of lattice correlation functions and the choice of reference matrix elements in renormalization. In addition, the Wilson-line ultraviolet divergences and short-distance logarithms must be treated consistently over the full $(z_1,z_2)$ plane before the momentum-space matching can be defined.

Recent theoretical developments have established the LaMET factorization and renormalization structure for baryon LCDAs~\cite{Deng:2023csv,Han:2023xbl,Han:2024ucv,Zhang:2025npd,Zhang:2026epr,Zhang:2026rql}. In our previous work, we demonstrated the feasibility of applying LaMET to baryon LCDAs~\cite{LatticeParton:2024vck} and explored the associated hybrid renormalization procedure~\cite{LatticePartonCollaborationLPC:2025vhd}. These developments make it possible to construct a complete framework for determining baryon LCDAs from lattice QCD.

In this work, we complete a baryon-LaMET framework for extracting baryon leading-twist LCDAs from lattice QCD within LaMET.
We first define the leading-twist baryon quasi-DAs for the $V$, $A$, and $T$ structures and analyze their coordinate-space symmetry properties.
We then derive the reduction formulas needed to extract the corresponding quasi-DAs from baryon two-point correlation functions, including the modified normalization required for antisymmetric amplitudes with vanishing local limits. 
A hybrid renormalization prescription is formulated on the two-dimensional $(z_1,z_2)$ plane, combining short-distance ratio renormalization with long-distance self-renormalization, with a structure-dependent implementation for the $V$, $A$, and $T$ amplitudes. 
The framework also includes a large-distance extrapolation strategy for the coordinate-space quasi-DA matrix elements based on the asymptotic long-distance expansion of Euclidean correlators. 
Finally, we derive the one-loop LaMET matching relation in the hybrid scheme, including the hybrid counterterms that remove the short-distance ultraviolet logarithms and render the two-dimensional matching convolution well defined.

As a numerical demonstration of the complete pipeline, we apply the framework to the $\Lambda$-baryon $A$-structure quasi-DA using seven $N_f=2+1$ lattice ensembles. The analysis includes hybrid renormalization, large-distance coordinate-space extrapolation, two-dimensional Fourier transform, perturbative matching, and a combined extrapolation to the continuum, physical-pion-mass, and infinite-momentum limits. The purpose of this numerical demonstration is to validate the baryon-LaMET framework and quantify the main sources of uncertainty in an end-to-end baryon LCDAs calculation. The complete physical determination of the leading-twist $\Lambda$-baryon $V$, $A$, and $T$ LCDAs, together with phenomenological applications, is presented in the companion Letter~\cite{LPC:2026lcj}.

The rest of this article is organized as follows. In Sec.~\ref{sec:Framework}, we introduce the leading-twist baryon LCDAs and their equal-time counterparts in LaMET, the baryon quasi-DAs. In Sec.~\ref{sec:Lattice_Simulation}, we describe the lattice setup, choices of interpolating operators and spinor projector, as well as the extraction of quasi-DA matrix elements. In Sec.~\ref{sec:Hybrid}, we present the hybrid renormalization prescription on the two-dimensional $(z_1,z_2)$ coordinate plane, for all the leading-twist amplitudes. In Sec.~\ref{sec:extrapolation}, we construct the coordinate-space large-distance extrapolation using the newly developed asymptotic ans\"atze of Euclidean correlators. In Sec.~\ref{sec:Matching}, we derive and implement the perturbative LaMET matching relation in the hybrid scheme. In Sec.~\ref{sec:Apply_Results}, we apply the full framework to the $\Lambda$-baryon $A$-structure LCDA and analyze the associated systematic uncertainties. We summarize in Sec.~\ref{sec:Summary}.

\section{Theoretical Formalism}\label{sec:Framework}
In this section, we set up the LaMET formalism for baryon leading-twist light-cone distribution amplitudes (LCDAs). Since LCDAs are defined from light-cone correlators, they cannot be computed directly in Euclidean lattice QCD. LaMET replaces them by equal-time spatial correlators evaluated in boosted baryon states. The corresponding quasi-distribution amplitudes (quasi-DAs) share the same infrared structure as the physical LCDAs and can be related to them at large hadron momentum through perturbative matching, up to power corrections suppressed by the hadron momentum.

Compared with meson LCDAs, baryon LCDAs require a genuinely two-dimensional treatment. They involve three valence quarks and depend on two independent momentum fractions $x_1$ and $x_2$, with the third fixed by momentum conservation $x_3=1-x_1-x_2$. The associated equal-time matrix elements therefore depend on two independent spatial separations. In addition, the leading-twist sector contains the $V$, $A$, and $T$ structures, whose symmetry properties depend on the baryon flavor content. These features affect the construction of quasi-DAs, their normalization, the renormalization and extrapolation procedure implemented in the lattice calculation.

\subsection{Baryon Leading-twist LCDAs}\label{sec:leading-twist_LCDAs}

\begin{figure}[htbp]
    \centering
    \includegraphics[width=0.85\linewidth]{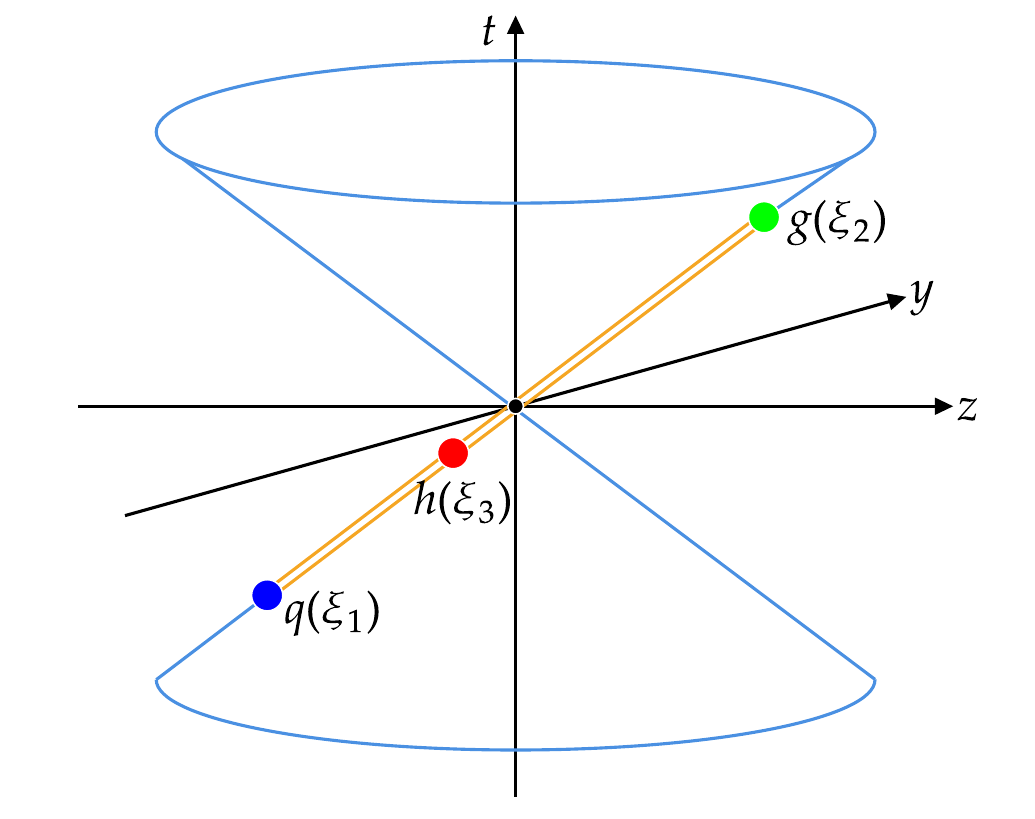}
    \caption{Schematic structure of the light-cone three-quark operator defining baryon leading-twist LCDAs. The three quark fields are located at $\xi_1$, $\xi_2$, and $\xi_3$ along the light-cone direction and are connected by Wilson lines to ensure gauge invariance.}
    \label{fig:structure_LCDA}
\end{figure}

The LCDAs of baryons can be defined as the baryon-to-vacuum matrix elements of gauge invariant light-like separated three-quark operators~\cite{Braun:1999te,Han:2024ucv,Braun:2000kw}:
\begin{equation}\label{eq:DA_def}
\begin{aligned}
    \epsilon^{ijk} \langle 0 | &q^{i'}_\alpha(\xi_1n) U_{i'i}(\xi_1n,\xi_0n)\\
    \times&g^{j'}_\beta(\xi_2n) U_{j'j}(\xi_2n,\xi_0n)\\
    \times&h^{k'}_\gamma(\xi_3n) U_{k'k}(\xi_3n,\xi_0n) | B(P) \rangle^R\ .
\end{aligned}
\end{equation}
Here $q,g,h$ denote the three valence-quark fields in the leading Fock component of the baryon, following the flavor convention summarized in Table~\ref{tab:valancequark}. The superscript $R$ denotes the renormalized matrix element. Gauge invariance is maintained by the Wilson lines $U$ from $\xi_0$ to $\xi$ along the light-cone:
\begin{equation}
    U(\xi, \xi_0) = \mathbb P \exp \left[ \rmi g \int_0^1\rmd \zeta(\xi-\xi_0)n\cdot A\big(\zeta\xi+(1-\zeta)\xi_0\big) \right]\ .
\end{equation}
The Greek indices $\alpha,\beta,\gamma,\dots$ denote Dirac spinor indices and the Latin indices $i,j,k,\dots$ denote color indices in the fundamental representation of ${\rm SU}(3)_c$. $\mathbb P$ is the path-order operator. We use the light-cone vectors $n=(1,0,0,-1)^\mu/\sqrt2$ and $\bar n^\mu=(1,0,0,1)/\sqrt2$, satisfying $n^2 = \bar n^2 = 0$. The $|B(P)\rangle$ is a baryon state with momentum $P^\mu = P^+ \bar n^\mu = (P^+,0,0,P^+)$ along the light-cone direction. Fig.~\ref{fig:structure_LCDA} schematically illustrates the light-cone definition of baryon leading-twist LCDAs.

\begin{table*}[ht]
    \centering
    \renewcommand{\arraystretch}{1.5}
    \setlength{\tabcolsep}{3mm}
    \begin{tabular}{c c c c c c c c c}
        \hline\hline
        Octet & $\rm p$ & $\rm n$ & $\Sigma^-$ & $\Sigma^0$ & $\Sigma^+$ & $\Xi^-$ & $\Xi^0$ & $\Lambda$ \\
        \hline
        $q,g,h$ & $\rm u\ u\ d$ & $\rm d\ d\ u$ & $\rm d\ d\ s$ & $\rm \frac{1}{\sqrt{2}}(u\ d\ s + d\ u\ s)$ & $\rm u\ u\ s$ & $\rm s\ s\ d$ & $\rm s\ s\ u$ & $\rm u\ d\ s$ \\
        \hline\hline
    \end{tabular}
    \caption{Valence quarks convention for octet baryons.}  
    \label{tab:valancequark}
\end{table*}

The general three-quark matrix element Eq.~\eqref{eq:DA_def} can be decomposed into 24 independent Lorentz structures~\cite{Braun:2000kw}. At leading-twist level, only three structures are relevant, of which we are focusing on in this work. They are conventionally denoted by $V$, $A$, and $T$, and are defined by~\cite{Chernyak:1984bm}:
\begin{equation}
\begin{aligned}
    &\langle 0 | f_\alpha(\xi_1n) g_\beta(\xi_2n) h_\gamma(\xi_3n) | B(P) \rangle^R\\
    =& \frac14 f_B \Big[ (\slashed P C)_{\alpha\beta}(\gamma^5 u_B)_\gamma \Phi^V_B(\xi_{\ell} n\cdot P,\mu)\\
    &\quad+ (\slashed P \gamma^5 C)_{\alpha\beta}(u_B)_\gamma \Phi^A_B(\xi_{\ell} n\cdot P,\mu) \Big]\\
    &+ \frac14 f^T_B (\rmi  \sigma_{\mu\nu} P^\nu C)_{\alpha\beta}(\gamma^\mu \gamma^5 u_B)_\gamma \Phi^T_B(\xi_{\ell} n\cdot P,\mu)\ .
\end{aligned}
\end{equation}
Here $\ell=1,2,3$, $C=\rmi  \gamma^2\gamma^0$ is the charge conjugation matrix, $u_B$ stands for the spinor of the baryon, and $\mu$ is the renormalization scale. We do not explicitly write out the Wilson lines and the color indices here and in the following context for notational simplicity. The $V$ and $A$ structures share the same decay constant $f_B$, while $T$ structure does not.

Therefore, the three leading-twist amplitudes can be projected from nonlocal matrix elements by choosing suitable Dirac-$\gamma$ structures:
\begin{equation}
\begin{aligned}
    &\ M^V_B(\xi_1,\xi_2,\xi_3;P,\mu) u_B \\
    =&\ \langle 0 | q^{\rm T}(\xi_1n) (C\slashed n) g(\xi_2n) h(\xi_3n) | B(P) \rangle^R \\
    =&\ -f_B \Phi^V_B(\xi_\ell n\cdot P,\mu) n\cdot P \gamma^5 u_B\ ,\\
    \\
    &\ M^A_B(\xi_1,\xi_2,\xi_3;P,\mu) u_B \\
    =&\ \langle 0 | q^{\rm T}(\xi_1n) (C\gamma^5\slashed n) g(\xi_2n) h(\xi_3n) | B(P) \rangle^R \\
    =&\ f_B \Phi^A_B(\xi_\ell n\cdot P,\mu) n\cdot P u_B\ ,\\
    \\
    &\ M^T_B(\xi_1,\xi_2,\xi_3;P,\mu) u_B \\
    =&\ \langle 0 | q^{\rm T}(\xi_1n) (\rmi  C \sigma_{\mu\nu} n^\nu ) g(\xi_2n) \gamma_\mu h(\xi_3n) | B(P) \rangle^R \\
    =&\ 2f^T_B \Phi^T_B(\xi_\ell n\cdot P,\mu) n\cdot P \gamma^5 u_B\ .
\end{aligned}
\end{equation}
In the following context, for convenience, we set $\xi_3=\xi_0=0$, which fixes the third quark at the Wilson-line junction and leaves only two independent light-cone separations $\xi_1$ and $\xi_2$.

With the conventions above, LCDAs in momentum space can be accessed through a two-dimensional Fourier transform of the matrix elements in coordinate space:
\begin{equation}
\begin{aligned}
    \phi_B^X(x_1,x_2;\mu) &= (n\cdot P)^2 \int \frac{ \rmd \xi_1}{2\pi} \frac{ \rmd \xi_2}{2\pi}\ \rme^{\rmi  (x_1\xi_1+x_2\xi_2) n\cdot P }\\
    &\quad \times\Phi_B^X(\xi_1n\cdot P,\xi_2n\cdot P;\mu)\ ,
\end{aligned}
\end{equation}
where $X=V,A,T$. The variables $x_1,x_2$ denote the longitudinal momentum fractions carried by the $q$ and $g$ quarks, while $x_3 = 1-x_1-x_2$ for the $h$ quark.The physical support is the triangular region $0\leq x_1,x_2,x_3\leq 1$. We use the normalization convention:
\begin{equation}
\begin{aligned}
    &\int [\rmd x]\ \phi_\Lambda^A(x_1,x_2)=1\ ,  \\
    &\int [\rmd x]\ \phi_\Lambda^{V,T}(x_1,x_2)=0\ , \\
    &\int [\rmd x]\ \phi_{B\neq\Lambda}^A(x_1,x_2)=0\ , \\
    &\int [\rmd x]\ \phi_{B\neq\Lambda}^{V,T}(x_1,x_2)=1\ ,
\end{aligned}
\end{equation}
where$\int [\rmd x]\equiv \int_0^1 \rmd x_1 \int_0^{1-x_1}\rmd x_2.$
Equivalently, in coordinate space this convention corresponds to
\begin{equation}\label{eq:norm_conv_lc}
\begin{aligned}
    \Phi_\Lambda^A(0,0) = 1\ , &\quad \Phi_\Lambda^{V,T}(0,0) = 0\ ,\\
    \Phi_{B\neq\Lambda}^A(0,0) = 0\ , &\quad \Phi_{B\neq\Lambda}^{V,T}(0,0) = 1\ .\\
\end{aligned}
\end{equation}
These local-limit relations will be important when defining the normalized quasi-DA matrix elements below.

\subsection{LaMET Expansion and Quasi-DAs}\label{sec:quasi}
Since LCDAs are defined through light-cone correlations, they are not directly accessible in the Euclidean lattice. In LaMET, instead of calculating the LCDAs directly, one calculates the corresponding equal-time spatial correlations in a hadron state boosted to a large but finite momentum $P^z$, known as quasi-distribution amplitudes (quasi-DAs). The resulting quasi-DAs have the same infrared structure as the corresponding LCDAs, while their ultraviolet difference can be treated perturbatively. The baryon LCDAs are then obtained from the quasi-DAs through the LaMET large-momentum expansion formula~\cite{Ji:2024oka}:
\begin{equation}\label{eq:LaMET}
\begin{aligned}
    \phi^X_B(x_1,x_2;\mu)
    =
    \int&\ \rmd y_1\rmd y_2\ \mathcal C^X(x_1,x_2;y_1,y_2;P^z,\mu)\\
    &\ \qquad\times\widetilde \phi^X_B(y_1,y_2;P^z,\mu)\\
    + \mathcal O\Bigg( \frac{\Lambda_{\rm QCD}^2}{(x_1P^z)^2}, &\frac{\Lambda_{\rm QCD}^2}{(x_2P^z)^2}, \frac{\Lambda_{\rm QCD}^2}{[(1-x_1-x_2)P^z]^2} \Bigg)\ .
\end{aligned}
\end{equation}
In this work we set the factorization and renormalization scales equal, $\mu_F=\mu_R=\mu$. The power corrections become enhanced near the endpoint regions where any of the momentum fractions $x_i$ approaches zero.

\begin{figure}[htbp]
    \centering
    \includegraphics[width=\linewidth]{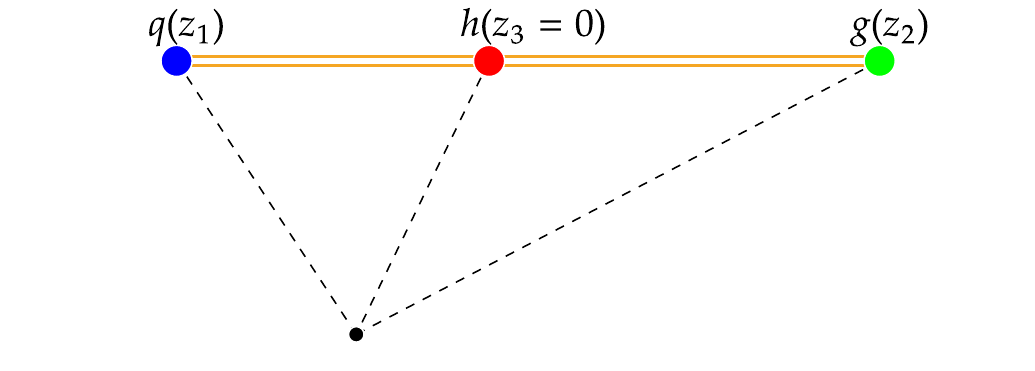}
    \caption{Structure of baryon leading-twist quasi-DAs. Three quarks are separated along the equal-time direction $n_z$, conected by Wilson lines.}
    \label{fig:structure_quasiDA}
\end{figure}

The momentum-space quasi-DAs are defined by the two-dimensional Fourier transform:
\begin{equation}
\begin{aligned}
    \widetilde \phi^X_B(y_1,y_2;P^z,\mu) &= (P^z)^2 \int \frac{ \rmd z_1}{2\pi} \frac{ \rmd z_2}{2\pi} \rme^{-\rmi  (y_1z_1+y_2z_2) P^z }\\
    &\quad \times\widetilde \Phi^X_B(z_1,z_2;P^z,\mu)\ .
\end{aligned}
\end{equation}
Here $z_1$ and $z_2$ are equal-time spatial separations along the longitudinal boost direction $n_z^\mu=(0,0,0,1)$, with the third quark fixed at $z_3=0$ for notational simplicity, as illustrated in Fig.~\ref{fig:structure_quasiDA}. The coordinate-space quasi-DAs are defined from matrix elements of spatially separated three-quark operators. A convenient choice for the leading-twist structures is:
\begin{equation}
\begin{aligned}\label{eq:quasi_matrix}
    &\ \widetilde M^V_B(z_1,z_2;P^z,\mu) u_B \\
    =&\ \langle 0 | q^{\rm T}(z_1) (C\gamma^t) g(z_2) \gamma^5 h(0) | B(P^z) \rangle^R \\
    =&\ -f_B \widetilde \Phi^V_B(z_1,z_2;P^z,\mu) P^z u_B\ , \\
    \\
    &\ \widetilde M^A_B(z_1,z_2;P^z,\mu) u_B \\
    =&\ \langle 0 | q^{\rm T}(z_1) (C\gamma^5\gamma^t) g(z_2) h(0) | B(P^z) \rangle^R \\
    =&\ f_B \widetilde \Phi^A_B(z_1,z_2;P^z,\mu) P^z u_B\ , \\
    \\
    &\ \widetilde M^T_B(z_1,z_2;P^z,\mu) u_B \\
    =&\ \langle 0 | q^{\rm T}(z_1) (\tfrac12 C [\gamma^t,\gamma^\mu] ) g(z_2) \gamma^5\gamma_\mu h(0) | B(P^z) \rangle^R \\
    =&\ 2f^T_B \widetilde \Phi^T_B(z_1,z_2;P^z,\mu) P^z u_B\ .
\end{aligned} 
\end{equation}

The coordinate-space quasi-DAs are normalized according to the same convention as the LCDAs:
\begin{equation}\label{eq:norm_conv_quasi}
\begin{aligned}
    \widetilde\Phi_\Lambda^A(0,0) = 1\ , &\quad \widetilde\Phi_\Lambda^{V,T}(0,0) = 0\ ,\\
    \widetilde\Phi_{B\neq\Lambda}^A(0,0) = 0\ , &\quad \widetilde\Phi_{B\neq\Lambda}^{V,T}(0,0) = 1\ .\\
\end{aligned}
\end{equation}
The local-limit relations in Eq.~\eqref{eq:norm_conv_quasi} determine how the coordinate-space quasi-DAs should be normalized in practice. For amplitudes with a nonzero local limit, the normalization can be taken with respect to the corresponding local matrix element. For amplitudes whose local limit vanishes, one must instead use a nonzero reference matrix element with the same baryon state. For the $\Lambda$ baryon, the $A$ amplitude has a nonzero local limit, while the $V$ and $T$ amplitudes are antisymmetric and vanish at $z_1=z_2=0$. We therefore define,
\begin{itemize}
    \item Normalization of $\Lambda$-baryon $A$ amplitude:
    \begin{equation}
    \begin{aligned}
        \widetilde\Phi_\Lambda^A(z_1,z_2;P^z) &= \frac{\widetilde M^A_\Lambda(z_1,z_2;P^z)}{f_\Lambda P^z} = \frac{\widetilde M^A_\Lambda(z_1,z_2;P^z)}{\widetilde M^A_\Lambda(0,0;P^z)} \\
        &\equiv \widehat M^A_\Lambda(z_1,z_2;P^z)\ ;
    \end{aligned}
    \end{equation}
    
    \item Normalization of $\Lambda$-baryon $V$ amplitude:
    \begin{equation}\label{eq:norm_V}
    \begin{aligned}
        \widetilde\Phi_\Lambda^V(z_1,z_2;P^z) &= \frac{\widetilde M^V_\Lambda(z_1,z_2;P^z)}{-f_\Lambda P^z} = -\frac{\widetilde M^V_\Lambda(z_1,z_2;P^z)}{\widetilde M^A_\Lambda(0,0;P^z)} \\
        &\equiv -\widehat M^V_\Lambda(z_1,z_2;P^z)\ ;
    \end{aligned}
    \end{equation}
    
    \item Normalization of $\Lambda$-baryon $T$ amplitude:
    \begin{equation}\label{eq:norm_T}
    \begin{aligned}
        \widetilde\Phi_\Lambda^T(z_1,z_2;P^z) &= \frac{\widetilde M^T_\Lambda(z_1,z_2;P^z)}{2f^T_\Lambda P^z} = \frac{f_\Lambda}{2f^T_\Lambda} \frac{\widetilde M^T_\Lambda(z_1,z_2;P^z)}{\widetilde M^A_\Lambda(0,0;P^z)} \\
        &\equiv \frac{f_\Lambda}{2f^T_\Lambda} \widehat M^T_\Lambda(z_1,z_2;P^z)\ .
    \end{aligned}
    \end{equation}
\end{itemize}
Thus the $\Lambda$-baryon $V$ and $T$ quasi-DAs are extracted using the local $\Lambda$-baryon $A$ matrix element as the reference normalization. This choice is required by the vanishing local limits of the antisymmetric amplitudes and does not modify their exchange-symmetry properties.

For the octet baryons except $\Lambda$, such as proton, the nonzero local limits occur in the $V$ and $T$ amplitudes, while the $A$ amplitude is antisymmetric. With the same notation, the corresponding normalized matrix elements are:
\begin{itemize}
    \item Normalization of proton $A$ amplitude:
    \begin{equation}
    \begin{aligned}
        \widetilde\Phi_{\rm p}^A(z_1,z_2;P^z) &= \frac{\widetilde M^A_{\rm p}(z_1,z_2;P^z)}{f_{\rm p} P^z} = -\frac{\widetilde M^A_{\rm p}(z_1,z_2;P^z)}{\widetilde M^V_{\rm p}(0,0;P^z)} \\
        &\equiv -\widehat M^A_{\rm p}(z_1,z_2;P^z)\ ;
    \end{aligned}
    \end{equation}
    
    \item Normalization of proton $V$ amplitude:
    \begin{equation}
    \begin{aligned}
        \widetilde\Phi_{\rm p}^V(z_1,z_2;P^z) &= \frac{\widetilde M^V_{\rm p}(z_1,z_2;P^z)}{-f_{\rm p} P^z} = \frac{\widetilde M^V_{\rm p}(z_1,z_2;P^z)}{\widetilde M^V_{\rm p}(0,0;P^z)} \\
        &\equiv \widehat M^V_{\rm p}(z_1,z_2;P^z)\ ;
    \end{aligned}
    \end{equation}
    
    \item Normalization of proton $T$ amplitude:
    \begin{equation}
    \begin{aligned}
        \widetilde\Phi_{\rm p}^T(z_1,z_2;P^z) &= \frac{\widetilde M^T_{\rm p}(z_1,z_2;P^z)}{2f^T_{\rm p} P^z} = \frac{\widetilde M^T_{\rm p}(z_1,z_2;P^z)}{\widetilde M^T_{\rm p}(0,0;P^z)} \\
        &\equiv \widehat M^T_{\rm p}(z_1,z_2;P^z)\ .
    \end{aligned}
    \end{equation}
\end{itemize}
After these normalized coordinate-space matrix elements are obtained from lattice correlation functions, the momentum-space quasi-DAs follow from the two-dimensional Fourier transform defined above.

\subsection{Symmetries of Octet Quasi-DAs in Coordinate Space}\label{sec:symmetries}
The baryon quasi-DAs are extracted from coordinate-space matrix elements on the two-dimensional $(z_1,z_2)$ plane. Their symmetry properties provide useful constraints on the lattice data and reduce the number of independent coordinate regions that need to be analyzed. In this subsection we summarize the two symmetry relations used in the numerical calculation, from the exchange symmetry of the first two momentum fractions, and the reality condition of the momentum-space quasi-DAs.

With the flavor convention in Table~\ref{tab:valancequark}, the exchange of the first two quark fields corresponds to $x_1\leftrightarrow x_2$ in momentum space and $z_1\leftrightarrow z_2$ in coordinate space. In the isospin-symmetric $N_f=2+1$ ensembles used in this work, this exchange gives definite symmetry or antisymmetry for the leading-twist structures. For octet baryons other than the $\Lambda$, one has:
\begin{equation}
\begin{aligned}
    \widetilde\phi_{B\neq\Lambda}^{V,T}(x_1,x_2)
    &= +\widetilde\phi_{B\neq\Lambda}^{V,T}(x_2,x_1)\ ,\\
    \widetilde\phi_{B\neq\Lambda}^{A}(x_1,x_2)
    &= -\widetilde\phi_{B\neq\Lambda}^{A}(x_2,x_1)\ .
\end{aligned}
\end{equation}
For the $\Lambda$ baryon, the symmetry pattern is reversed:
\begin{equation}
\begin{aligned}
    \widetilde\phi_{\Lambda}^{V,T}(x_1,x_2)
    &= -\widetilde\phi_{\Lambda}^{V,T}(x_2,x_1)\ ,\\
    \widetilde\phi_{\Lambda}^{A}(x_1,x_2)
    &= +\widetilde\phi_{\Lambda}^{A}(x_2,x_1)\ ,
\end{aligned}
\end{equation}
equivalently, the coordinate-space quasi-DAs satisfy:
\begin{equation}
\begin{aligned}
    \widetilde\Phi_{B\neq\Lambda}^{V,T}(z_1,z_2)
    &= +\widetilde\Phi_{B\neq\Lambda}^{V,T}(z_2,z_1)\ ,\\
    \widetilde\Phi_{B\neq\Lambda}^{A}(z_1,z_2)
    &= -\widetilde\Phi_{B\neq\Lambda}^{A}(z_2,z_1)\ ,\\
    \widetilde\Phi_{\Lambda}^{V,T}(z_1,z_2)
    &= -\widetilde\Phi_{\Lambda}^{V,T}(z_2,z_1)\ ,\\
    \widetilde\Phi_{\Lambda}^{A}(z_1,z_2)
    &= +\widetilde\Phi_{\Lambda}^{A}(z_2,z_1)\ .
\end{aligned}
\end{equation}
These relations follow from the spin-flavor structures of the interpolating operators and are observed configuration by configuration in the lattice data. A derivation at the contraction level is given in Appendix~\ref{app:operator_exchange_symmetry}.

The second constraint follows from the reality of the momentum-space quasi-DAs. Since the physical quasi-DAs are real functions of the momentum fractions:
\begin{equation}
\begin{aligned}
    {\rm Re}\ \widetilde\phi_B^X(x_1,x_2)
    &= \widetilde\phi_B^X(x_1,x_2)\ ,\\
    {\rm Im}\ \widetilde\phi_B^X(x_1,x_2)
    &= 0\ ,
\end{aligned}
\end{equation}
their coordinate-space counterparts obey the Hermiticity relation:
\begin{equation}
    \widetilde\Phi_B^X(-z_1,-z_2)
    = \left[\widetilde\Phi_B^X(z_1,z_2)\right]^*\ ,
\end{equation}
equivalently:
\begin{equation}
\begin{aligned}
    {\rm Re}\ \widetilde\Phi_B^X(-z_1,-z_2)
    &= +{\rm Re}\ \widetilde\Phi_B^X(z_1,z_2)\ ,\\
    {\rm Im}\ \widetilde\Phi_B^X(-z_1,-z_2)
    &= -{\rm Im}\ \widetilde\Phi_B^X(z_1,z_2)\ .
\end{aligned}
\end{equation}
Unlike the exchange symmetry, this relation is realized only after statistical averaging.

\begin{figure}[htbp]
    \centering
    \includegraphics[width=\linewidth]{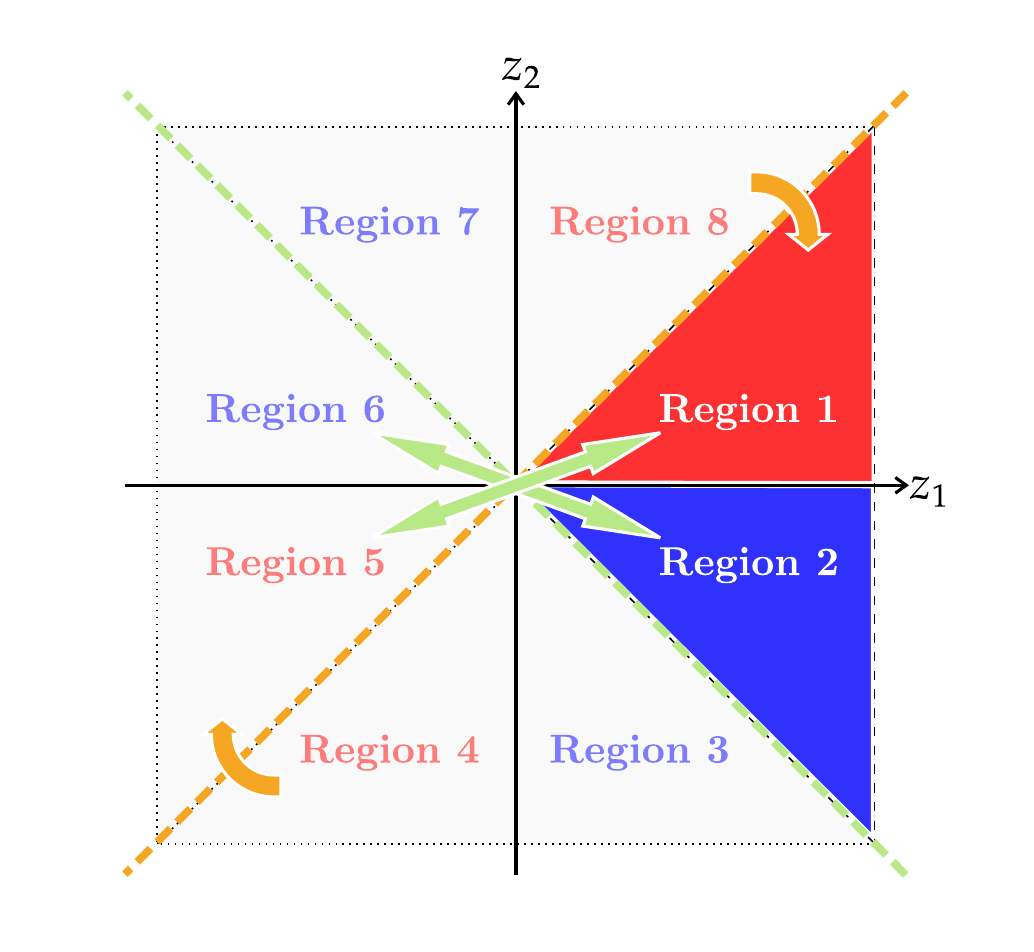}
    \caption{The region partition on the $(z_1,z_2)$ plane for quasi-DAs. According to the symmetry properties in coordinate space, only 2 regions (Region 1 and Region 2) are independent. These regions form an independent quarter region $z_1>|z_2|$, bounded by $z_1=\pm z_2$.}
    \label{fig:symmetry_region_division}
\end{figure}

Combining the exchange symmetry with the Hermiticity relation determines the symmetry structure on the full $(z_1,z_2)$ plane. As illustrated in Fig.~\ref{fig:symmetry_region_division}, the $(z_1,z_2)$ plane can be divided into eight regions, among which only two, the Region 1 and 2, are independent under these relations. In the following analysis, the hybrid renormalization and large-distance extrapolation are performed in the independent quarter region $z_1>|z_2|$, bounded by $z_1=\pm z_2$, and the full coordinate-space matrix elements are then reconstructed from this region using the exchange symmetry and Hermiticity relation.

\section{Lattice Simulation}\label{sec:Lattice_Simulation}
The lattice calculation in this work is designed to demonstrate and validate the complete baryon-LaMET analysis pipeline. The focus of this article is the formalism, renormalization, large-distance extrapolation, matching, and systematic treatment required to extract baryon LCDAs in the physical limit.
In this section, we describe the lattice ensembles used in the calculation, the signal-improvement techniques, the choices of interpolating operators and projectors, and the correlation-function analysis that provides the input to the subsequent stages of the pipeline.

\subsection{Lattice Setup}

\begin{table*}[ht]
    \centering
    \renewcommand{\arraystretch}{1.5}
      \setlength{\tabcolsep}{3mm}
    \begin{tabular}{c c c c c c c}
        \hline\hline
        Ensembles & $a$ (fm) & $m_\pi$ (MeV) & Volume & $n_{\rm cfg}$ & $n_{\rm src}$ & $P^z$ (GeV)  \\
         \hline
        C24P29 & 0.1052 & 292.3 & $24\times72 $ & 864 & $4\times9 $ & 0, 1.96, 2.45, 2.94 \\
        C32P23 & 0.1052 & 227.9 & $32\times64 $ & 954 & $4\times8 $ & 0, 1.84, 2.21, 2.57, 2.94 \\
        C48P14 & 0.1052 & 136.4 & $48\times96 $ & 302 & $4\times16$ & 0, 1.96, 2.45, 2.94 \\
        F32P30 & 0.0775 & 300.4 & $32\times96 $ & 777 & $4\times8 $ & 0, 2.00, 2.49, 2.99 \\
        F32P21 & 0.0775 & 210.3 & $32\times64 $ & 459 & $4\times16$ & 0, 2.00, 2.49, 2.99 \\
        G36P29 & 0.0689 & 297.2 & $36\times108$ & 656 & $6\times8 $ & 0, 2.00, 2.50, 3.00 \\
        H48P32 & 0.0520 & 316.6 & $48\times144$ & 550 & $6\times9 $ & 0, 1.98, 2.48, 2.98 \\
        \hline\hline
    \end{tabular}
    \caption{Lattice ensembles used in this work. The table lists the lattice spacing $a$, pion mass $m_\pi$, lattice volume, number of configurations $n_{\rm cfg}$, number of sources $n_{\rm src}$, and hadron momenta $P^z$.}  
    \label{tab:Ensembles}
\end{table*}

The numerical calculation is performed on seven $N_f=2+1$ lattice ensembles generated by the CLQCD Collaboration, using stout-smeared clover fermions coupled with the Symanzik-improved gauge action~\cite{CLQCD:2023sdb,CLQCD:2024yyn,Zhang:2021oja}. The ensembles cover pion masses from $m_\pi\approx 317~{\rm MeV}$ to $136~{\rm MeV}$ and lattice spacings from $a\approx 0.105~{\rm fm}$ down to $0.052~{\rm fm}$. The boosted baryon matrix elements are computed at several hadron momenta up to $P^z\approx 3~{\rm GeV}$. The ensemble parameters are summarized in Table~\ref{tab:Ensembles}.

\begin{figure}[htbp]
    \centering
    \includegraphics[width=\linewidth]{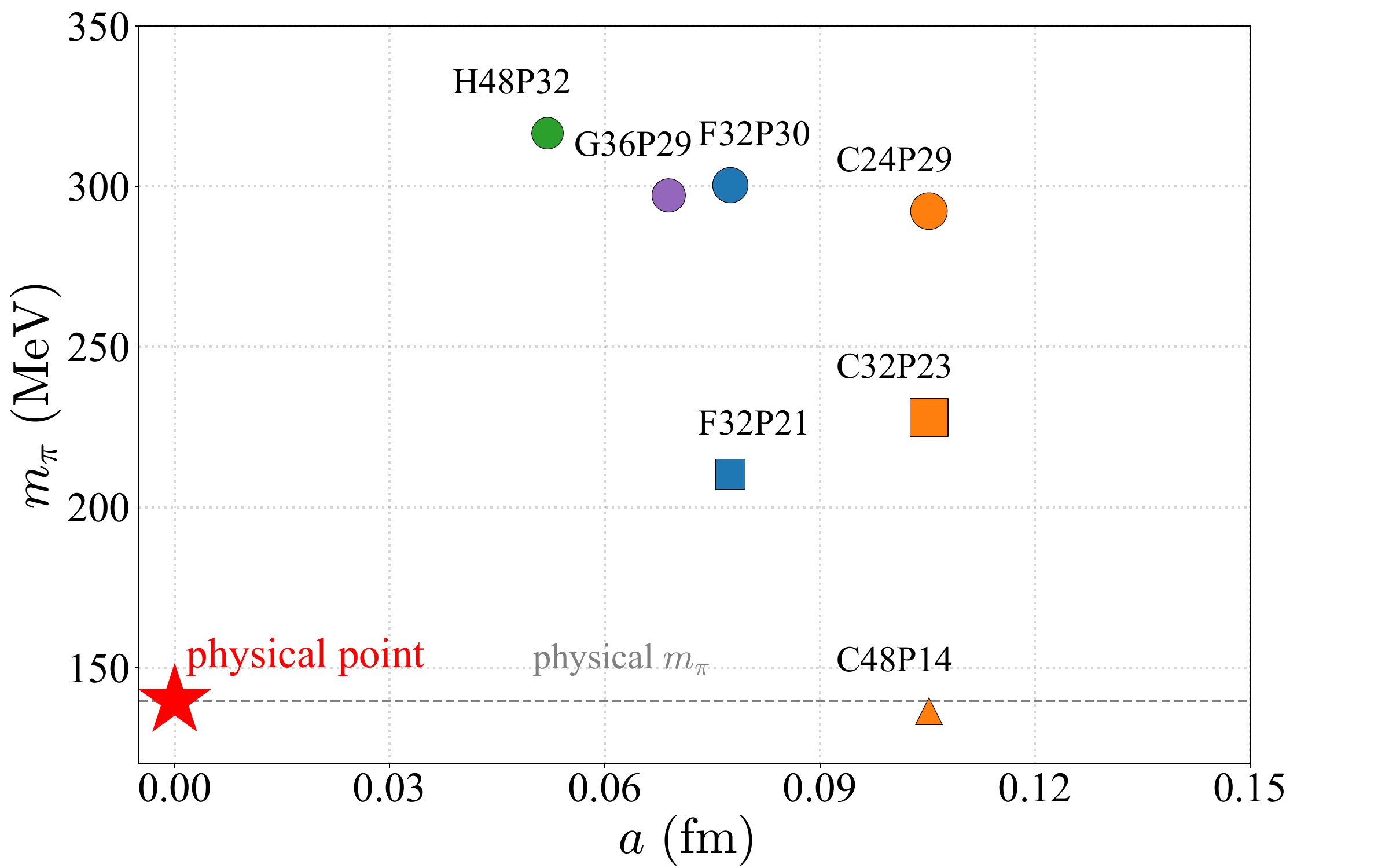}
    \caption{Distribution of the lattice ensembles in the $a$--$m_\pi$ plane. The red star marks the physical point, and the marker size indicates the number of configurations used for each ensemble.}
    \label{fig:ensemble_used}
\end{figure}

Fig.~\ref{fig:ensemble_used} shows the distribution of the ensembles in the $a$--$m_\pi$ plane. The use of several lattice spacings allows the lattice-spacing dependence to be extracted, while the range of pion masses, including one ensemble close to the physical pion mass, provides the lever arm for the physical-pion-mass extrapolation. Together with the multiple hadron momenta listed in Table~\ref{tab:Ensembles}, this setup enables the combined continuum, physical-pion-mass, and infinite-momentum extrapolation used in the final stage of the analysis.

\subsection{Signal Improvement Smearing}
To improve the signal quality of boosted-baryon matrix elements, we employ momentum smearing~\cite{Bali:2016lva} for the quark propagators and hypercubic (HYP) smearing~\cite{Hasenfratz:2001hp,DeGrand:2002vu} for the spatial gauge links in the nonlocal operators.
Momentum smearing enhances the overlap with baryon states carrying large momentum and is therefore important for the LaMET calculation, where momenta up to about $3~{\rm GeV}$ are used.
HYP smearing reduces statistical fluctuations associated with extended gauge links. This is particularly useful for baryon quasi-DAs, whose coordinate-space matrix elements depend on two independent separations, $z_1$ and $z_2$, and therefore suffer stronger signal degradation at large spatial separations than the corresponding mesonic observables.

The use of HYP smearing must be combined with a consistent renormalization procedure, because it modifies the ultraviolet behavior of the Wilson-line operator. In particular, the coefficient of the Wilson-line linear divergence depends on the number of HYP-smearing steps.
In our previous study of hybrid renormalization~\cite{LatticePartonCollaborationLPC:2025vhd}, we compared calculations with zero, one, and two HYP-smearing steps.
The linear-divergence coefficient changes with the smearing level, while the corresponding renormalized matrix elements agree within statistical uncertainties. This behavior indicates that the dominant effect of HYP smearing can be absorbed into the renormalization factor of hybrid renormalization.

Based on this observation, we use one-step HYP smearing in the present calculation. This choice provides a practical balance between improving the signal and limiting the modification of the short-distance lattice operator. In practice, going from unsmeared links to one HYP-smearing step improved the signal-to-noise ratio by approximately a factor of two~\cite{LatticePartonCollaborationLPC:2025vhd}, whereas the additional improvement from a second HYP-smearing step was much smaller. The remaining signal-improvement strategies, associated with the source-side and sink-side interpolating operators, are discussed below.

\subsection{Two-point Correlation Functions}
The quasi-DA matrix elements are extracted from baryon two-point correlation functions with nonlocal sink-side interpolating operators. We denote the Euclidean-space source position by $x_0^\mu=(\tau_0, \vec x_0)$ and the sink position by $x^\mu=(\tau_0+\tau, \vec x_0+\vec x)$. The Euclidean two-point correlation with spatial momentum $\vec P$ is defined as:
\begin{equation}
\begin{aligned}\label{eq:def2pt}
     &\ C_{\rm 2pt}(\tau,\vec P;z_1,z_2) \\
    =&\ \sum_{\vec x} \rme^{-\rmi  \vec P\cdot\vec x} \langle \mathbb T^{\gamma'\gamma}  {\mathcal O}^{\rm snk}_{\gamma} (\vec x_0 + \vec x,\tau_0 + \tau;z_1,z_2) \overline {\mathcal O}^{\rm src}_{\gamma'} (\vec x_0,\tau_0) \rangle\ .
\end{aligned}
\end{equation}
Here ${\mathcal O}^{\rm snk}_\gamma$ is the nonlocal sink-side interpolating operator associated with the quasi-DA structure, $\overline{\mathcal O}^{\rm src}_{\gamma'}$ is the source-side interpolating operator that creates the external baryon state, and $\mathbb T^{\gamma'\gamma}$ is the spin ``projector''. Their explicit choices are described below.

\subsubsection{Sink-side Interpolators: Relation to DA Structures}
The sink-side interpolators are chosen to project the leading-twist quasi-DA matrix elements defined in Sec.~\ref{sec:quasi}. Since the leading-twist baryon LCDAs contain three structures $V$, $A$, and $T$, we use three corresponding nonlocal sink-side interpolating operators. Following Eq.~\eqref{eq:quasi_matrix}, they are:
\begin{equation}
\begin{aligned}\label{eq:inter_sink}
    &\ \mathcal O^{{\rm snk},A}(\vec x,\tau;z_1,z_2) \\
    =&\ q^{\rm T}(x+z_1 n_z) \left( C\gamma^5\gamma^t \right) g(x+z_2 n_z) h(x)\ ,\\
    \\
    &\ \mathcal O^{{\rm snk},V}(\vec x,\tau;z_1,z_2) \\
    =&\ q^{\rm T}(x+z_1 n_z) \left( C\gamma^t \right) g(x+z_2 n_z) \gamma^5 h(x)\ ,\\
    \\
    &\ \mathcal O^{{\rm snk},T}(\vec x,\tau;z_1,z_2) \\
    =&\ q^{\rm T}(x+z_1 n_z) ( \tfrac12 C [\gamma^t,\gamma^{x,y}] ) g(x+z_2 n_z) \gamma^5\gamma_{x,y} h(x)\ .\\
\end{aligned}
\end{equation}
The quark flavors $q,g,h$ follow the convention inTable~\ref{tab:valancequark}. With these sink-side operators, the ground-state contribution to the two-point correlation is proportional to $\widetilde M_B^X(z_1,z_2;P^z)$ with $X=V,A,T$. The desired quasi-DA matrix elements can therefore be obtained from the corresponding two-point correlations after the reduction procedure described in Sec.~\ref{sec:quasi}.

The sink-side operators must also respect the exchange symmetries discussed in Sec.~\ref{sec:symmetries}. For the $\Lambda$ baryon, the $A$ structure is symmetric under $z_1\leftrightarrow z_2$, whereas the $V$ and $T$ structures are antisymmetric. This symmetry pattern is reflected in the local limit: the $\Lambda$-baryon $A$ amplitude has a nonzero local matrix element, while the $\Lambda$-baryon $V$ and $T$ amplitudes vanish at $z_1=z_2=0$. As a result, the three structures can be computed within the same two-point framework, but their normalization and reduction procedures are not identical.
The operator-level derivation of these symmetry properties are given in Appendix~\ref{app:operator_exchange_symmetry} and the numerical verification can be checked in Sec.~\ref{sec:hybridVAT}.

\subsubsection{Source-side Interpolator: Conventional and Kinematically Enhanced Choices}
The source-side interpolator is used to create the desired external baryon state. Since the quasi-DA structure is fixed by the nonlocal sink-side interpolating operator, the source-side interpolating operator only needs to have a nonzero overlap with the baryon ground state and the correct quantum numbers. A conventional choice for the $\Lambda$ and proton source-side interpolators is:
\begin{equation}
\begin{aligned}\label{eq:inter_src_ori}
    \mathcal O^{\rm src}_\Lambda &= \frac{ 2u^{\rm T} ( C\gamma^5 ) d s + u^{\rm T} ( C\gamma^5 ) s d + s^{\rm T} ( C\gamma^5 ) d u }{\sqrt6}\ ,\\
    \mathcal O^{\rm src}_{\rm p} &= u^{\rm T} ( C\gamma^5 ) d u\ .
\end{aligned}
\end{equation}
This choice has the correct flavor and spin quantum numbers and is widely used in baryon lattice calculations.

In LaMET calculations, however, the baryon must be boosted to large momentum. The signal-to-noise ratio of baryon correlators deteriorates rapidly as $P^z$ increases, so improving the overlap with the boosted ground state is important. Inspired by Ref.~\cite{Zhang:2025hyo,Reitinger:2026hta}, we use a kinematically enhanced source-side interpolator by inserting an additional $\gamma^t$ matrix aligned with the boost direction:
\begin{equation}
\begin{aligned}\label{eq:inter_src}
    \mathcal O^{\rm src,KE}_\Lambda &= \frac{ 2 u^{\rm T} ( C\gamma^5\gamma^t ) d s + u^{\rm T} ( C\gamma^5\gamma^t ) s d + s^{\rm T} ( C\gamma^5\gamma^t ) d u }{\sqrt6}\ ,\\
    \mathcal O^{\rm src,KE}_{\rm p} &= u^{\rm T} ( C\gamma^5\gamma^t ) d u\ .
\end{aligned}
\end{equation}
This modification preserves the baryon quantum numbers and does not change the quasi-DA structure selected by the sink-side interpolating operator. It only changes the overlap factor between the source-side interpolating operator and the boosted baryon state, which is canceled in the following reduction procedure. As shown in Ref.~\cite{LatticePartonCollaborationLPC:2025vhd}, the kinematically enhanced source-side interpolator improves the signal-to-noise ratio of baryon two-point correlation at $P^z\approx 2.5~{\rm GeV}$ by approximately a factor of two, with a larger improvement expected at higher momenta.

\subsubsection{Projector Choice}
A spin projector is required to extract the desired component of the baryon two-point correlation function. In zero-momentum baryon spectroscopy, the parity projector $(1+\gamma^t)/2$ is commonly used to enhance the positive-parity contribution and suppress excited-state contamination from negative-parity states. For boosted baryons, however, parity is not a good quantum number in the same way as at rest, and states associated with different rest-frame parities can mix in a moving frame. Therefore, $(1+\gamma^t)/2$ is not sufficient to isolate the desired ground state leading-twist structures at nonzero momentum.

\begin{figure}[htbp]
    \centering
    \subfloat[\ $\lambda_1=3$]{
        \hspace{-1cm}\includegraphics[width=0.93\linewidth]{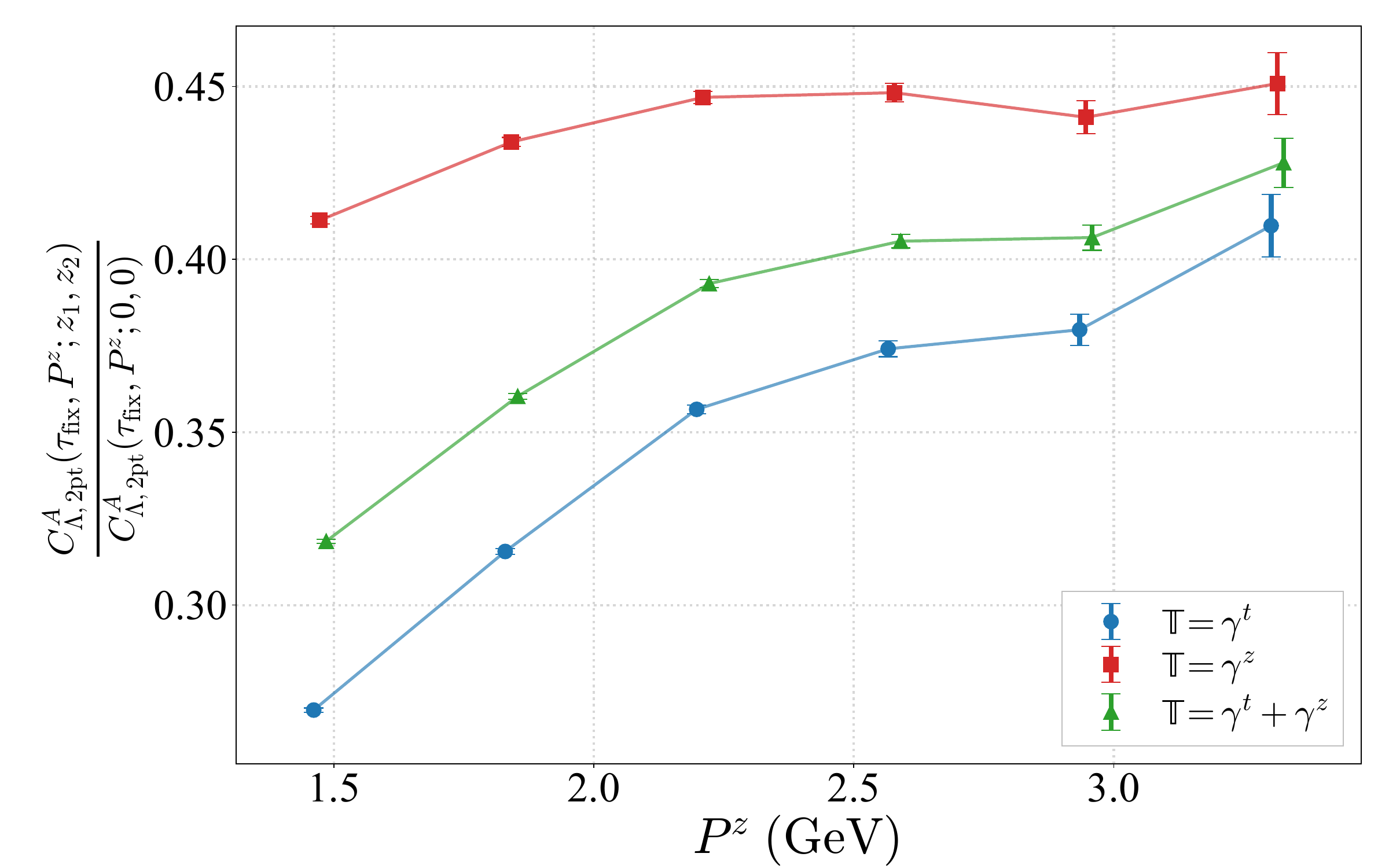}
        }\\
    \subfloat[\ $\lambda_1=\lambda_2=2$]{
        \hspace{-1cm}\includegraphics[width=0.93\linewidth]{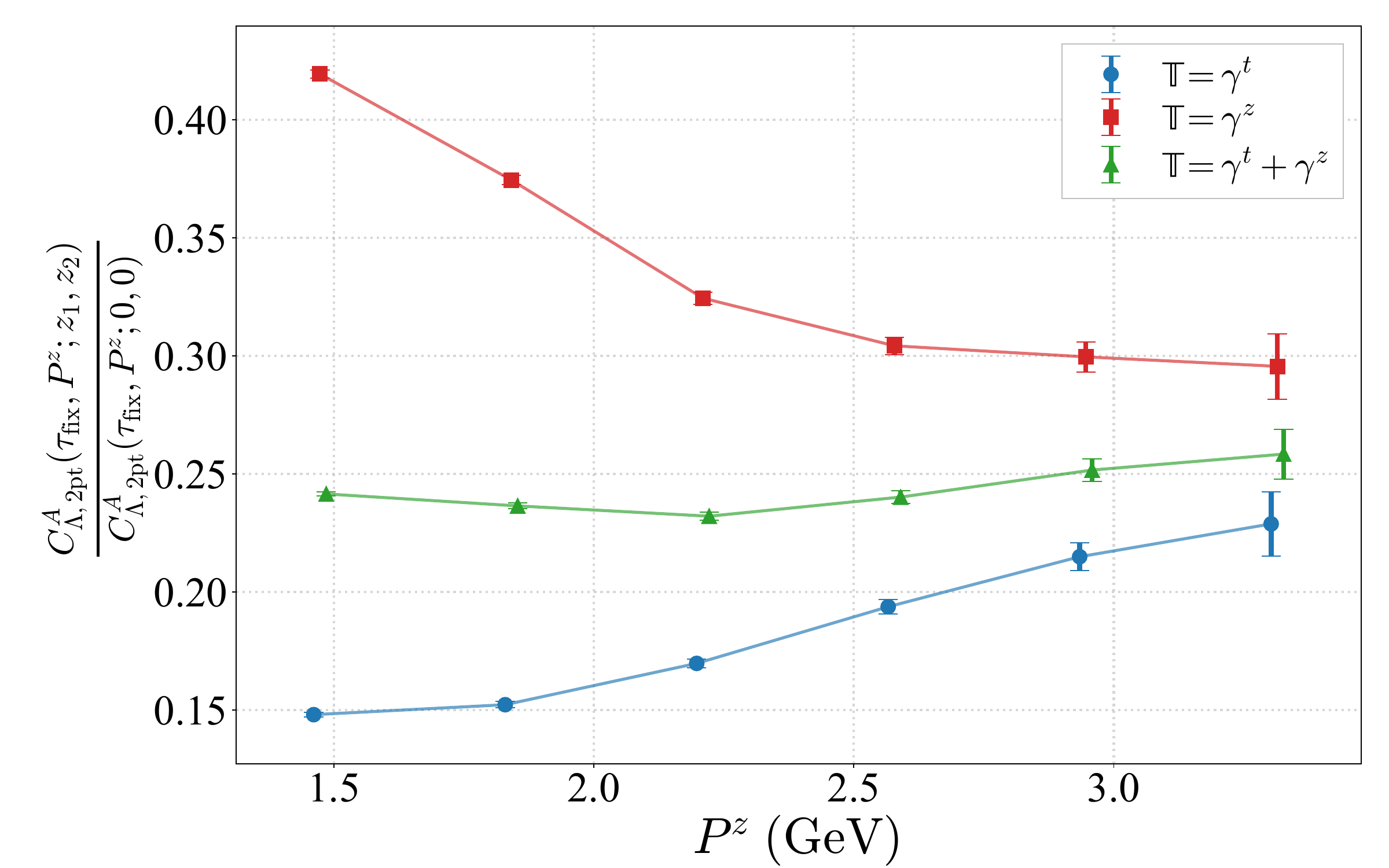}
        }
    \caption{Projector dependence of the real part of the normalized nonlocal $\Lambda$-baryon $A$-structure matrix elements on the C32P23 ensemble. The results are shown as functions of $P^z$ at fixed $\lambda=zP^z$, along the $z_2=0$ slice at $\lambda_1=3$ and the diagonal slice $z_1=z_2$ at $\lambda_1=\lambda_2=2$, comparing $\mathbb T=\gamma^t$, $\gamma^z$, and $\gamma^t+\gamma^z$.}
    \label{fig:lambdaA_projector_splitting_diag}
\end{figure}

In this work we choose the spin projector according to the leading-twist light-cone structures. In the light-cone decomposition, the leading contribution is associated with the projection ${\rm tr}\left[u_B(n\cdot P)\slashed{\bar n}u_B(n\cdot P)\right]$. Motivated by this structure, we use
\begin{equation}\label{eq:projector}
    \mathbb T=\slashed{\bar n}=\gamma^t+\gamma^z\ .
\end{equation}
At finite momentum, this choice reduces the sensitivity to subleading-power contributions compared with using $\gamma^t$ or $\gamma^z$ alone. The remaining finite-momentum dependence is controlled by using several values of $P^z$ and by carrying out the infinite-momentum extrapolation.

As a numerical check, Fig.~\ref{fig:lambdaA_projector_splitting_diag} compares the normalized nonlocal $\Lambda$-baryon $A$-structure matrix elements obtainedwith the projectors $\mathbb T=\gamma^t$, $\gamma^z$, and $\gamma^t+\gamma^z$. The comparison is performed on the C32P23 ensemble for the $z_2=0$ slice at $\lambda_1=3$ and the diagonal slice $z_1=z_2$ at $\lambda_1=\lambda_2=2$. The source--sink separations are averaged over the fixed Euclidean time slices $\tau_{\rm fix}/a=5,6,7$, corresponding to $0.526$, $0.631$, and $0.736~{\rm fm}$, respectively.

At smaller momenta, the separate $\gamma^t$ and $\gamma^z$ projectors exhibit a visible splitting, indicating sizable residual finite-momentum effects. As $P^z$ increases, the three projector choices become more consistent, while the combined projector $\mathbb T=\gamma^t+\gamma^z$ reaches a stable behavior more rapidly. This pattern supports the use of $\mathbb T=\slashed{\bar n}$ as the default projector for the finite-momentum quasi-DA matrix elements.

\subsection{Extraction of Quasi-DAs}

\subsubsection{Reduction Formula for \texorpdfstring{$A$}{A} and \texorpdfstring{$V$}{V}, \texorpdfstring{$T$}{T}}\label{sec:reduction}
We now describe how the normalized coordinate-space quasi-DA matrix elements are extracted from the two-point correlations defined in Eq.~\eqref{eq:def2pt}. For the $\Lambda$ baryon, the $A$ structure is symmetric and has a nonzero local limit, whereas the $V$ and $T$ amplitudes are antisymmetric and vanish at $z_1=z_2=0$. Therefore the reduction formula must be chosen separately for the symmetric and antisymmetric structures.

With the sink-side and source-side interpolators given in Eq.~\eqref{eq:inter_sink} and Eq.~\eqref{eq:inter_src}, and with the hadron momentum chosen as $\vec P=(0,0,P^z)$, the two-point correlation has the spectral decomposition:
\begin{equation}\label{eq:2pt}
\begin{aligned}
    C_{B,\rm 2pt}^X(\tau,P^z;z_1,z_2)
    =&\ \sum_k \frac{\rme^{-E_k\tau}}{2E_k} \langle E_k,P^z | \overline{\mathcal O}_{B,\gamma'}^{\rm src,KE} \mathbb T^{\gamma'\gamma} | 0 \rangle \\
    &\ \times \langle 0 | {\mathcal O}^{{\rm snk},X}_{\gamma}(z_1,z_2) | E_k,P^z \rangle\ ,
\end{aligned}
\end{equation}
where $X=V,A,T$. At large Euclidean time $\tau$, the ground-state contribution takes the form:
\begin{equation}\label{eq:reduction}
\begin{aligned}
    &\ C_{B,\rm 2pt}^X(\tau,P^z;z_1,z_2) \\
    \simeq&\ \widetilde M_B^X(z_1,z_2;P^z)\left[ \langle B(P^z) | \overline{\mathcal O}_{B,\gamma'}^{\rm src,KE} \mathbb T^{\gamma'\gamma}|0\rangle u_{B,\gamma}\right] \\
    &\ \times \frac{\rme^{-E_0\tau}}{2E_0} \left[ 1 + C_{B,1}^X(z_1,z_2;P^z)\ \rme^{-\Delta E_1\tau} +\cdots \right]\ .
\end{aligned}
\end{equation}
Here $\widetilde M_B^X(z_1,z_2;P^z)$ is the quasi-DA matrix element defined in Eq.~\eqref{eq:quasi_matrix}. In ratios of two-point correlations with the same source-side interpolator and spin projector, the source overlap and the leading Euclidean-time dependence cancel, leaving the desired normalized matrix element up to excited-state corrections.

For the $\Lambda$-baryon $A$ structure, the local matrix element is nonzero. Taking the ground state, the standard reduction is therefore:
\begin{itemize}
    \item $\Lambda$-baryon $A$ amplitude $\widetilde\Phi_\Lambda^A$:
    \begin{equation}\label{eq:ratio_A}
    \begin{aligned}
        &\ \frac{C^A_{\Lambda,\rm 2pt}(\tau,P^z;z_1,z_2)}{C^A_{\Lambda,\rm 2pt}(\tau,P^z;0,0)} \simeq \frac{f_\Lambda \widetilde \Phi^A_\Lambda(z_1,z_2;P^z) P^z}{f_\Lambda P^z } \\
        =&\  \widehat M^A_\Lambda(z_1,z_2;P^z) = \widetilde\Phi_\Lambda^A(z_1,z_2;P^z)\ .
    \end{aligned}
    \end{equation}
\end{itemize}
This relation follows directly from the normalization convention $\widetilde\Phi_\Lambda^A(0,0)=1$ in Sec.~\ref{sec:quasi}.

For the $\Lambda$-baryon $V$ and $T$ structures, the corresponding local matrix elements vanish due to antisymmetry. Therefore, $C_{\Lambda,\rm 2pt}^{V,T}(\tau,P^z;0,0)=0$, and the same self-reference ratio procedure as used for the $A$ structure cannot be applied directly. Guided by the leading-twist definition in Eq.~\eqref{eq:quasi_matrix} and the normalization conventions in Eqs.~\eqref{eq:norm_V}--\eqref{eq:norm_T}, we define the normalized $V$ and $T$ quasi-DA matrix elements by dividing the nonlocal correlation functions $C_{\Lambda,\rm 2pt}^{V,T}(\tau,P^z;z_1,z_2)$ by the local, non-vanishing $C^{A}_{\Lambda,\rm 2pt}(\tau,P^z;0,0)$:
\begin{itemize}
    \item $\Lambda$-baryon $V$ amplitude $\widetilde\Phi_\Lambda^V$:
    \begin{equation}\label{eq:ratio_V}
    \begin{aligned}
        &\ \frac{C^V_{\Lambda,\rm 2pt}(\tau,P^z;z_1,z_2)}{C^A_{\Lambda,\rm 2pt}(\tau,P^z;0,0)} \simeq \frac{-f_\Lambda \widetilde \Phi^V_\Lambda(z_1,z_2;P^z) P^z}{f_\Lambda P^z}\\
        =&\ \widehat M^V_\Lambda(z_1,z_2;P^z) = -\widetilde\Phi_\Lambda^V(z_1,z_2;P^z)\ ;
    \end{aligned}
    \end{equation}
    
    \item $\Lambda$-baryon $T$ amplitude $\widetilde\Phi_\Lambda^T$:
    \begin{equation}\label{eq:ratio_T}
    \begin{aligned}
        &\frac{C^T_{\Lambda,\rm 2pt}(\tau,P^z;z_1,z_2)}{C^A_{\Lambda,\rm 2pt}(\tau,P^z;0,0)} \simeq \frac{2f^T_\Lambda \widetilde \Phi^T_\Lambda(z_1,z_2;P^z) P^z }{f_\Lambda P^z}\\
        =&\ \widehat M^T_\Lambda(z_1,z_2;P^z)= \frac{2f_\Lambda^T}{f_\Lambda} \widetilde\Phi_\Lambda^T(z_1,z_2;P^z)\ .
    \end{aligned}
    \end{equation}
\end{itemize}
The same source-side interpolator and spin projector are used in the numerator and denominator, so the source overlap cancels in these ratios in the ground-state limit. Therefore, these reduction enable a direct determination for the $V$ and $T$ structures of the $\Lambda$-baryon quasi-DA in this work.

\subsubsection{Two-state Fit with Model Average}
At finite source-sink separation, the ratios in Eqs.~\eqref{eq:ratio_A}--\eqref{eq:ratio_T} contain residual excited-state contamination. We describe the Euclidean-time $\tau$ dependence using the two-state fit form:
\begin{equation}\label{eq:2_state_fit}
\begin{aligned}
    &\ \frac{C_{\rm 2pt}(\tau,P^z;z_1,z_2)}{C_{\rm 2pt}(\tau,P^z;0,0)} \\
    =&\ \widehat M(z_1,z_2;P^z)\left[1+
    \Delta C_1(z_1,z_2;P^z)\ \rme^{-\Delta E_1(z_1,z_2;P^z) \tau}\right]\ ,
\end{aligned}
\end{equation}
where $\widehat M(z_1,z_2;P^z)$ is the desired ground-state normalized matrix element. The parameters $\Delta C_1$ and $\Delta E_1$ describe the leading excited-state contribution.

The fitted ground-state matrix element could depend on different fit ranges $[\tau^{(\kappa)}_{\min},\tau^{(\kappa)}_{\max}]$ with lattice spacing $a$. To estimate this fit-range dependence, we use the model-averaging procedure~\cite{Jay:2020jkz}. For each fit range $\kappa$, the weight is assigned as:
\begin{equation}\label{eq:weight}
    w_\kappa \propto
    \exp\left(-\frac{1}{2}\chi_\kappa^2-\frac{\tau_{\min}^{(\kappa)}}{a}\right)\ ,
\end{equation}
with the normalization $\sum_\kappa w_\kappa=1$. The model-averaged matrix element and its variance are then:
\begin{equation}\label{eq:model_ave}
    \overline M
    = \sum_\kappa w_\kappa \widehat M_\kappa\ ,\qquad
    \sigma_{\overline M}^2
    = \sum_\kappa w_\kappa\left(\widehat M_\kappa^2+\sigma_\kappa^2\right)-\overline M^2\ ,
\end{equation}
where $\widehat M_\kappa$ and $\sigma_\kappa$ are the fitted matrix element and its statistical uncertainty from the fit range $\kappa$. This procedure incorporates the spread among acceptable fit ranges into the quoted uncertainty of the extracted coordinate-space matrix element.

\subsection{Dispersion Relation}\label{dispersion}
As a consistency check of the boosted-baryon calculation, we examine the dispersion relation of the ground-state energies extracted from the two-point correlation functions. This check is relevant for the LaMET analysis because the quasi-DA matrix elements are computed with external baryon states carrying momenta up to about $3~{\rm GeV}$.

The temporal dependence of the two-point correlation is described by the spectral decomposition:
\begin{equation}
    C_{\rm 2pt}(\tau,P^z)=
    \sum_k C'_k(P^z)\ \rme^{-E_k(P^z)\tau}\ ,
\end{equation}
where $C'_k$ is the overlap factor for the $k$-th energy eigenstate. From this, the ground-state energy $E_0$ can be extracted using a standard two-state fit procedure. 

In the continuum limit, the relativistic dispersion relation for ground state is given by $E_0^2 = m_0^2 + (P^z)^2$. On the lattice, however, discretization effects can lead to deviations from this relation, especially at large momenta. To quantify such effects, we parameterize the lattice dispersion relation as:
\begin{equation}\label{eq:dispersion_fit}
    E^2 = m_{0,\rm ens}^2 + c_0 (P^z)^2 + c_1 a^2 (P^z)^4\ ,
\end{equation}
where $m_{0,{\rm ens}}$ is an ensemble-dependent rest mass, while the coefficients $c_0$ and $c_1$ encode deviations from the continuum relativistic expectation. In the continuum limit, one expects $c_0 \to 1$ and $c_1 \to 0$.

\begin{figure}[htbp]
    \centering
    \includegraphics[width=\linewidth]{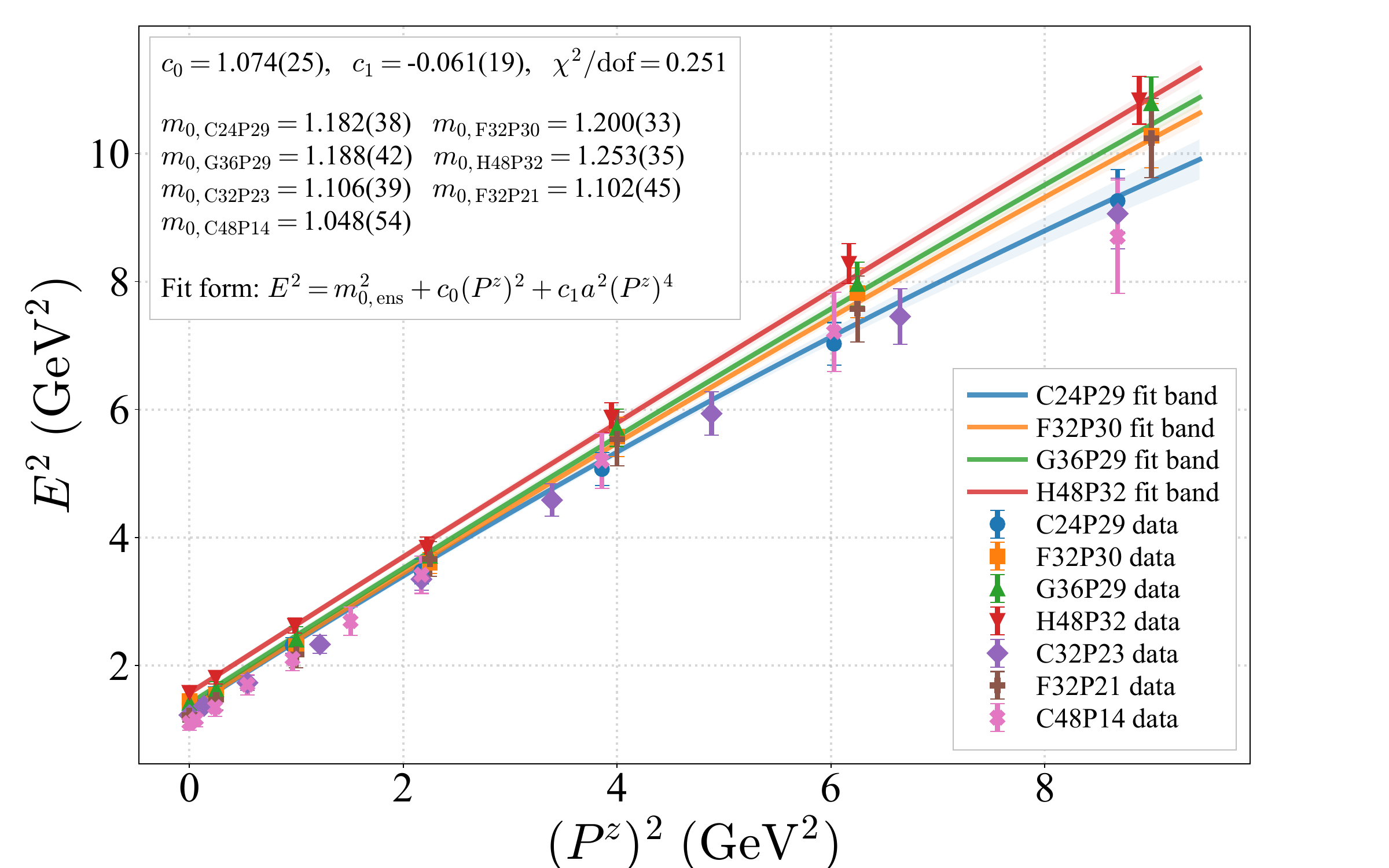}
    \caption{Combined fit of the lattice dispersion relation using all seven ensembles. The data points with error bars show the extracted ground-state energies, while the fit bands are displayed for the four ensembles with pion masses near $300~{\rm MeV}$: C24P29, F32P30, G36P29, and H48P32.}
    \label{fig:dispersion_relation_fit}
\end{figure}

In numerical analysis, the dispersion relation is combined fitted using all seven ensembles with the form in Eq.~\eqref{eq:dispersion_fit}, where each ensemble has an independent $m_{0,{\rm ens}}$ and the coefficients $c_0$ and $c_1$ are common to all ensembles. The resulting fit gives:
\begin{equation}
    c_0 = 1.074(25)\ , \quad c_1 = -0.061(19)\ ,
\end{equation}
with $\chi^2/{\rm d.o.f.} = 0.251$.

As shown in Fig.~\ref{fig:dispersion_relation_fit}, the data points are displayed for all seven ensembles, while the fit bands are shown for the four ensembles with pion masses close to $300~{\rm MeV}$: C24P29, F32P30, G36P29, and H48P32. The fitted coefficient $c_0$ is close to the continuum expectation, and the $c_1$ term indicates a moderate momentum-dependent discretization effect over the momentum range used in this work.
Overall, the extracted energies follow the expected approximately relativistic dispersion relation in the kinematic region relevant for the quasi-DA analysis. A mild deviation is visible on the coarsest ensembles, such as $a\approx 0.105~{\rm fm}$ with the largest momenta around $3~{\rm GeV}$. Such momentum-dependent lattice artifacts are taken into account in the subsequent physical-limit extrapolation in Sec.~\ref{sec:extrap_phys}.

\section{Hybrid renormalization}\label{sec:Hybrid}
The lattice-calculated quasi-DA matrix elements are bare quantities regularized by the lattice cutoff. For the nonlocal three-quark operators considered here, the Wilson-line self energies generate linear ultraviolet divergences, while short-distance logarithms arise when the relevant coordinate separations become small.
In the baryon case, the coordinate dependence is genuinely two-dimensional: the matrix element depends on $z_1$ and $z_2$, and the perturbative short-distance structures involve the separations $|z_1|$, $|z_2|$, and $|z_1-z_2|$. Therefore the renormalization prescription must be formulated on the full $(z_1,z_2)$ plane.

In this section, we construct the extension of hybrid renormalization scheme~\cite{Ji:2020brr} used for the large-momentum matrix elements of baryon quasi-DAs. The scheme combines the advantages of two prescriptions. At short distances, a ratio to the corresponding zero-momentum matrix elements removes the short-distance logarithms efficiently. At large distances, however, the same ratio would introduce uncontrolled infrared contamination from the zero-momentum matrix elements. We therefore use the self-renormalization~\cite{LatticePartonLPC:2021gpi} in the long-distance region, where the Wilson-line linear divergences are removed without dividing by long-distance nonlocal matrix elements. The hybrid prescription combines these two treatments on the two-dimensional coordinate plane and preserves the exchange symmetries of the $V$, $A$, and $T$ amplitudes.

\subsection{Self Renormalization of Baryon Quasi-DA Matrix Elements}\label{sec:self_formalism}

Throughout this section, we denote the normalized bare lattice matrix element by $\widehat M_{\rm bare}(z_1,z_2;P^z,a)$, where the normalization follows the convention introduced in Sec.~\ref{sec:quasi} and Sec.~\ref{sec:reduction}. 
A simple ratio prescription is:
\begin{equation}
    \widehat M_{\rm Ratio}(z_1,z_2;P^z,a) = \frac{\widehat M_{\rm bare}(z_1,z_2;P^z,a)}{\widehat M_{\rm bare}(z_1,z_2;0,a)}\ .
\end{equation}
The numerator and denominator contain the same Wilson-line geometry, so the Wilson-line linear divergences cancel in the ratio. This prescription is well suited to the short-distance region, where the zero-momentum matrix element is perturbatively controlled. At large separations, however, the zero-momentum matrix element contains long-distance hadronic physics, and using it as a denominator over the entire $(z_1,z_2)$ plane would introduce indeterminate‌ infrared contamination.

To remove ultraviolet divergences without introducing ‌indeterminate‌ infrared effects, we employ the self-renormalization procedure at long distances. The normalized bare matrix element is parametrized as~\cite{Han:2023xbl}:
\begin{equation}
\begin{aligned}
    &\ \widehat M_{\rm bare}(z_1,z_2;P^z,a)\\
    =&\ \exp\Bigg[ \frac{k \tilde z}{a\ln(a \Lambda_{\rm QCD}^{\rm latt})} + g(z_1,z_2;P^z) + \frac{\gamma_0}{b_0} \ln\frac{\ln(a \Lambda_{\rm QCD}^{\rm latt})}{\ln(\mu / \Lambda_{\rm QCD}^{\rm \overline{MS}})}  \\
    &\ \qquad+ \ln\left( 1 + \frac{d}{\ln(a \Lambda_{\rm QCD}^{\rm latt})} \right) + f(z_1,z_2) a^2 \Bigg]\ ,
\end{aligned}
\end{equation}
here $k$ is the coefficient of the Wilson-line linear divergence,  $f(z_1,z_2)a^2$ parametrizes the leading discretization effect. The logarithmic terms describe the residual cutoff dependence associated with the perturbative scheme conversion. The effective Wilson-line length is:
\begin{equation}\label{eq:eff_length}
    \tilde z =
    \begin{cases}
        \max(|z_1|,|z_2|)\ , & z_1z_2 \ge 0\ ,\\
        |z_1-z_2|\ , & z_1z_2 < 0\ .
    \end{cases}
\end{equation}

The $g(z_1,z_2;P^z)$ denotes the effects from non-perturbative physics, including renormalons. At short distance region $a \ll z_i < 1/\mu$, it relates to the results in perturbative $\rm \overline{MS}$ scheme through following matching relation:
\begin{equation} 
    g(z_1,z_2;P^z) = \ln \widehat M_{\rm \overline{MS}, pert}(z_1,z_2;P^z,\mu) + m_0 \tilde z\ ,
\end{equation}
where $m_0$ is the residual mass parameter. In one-loop perturbative calculation, the zero-momentum matrix elements in $\rm \overline{MS}$ scheme for $V$, $A$, and $T$ follow~\cite{Han:2024ucv}:
\begin{equation}\label{eq:pert_res_VA}
\begin{aligned}
    &\ \widehat M_{\rm \overline{MS}, pert}^{(1),V/A}(z_1,z_2;P^z=0,\mu) \\
    =&\ 1 + \frac{\alpha_sC_F}{2\pi} \Bigg( \frac78\ln\frac{z_1^2\mu^2\rme^{2\gamma_E}}{4} + \frac78\ln\frac{z_2^2\mu^2\rme^{2\gamma_E}}{4} \\
    &\ \qquad\qquad + \frac34\ln\frac{(z_1-z_2)^2\mu^2\rme^{2\gamma_E}}{4} + 4 \Bigg)\ ,
\end{aligned}
\end{equation}
\begin{equation}\label{eq:pert_res_T}
\begin{aligned}
        &\ \widehat M_{\rm \overline{MS}, pert}^{(1),T}(z_1,z_2;P^z=0,\mu) \\
        =&\ 1 + \frac{\alpha_sC_F}{2\pi} \Bigg( \frac78\ln\frac{z_1^2\mu^2\rme^{2\gamma_E}}{4} + \frac78\ln\frac{z_2^2\mu^2\rme^{2\gamma_E}}{4} \\
        &\ \qquad\qquad + \frac12\ln\frac{(z_1-z_2)^2\mu^2\rme^{2\gamma_E}}{4} + \frac{13}{4} \Bigg)\ .
\end{aligned}
\end{equation}
The difference between the $V/A$ and $T$ expressions will be relevant for the treatment of different structures in Sec.~\ref{sec:hybridVAT}.

The self-renormalization factor is then defined as:
\begin{equation}
\begin{aligned}\label{eq:self_Z}
    &\ Z_{\rm Self}(z_1,z_2;a,\mu) \\
    =&\ \exp\Bigg[ \left( \frac{k}{a\ln(a \Lambda_{\rm QCD}^{\rm latt})}-m_0 \right)\tilde z + \frac{\gamma_0}{b_0} \ln\frac{\ln(a \Lambda_{\rm QCD}^{\rm latt})}{\ln(\mu / \Lambda_{\rm QCD}^{\rm \overline {MS}})} \\
    &\ \qquad+ \ln\left( 1 + \frac{d}{\ln(a \Lambda_{\rm QCD}^{\rm latt})} \right) + f(z_1,z_2) a^2 \Bigg]\ .
\end{aligned}
\end{equation}
The parameters $k$, $m_0$, and $d$ are extracted from zero-momentum matrix elements through the two-step fitting procedure described in Sec.~\ref{sec:hybrid_implement}. Dividing the bare lattice matrix elements by this factor gives the self-renormalized matrix elements:
\begin{equation}\label{eq:Self_Re}
    \widehat{M}_{\rm Self}\left(z_1, z_2 ; P^z, \mu\right) = \frac{\widehat{M}_{\rm bare}\left(z_1, z_2 ; P^z, a\right)}{Z_{\rm Self}\left(z_1, z_2 ; a, \mu\right)}\ .
\end{equation}
The self-renormalized matrix elements remove the Wilson-line linear divergences and converts the lattice result to the $\overline{\rm MS}$ scheme. It will serve as the building block for the hybrid prescription introduced in the next subsection.

\subsection{Hybrid Prescription on Two-dimensional Coordinate Plane} \label{sec:hybrid_2D}

The self-renormalization prescription removes the Wilson-line linear divergences and converts the lattice matrix element to the $\overline{\rm MS}$ scheme. However, applying it uniformly over the full $(z_1,z_2)$ plane is not optimal. At short distances, the perturbative logarithms in Eqs.~\eqref{eq:pert_res_VA}--\eqref{eq:pert_res_T} associated with $|z_1|$, $|z_2|$ and $|z_1-z_2|$ generate sharp structures in the self-renormalized matrix element and in the corresponding matching kernel. In such regions, the ratio prescription is more suitable, since the zero-momentum matrix elements are numerically smooth and properly cancel the short-distance logarithmic dependence. The hybrid scheme therefore combines these two prescriptions: it adopts the ratio scheme at short distances and self-renormalization at large distances.

\begin{figure}[htbp]
    \centering
    \includegraphics[width=\linewidth]{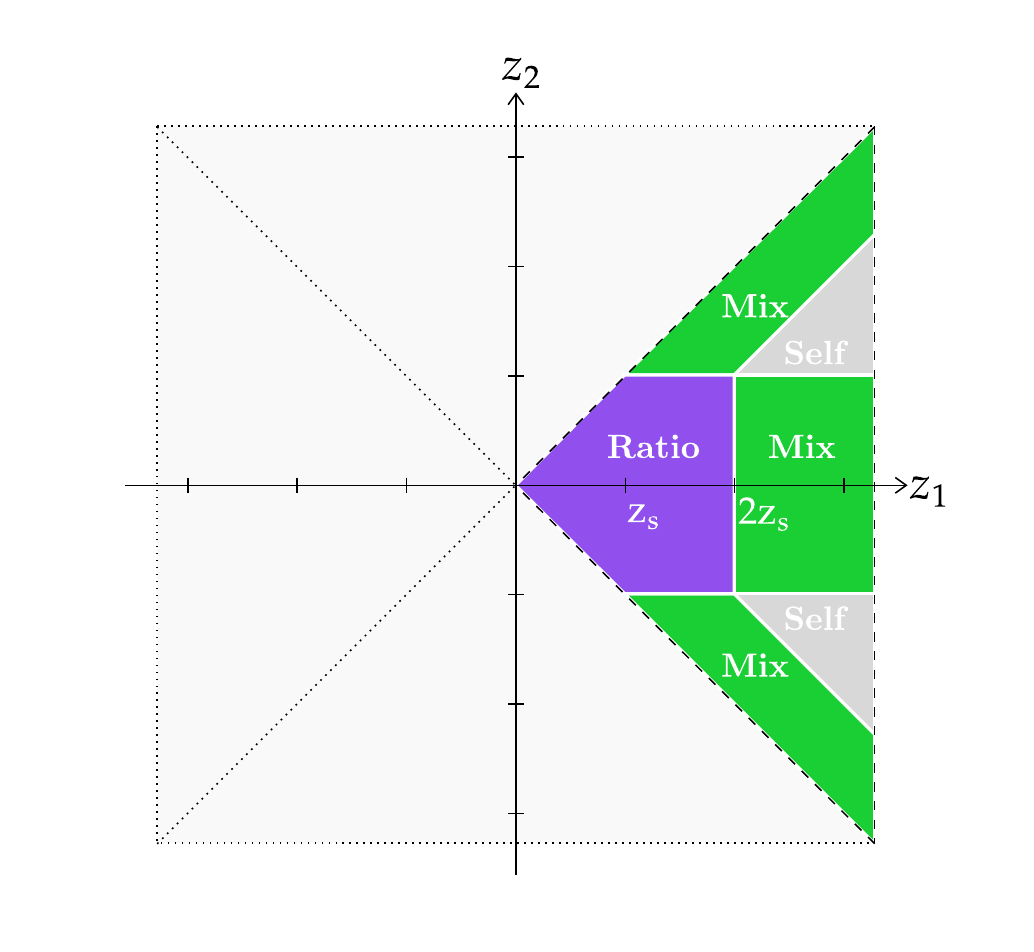}
    \caption{The region partition for the hybrid renormalization of bare matrix elements in the independent quarter region $z_1>|z_2|$, bounded by $z_1=\pm z_2$. The purple region is treated with the ratio scheme, the gray regions with self-renormalization, and the green regions with the mixing prescription.}
    \label{fig:hybrid_region_division}
\end{figure}

For baryon quasi-DAs, this construction must be implemented on the two-dimensional coordinate plane. The physical short-distance structures are associated with the three separations $|z_1|$, $|z_2|$, and $|z_1-z_2|$. In practice, after using the $z_1 \leftrightarrow z_2$ exchange symmetry and Hermiticity relations discussed in Sec.~\ref{sec:symmetries}, we implement the prescription on the independent quarter region $z_1>|z_2|$, bounded by $z_1=\pm z_2$. In this symmetry-reduced domain, the distance $|z_1+z_2|$ appears as an auxiliary boundary distance. It is not an additional Wilson-line short-distance singularity, but is used to define the hybrid regions consistently after symmetry reconstruction of the full $(z_1,z_2)$ plane.

The resulting region division is illustrated in Fig.~\ref{fig:hybrid_region_division}. The purple region is treated with the ratio prescription, the gray regions with the self-renormalization, and the green regions with a mixed prescription that projects the short-distance coordinate to the corresponding boundary point while keeping the long-distance dependence self-renormalized. The switching distance is denoted by $z_s$.

With this convention, different regions are defined as:
\begin{itemize}
    \item Short-distance region (Ratio):
    \begin{equation}
    \begin{aligned}
            S_{\rm short} =&\ \theta(2z_s-|z_1|)\theta(z_s-|z_2|) \\
            & + \theta(z_s-|z_1|)\theta(2z_s-|z_2|)\theta(|z_2|-z_s)\ ;
    \end{aligned}
    \end{equation}
    
    \item Long-distance regions (Self-renormalization):
    \begin{equation}
    \begin{aligned}
            S_{\rm long} = &\ \theta(|z_1|-z_s)\theta(|z_2|-z_s)\\
            & \times\theta(|z_1-z_2|-z_s)\theta(|z_1+z_2|-z_s)\ ;
    \end{aligned}
    \end{equation}
    
    \item Mixing regions: \\
    The remaining regions, where at least one of the distances is short while the others are long. In these regions, the short-distance dependence is removed by the ratio prescription, while the long-distance part is self-renormalized, and the matrix elements are fixed by requiring continuity at the corresponding boundary points.
\end{itemize}

The renormalized matrix elements in the hybrid scheme are given by
\begin{equation}
    \widehat M_{\rm Hybrid}(z_1,z_2;P^z,\mu) = \frac{\widehat M_{\rm Self}(z_1,z_2;P^z,\mu)}{\widehat D_{\rm Hybrid}(z_1,z_2;0,\mu)}\ ,
\end{equation}
where $\widehat D_{\rm Hybrid}(z_1,z_2;0,\mu)$ is a piecewise denominator implementing the ratio, self-renormalization, and boundary continuity prescriptions. Explicitly:

\begin{widetext}
\begin{equation}\label{eq:denomi_hybrid}
\begin{aligned}
    &\ \widehat D_{\rm Hybrid}(z_1,z_2;0,\mu) \\
    =&\ \widehat M_{\rm Self}\big(z_1,z_2;0,\mu \big) \times \Big( \theta(2z_s-|z_1|)\theta(z_s-|z_2|) + \theta(z_s-|z_1|)\theta(2z_s-|z_2|)\theta(|z_2|-z_s) \Big)\\
    & +\widehat M_{\rm Self} \big(z_1,{\rm sign}(z_2)2z_s;0,\mu \big) \times \theta(z_s-|z_1|)\theta(|z_2|-2z_s)\\
    & +\widehat M_{\rm Self} \big({\rm sign}(z_1)2z_s,z_2;0,\mu \big) \times \theta(|z_1|-2z_s)\theta(z_s-|z_2|)\\
    & +\widehat M_{\rm Self} \big(z_s+(z_1-z_2)\theta(z_1-z_2),z_s+(-z_1+z_2)\theta(-z_1+z_2);0,\mu \big) \times \theta(|z_1|-z_s)\theta(|z_2|-z_s)\theta(z_s-|z_1-z_2|)\\
    & +\widehat M_{\rm Self} \big(z_s+(z_1+z_2)\theta(z_1+z_2),-z_s+(z_1+z_2)\theta(-z_1-z_2);0,\mu \big) \times \theta(|z_1|-z_s)\theta(|z_2|-z_s)\theta(z_s-|z_1+z_2|)\\
    & +\widehat M_{\rm Self} \big({\rm sign}(z_1)2z_s,{\rm sign}(z_2)2z_s;0,\mu \big) \times \theta(|z_1|-z_s)\theta(|z_2|-z_s)\theta(|z_1-z_2|-z_s)\theta(|z_1+z_2|-z_s)\ .\\
\end{aligned}
\end{equation}
\end{widetext}
This construction removes the short-distance logarithmic structures through the ratio prescription, while avoiding the use of zero-momentum matrix elements at long distances. At the same time, the boundary prescription provides a continuous interpolation across the short-distance, long-distance, and mixing regions, making the resulting coordinate-space matrix elements suitable for the subsequent Fourier transform.

\subsection{Structure Dependent Implementation for \texorpdfstring{$V$}{V}, \texorpdfstring{$A$}{A}, and \texorpdfstring{$T$}{T}}\label{sec:hybridVAT}

\begin{figure*}[t]
    \centering
    \subfloat[\ $\Lambda$-baryon $A$]{
        \makebox[0.31\textwidth][l]{
            \includegraphics[scale=0.24]{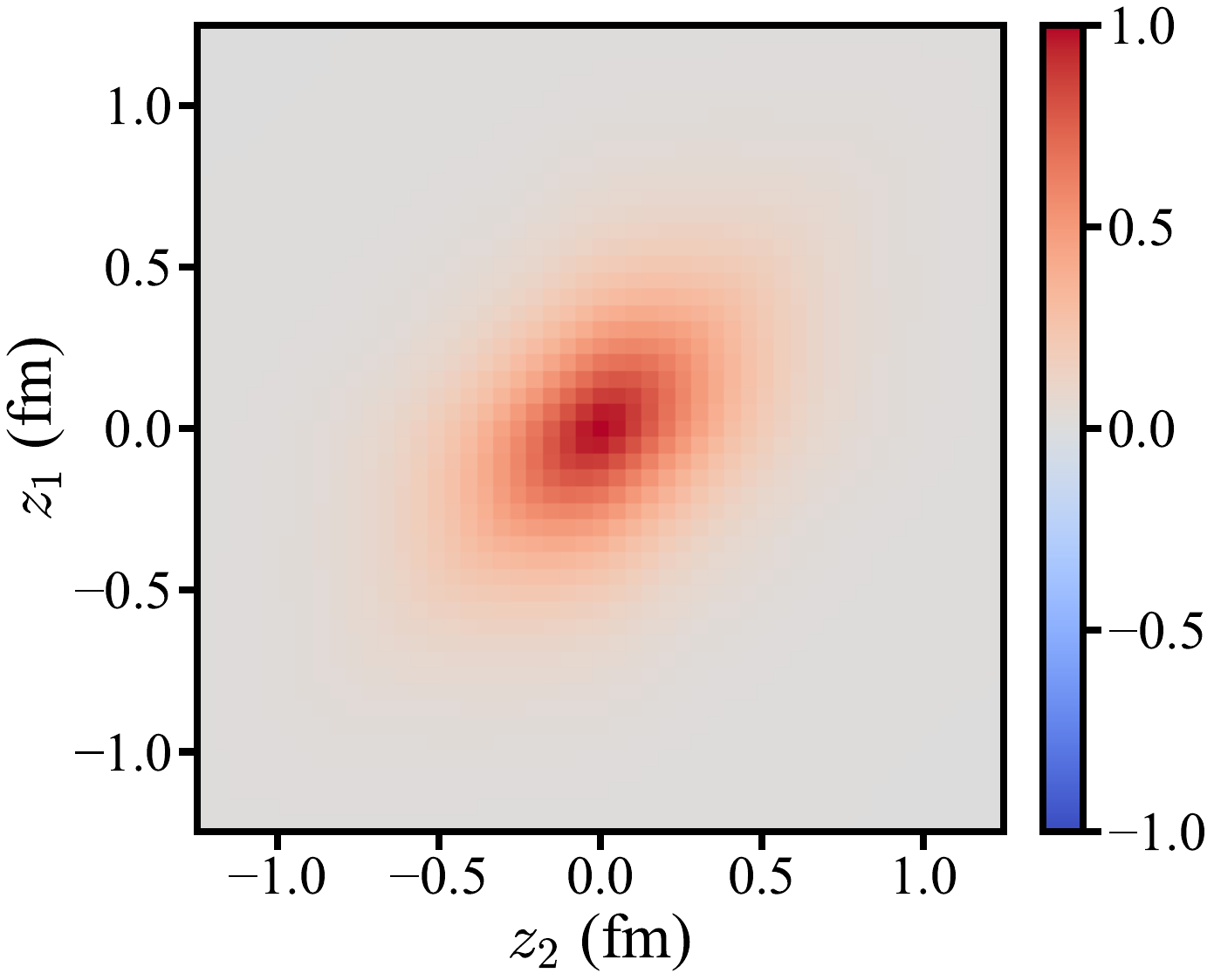}
            }
        }
    \subfloat[\ $\Lambda$-baryon $V$]{
        \makebox[0.31\textwidth][l]{
            \includegraphics[scale=0.24]{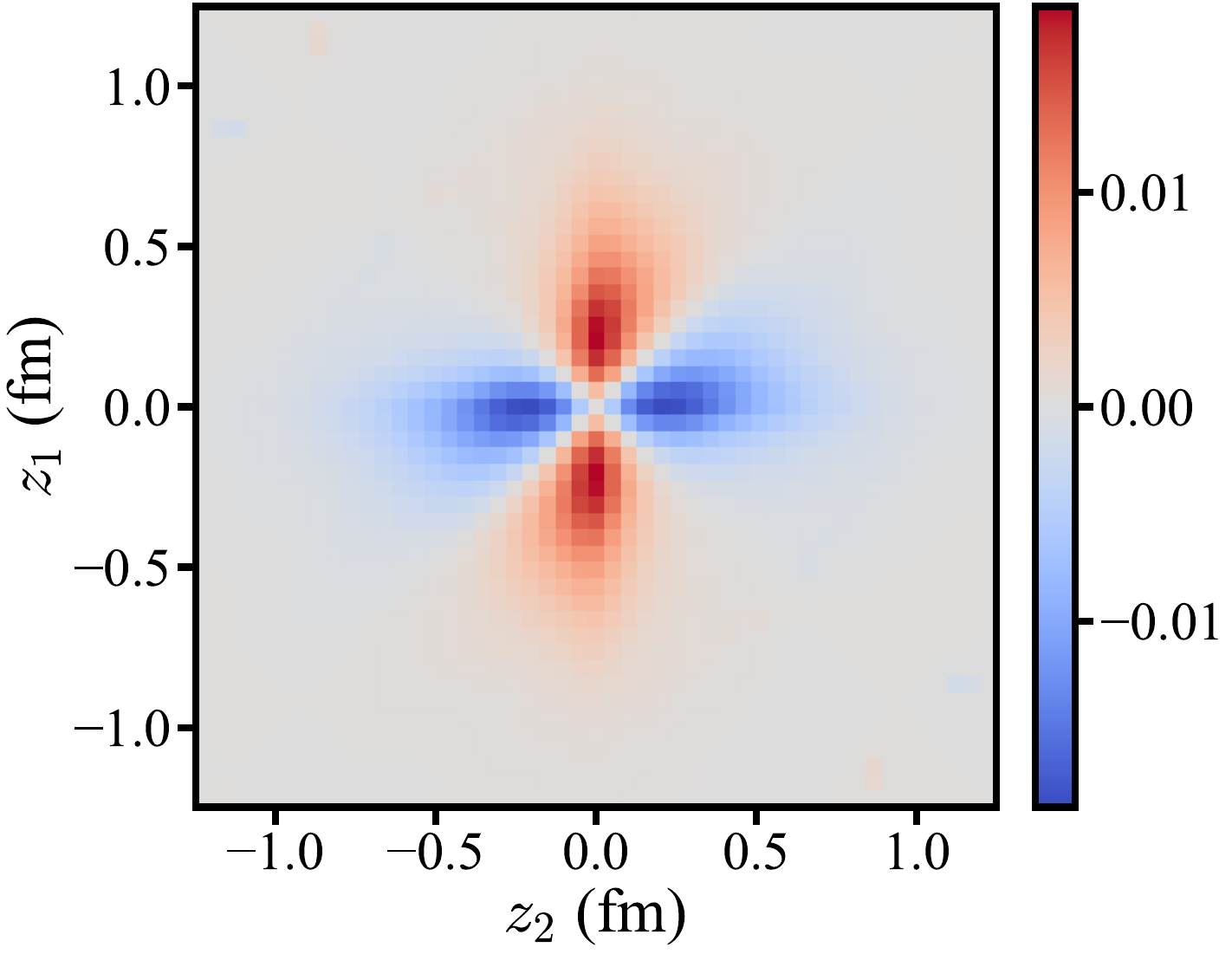}
            }
        }
    \subfloat[\ $\Lambda$-baryon $T$]{
        \makebox[0.31\textwidth][l]{
            \includegraphics[scale=0.24]{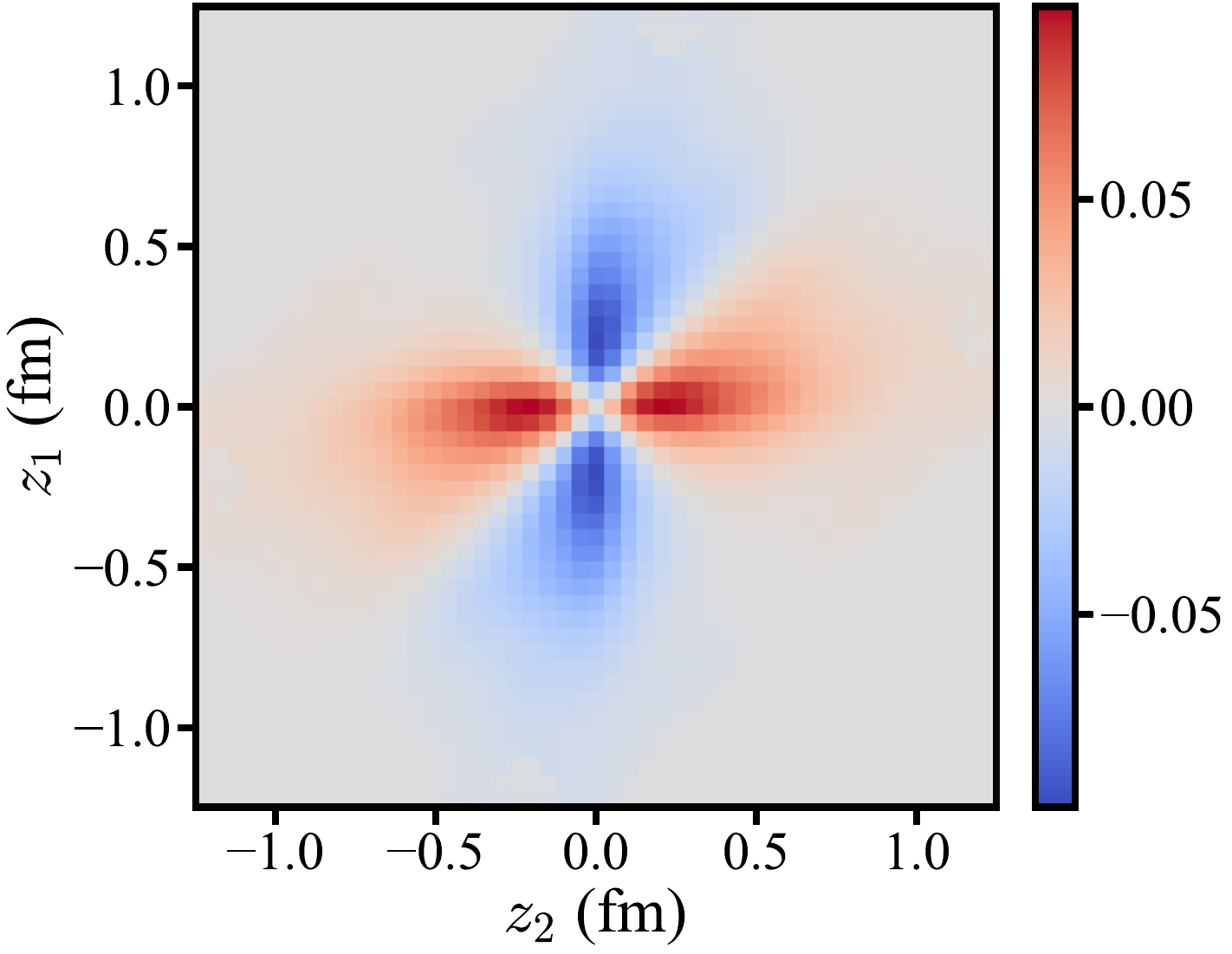}
            }
        }\\
    \subfloat[\ proton $A$]{
        \makebox[0.31\textwidth][l]{
            \includegraphics[scale=0.24]{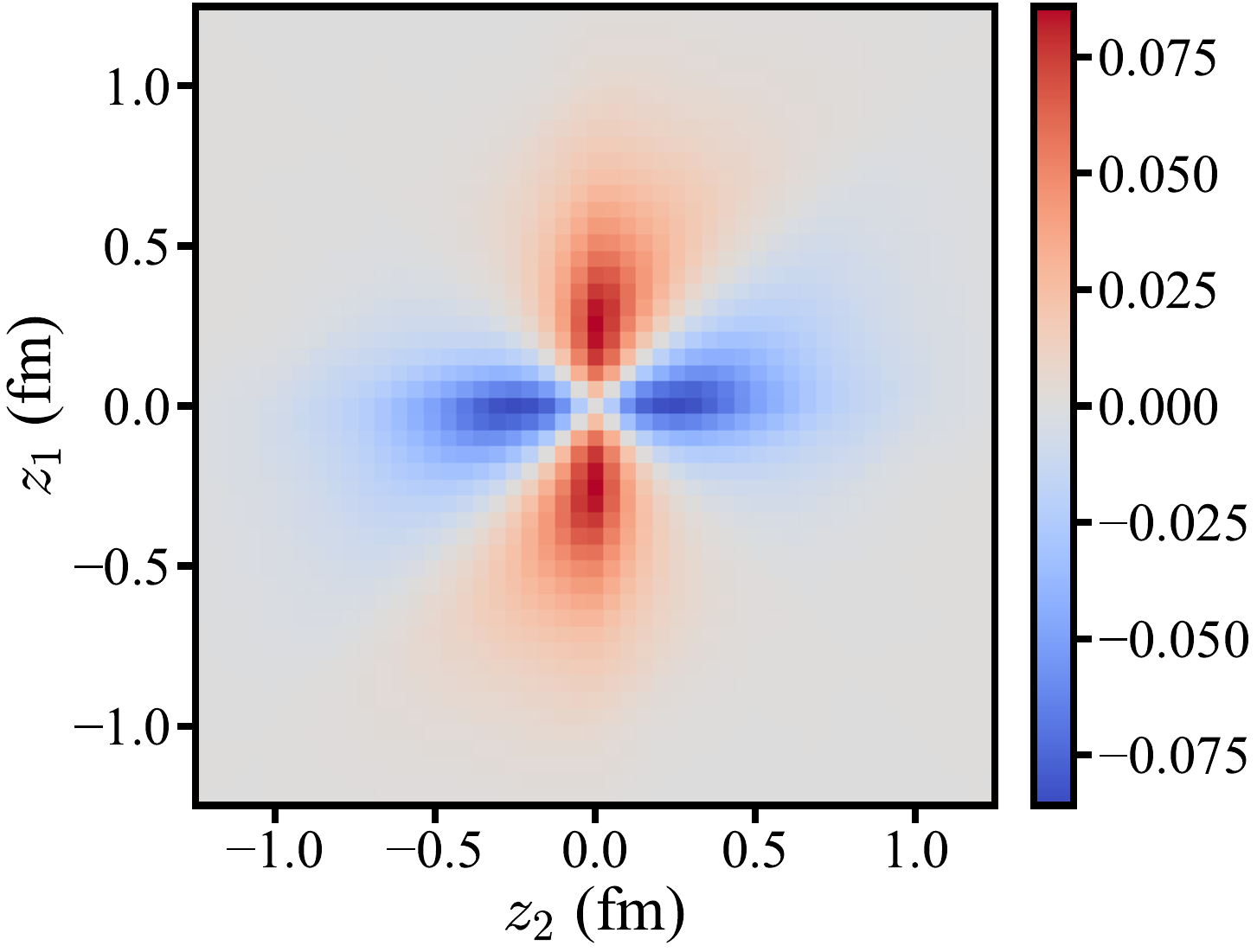}
            }
        }
    \subfloat[\ proton $V$]{
        \makebox[0.31\textwidth][l]{
            \includegraphics[scale=0.24]{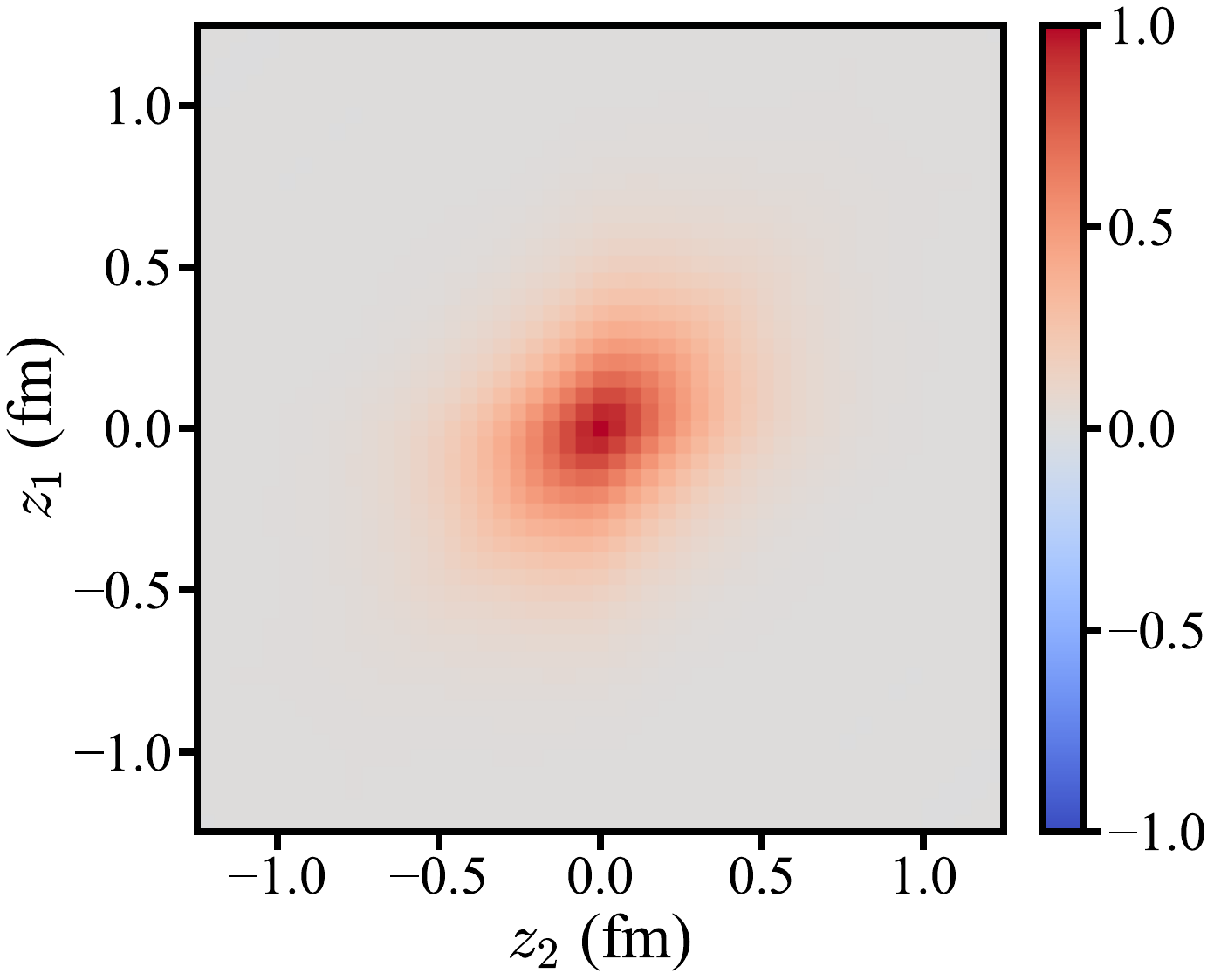}
            }
        }
    \subfloat[\ proton $T$]{
        \makebox[0.31\textwidth][l]{
            \includegraphics[scale=0.24]{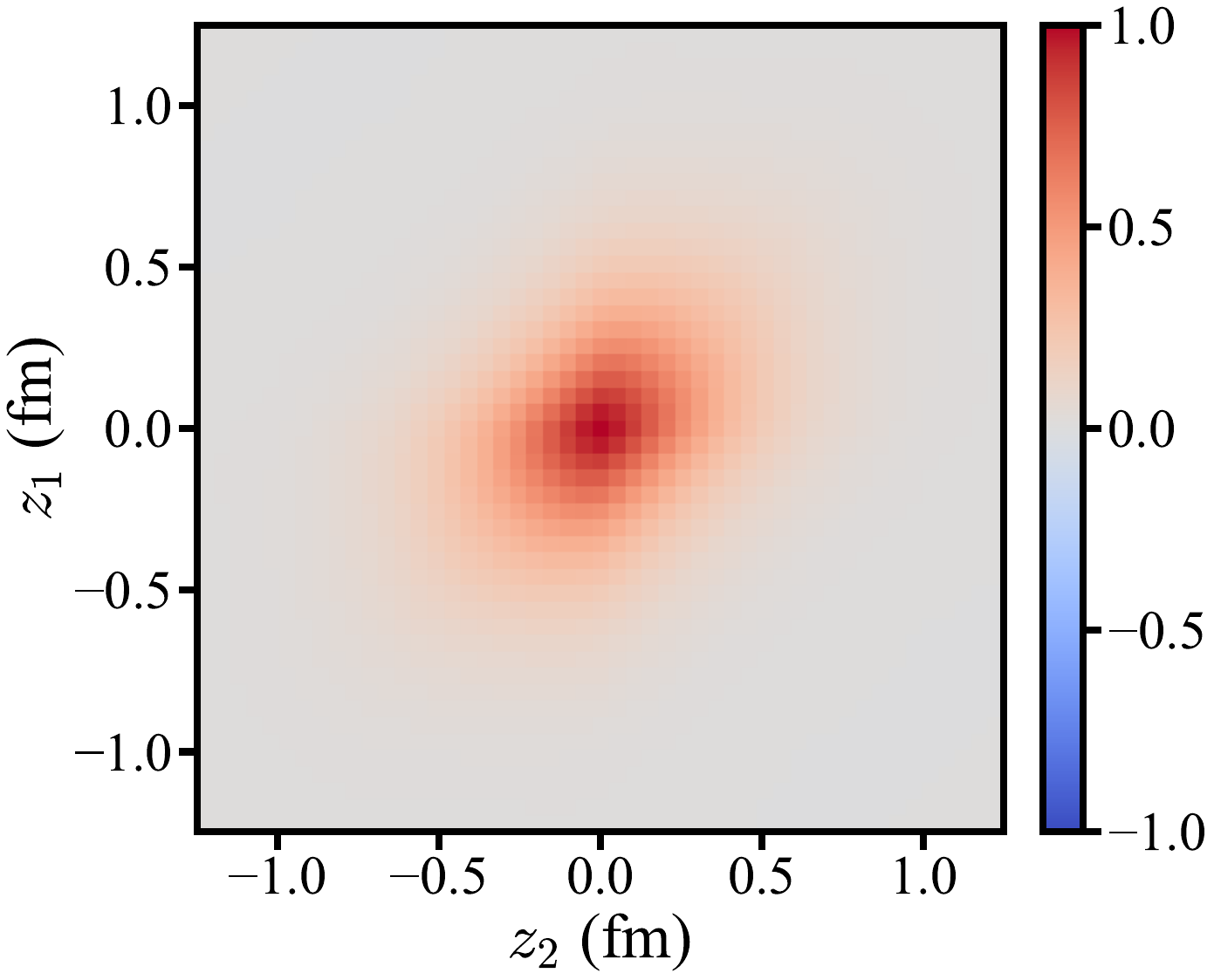}
            }
        }
    \caption{Real part of the bare zero-momentum coordinate-space matrix elements $\widehat M_{\rm bare}(z_1,z_2;0,a)$ before renormalization on the H48P32 ensemble. The upper row shows the $A$, $V$, and $T$ structures of the $\Lambda$ baryon, while the lower row shows the corresponding proton structures. The heat maps illustrate the symmetry patterns used to choose the reference matrix elements for the ratio and self-renormalization procedures.}
    \label{fig:before_renorm_heatmap_h48p32}
\end{figure*}

The hybrid renormalization scheme must be implemented consistently with the $z_1\leftrightarrow z_2$ exchange symmetry of each leading-twist structure. 
For the $\Lambda$ baryon, the $A$ amplitude is symmetric under $z_1\leftrightarrow z_2$, whereas the $V$ and $T$ amplitudes are antisymmetric. For the proton, the $V$ and $T$ amplitudes are symmetric, while the $A$ amplitude is antisymmetric. These symmetry properties are visible in the bare zero-momentum matrix elements shown in Fig.~\ref{fig:before_renorm_heatmap_h48p32}.

These symmetries constrain the reference matrix elements used in the hybrid renormalization scheme. First, the renormalization procedure must preserve the exchange symmetry of the target matrix element on the $(z_1,z_2)$ plane. Second, the zero-momentum matrix elements used to determine the self-renormalization factors $Z_{\rm Self}$ must be compatible with the one-loop perturbative expressions Eqs.~\eqref{eq:pert_res_VA}--\eqref{eq:pert_res_T}, which are symmetric under $z_1\leftrightarrow z_2$. Antisymmetric zero-momentum matrix elements are therefore not suitable as direct references for the self-renormalization fit.
    
These requirements lead us to use symmetric reference matrix elements for both the ratio prescription and the self-renormalization fit. Since the self-renormalization factor is associated with the Wilson-line ultraviolet structures and scheme conversion, it is expected to be largely independent of the external baryon states and only weakly dependent on the Dirac-$\gamma$ structures~\cite{LatticePartonLPC:2021gpi,Zhang:2024omt}. The relevant structure dependence enters through the short-distance perturbative expressions: the $T$ amplitude uses Eq.~\eqref{eq:pert_res_T}, while the $V/A$ amplitudes use Eq.~\eqref{eq:pert_res_VA}. The implementation details in this work are summarized as follows:
\begin{itemize}
    \item For the $\Lambda$-baryon $A$ structure:
    \begin{itemize}
        \item Ratio prescription: divide the large-momentum  matrix element of $\Lambda$ $A$ structure $\widehat M_{\Lambda,\rm bare}^A(z_1,z_2;P^z,a)$ by the zero-momentum one of $\Lambda$ $A$ structure $\widehat M_{\Lambda,\rm bare}^A(z_1,z_2;0,a)$;
        \item Self-renormalization: $Z^{A}_{\rm Self}$ by fitting the zero-momentum matrix  element of $\Lambda$ $A$ structure $\widehat M_{\Lambda,\rm bare}^A(z_1,z_2;0,a)$ with perturbative $\widehat M_{\rm \overline{MS},pert}^{(1),V/A}(z_1,z_2;0,\mu)$ in Eq.~\eqref{eq:pert_res_VA}.
    \end{itemize}

    \item For the $\Lambda$-baryon  $V$structure:
    \begin{itemize}
        \item Ratio prescription: divide the large-momentum matrix element of $\Lambda$ $V$ structure $\widehat M_{\Lambda,\rm bare}^V(z_1,z_2;P^z,a)$ by the zero-momentum one of $\Lambda$ $A$ structure $\widehat M_{\Lambda,\rm bare}^A(z_1,z_2;0,a)$;
        \item Self-renormalization: $Z^{V}_{\rm Self} = Z^{A}_{\rm Self}$ from $\Lambda$ $A$ structure.
    \end{itemize}

    \item For the $\Lambda$-baryon $T$ structure:
    \begin{itemize}
        \item Ratio prescription: divide the large-momentum matrix element of $\Lambda$ $T$ structure $\widehat M_{\Lambda,\rm bare}^T(z_1,z_2;P^z,a)$ by the zero-momentum one of proton $T$ structure $\widehat M_{\rm p, bare}^T(z_1,z_2;0,a)$;
        \item Self-renormalization: $Z^{T}_{\rm Self}$ by fitting the zero-momentum matrix element of proton $T$ structure $\widehat M_{\rm p, bare}^T(z_1,z_2;0,a)$ with perturbative $\widehat M_{\rm \overline{MS},pert}^{(1),T}(z_1,z_2;0,\mu)$ in Eq.~\eqref{eq:pert_res_T}.
    \end{itemize}
\end{itemize}
This prescription preserves the required exchange symmetry of each target matrix element while using symmetric zero-momentum references compatible with the perturbative expressions.

\subsection{Numerical Implementation and Checks}\label{sec:hybrid_implement}
We now describe the numerical implementation of the self-renormalization fit and the hybrid renormalization procedure. The analysis is performed on all seven ensembles listed in Table~\ref{tab:Ensembles}. To compare matrix elements at different lattice spacings, the bare zero-momentum matrix elements are interpolated onto a common coordinate grid with spacing $a_0=0.05~{\rm fm}$. The four ensembles with similar pion masses around $300~{\rm MeV}$ and different lattice spacings, C24P29, F32P30, G36P29, and H48P32, are used in the self-renormalization fit. The remaining three ensembles with different pion masses, F32P21, C48P14, and C32P23, are used as consistency checks.

As discussed in Sec.~\ref{sec:hybridVAT}, the $\Lambda$-baryon $A$ and $V$ matrix elements are renormalized using the zero-momentum $\Lambda$-baryon $A$ matrix element as the reference. The $\Lambda$-baryon $T$ matrix element is renormalized using the zero-momentum proton $T$ matrix element as the reference, in order to match the corresponding short-distance perturbative structure. The numerical implementation of these two reference structures is described below.

\begin{figure}[htbp]
    \centering
    \subfloat[\ $\Lambda$-baryon $A$ matrix elements]{
        \centering
        \includegraphics[width=\linewidth]{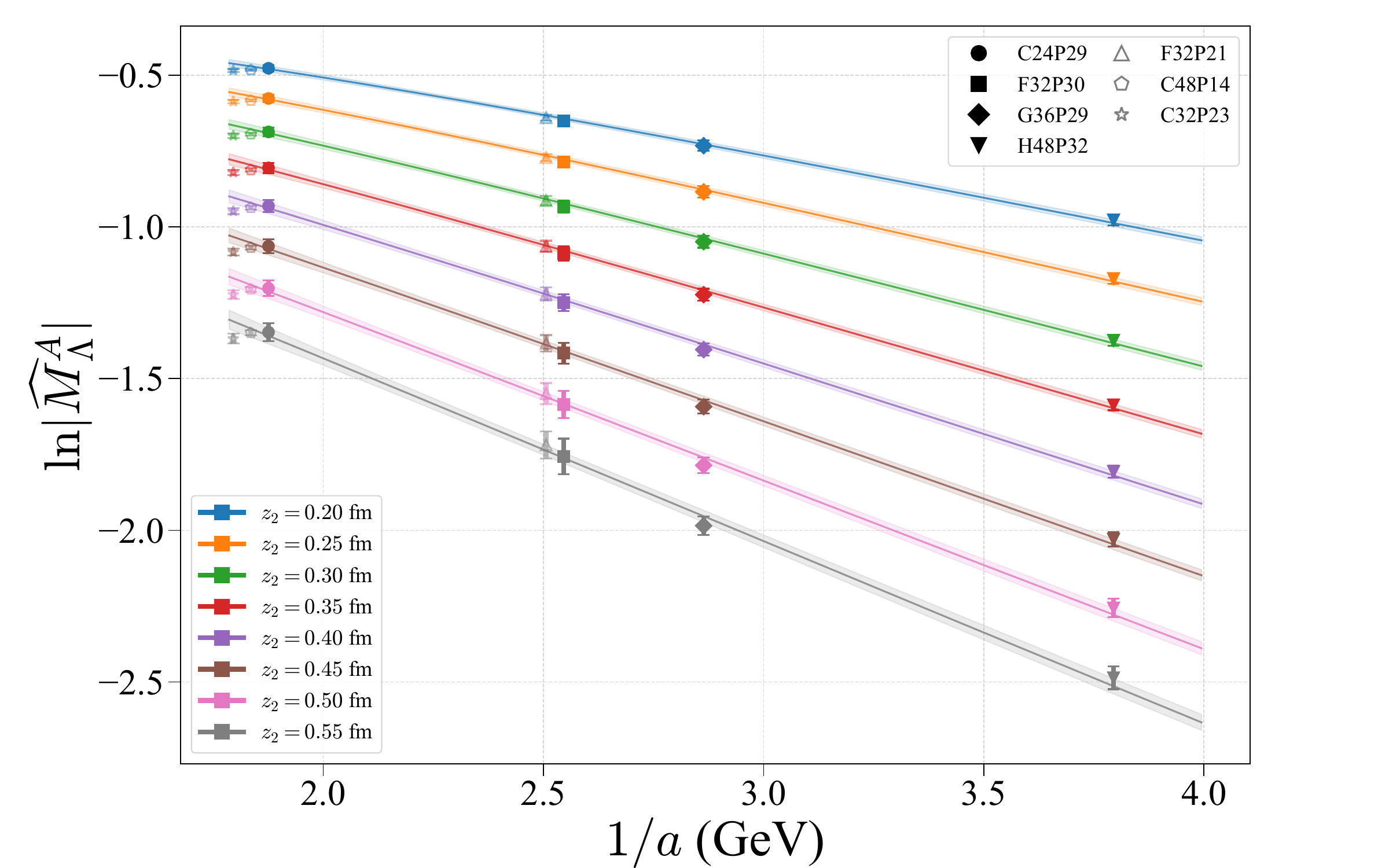}
        }\\
    \subfloat[\ proton $T$ matrix elements]{
        \centering
        \includegraphics[width=\linewidth]{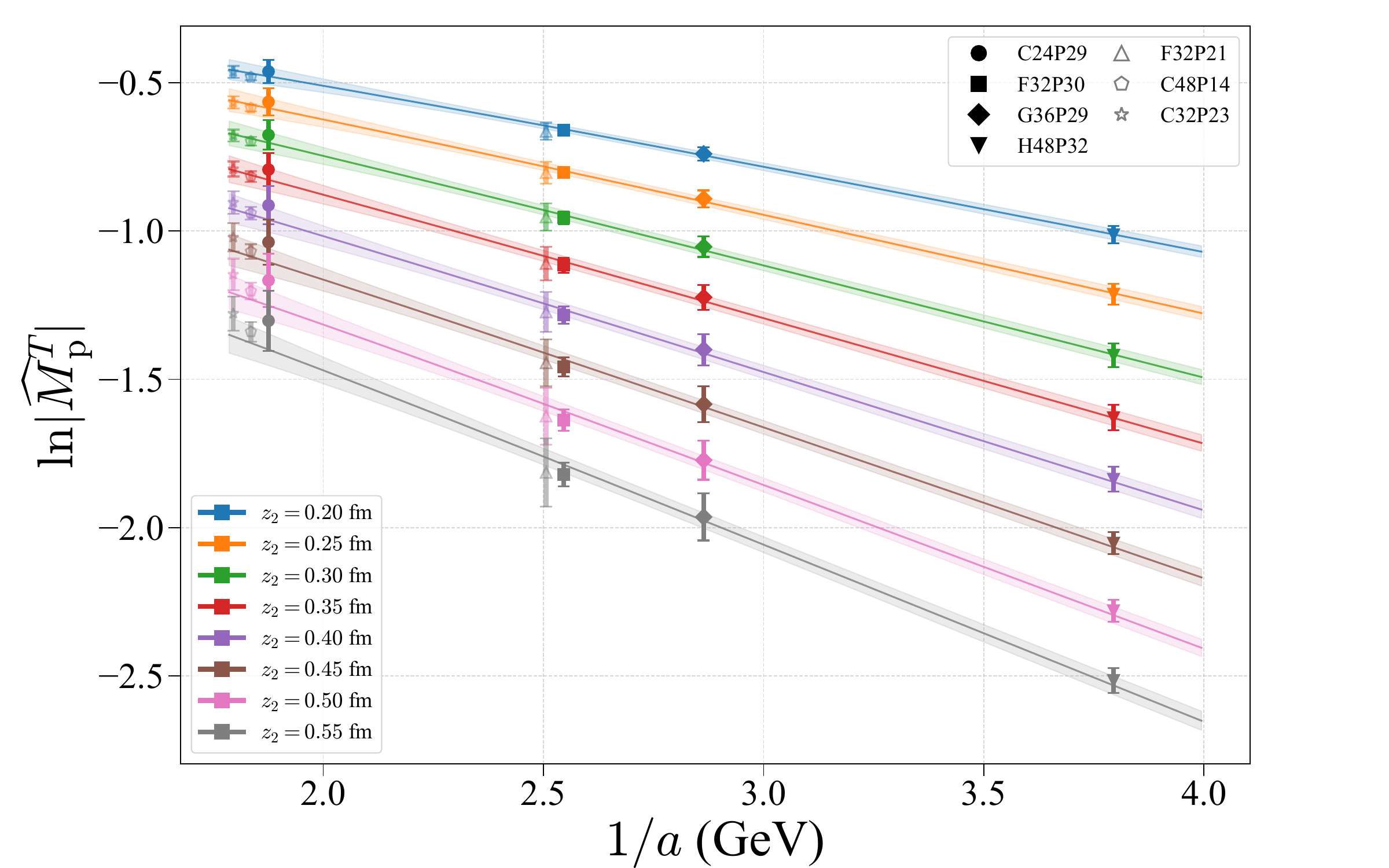}
        }
    \caption{Extraction of the linear divergence coefficient $k$ from the lattice-spacing dependence of zero-momentum matrix elements, shown for fixed $z_1 = -0.20~{\rm fm}$ with varying $z_2$. Filled markers correspond to the four fit ensembles (C24P29, F32P30, G36P29, H48P32), and open markers to the three consistency-check ensembles (F32P21, C48P14, C32P23). These colored bands show the global fit result.}
    \label{fig:hybrid_a_dependence}
\end{figure}

\subsubsection{Step 1: Extraction of the Linear Divergence}
The first step of the self-renormalization procedure is to determine the Wilson-line linear-divergence coefficient $k$ from the lattice-spacing dependence of the zero-momentum matrix elements. We use data in the medium- and long-distance region satisfying
\begin{equation}
    |z_1|,\ |z_2|,\ |z_1-z_2| \geq 4a_0 = 0.20~{\rm fm}\ ,
\end{equation}
where $a_0=0.05~{\rm fm}$ is the common interpolation-grid spacing introduced above. In this region, the lattice-spacing dependence is dominated by the Wilson-line linear divergence. The lattice QCD scale is fixed to $\Lambda_{\rm QCD}^{\rm latt}=0.2~{\rm GeV}$, following Ref.~\cite{LatticePartonCollaborationLPC:2025vhd}.

The $k$ values from the four fit ensembles, extracted at the representative scale $\mu=2.0~{\rm GeV}$ are:
\begin{itemize}
    \item $\Lambda$-baryon $A$: $k = 0.7778(22)$, $\chi^2/{\rm d.o.f.} = 0.293$,
    \item proton $T$: $k = 0.7810(22)$, $\chi^2/{\rm d.o.f.} = 0.329$.
\end{itemize}
The agreement between the values extracted from the $\Lambda$-baryon $A$ and proton $T$ reference matrix elements is consistent with the expectation that the linear divergence is governed by the Wilson-line self energy and is largely independent of the external baryon states and Dirac-$\gamma$ structures. This is also consistent with our previous study~\cite{LatticePartonCollaborationLPC:2025vhd}.

Fig.~\ref{fig:hybrid_a_dependence} illustrates the lattice-spacing dependence of the zero-momentum matrix elements. The data are plotted as $\ln|\widehat M_{\rm bare}|$ versus $1/a$, at fixed $z_1=-0.20~{\rm fm}$ with $z_2$ ranging from $0.20$ to $0.55~{\rm fm}$. The four fit ensembles, shown by filled markers, are well described by the fitted bands. The three consistency-check ensembles, shown by open markers, also lie close to the same bands, providing an additional check of the extracted linear-divergence coefficient.

\subsubsection{Step 2: Short-distance Fitting to \texorpdfstring{$\overline{\rm MS}$}{MS-bar} Scheme}

After the determination of the linear divergence coefficient $k$, the second step is to extract the residual self-renormalization parameters $m_0$ and $d$. This is done by matching the short-distance lattice matrix elements to the one-loop zero-momentum perturbative expressions in the $\overline{\rm MS}$ scheme. The $\Lambda$-baryon $A$ reference is matched to the $V/A$ perturbative expression, while the proton $T$ reference is matched to the $T$ perturbative expression.

Following Ref.~\cite{LatticePartonCollaborationLPC:2025vhd}, the fit is restricted to the region with $z_1z_2<0$. This choice avoids the sharp perturbative structure associated with the short-distance singularity near $|z_1-z_2|=0$. The fits are performed independently at three renormalization scales,
$\mu=\sqrt{2}$, $2$, and $2\sqrt{2}~{\rm GeV}$,
using the four-flavor running coupling with $\Lambda_{\overline{\rm MS}}^{N_f=4}=0.322~{\rm GeV}$. The central analysis uses $\mu=2~{\rm GeV}$, while the two additional scales are used to quantify the residual scale dependence of the one-loop scheme conversion.

The extracted parameters are summarized in Table~\ref{tab:hybrid_m0_d_three_scales}. The parameter $m_0$ is quoted in GeV, while $d$ is dimensionless and follows the convention in Eq.~\eqref{eq:self_Z}.

\begin{table}[htbp]
    \centering
    \renewcommand{\arraystretch}{1.15}
    \setlength{\tabcolsep}{7.5pt}
    \begin{tabular}{c c c c}
        \hline\hline
        $\mu~({\rm GeV})$ & Structures & $m_0~({\rm GeV})$ & $d$ \\
        \hline
        \multirow{2}{*}{$\sqrt{2}$}
        & $\Lambda$-baryon $A$ & $0.1307(34)$ & $-0.1798(25)$ \\
        & proton $T$ & $0.1392(46)$ & $-0.1283(33)$ \\
        \hline
        \multirow{2}{*}{$2$}
        & $\Lambda$-baryon $A$ & $0.2159(35)$ & $-0.1718(28)$ \\
        & proton $T$ & $0.2242(47)$ & $-0.1222(35)$ \\
        \hline
        \multirow{2}{*}{$2\sqrt{2}$}
        & $\Lambda$-baryon $A$ & $0.2654(38)$ & $-0.1375(32)$ \\
        & proton $T$ & $0.2725(49)$ & $-0.0867(39)$ \\
        \hline\hline
    \end{tabular}
    \caption{Residual self-renormalization parameters $m_0$ and $d$ from the Step-2 short-distance fits at $\mu=\sqrt{2}$, $2$ and $2\sqrt{2}~{\rm GeV}$.}
    \label{tab:hybrid_m0_d_three_scales}
\end{table}

At the central scale $\mu=2~{\rm GeV}$, the parameters used in the main analysis are
\begin{itemize}
    \item $\Lambda$-baryon $A$: $m_0 = 0.216(4)~{\rm GeV}$, $d = -0.172(3)$,
    \item proton $T$: $m_0 = 0.224(5)~{\rm GeV}$, $d = -0.122(4)$.
\end{itemize}
The values of $m_0$ extracted from the two reference structures are close at each scale, consistent with the interpretation of $m_0$ as a residual mass parameter associated primarily with the Wilson-line self energy~\cite{LatticePartonLPC:2021gpi,Zhang:2024omt}. The difference in $d$ reflects the remaining short-distance scheme dependence and the different one-loop perturbative structures in the $V/A$ and $T$ amplitudes.

\begin{figure*}[t]
    \centering
    \subfloat[\ $\Lambda$-baryon $A$, $\mu = \sqrt{2}~{\rm GeV}$]{
        \includegraphics[width=0.31\textwidth]{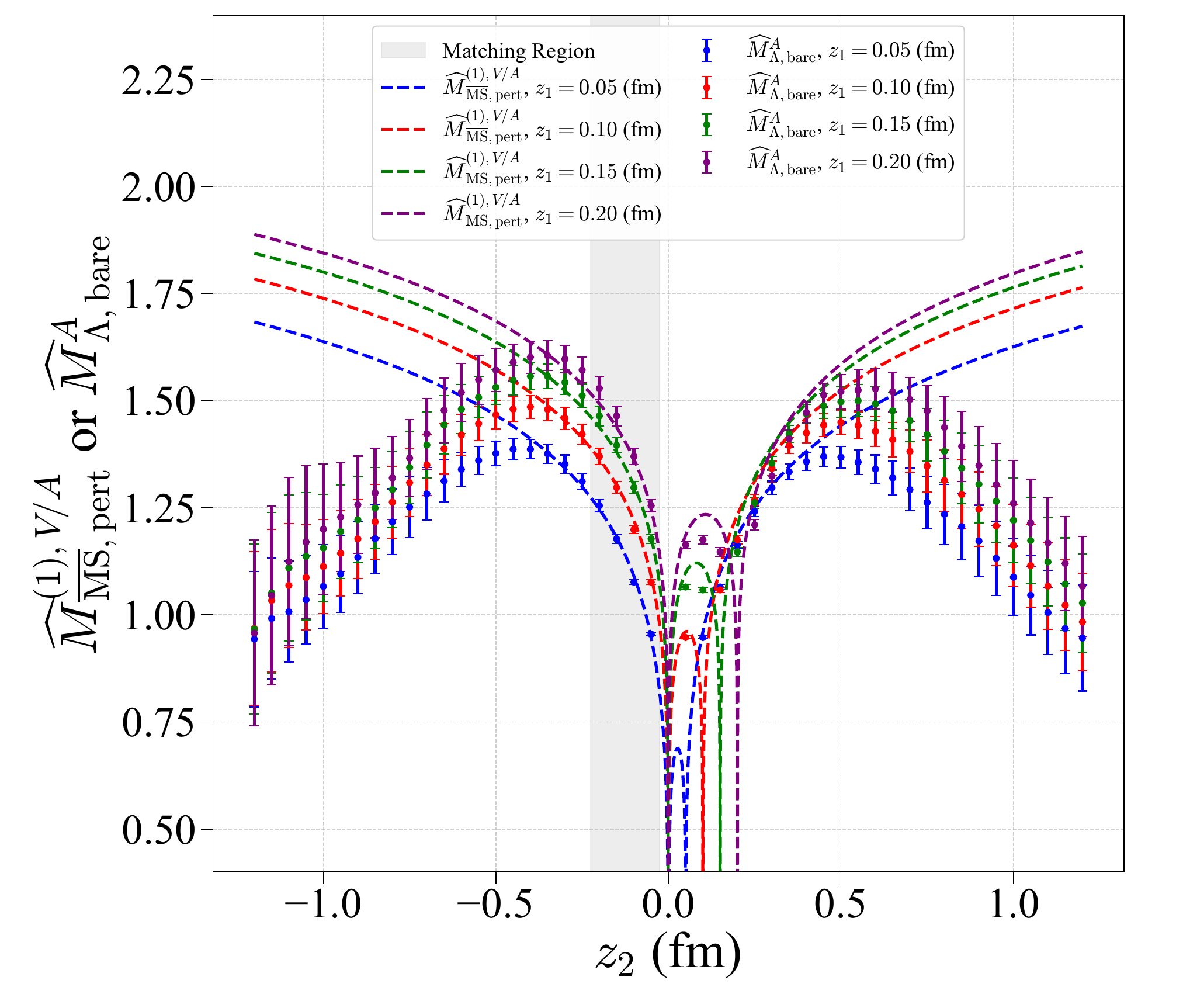}
        }
    \subfloat[\ $\Lambda$-baryon $A$, $\mu = 2~{\rm GeV}$]{
        \includegraphics[width=0.31\textwidth]{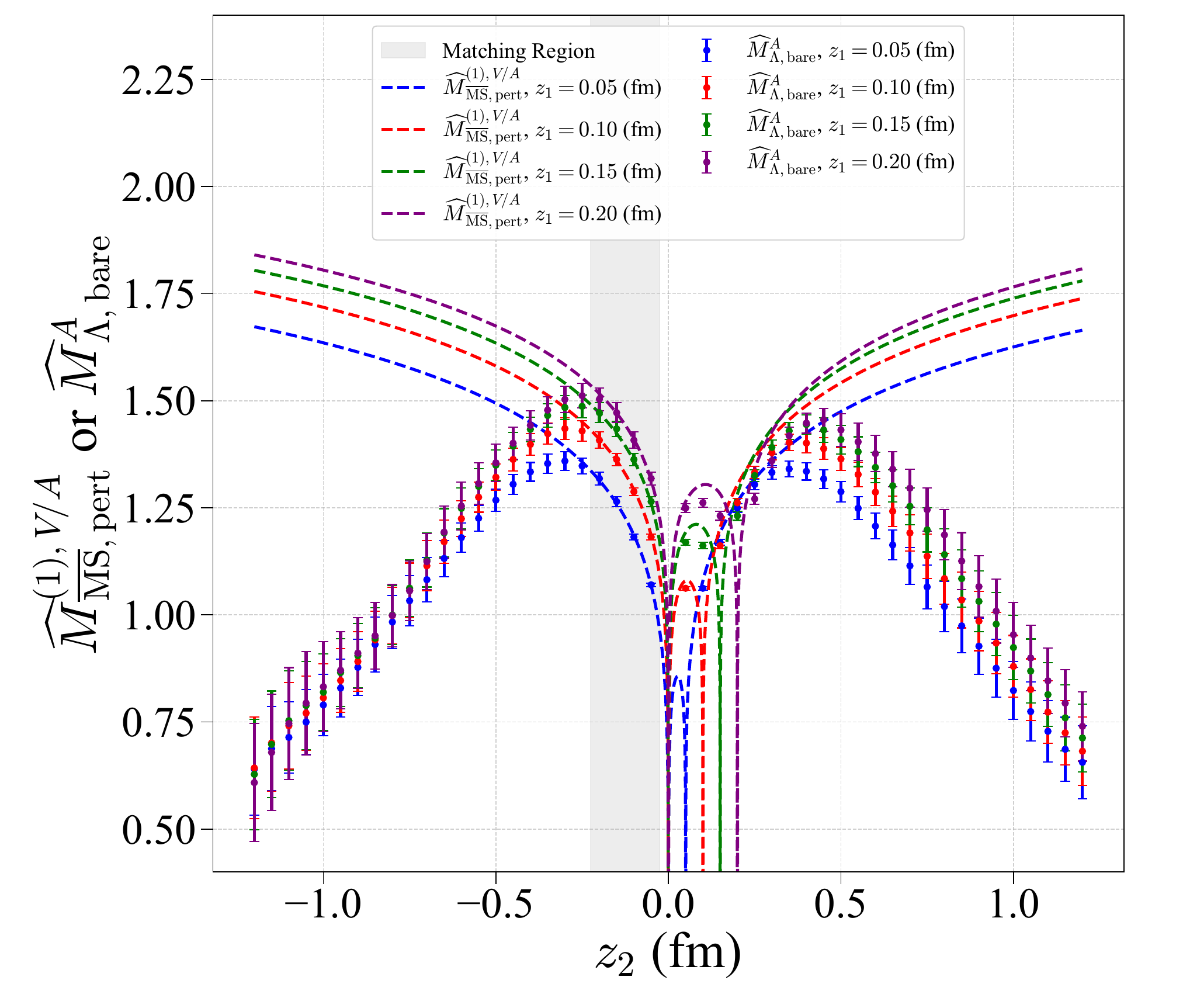}
        }
    \subfloat[\ $\Lambda$-baryon $A$, $\mu = 2\sqrt{2}~{\rm GeV}$]{
        \includegraphics[width=0.31\textwidth]{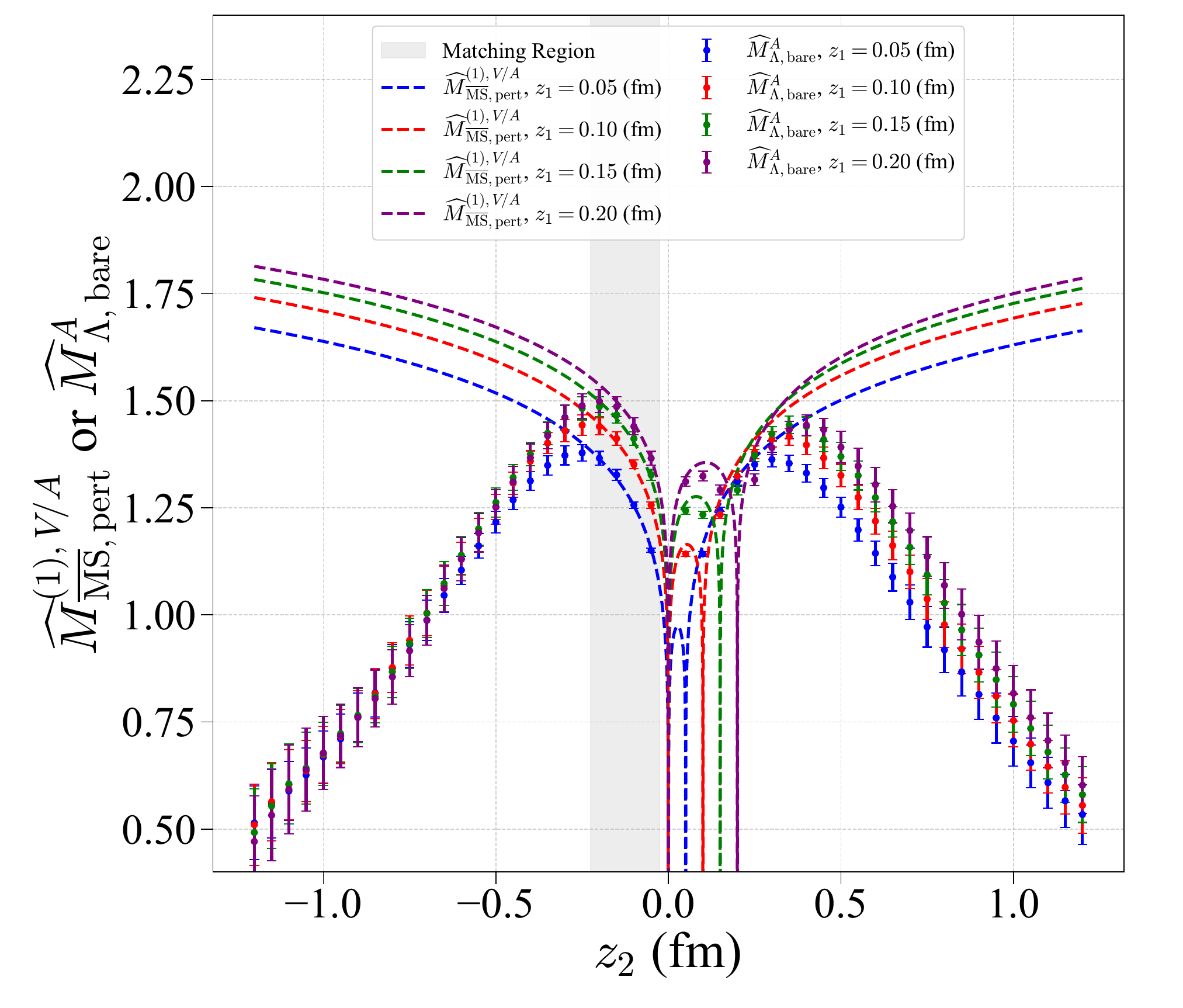}
        }\\
    \subfloat[\ proton $T$, $\mu = \sqrt{2}~{\rm GeV}$]{
        \includegraphics[width=0.31\textwidth]{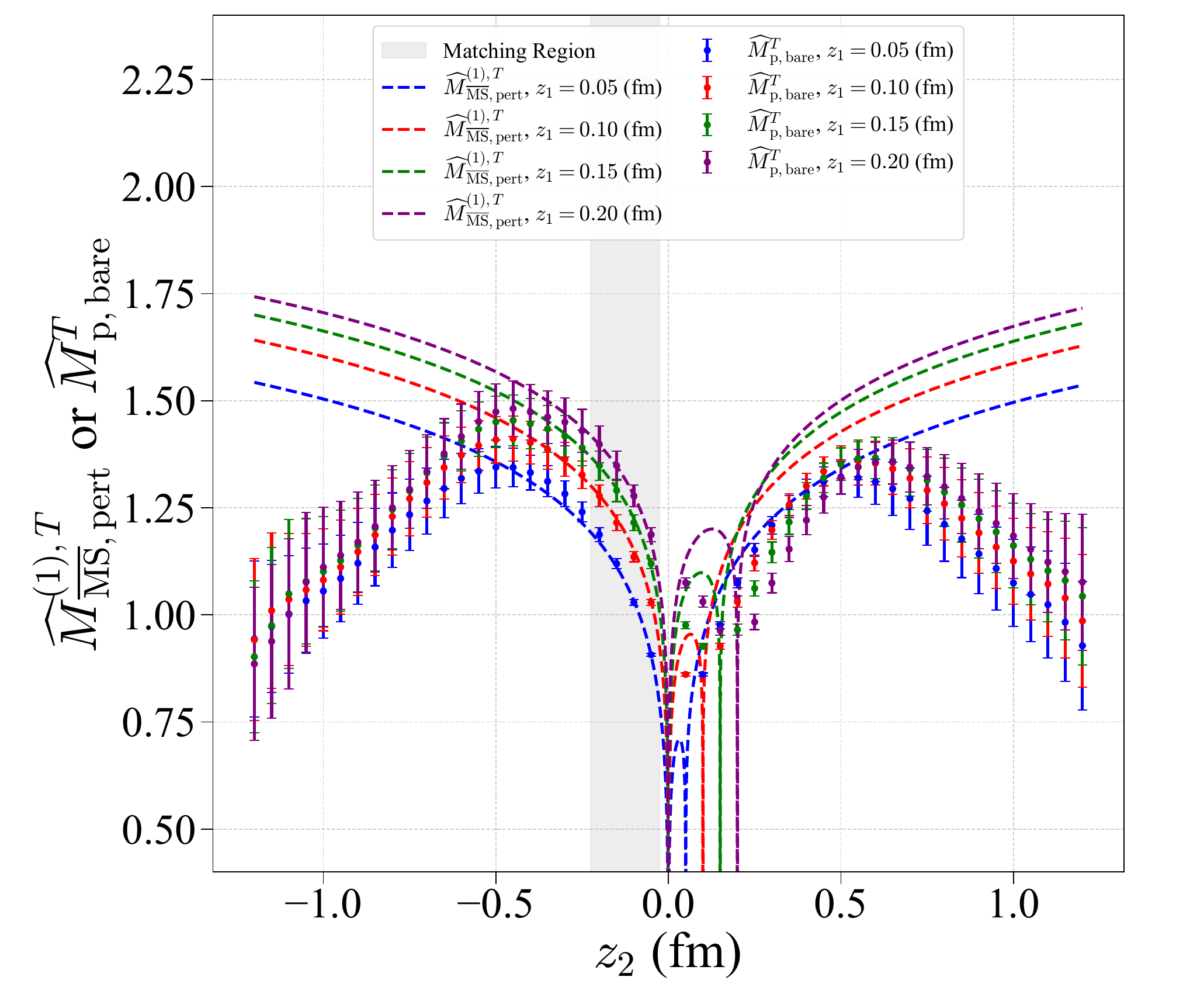}
        }
    \subfloat[\ proton $T$, $\mu = 2~{\rm GeV}$]{
        \includegraphics[width=0.31\textwidth]{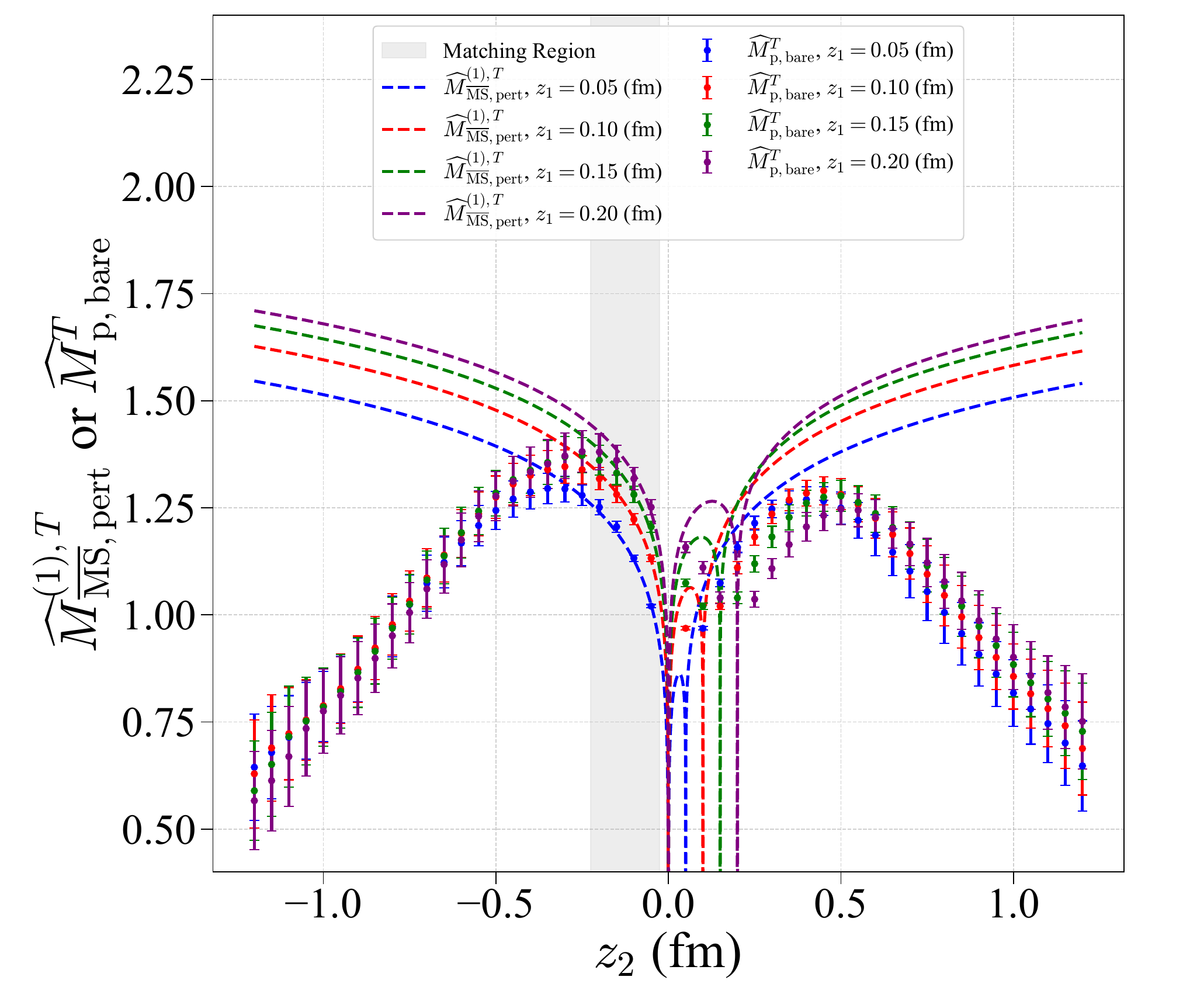}
        }
    \subfloat[\ proton $T$, $\mu = 2\sqrt{2}~{\rm GeV}$]{
        \includegraphics[width=0.31\textwidth]{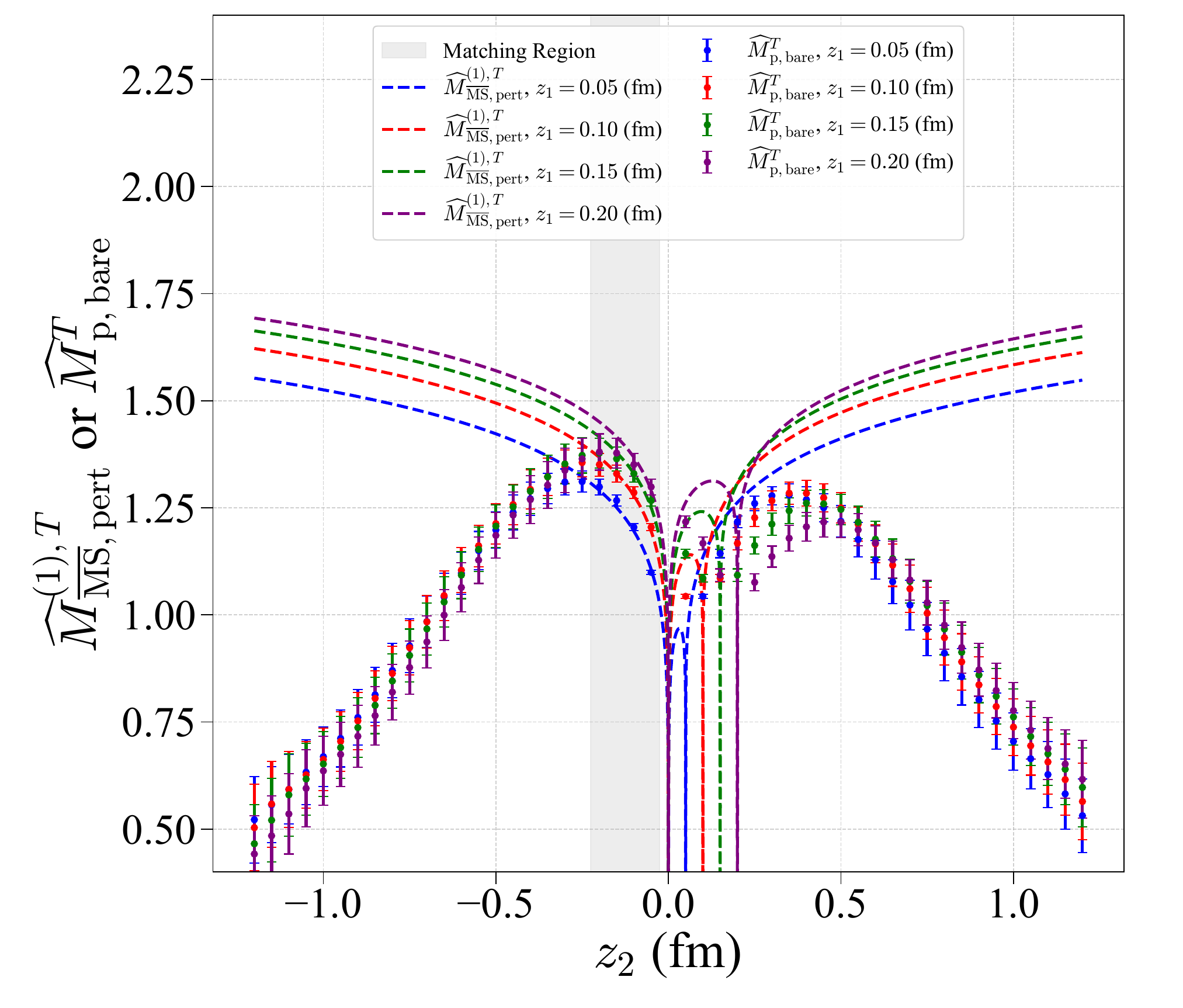}
        }
    \caption{Second-step fitting to the $\overline{\rm MS}$ scheme perturbative expressions at three renormalization scales $\mu = \sqrt{2}$, $2$, and $2\sqrt{2}~{\rm GeV}$. The fitted lattice-side short-distance data (points with error bars) are compared with the one-loop perturbative $\overline{\rm MS}$ expressions (dashed curves) for the $\Lambda$-baryon $A$ amplitude (top row) and proton $T$ amplitude (bottom row). The gray band indicates the chosen short-distance fitting window. }
    \label{fig:hybrid_msbar_matching}
\end{figure*}

The quality of the short-distance fits is shown in Fig.~\ref{fig:hybrid_msbar_matching}. The fitted lattice short-distance quantity is compared with the corresponding one-loop $\overline{\rm MS}$ expression at the three different scales. Within the selected fitting window, the lattice data are described by the perturbative curves for both $\Lambda$-baryon $A$ and proton $T$ reference amplitudes. This agreement supports the extraction of the residual self-renormalization parameters and indicates that the renormalization factor provides a stable conversion of the lattice matrix elements to the $\overline{\rm MS}$ scheme. 

\subsubsection{Implementation of Hybrid Renormalization}
\begin{figure*}[t]
    \centering
    \subfloat[\ Bare, $\Lambda$-baryon $A$ matrix elements]{
        \centering
        \includegraphics[width=0.47\textwidth]{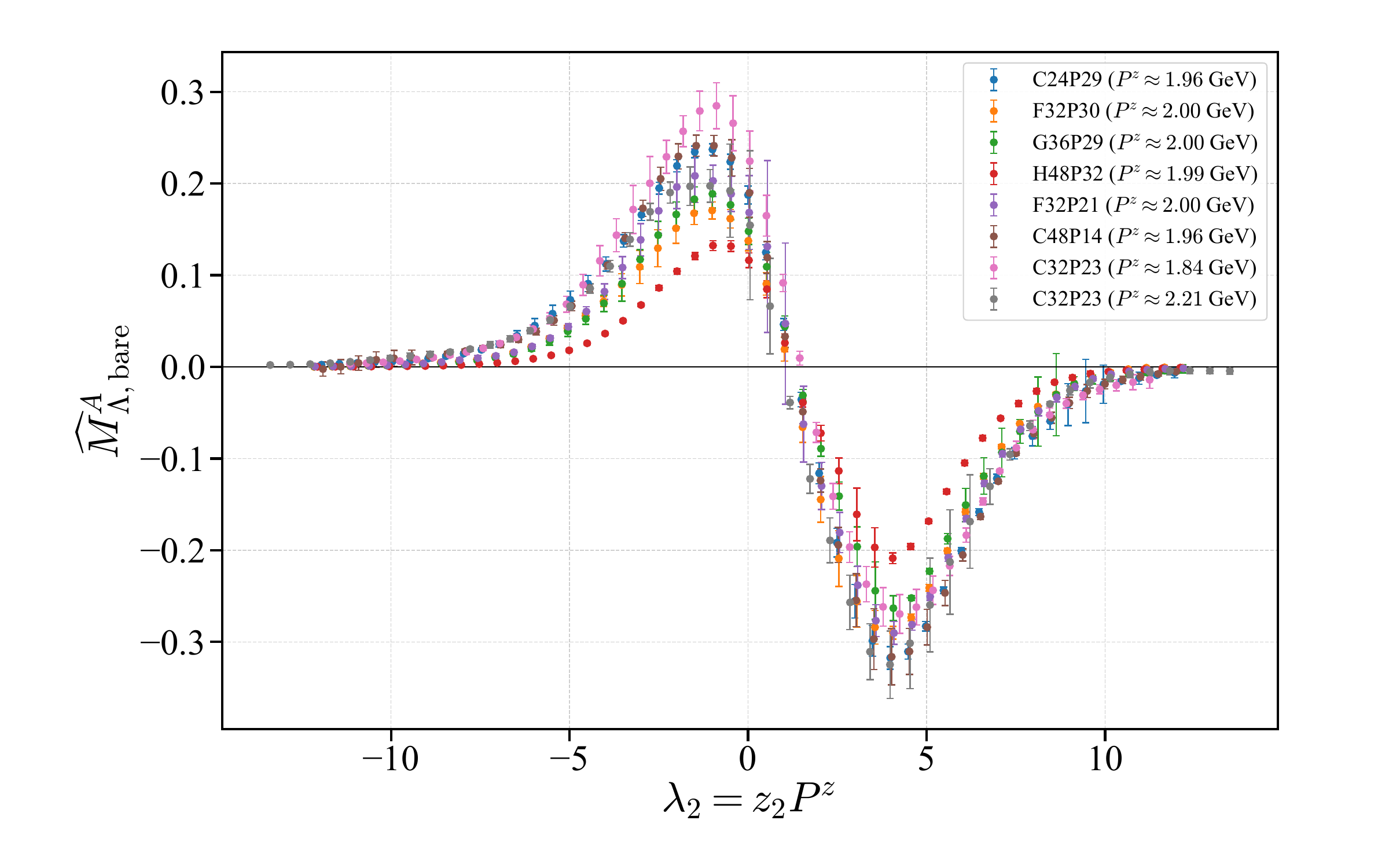}
        \label{subfig:bare_matrix_elem}
        }
    \subfloat[\ Hybrid, $\Lambda$-baryon $A$ matrix elements]{
        \centering
        \includegraphics[width=0.47\textwidth]{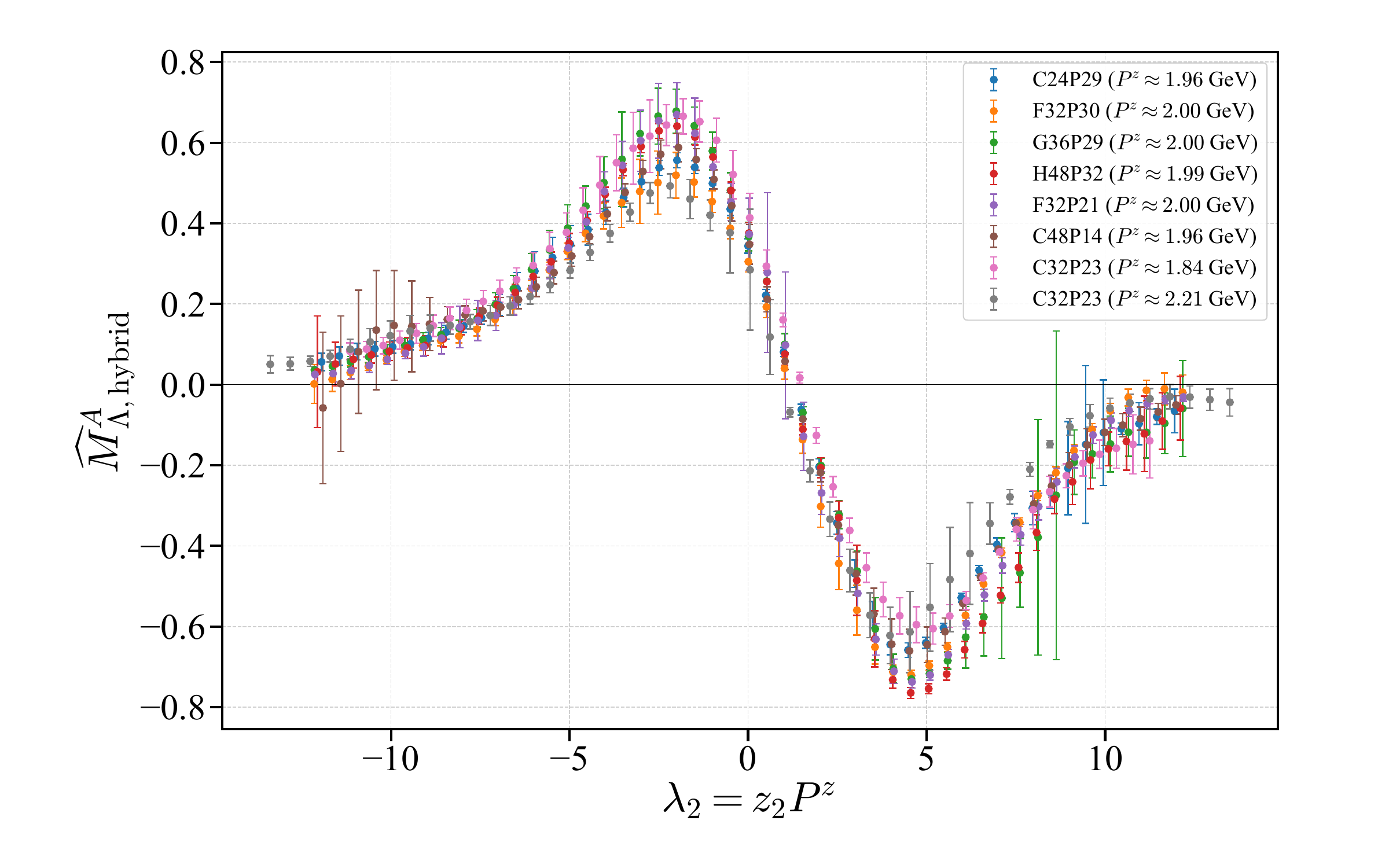}
        \label{subfig:hybrid_matrix_elem}
        }
    \caption{Comparison of bare (left) and hybrid-renormalized (right) $\Lambda$-baryon $A$ matrix elements at the $P^z \approx 2.0~{\rm GeV}$ momentum group, with fixed $z_1 = 0.40~{\rm fm}$, shown as a function of $\lambda_2 = z_2 P^z$. All seven ensembles are displayed. The hybrid renormalization effectively removes the lattice-spacing-dependent linear divergences, yielding consistent results across ensembles.}
    \label{fig:hybrid_bare_vs_hybrid}
\end{figure*}

With the self-renormalization factor $Z_{\rm Self}(z_1,z_2;a,\mu)$ determined, we apply hybrid renormalization to the large-momentum matrix elements following the procedure described in Sec.~\ref{sec:hybrid_2D}. To demonstrate its effectiveness, we present a comparison in Fig.~\ref{fig:hybrid_bare_vs_hybrid} between the bare matrix element of $\Lambda$-baryon $A$ structure $\widehat M^A_{\Lambda,\rm bare}(z_1,z_2;P^z,a)$ and the hybrid-renormalized one $\widehat M_{\Lambda,\rm hybrid}^A(z_1,z_2;P^z,\mu)$, at the $P^z \approx 2.0~{\rm GeV}$ momentum, with a fixed slice at $z_1 = 0.40~{\rm fm}$.
In the bare case Fig.~\ref{subfig:bare_matrix_elem}, significant discrepancies are visible among the seven ensembles, reflecting the large linear divergences. After hybrid renormalization Fig.~\ref{subfig:hybrid_matrix_elem}, the results from different ensembles collapse onto a consistent curve within uncertainties. This demonstrates the effectiveness of the hybrid renormalization scheme in removing the dominant ultraviolet lattice divergences before the large-distance extrapolation and Fourier transform.

\subsubsection{Residual Renormalization-scale Dependence}

\begin{figure}[htbp]
    \centering
    \subfloat[]{
        \centering
        \includegraphics[width=\linewidth]{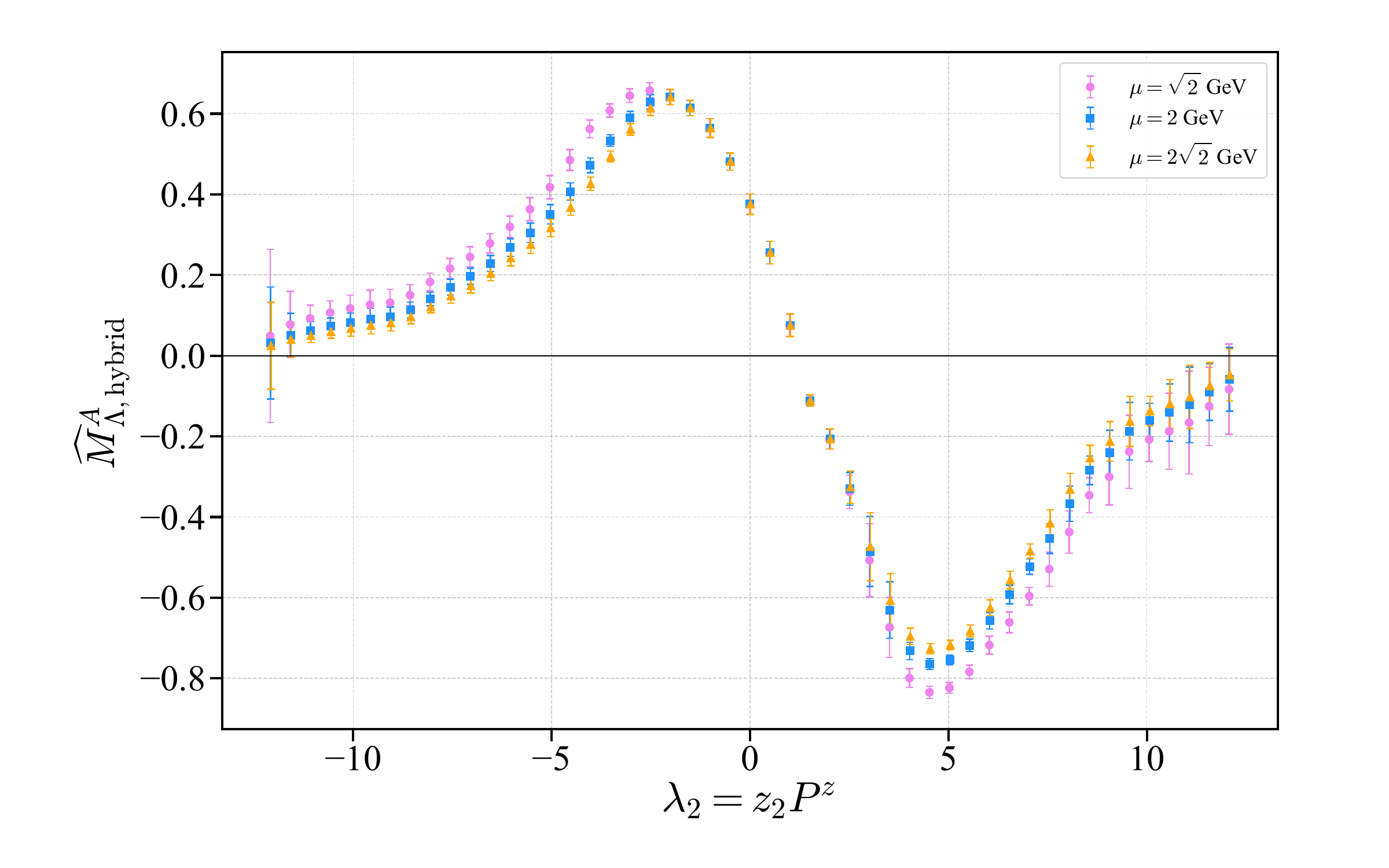}
        \label{subfig:mu_coordinate_z1}
        }\\
    \subfloat[]{
        \centering
        \includegraphics[width=\linewidth]{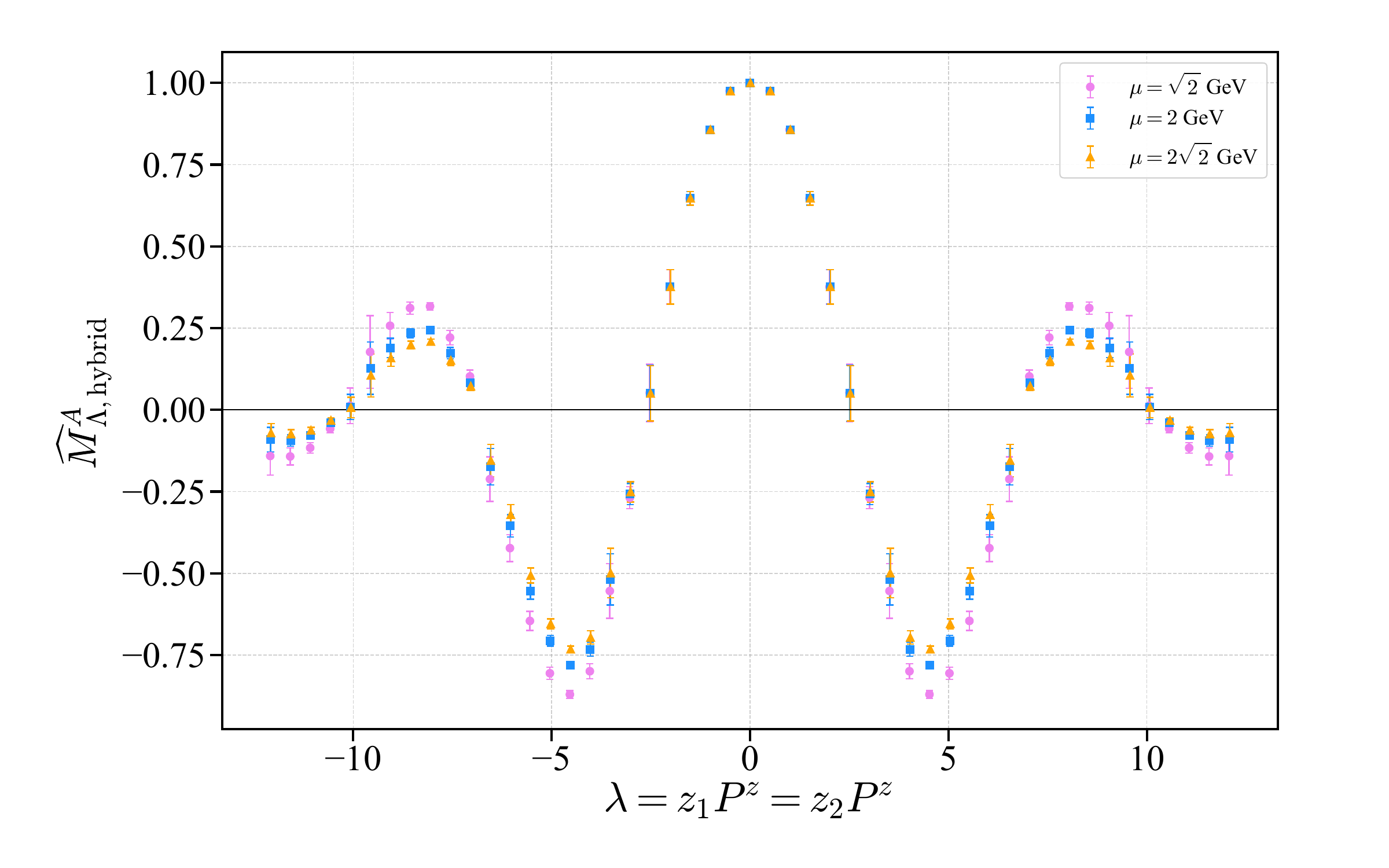}
        \label{subfig:mu_coordinate_z1-z2}
        }
    \caption{Residual renormalization-scale dependence of the hybrid-renormalized $\Lambda$ $A$ matrix element on the H48P32 ensemble at $P^z \approx 2.0~{\rm GeV}$. The three curves correspond to $\mu = \sqrt{2}$, $2$, and $2\sqrt{2}~{\rm GeV}$. Upper: fixed $z_1 = 0.40~{\rm fm}$. Lower: diagonal slice $z_1 = z_2$.}
    \label{fig:hybrid_mu_dependence}
\end{figure}

The self-renormalization parameters $m_0$ and $d$ depend on the $\overline{\rm MS}$ scale $\mu$ at finite order in perturbation theory. To quantify the residual scale dependence of the hybrid-renormalized matrix elements, Fig.~\ref{fig:hybrid_mu_dependence} compares the $\Lambda$-baryon $A$ matrix elements on the H48P32 ensemble at $P^z\approx 2.0~{\rm GeV}$, obtained with $\mu=\sqrt{2}$, $2$, and $2\sqrt{2}~{\rm GeV}$. Two representative coordinate-space slices are shown: fixed $z_1=0.40~{\rm fm}$ Fig.~\ref{subfig:mu_coordinate_z1} and the diagonal line $z_1=z_2$ Fig.~\ref{subfig:mu_coordinate_z1-z2}. 
The spread among the results from different scales is used to estimate the residual perturbative uncertainty associated with the renormalization and subsequent one-loop matching procedure.

\section{Large-\texorpdfstring{$\lambda$}{lambda} Extrapolation}\label{sec:extrapolation}

The renormalized quasi-DAs are obtained on the lattice as coordinate-space matrix elements at a finite region of spatial separations $(z_1,z_2)$, and the signal deteriorates rapidly at large spatial separations. However, the momentum-space quasi-DAs are defined through a Fourier transform of these coordinate-space distributions, requiring knowledge of the matrix elements over the whole $(z_1,z_2)$ plane. A direct Fourier transform of the truncated lattice data would therefore suffer from sizable finite-range effects and artificial oscillations in momentum-space, especially in the small-momentum-fraction regions. To reduce these artifacts, it is necessary to supplement the lattice data with a physically motivated description at the large quasi light-cone distance $\lambda = z P^z$, before performing the Fourier transform.

\subsection{Asymptotic Forms for Quasi-DAs}\label{sec:asym-ext}

Previous LaMET calculations have controlled the large-distance tail of quasi-correlators through large-$\lambda$ extrapolations. In those applications, the tail was modeled using the endpoint behavior of the corresponding light-cone correlators, such as the Regge behavior at small momentum fractions and the quark-counting behavior near the endpoints~\cite{Ji:2020brr}. These endpoint constraints imply an algebraic falloff of the coordinate-space correlators and have been used, for example, in hybrid-renormalization analyses and calculations of pion quasi-DAs~\cite{LatticeParton:2022zqc} to reduce finite-range artifacts in the Fourier transform.

In this work, instead, we adopt a different strategy based on the newly developed asymptotic large-distance expansion of Euclidean correlators~\cite{Ji:2026vir}. This approach provides a more direct coordinate-space description of quasi-distributions at finite $P^z$. Based on the heavy-quark effective theory (HQET) reduction and a dispersive analysis, the generic large-distance behavior for a single spatial separation $z$ can be written as:
\begin{equation}\label{eq:asym_formula}
\begin{aligned}
    &\ \widetilde M(z;P^z) = \sum_{\Lambda^{J^P}} \rme^{-\Lambda^{J^P}|z|}\\
    &\quad\times\left[ \left( \mathcal H_1 \rme^{-\rmi zP^z} + \mathcal H_2 \right) + \left( \mathcal H_1' \rme^{-\rmi zP^z} + \mathcal H_2' \right) \frac{1}{|z|} + \cdots \right]\ .
\end{aligned}
\end{equation}
Here the sum runs over intermediate states in the HQET description of the Wilson line, with each state characterized by a binding energy $\Lambda^{J^P}$. The superscript $J^P$ labels the spin and parity of the corresponding heavy--light intermediate state.
States with larger binding energies are therefore exponentially suppressed at large $|z|$. 
Given the present statistical precision, we retain only the lowest mesonic channels, $\Lambda^{0^-}$ and $\Lambda^{0^+}$, while baryonic channels, expected to have larger binding energies, are neglected as more strongly suppressed at large distances~\cite{Ji:2026vir}.

For each fixed intermediate state, the prefactor multiplying the exponential factor admits an asymptotic expansion in inverse powers of the large-distance separation $|z|$. The coefficients $\mathcal H_1$, $\mathcal H_2$, $\mathcal H_1'$, $\mathcal H_2'$, together with the higher-order analogous coefficients, depend on the intermediate channel and encode the corresponding overlap and matrix-element information. In constructing the large-distance extrapolation, we denote the ansatz that keeps only the leading $|z|^0$ term as the leading-asymptotic $(\mathrm{LA})$ form, while the ansatz that also includes the first subleading correction proportional to $1/|z|$ is referred to as the next-to-leading-asymptotic $(\mathrm{NLA})$ form.

For baryon quasi-DAs, the coordinate-space matrix elements depend on two independent spatial separations. The relevant large-distance expansion variables are therefore $|z_1|$, $|z_2|$, and $|z_1-z_2|$. Consequently, the large-distance behavior cannot be described by a single universal asymptotic form over the full $(z_1,z_2)$ plane. 

\begin{figure}[htbp]
    \centering
    \includegraphics[width=\linewidth]{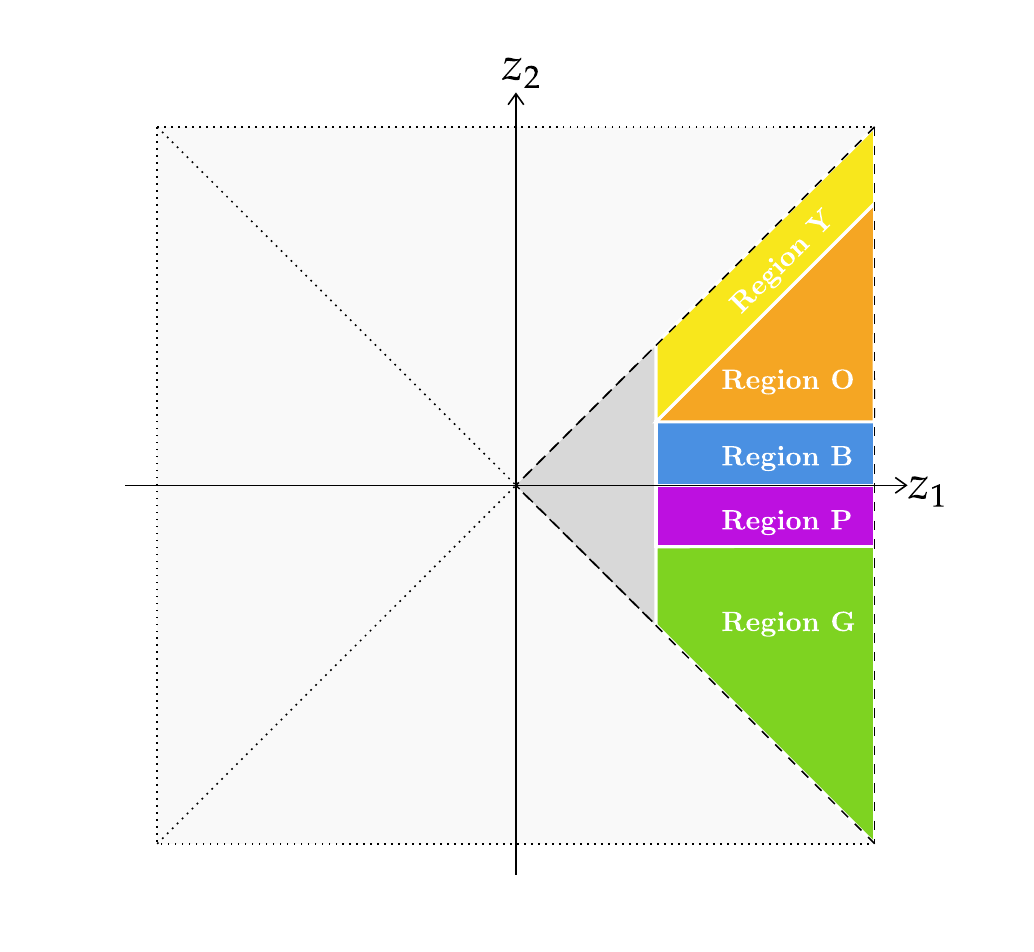}
    \caption{The region partition for large-$\lambda$ extrapolation of quasi-DAs in the independent quarter region $z_1>|z_2|$, bounded by $z_1=\pm z_2$. In the different coloed regions we derived corresponding asymptotic ans\"atze.}
    \label{fig:asymptotic_region_division}
\end{figure}

We therefore, after considering the $z_1 \leftrightarrow z_2$ exchange symmetry and Hermiticity relations discussed in Sec.~\ref{sec:symmetries}, partition the independent quarter region $z_1>|z_2|$ that bounded by $z_1=\pm z_2$, according to which of those separations becomes large, and use the corresponding asymptotic ansatz in each sub-region. This choice avoids fitting all symmetry-related regions independently, while retaining the information needed to reconstruct the full $(z_1,z_2)$ plane. As shown in Fig.~\ref{fig:asymptotic_region_division}, the triangular region is divided into six sub-regions. In each region, except the short-distance gray region, the appropriate asymptotic forms of baryon quasi-DAs are derived according to the large-distance expansion of $|z_1|$, $|z_2|$, and $|z_1-z_2|$. 

Because the $A$ amplitude and the $V/T$ amplitudes have different exchange symmetries and different large-distance behaviors, we use different extrapolation ans\"atze for these different amplitudes, as described below.

\subsubsection{Extrapolation Ans\"atze for the \texorpdfstring{$\Lambda$}{Lambda} \texorpdfstring{$A$}{A} Amplitude}

For the symmetric $A$ amplitude of $\Lambda$, the large-distance signal is sufficiently precise to support stable fits with the $\mathrm{NLA}$ ansatz. In the central analysis we use $\mathrm{NLA}$ forms constructed from the lowest mesonic binding energy $\Lambda^{0^-}$. The explicit $\mathrm{NLA}$ expressions in the different regions are given by:

\begin{itemize}

    \item \textbf{Region G}: $z_1z_2<0$, $|z_1|\to\infty$, $|z_2|\to\infty$ and $|z_1-z_2|\to\infty$,
    \begin{equation}
    \begin{aligned}
        &\ \widetilde A_\Lambda^{\rm NLA}(z_1,z_2;P^z) \\
        =&\ \rme^{-\Lambda^{0^-}|z_1|}
            \rme^{-\Lambda^{0^-}|z_2|}
        \left[
        \begin{aligned}
            &\mathcal G_A(\rmi\hat z_1,\rmi\hat z_2,P^z)\\
            +&\frac{
            \mathcal G_A^{(1)}(\rmi\hat z_1,\rmi\hat z_2,P^z)
            }{|z_1|}\\
            +&\frac{
            \mathcal G_A^{(2)}(\rmi\hat z_1,\rmi\hat z_2,P^z)
            }{|z_2|}
        \end{aligned}
        \right]\ .
    \end{aligned}
    \end{equation}
    
    \item \textbf{Region P}: $z_1z_2<0$, $|z_1|\to\infty$, $|z_1-z_2|\to\infty$ while $|z_2|$ keeps finite,
    \begin{equation}
    \begin{aligned}
        &\ \widetilde A_\Lambda^{\rm NLA}(z_1,z_2;P^z) \\
        = &\ \rme^{-\Lambda^{0^-}|z_2|}
             \rme^{\rmi z_1P^z}
        \left[
        \begin{aligned}
            &\mathcal P_{A,1}(\rmi z_1,\rmi \hat z_2,P^z)\\
            +&\frac{
            \mathcal P_{A,1}^{(1)}(\rmi z_1,\rmi \hat z_2,P^z)
            }{|z_1|}
        \end{aligned}
        \right] \\
        & + \rme^{-\Lambda^{0^-}|z_2|}
        \left[
        \begin{aligned}
            &\mathcal P_{A,2}(\rmi z_1,\rmi \hat z_2,P^z)\\
            +&\frac{
            \mathcal P_{A,2}^{(1)}(\rmi z_1,\rmi \hat z_2,P^z)
            }{|z_1|}
        \end{aligned}
        \right]\ .
    \end{aligned}
    \end{equation}
    
    \item \textbf{Region B}: $z_1z_2>0$, $|z_1|\to\infty$, $|z_1-z_2|\to\infty$ while $|z_2|$ keeps finite,
    \begin{equation}
    \begin{aligned}
        &\ \widetilde A_\Lambda^{\rm NLA}(z_1,z_2;P^z) \\
        = &\ \rme^{-\Lambda^{0^-}|z_2-z_1|}
             \rme^{\rmi z_1P^z}
        \left[
        \begin{aligned}
            &\mathcal B_{A,1}(-\rmi z_1,\rmi \hat z_2,P^z)\\
            +&\frac{
            \mathcal B_{A,1}^{(1)}(-\rmi z_1,\rmi \hat z_2,P^z)
            }{|z_1|}
        \end{aligned}
        \right] \\
        & + \rme^{-\Lambda^{0^-}|z_2-z_1|}
        \left[
        \begin{aligned}
            &\mathcal B_{A,2}(-\rmi z_1,\rmi \hat z_2,P^z)\\
            +&\frac{
            \mathcal B_{A,2}^{(1)}(-\rmi z_1,\rmi \hat z_2,P^z)
            }{|z_1|}
        \end{aligned}
        \right]\ .
    \end{aligned}
    \end{equation}
    
    \item \textbf{Region O}: $z_1z_2>0$, $|z_1|\to\infty$, $|z_2|\to\infty$ and $|z_1-z_2|\to\infty$,
    \begin{equation}
    \begin{aligned}
        &\ \widetilde A_\Lambda^{\rm NLA}(z_1,z_2;P^z) \\
        =&\ \rme^{-\Lambda^{0^-}|z_1|}
            \rme^{-\Lambda^{0^-}|z_2-z_1|}
            \rme^{\rmi z_1P^z}
        \left[
        \begin{aligned}
            & \mathcal O_A(\rmi \hat z_1,\rmi \hat z_2,P^z)\\
            +&\frac{
            \mathcal O_A^{(1)}(\rmi \hat z_1,\rmi \hat z_2,P^z)
            }{|z_1|}\\
            +&\frac{
            \mathcal O_A^{(12)}(\rmi \hat z_1,\rmi \hat z_2,P^z)
            }{|z_1-z_2|}
        \end{aligned}
        \right]\ .
    \end{aligned}
    \end{equation}
    
    \item \textbf{Region Y}: $z_1z_2>0$, $|z_1|\to\infty$, $|z_2|\to\infty$ while $|z_1-z_2|$ keeps finite,
    \begin{equation}
    \begin{aligned}
        &\ \widetilde A_\Lambda^{\rm NLA}(z_1,z_2;P^z) \\
        = &\ \rme^{-\Lambda^{0^-}|z_1|}
             \rme^{\rmi z_2P^z}
        \left[
        \begin{aligned}
            & \mathcal Y_{A,1}(\rmi \hat z_1,\rmi (z_2-z_1),P^z)\\
            +& \frac{
            \mathcal Y_{A,1}^{(2)}(\rmi \hat z_1,\rmi (z_2-z_1),P^z)
            }{|z_2|}
        \end{aligned}
        \right] \\
        & + \rme^{-\Lambda^{0^-}|z_1|}
            \rme^{\rmi z_1P^z}
        \left[
        \begin{aligned}
            & \mathcal Y_{A,2}(\rmi \hat z_1,\rmi (z_2-z_1),P^z)\\
            +& \frac{
            \mathcal Y_{A,2}^{(2)}(\rmi \hat z_1,\rmi (z_2-z_1),P^z)
            }{|z_2|}
        \end{aligned}
        \right]\ .
    \end{aligned}
    \end{equation}

\end{itemize}
Here $\hat z_i \equiv z_i/|z_i|$ denotes the sign of the corresponding coordinate separation and is used to distinguish different asymptotic directions. The functions $\mathcal G_A$, $\mathcal P_{A,i}$, $\mathcal B_{A,i}$, $\mathcal O_A$, and $\mathcal Y_{A,i}$, denote the coefficients in the asymptotic expansion and are fitted from the lattice data in the corresponding regions.

We take the $\mathrm{NLA}$ result as the central value for the $\Lambda$ $A$ amplitude. To estimate the uncertainty associated with neglected higher-asymptotic terms in the large-distance expansion, we repeat the analysis with the corresponding $\mathrm{LA}$ ans\"atze, constructed with the same mesonic binding-energy content. The difference between the $\mathrm{NLA}$ and $\mathrm{LA}$ results is included as the systematic uncertainty from the large-$\lambda$ extrapolation.

\subsubsection{Extrapolation Ans\"atze for the \texorpdfstring{$\Lambda$}{Lambda} \texorpdfstring{$V$}{V} and \texorpdfstring{$T$}{T} Amplitudes}

For the $V$ and $T$ amplitudes, the large-distance signals are less precise than in the $A$ amplitude. Introducing the additional parameters of the $\mathrm{NLA}$ ansatz would not be sufficiently constrained by the current data and can lead to unstable fits. We therefore use the $\mathrm{LA}$ ansatz for the $V$ and $T$ amplitudes, given by:
\begin{itemize}
    
    \item \textbf{Region G}: $z_1z_2<0$, $|z_1|\to\infty$, $|z_2|\to\infty$ and $|z_1-z_2|\to\infty$,
    \begin{equation}
    \begin{aligned}
        &\ \widetilde V_\Lambda^{\rm LA}/\widetilde T_\Lambda^{\rm LA}(z_1,z_2;P^z) \\
        =&\ \rme^{-\Lambda^{0^-}|z_1|} \rme^{-\Lambda^{0^+}|z_2|} \mathcal G_{V/T}(\rmi \hat z_1,\rmi \hat z_2,P^z)\\
        & - \rme^{-\Lambda^{0^+}|z_1|} \rme^{-\Lambda^{0^-}|z_2|} \mathcal G_{V/T}^*(\rmi \hat z_1,\rmi \hat z_2,P^z) \ ;
    \end{aligned}
    \end{equation}

    \item \textbf{Region P}: $z_1z_2<0$, $|z_1|\to\infty$, $|z_1-z_2|\to\infty$ while $|z_2|$ keep finite,
    \begin{equation}
    \begin{aligned}
        &\ \widetilde V_\Lambda^{\rm LA}/\widetilde T_\Lambda^{\rm LA}(z_1,z_2;P^z) \\
        = &\ \rme^{-\Lambda^{0^-}|z_2|} \rme^{\rmi z_1P^z} \mathcal P_{V/T,1}(\rmi z_1,\rmi \hat z_2,P^z)\\
        & + \rme^{-\Lambda^{0^-}|z_2|} \mathcal P_{V/T,2}(\rmi z_1,\rmi \hat z_2,P^z)\ ;
    \end{aligned}
    \end{equation}
    
    \item \textbf{Region B}: $z_1z_2>0$, $|z_1|\to\infty$, $|z_1-z_2|\to\infty$ while  $|z_2|$ keep finite,
    \begin{equation}
    \begin{aligned}
        &\ \widetilde V_\Lambda^{\rm LA}/\widetilde T_\Lambda^{\rm LA}(z_1,z_2;P^z) \\
        = &\ \rme^{-\Lambda^{0^-}|z_2-z_1|} \rme^{\rmi z_1P^z} \mathcal B_{V/T,1}(-\rmi z_1,\rmi \hat z_2,P^z)\\
        & + \rme^{-\Lambda^{0^-}|z_2-z_1|} \mathcal B_{V/T,2}(-\rmi z_1,\rmi \hat z_2,P^z)\ ;
    \end{aligned}
    \end{equation}

    \item \textbf{Region O}: $z_1z_2>0$, $|z_1|\to\infty$, $|z_2|\to\infty$ and $|z_1-z_2|\to\infty$,
    \begin{equation}
    \begin{aligned}
        &\ \widetilde V_\Lambda^{\rm LA}/\widetilde T_\Lambda^{\rm LA}(z_1,z_2;P^z) \\
        = &\ \rme^{-\Lambda^{0^-}|z_1|} \rme^{-\Lambda^{0^-}|z_2-z_1|} \rme^{\rmi z_1P^z} \mathcal O_{V/T}(\rmi \hat z_1,\rmi \hat z_2,P^z)\ ;
    \end{aligned}
    \end{equation}

    \item \textbf{Region Y}: $z_1z_2>0$, $|z_1|\to\infty$, $|z_2|\to\infty$ while $|z_1-z_2|$ keep finite,
    \begin{equation}
    \begin{aligned}
        &\ \widetilde V_\Lambda^{\rm LA}/\widetilde T_\Lambda^{\rm LA}(z_1,z_2;P^z) \\
        = &\ \rme^{-\Lambda^{0^-}|z_1|} \rme^{\rmi z_2P^z} \mathcal Y_{V/T,1}(\rmi \hat z_1,\rmi (z_2-z_1),P^z)\\
        & + \rme^{-\Lambda^{0^-}|z_1|} \rme^{\rmi z_1P^z} \mathcal Y_{V/T,2}(\rmi \hat z_1,\rmi (z_2-z_1),P^z)\ .
    \end{aligned}
    \end{equation}
\end{itemize}

In both $A$ and $V/T$ cases, the gray region is short-distance-like for all three coordinate separations, so the original lattice data are always used directly. The asymptotic behavior in other sectors can be inferred from the symmetry properties of the coordinate-space quasi-DAs given in Sec.~\ref{sec:symmetries}.

\subsection{Implementation of Large-\texorpdfstring{$\lambda$}{lambda} Extrapolation }

\begin{figure}[htbp]
    \centering
    \includegraphics[width=\linewidth]{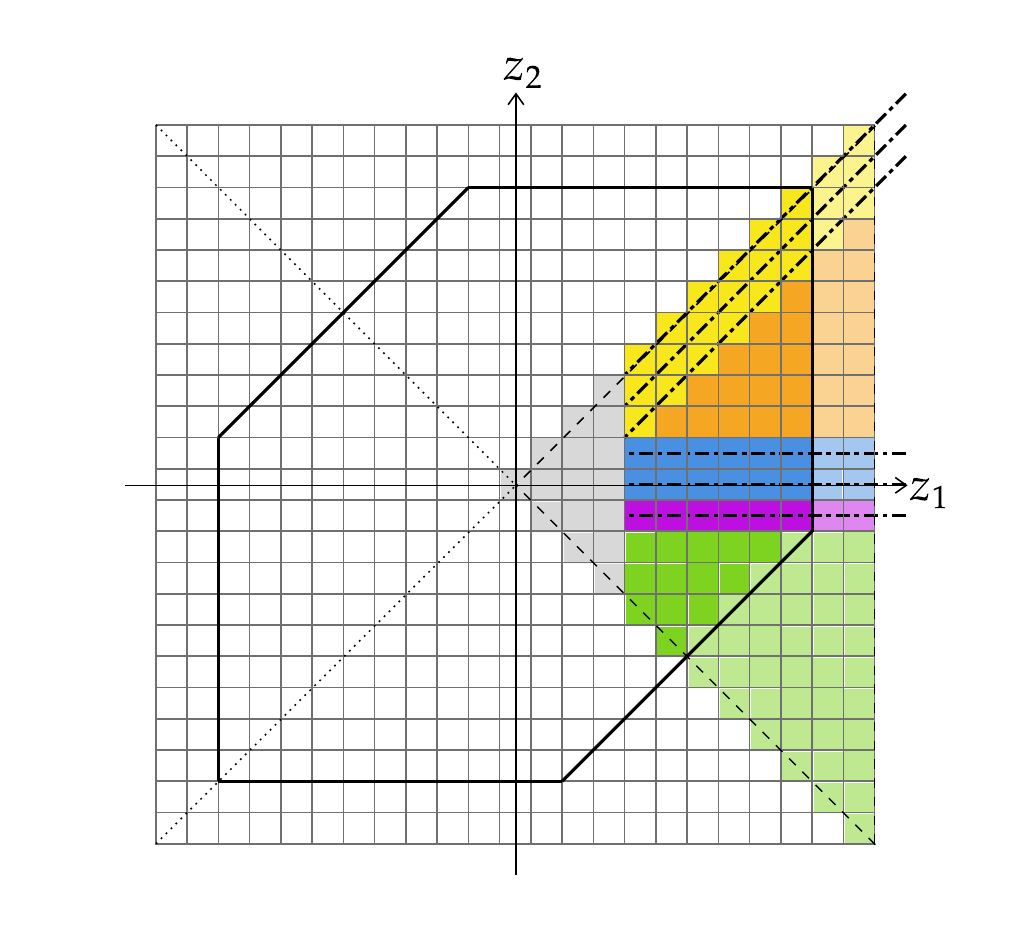}
    \caption{
    Practical region partition for the numerical implementation of the large-$\lambda$ extrapolation. The gray region denotes the retained original data, and the colored regions are fitted with the corresponding ans"atze. Only the fully colored points inside the black hexagon are included in the fits. The black dashed lines indicate line-by-line fits at fixed short $|z_2|$ or $|z_1-z_2|$.
    }
    \label{fig:extrapolation_region_division}
\end{figure}

We now apply the large-$\lambda$ extrapolation to the hybrid-renormalized coordinate-space quasi-DA matrix elements. Following the region partition used to derive the asymptotic forms in the previous subsection, the practical partition of the lattice data used in the numerical analysis is shown in Fig.~\ref{fig:extrapolation_region_division}. The gray central region denotes the coordinate range where the original lattice data are retained directly. The colored regions are extrapolated using the corresponding asymptotic ans"atze described above: the yellow region lies close to the diagonal $|z_1-z_2|\simeq 0$, the blue and purple regions are close to the boundary $|z_2|\simeq 0$, and the orange and green regions correspond to large-distance sectors away from the short-distance boundaries.

Since the signal quality of the nonlocal baryon matrix elements deteriorates at large effective Wilson-line length $\tilde z$ defined in Eq.~\eqref{eq:eff_length}, we restrict the fits to the fully colored points inside the black hexagon in Fig.~\ref{fig:extrapolation_region_division}, where the coordinate-space matrix elements are better constrained. The selected fit region nevertheless extends beyond $1~{\rm fm}$ along the $z_1$ direction and therefore retains direct sensitivity to the large-distance behavior.
The faded points outside the hexagon are shown only for reference and are excluded from the fitting procedure.

The fitting procedure is carried out in two stages. In the first stage, the orange and green regions, where all relevant separations are in the large-distance regime, are fitted simultaneously to determine the common asymptotic binding energies $\Lambda^{J^P}$. In the second stage, using the large-distance parameters constrained in the first stage, the yellow, blue, and purple regions are fitted line by line along the black dashed lines shown in Fig.~\ref{fig:extrapolation_region_division}. These fits are performed at different fixed short distances, such as $|z_2|$ or $|z_1-z_2|$, in order to determine the coefficient functions associated with the remaining finite-distance dependence. This two-stage strategy uses the genuinely large-distance regions to constrain the universal asymptotic behavior, while allowing the boundary regions to retain their dependence on the short coordinate separations.

\begin{figure*}[tbp]
    \centering
    \subfloat[$z_1-z_2=0$, $z_1+z_2\geq 0$]{
        \centering
        \includegraphics[width=0.4\textwidth]{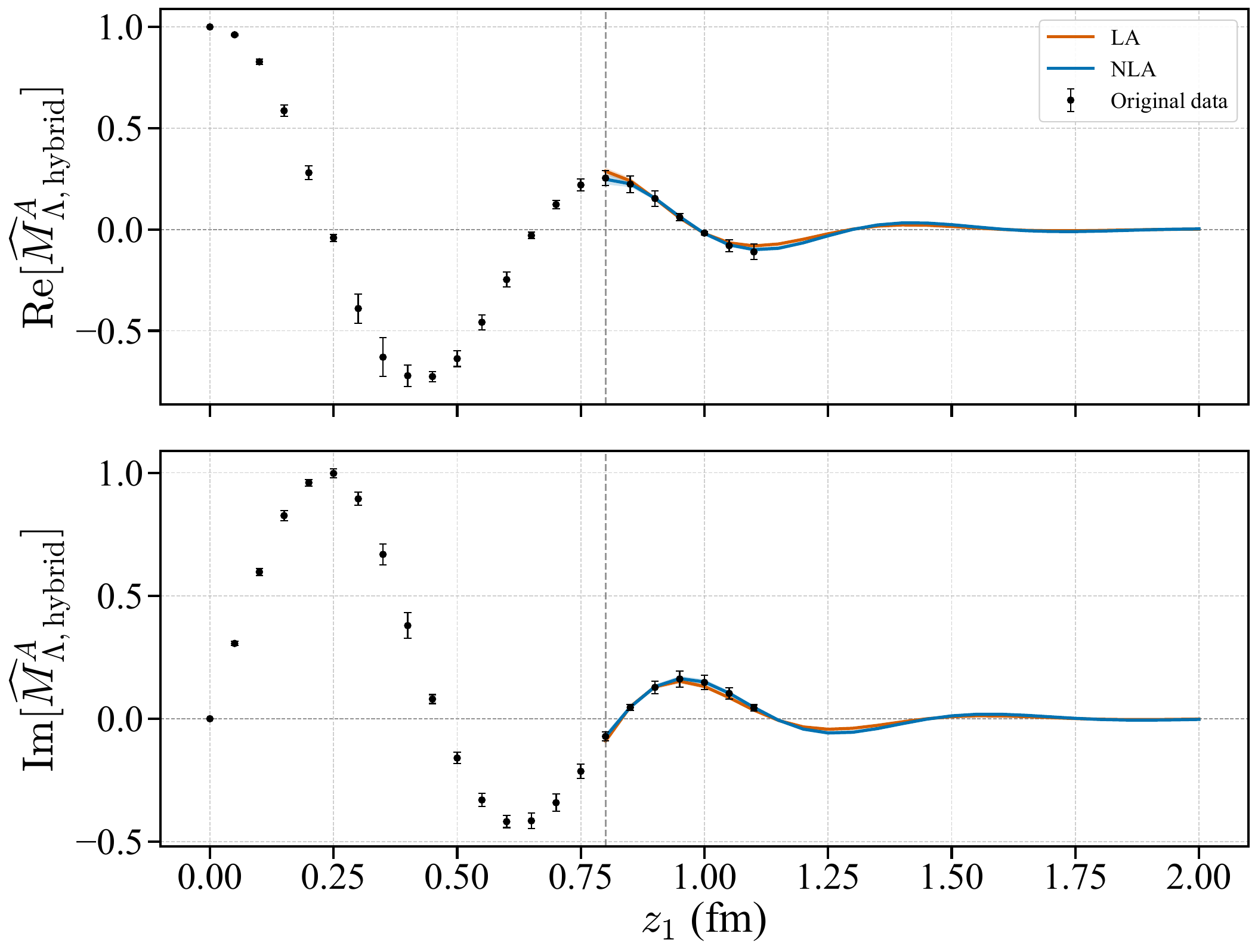}
        }\hspace{0.05\textwidth}
    \subfloat[$z_2=1$ grid unit, $z_1\geq 0$]{
        \centering
        \includegraphics[width=0.4\textwidth]{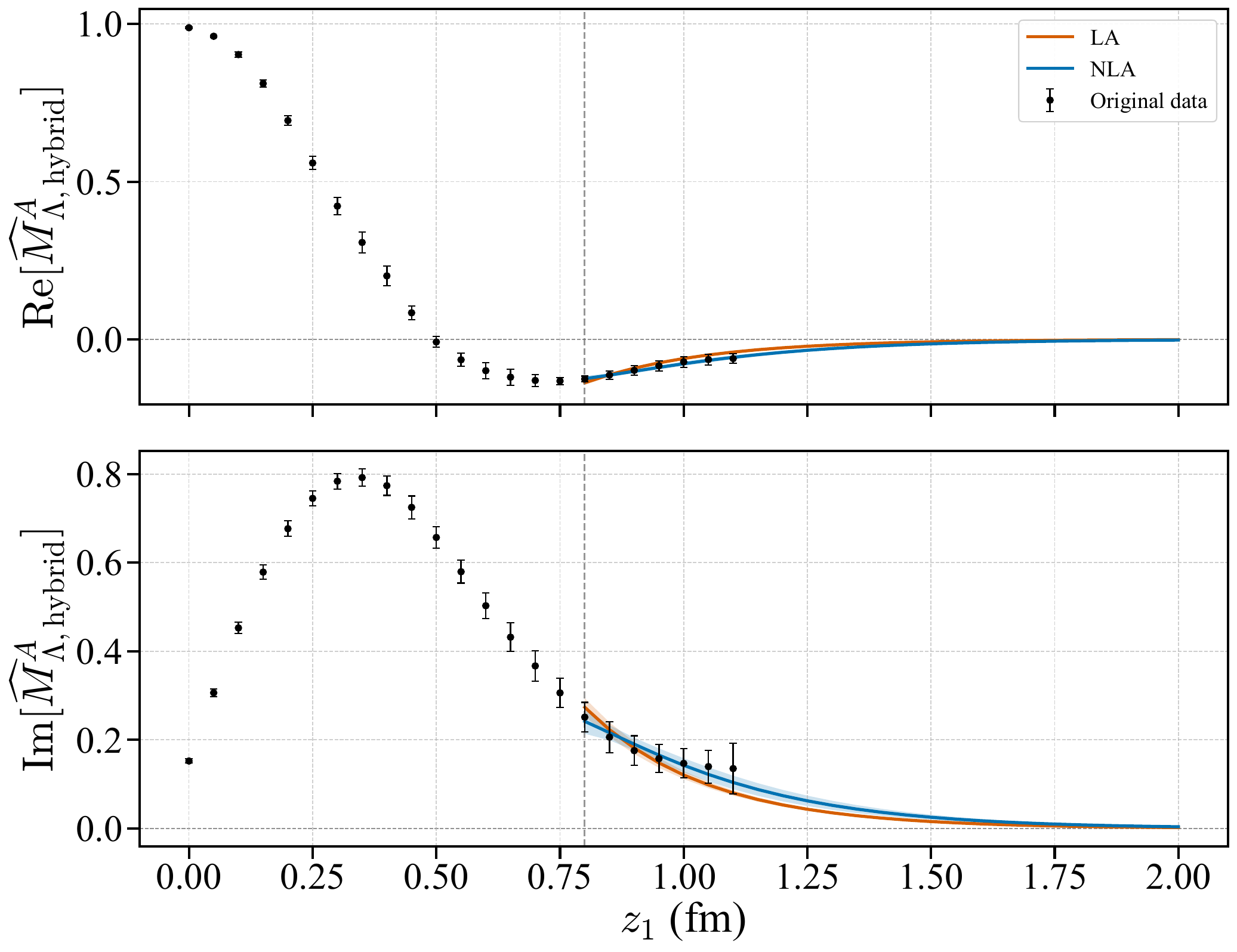}
        }
    \caption{Representative line-cut extrapolation results for the $\Lambda$ $A$ quasi-DA matrix element on the F32P30 ensemble at momentum $P^z\approx2.0~{\rm GeV}$. The two panels compare the $\mathrm{LA}$ and $\mathrm{NLA}$ extrapolation ans\"atze on different coordinate-space cuts. The black points show the retained hybrid-renormalized data, and the colored bands show the extrapolated results with statistical uncertainties.}
    \label{fig:extrapolation_linecuts_A}
\end{figure*}

For the $A$, $V$, and $T$ amplitudes, we employ the same two-stage fitting strategy described above. The mesonic binding-energy information is transferred among the three structures in a controlled way, as justified by the common mesonic intermediate states in the HQET description of the Wilson line. In practice, the $A$ amplitude has the best signal quality and is first used to determine the negative-parity mesonic binding energy $\Lambda^{0^-}$. In the $V$-amplitude fit, this value is used as a Bayesian prior, while the positive-parity mesonic binding energy $\Lambda^{0^+}$ is determined from the $V$ data. Since the $T$ amplitude has an even weaker large-distance signal, the mesonic binding energies obtained from the $A$ and $V$ fits are used as Bayesian priors in the $T$ fit. Apart from these shared binding-energy inputs, all coefficient functions entering the asymptotic forms are fitted independently for the $A$, $V$, and $T$ amplitudes.

For the $A$ amplitude, we compare the leading-asymptotic ansatz $(\mathrm{LA})$ with the next-to-leading-asymptotic ansatz $(\mathrm{NLA})$. For the $V$ and $T$ amplitudes, where the present data do not support stable $\mathrm{NLA}$ fits, we use the $\mathrm{LA}$ ansatz for the representative results.

Fig.~\ref{fig:extrapolation_linecuts_A} shows two representative line cuts for the $\Lambda$ $A$-amplitude matrix element on the F32P30 ensemble at momentum $P^z\approx2.0~{\rm GeV}$. The black scatters are the retained hybrid-renormalized lattice data, while the colored bands show the extrapolated results. The $\mathrm{LA}$ and $\mathrm{NLA}$ ans\"atze give compatible descriptions on these cuts within the displayed uncertainties, and the extrapolated bands join smoothly onto the retained coordinate-space data.

\begin{figure*}[tbp]
    \centering
    \subfloat[$\Lambda$ $V$ quasi-DA, $z_1-z_2=3$ grid units]{
        \centering
        \includegraphics[width=0.4\textwidth]{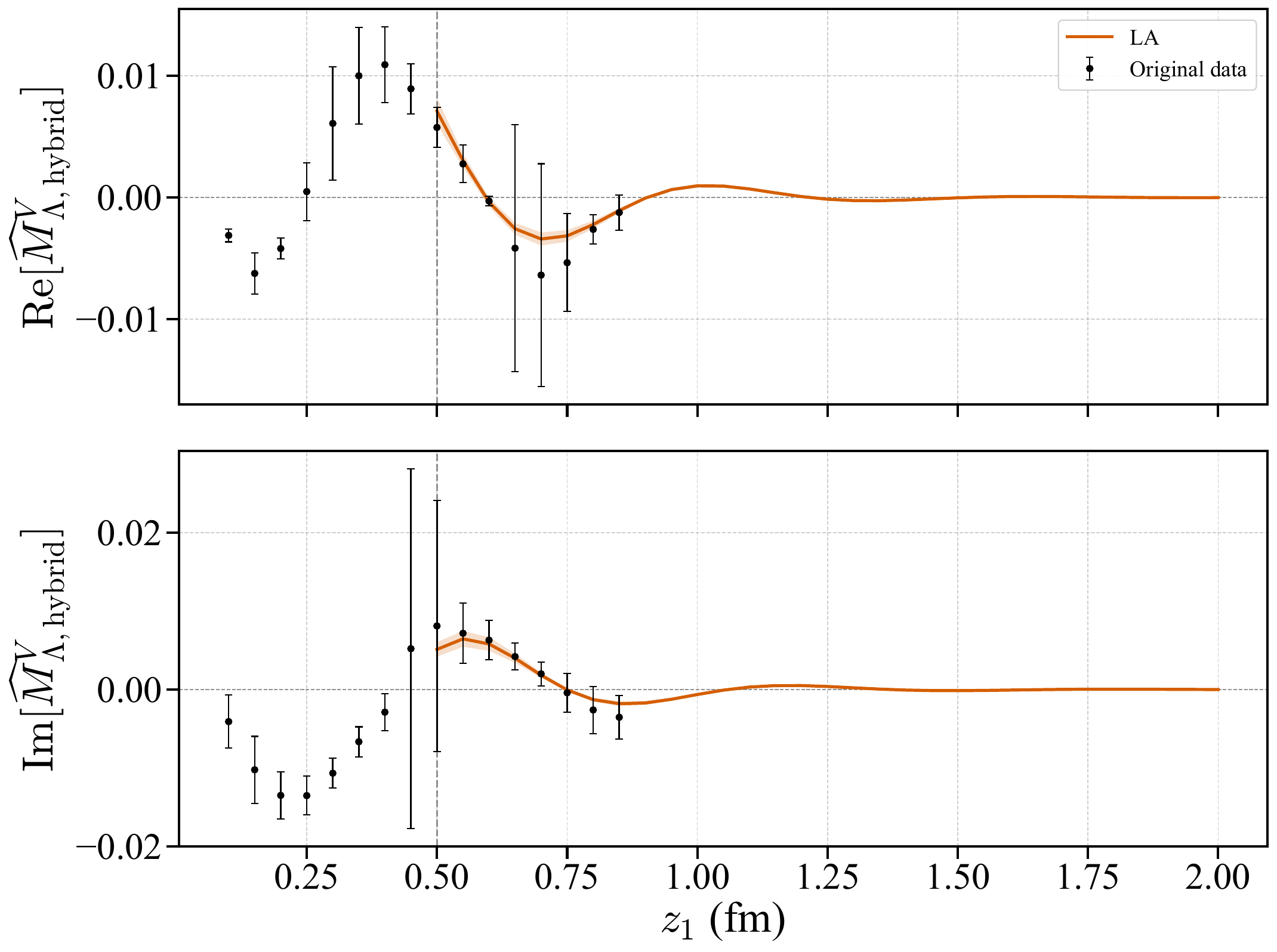}
        }\hspace{0.05\textwidth}
    \subfloat[$\Lambda$ $T$ quasi-DA, $z_1-z_2=3$ grid units]{
        \centering
        \includegraphics[width=0.4\textwidth]{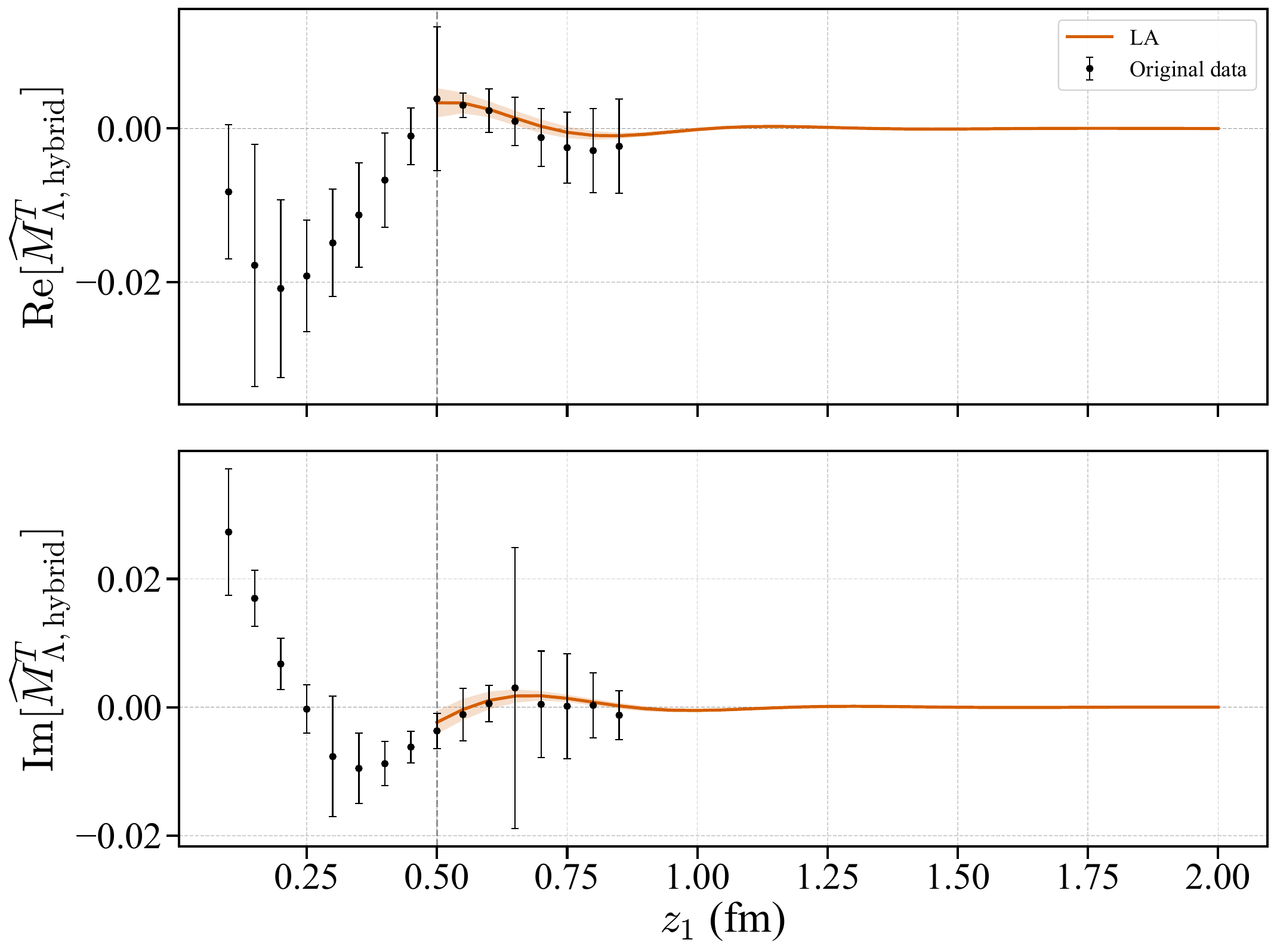}
        }
    \caption{Representative line-cut extrapolation results for the $\Lambda$ $V$ and $T$ quasi-DA matrix elements on the F32P30 ensemble at momentum $P^z \approx 2.0~{\rm GeV}$, using the \rm{LA} ansatz. The black points show the retained hybrid-renormalized data, and the bands show the large-distance extrapolation with statistical uncertainties.}
    \label{fig:extrapolation_linecuts_VT}
\end{figure*}

The same procedure is performed for the $V$ and $T$ amplitudes. Representative line cuts are shown in Fig.~\ref{fig:extrapolation_linecuts_VT}. In analyses of these two structures, the $\mathrm{LA}$ extrapolation is used as the central ansatz. The fitted curves describe the retained data on the selected cuts within uncertainties and provide a smooth continuation into the large-distance coordinate region.

\begin{figure*}[htbp]
    \centering
    \includegraphics[width=\textwidth]{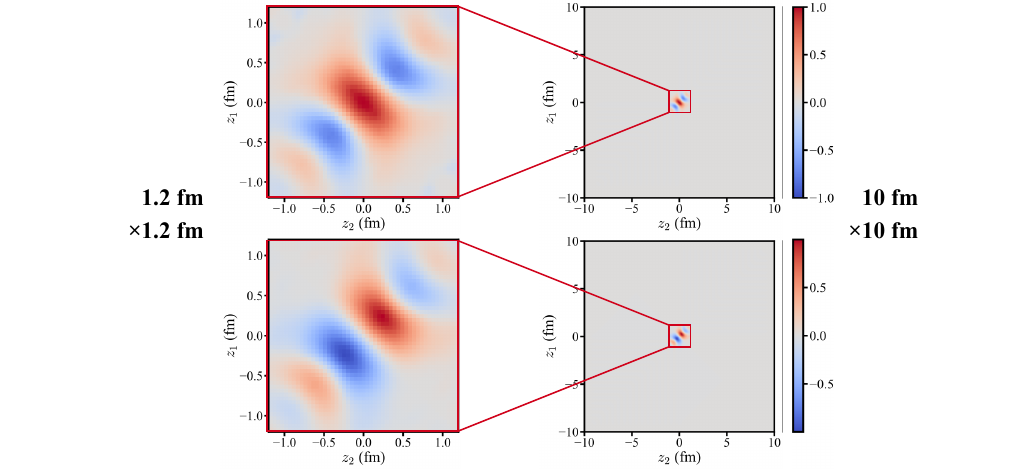}
    \caption{Overview of the original and extrapolated coordinate-space $\Lambda$ $A$ amplitude matrix element on the F32P30 ensemble at momentum $P^z \approx 2.0~{\rm GeV}$. The left column shows the original hybrid-renormalized data and the right column shows the extrapolated result; the upper and lower panels correspond to the real and imaginary parts.}
    \label{fig:extrapolation_heatmap_overview}
\end{figure*}

Finally, Fig.~\ref{fig:extrapolation_heatmap_overview} gives a two-dimensional overview of the original and extrapolated coordinate-space matrix elements for the $\Lambda$ $A$ amplitude. The left-side panels show the original hybrid-renormalized data, while the right-side panels show the result after filling the enlarged coordinate-space range needed for the subsequent Fourier transform, through the large-$\lambda$ extrapolation. The top and bottom panels display the real and imaginary parts, respectively. The comparison illustrates how the finite original data coverage is extended to a larger coordinate-space domain while maintaining a smooth continuation from the region constrained by the lattice data.

\section{LaMET Matching under Hybrid Scheme}\label{sec:Matching}
After the hybrid renormalization and the large-$\lambda$ extrapolation, the coordinate-space quasi-DAs are Fourier transformed to momentum space. The resulting quasi-DAs are still defined at finite hadron momentum $P^z$ and in the hybrid renormalization scheme. To obtain the physical LCDAs in the $\overline{\rm MS}$ scheme, we apply the LaMET factorization formula in Eq.~\eqref{eq:LaMET}. At next-to-leading order (NLO), the matching relation reads:
\begin{equation}\label{eq:matching_ori}
\begin{aligned}
    &\ \phi_B^X(x_1,x_2;\mu) \\
    =&\ \int \rmd y_1\rmd y_2\ \mathcal C^X(x_1,x_2;y_1,y_2;P^z,\mu)\ \widetilde \phi_B(y_1,y_2;P^z,\mu)\\
    &\ + \mathcal O\left( \frac{\Lambda_{\rm QCD}^2}{(x_1P^z)^2}, \frac{\Lambda_{\rm QCD}^2}{(x_2P^z)^2}, \frac{\Lambda_{\rm QCD}^2}{[(1-x_1-x_2)P^z]^2} \right)\\
    =&\ \widetilde \phi_B^X(x_1,x_2;P^z,\mu) \\
    &\ + \frac{\alpha_sC_F}{2\pi}\int \rmd y_1\rmd y_2\ \mathcal C^{(1),X}(x_1,x_2;y_1,y_2;P^z,\mu)\ \\ 
    &\ \qquad\qquad\times \widetilde \phi_B^X(y_1,y_2;P^z,\mu)\\
    &\ + \mathcal O\left( \frac{\Lambda_{\rm QCD}^2}{(x_1P^z)^2}, \frac{\Lambda_{\rm QCD}^2}{(x_2P^z)^2}, \frac{\Lambda_{\rm QCD}^2}{[(1-x_1-x_2)P^z]^2} \right)\ .
\end{aligned}
\end{equation}
Here $X=V,A,T$. The hybrid-renormalized quasi-DAs at finite hadron momentum $P^z$ are denoted by $\widetilde \phi_B^X(y_1,y_2;P^z,\mu)$, where the renormalization scale $\mu$ is chosen to coincide with the factorization scale. The matched LCDAs in the $\overline{\rm MS}$ scheme are denoted by $\phi_B^X(x_1,x_2;\mu)$. The power corrections are suppressed at large hadron momentum $P^z$, but can be enhanced in the endpoint regions where one of the three parton momentum fractions, $x_1$, $x_2$, or $x_3=1-x_1-x_2$, becomes small.

Because the quasi-DAs entering Eq.~\eqref{eq:matching_ori} are renormalized in the hybrid scheme, the one-loop matching kernel differs from the standard $\overline{\rm MS}$ kernel by hybrid counterterms. The current matching kernel can be written as:
\begin{equation}
\begin{aligned}\label{eq:matching_kernel}
    &\ \mathcal C^{(1),X}(x_1,x_2;y_1,y_2;P^z,\mu)\\
    =&\ \Big[ \mathcal C^{(1),X}_{\overline{\rm MS}}(x_1,x_2;y_1,y_2;P^z,\mu)\\
    &\  + \delta \mathcal C^{(1),X}_{\rm H}(x_1,x_2;y_1,y_2;P^z,\mu) \Big]_\oplus\ .
\end{aligned}
\end{equation}
The symbol $\oplus$ denotes the double-plus distribution for the two momentum fractions, defined as:
\begin{equation}\label{eq:double_plus}
\begin{aligned}
    \big[g(&x_1,x_2;y_1,y_2)\big]_{\oplus} \equiv g(x_1,x_2;y_1,y_2) \\
    &\ -\delta(x_1-y_1)\delta(x_2-y_2)
    \int \rmd t_1 \rmd t_2\ g(t_1,t_2;y_1,y_2)\ .
\end{aligned}
\end{equation}

\subsection{Matching Kernel in \texorpdfstring{$\overline{\rm MS}$}{MSbar} Scheme}\label{sec:MSbar_kernel}

The $\mathcal C^{(1),X}_{\overline{\rm MS}}$ are the standard one-loop LaMET matching kernels in the $\overline{\rm MS}$ scheme. For the $V$, $A$, and $T$ amplitudes, their explicit forms are given by~\cite{Han:2023xbl}:
\begin{equation}\label{eq:MSbar_kernel_VA}
\begin{aligned}
    &\ \mathcal C^{(1), V/A}_{\overline{\rm MS}}(x_1,x_2;y_1,y_2;P^z,\mu)\\
    =&\ -\frac14\Bigg[\mathcal C_2(x_1,x_2;y_1,y_2;P^z,\mu)\delta (x_2-y_2)\\
    &\ +\mathcal C_3(x_1,x_2;y_1,y_2;P^z,\mu)\delta (x_3-y_3)\\
    &\ +\{x_1\leftrightarrow x_2,y_1\leftrightarrow y_2\}\Bigg]_\oplus\ ,
\end{aligned}
\end{equation}
\begin{equation}\label{eq:MSbar_kernel_T}
\begin{aligned}
    &\ \mathcal C^{(1), T}_{\overline{\rm MS}}(x_1,x_2;y_1,y_2;P^z,\mu)\\
    =&\ -\frac14\Bigg[\mathcal C_2(x_1,x_2;y_1,y_2;P^z,\mu)\delta (x_2-y_2)\\
    &\ +(\mathcal C_3-\mathcal C_5)(x_1,x_2;y_1,y_2;P^z,\mu)\delta (x_3-y_3)\\
    &\ +\{x_1\leftrightarrow x_2,y_1\leftrightarrow y_2\}\Bigg]_\oplus\ ,
\end{aligned}
\end{equation}
where $x_3=1-x_1-x_2$ and $y_3=1-y_1-y_2$. The complete expressions for $\mathcal C_2$, $\mathcal C_3$, and $\mathcal C_5$ are collected in Appendix~\ref{app:kernel_MSbar}.

A subtlety arises when the $\overline{\rm MS}$ matching kernel is written directly as a momentum-space double-plus distribution. In coordinate space, the one-loop matrix elements contain short-distance logarithms of the form $\ln(\mu^2 z^2)$, as in Eqs.~\eqref{eq:pert_res_VA}--\eqref{eq:pert_res_T}. Upon Fourier transformation, these logarithms generate
slowly decaying ultraviolet tails in the quasi momentum-fraction variables, through the double-plus prescription Eq.~\eqref{eq:double_plus}.

This issue is particularly relevant for baryon quasi-DAs because the matching convolution is genuinely two-dimensional. As discussed in Sec.~\ref{sec:self_formalism}, The short-distance logarithms $\ln(\mu^2 z_1^2)$, $\ln(\mu^2 z_2^2)$, and $\ln[\mu^2(z_1-z_2)^2]$ are associated with three coordinate separations $z_1$, $z_2$, and $z_1-z_2$, respectively. These logarithms are mapped into ultraviole tails in the quasi momentum-fraction variables $y_1$ and $y_2$. Without an additional subtraction, the virtual contribution becomes logarithmically sensitive to the large momentum fractions, and the momentum-space kernel does not define a finite distribution.

This provides an additional matching-level consequence of the hybrid scheme beyond the perspective of renormalization introduced in Sec.~\ref{sec:Hybrid}. In the short-distance region, the ratio-type subtraction cancels the ultraviolet logarithms associated with the limits $z_1\to 0$, $z_2\to 0$, and $(z_1-z_2)\to 0$. The corresponding perturbative subtraction appears as hybrid counterterms in the matching kernel. After this subtraction, the virtual contribution in the double-plus distribution becomes finite, and the matching kernel defines a well-behaved distribution that can be convoluted with the quasi-DAs.

\subsection{Counterterms in the Hybrid Scheme Matching}\label{sec:hybrid_kernel}

As discussed in the previous subsection, the short-distance logarithms in the $\overline{\rm MS}$ matching kernel lead to logarithmic ultraviolet tails in momentum-fraction space. In the hybrid scheme, these logarithms are removed by the ratio-type subtraction in the short-distance region of coordinate space. Perturbatively, this subtraction induces additional counterterms in the momentum-space matching kernel. These counterterms are obtained from the perturbative counterparts of the coordinate-space hybrid denominators:
\begin{equation}
\begin{aligned}
    &\ \delta \mathcal C^{(1),X}_{\rm H}(x_1,x_2;y_1,y_2;P^z,\mu) \\
    =&\  (P^z)^2 \int \frac{ \rmd z_1}{2\pi} \frac{ \rmd z_2}{2\pi}\ \rme^{ \rmi  [(x_1-y_1)z_1+(x_2-y_2)z_2] P^z }\\
    &\ \times \delta \widehat M_{\rm H,pert}^{(1),X}(z_1,z_2;0,\mu) \ .
\end{aligned}
\end{equation}
Here $\delta \widehat M_{\rm H,pert}^{(1),X}(z_1,z_2;0,\mu)$ are constructed using the same piecewise prescription as the hybrid denominators in Eq.~\eqref{eq:denomi_hybrid}, with the zero-momentum self-renormalized lattice matrix elements $\widehat M_{\rm Self}^X(z_1,z_2;0,\mu)$ replaced by the corresponding one-loop results in the $\overline{\rm MS}$ scheme $\widehat M_{\rm \overline{MS}, pert}^{(1),X}(z_1,z_2;0,\mu)$, given in Eqs.~\eqref{eq:pert_res_VA}--\eqref{eq:pert_res_T}. Thus, the counterterms reproduce, at one-loop order, the same short-distance subtraction that has been applied non-perturbatively in the hybrid-renormalized lattice matrix elements.

Because the hybrid denominators are defined piecewise on the two-dimensional $(z_1,z_2)$ plane, their Fourier transform decomposes into contributions from the short-distance, long-distance, and mixing regions, as discussed in Sec.\ref{sec:hybrid_2D}. Performing the Fourier transform region by region gives:
\begin{equation}\label{eq:hyb_count_VA}
\begin{aligned}
    \delta & \mathcal C^{(1),V/A}_{\rm H} = (P^z)^2 \frac{\alpha_s C_F}{2\pi} \Bigg[ I^{V/A}_{\rm H} (\Delta_1,\Delta_2) + I^{V/A}_{\rm HSI} (\Delta_1,\Delta_2) \\
    +&\  I^{V/A}_{\rm HSII} (\Delta_1,\Delta_2) + I^{V/A}_{\rm HSIII} (\Delta_1,\Delta_2) + I^{V/A}_{\rm HSIV} (\Delta_1,\Delta_2) \\
    +&\ I^{V/A}_{\rm S} (\Delta_1,\Delta_2) + \delta(\Delta_1) \delta(\Delta_2) \left(\frac{5}{2} \ln\frac{\mu^2 \rme^{2 \gamma_E}}{4} + 4\right) \Bigg]\ ,
\end{aligned}
\end{equation}
\begin{equation}\label{eq:hyb_count_T}
\begin{aligned}
    \delta & \mathcal C^{(1),T}_{\rm H} = (P^z)^2 \frac{\alpha_s C_F}{2\pi} \Bigg[ I^{T}_{\rm H} (\Delta_1,\Delta_2) + I^{T}_{\rm HSI} (\Delta_1,\Delta_2) \\
    +&\ I^{T}_{\rm HSII} (\Delta_1,\Delta_2) + I^{T}_{\rm HSIII} (\Delta_1,\Delta_2) + I^{T}_{\rm HSIV} (\Delta_1,\Delta_2) \\
    +&\ I^{T}_{\rm S} (\Delta_1,\Delta_2) + \delta(\Delta_1) \delta(\Delta_2) \left(\frac{9}{4} \ln\frac{\mu^2 \rme^{2 \gamma_E}}{4} + \frac{13}{4}\right) \Bigg]\ ,
\end{aligned}
\end{equation}
where $\Delta_1 \equiv (x_1-y_1)P^z$ and $\Delta_2 \equiv (x_2-y_2)P^z$. The explicit expressions for $I_{\rm H}$, $I_{\rm HSI}$, $I_{\rm HSII}$, $I_{\rm HSIII}$, $I_{\rm HSIV}$, and $I_{\rm S}$ are collected in Appendix~\ref{app:hybrid_count}.

A useful consistency check is the large-momentum-fraction behavior of the hybrid-subtracted kernel. Terms proportional to $1/|x_1|$ or $1/|x_2|$ in the limits $|x_1|\gg 1$ or $|x_2|\gg 1$ would indicate uncanceled short-distance logarithms and would make the double-plus distribution ill defined. We have verified that these terms cancel after the hybrid counterterms are included. This cancellation is the momentum-space counterpart of removing the coordinate-space logarithms $\ln(\mu^2 z_1^2)$, $\ln(\mu^2 z_2^2)$, and $\ln[\mu^2(z_1-z_2)^2]$, associated with the short-distance singularities $z_1\to 0$, $z_2\to 0$, and $(z_1-z_2)\to 0$, respectively.

In practical implementation, the hybrid counterterms can be included either in momentum space through $\delta \mathcal C_{\rm H}^{(1)}$ or, equivalently, in coordinate space through the perturbative hybrid denominator $\delta \widehat M_{\overline{\rm MS}}^{(1)}$. In this work we use the coordinate-space implementation. This avoids introducing an additional two-dimensional convolution with $\delta \mathcal C_{\rm H}^{(1)}$ on top of the LaMET matching convolution in momentum space. Concretely, we implement the perturbative counterpart of the hybrid denominator in coordinate space, then perform the Fourier transform and apply the remaining $\overline{\rm MS}$ matching kernel in momentum space.

\subsection{Numerical Illustration of The Matching Procedure}

\begin{figure}[htbp]
    \centering
    \subfloat{
        \centering
        \includegraphics[width=\linewidth]{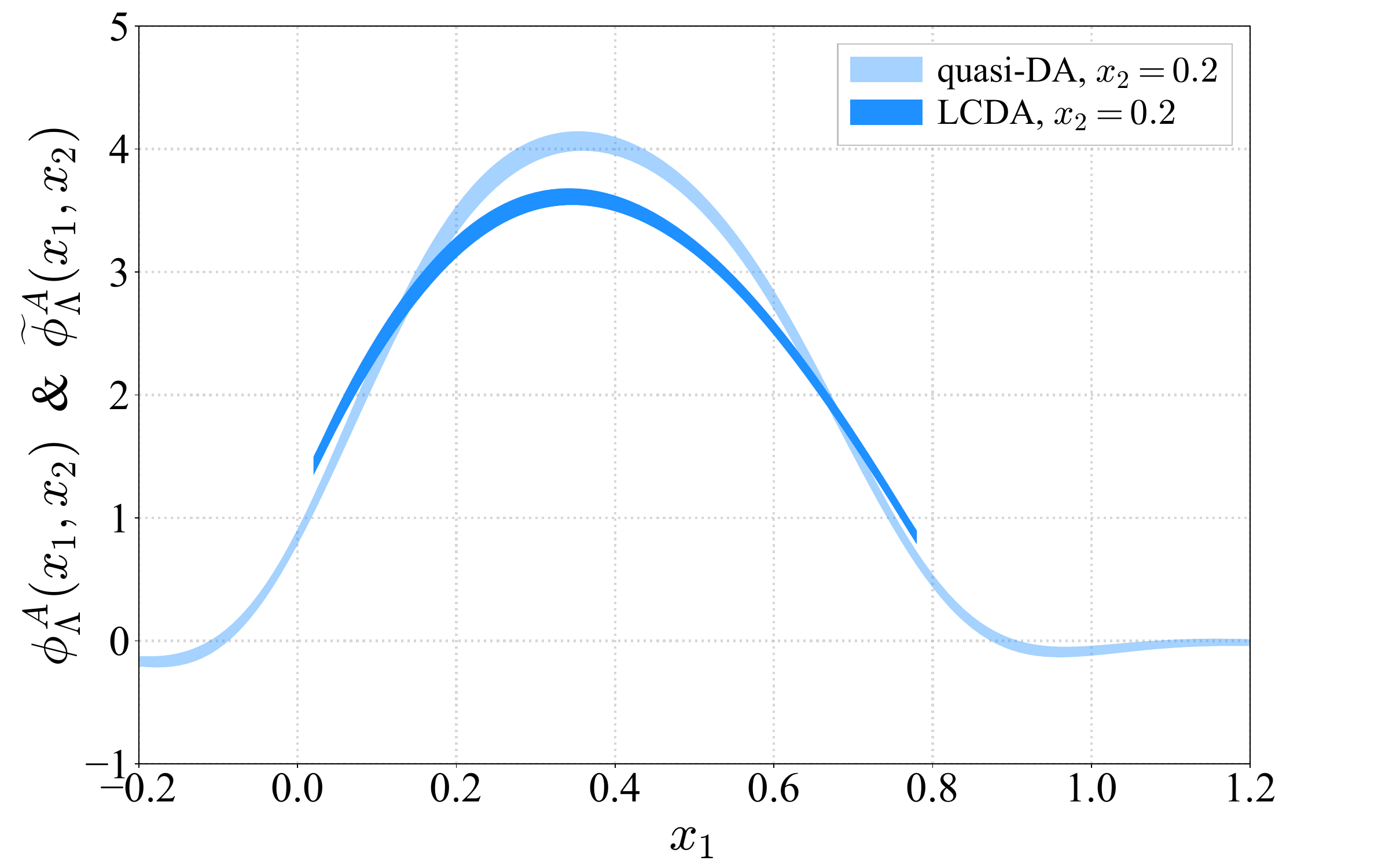}
        }\\
    \subfloat{
        \centering
        \includegraphics[width=\linewidth]{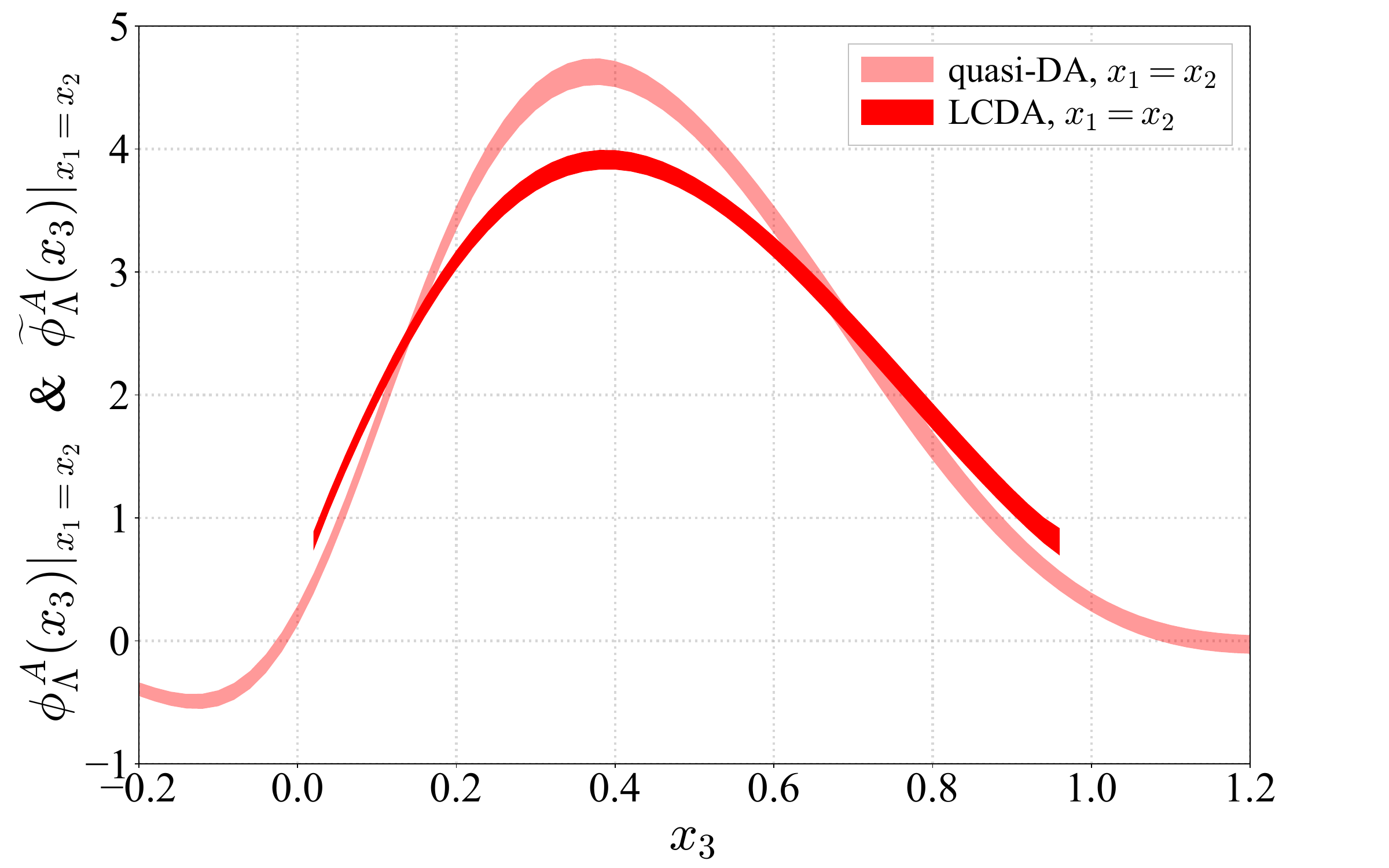}
        }
    \caption{Comparison of quasi-DAs and LCDAs on the H48P32 ensemble with $P^z\approx3.0{\rm\ GeV}$. The upper panel shows the shapes of DAs depend on $x_1$ when fix $x_2=0.2$, while the lower panel shows the shapes depend on $x_3$ when fix $x_1=x_2$.}
    \label{fig:matching_xcut}
\end{figure}

\begin{figure}[htbp]
    \centering
    \includegraphics[width=\linewidth]{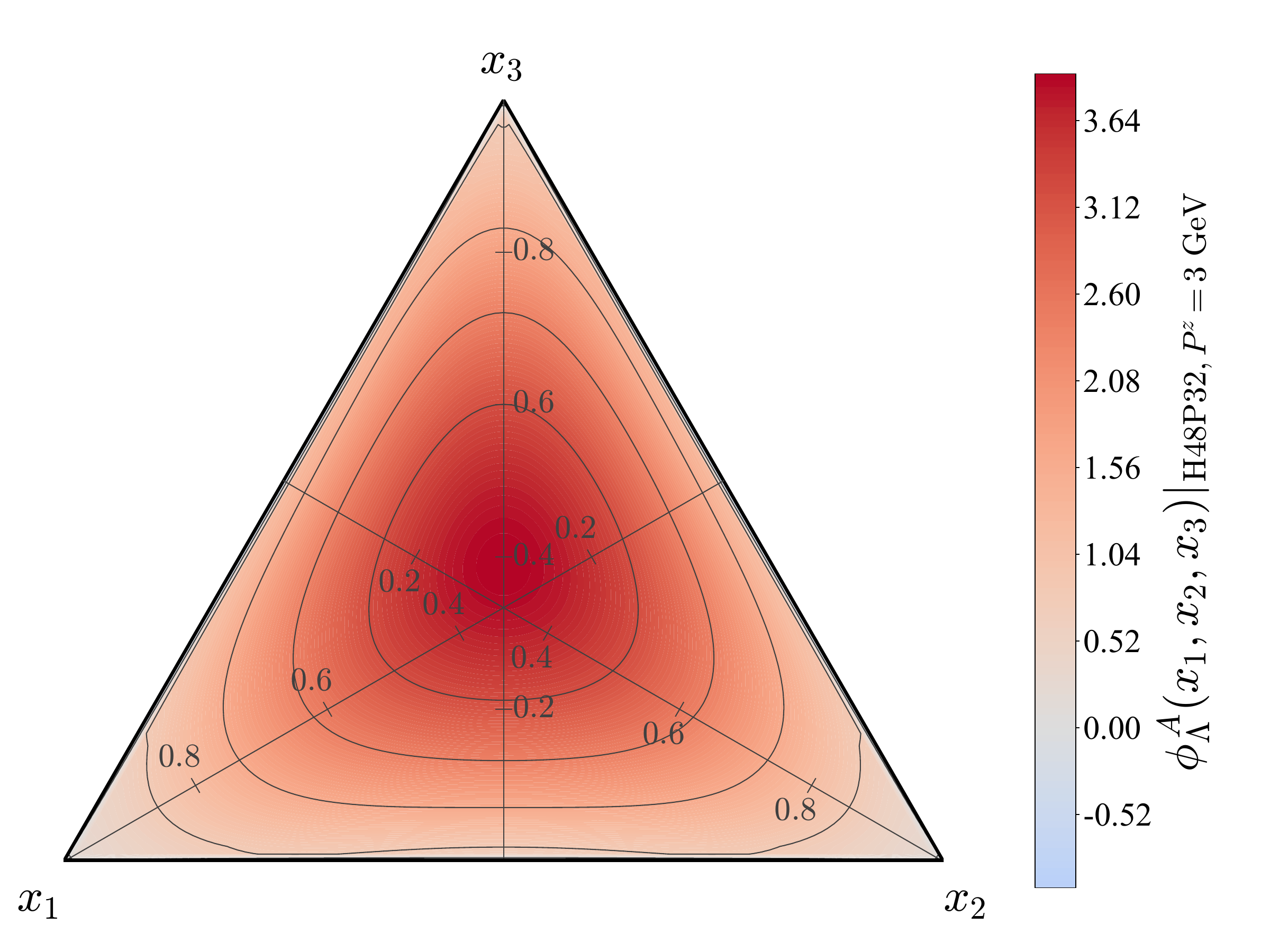}
    \caption{Central value of two-dimensional LCDA in physical momentum-fraction triangle region, calculated on the H48P32 ensemble with $P^z\approx3{\rm\ GeV}$.}
    \label{fig:matching_heatmap}
\end{figure}

After implementing the LaMET matching with hybrid counterterms, we examine its numerical impact on the baryon quasi-DAs. As a representative example, Fig.~\ref{fig:matching_xcut} compares the hybrid-renormalized quasi-DA before LaMET matching with the corresponding matched LCDA on the H48P32 ensemble at $P^z\approx 3~{\rm GeV}$. Since baryon DAs depend on two independent momentum fractions subject to $x_1+x_2+x_3=1$, we display the comparison through one-dimensional slices. In the upper panel, $x_2$ is fixed at $0.2$ and the dependence on $x_1$ is shown. In the lower panel, the diagonal direction $x_1=x_2$ is shown as a function of $x_3=1-x_1-x_2$. These slices illustrate that the one-loop matching produces a momentum-fraction-dependent modification of the quasi-DAs, rather than a simple overall rescaling. Fig.~\ref{fig:matching_heatmap} shows the same matched LCDA as a two-dimensional heat map in the physical triangular region. The three edges of the triangle correspond to the endpoint limits $x_1=0$, $x_2=0$, and $x_3=0$, while the interior satisfies $0\leq x_1,x_2,x_3\leq 1$ and $x_1+x_2+x_3=1$. This representation provides a direct visualization of the full two-dimensional momentum-fraction dependence after matching. 

\begin{figure}[htbp]
    \centering
    \subfloat{
        \centering
        \includegraphics[width=\linewidth]{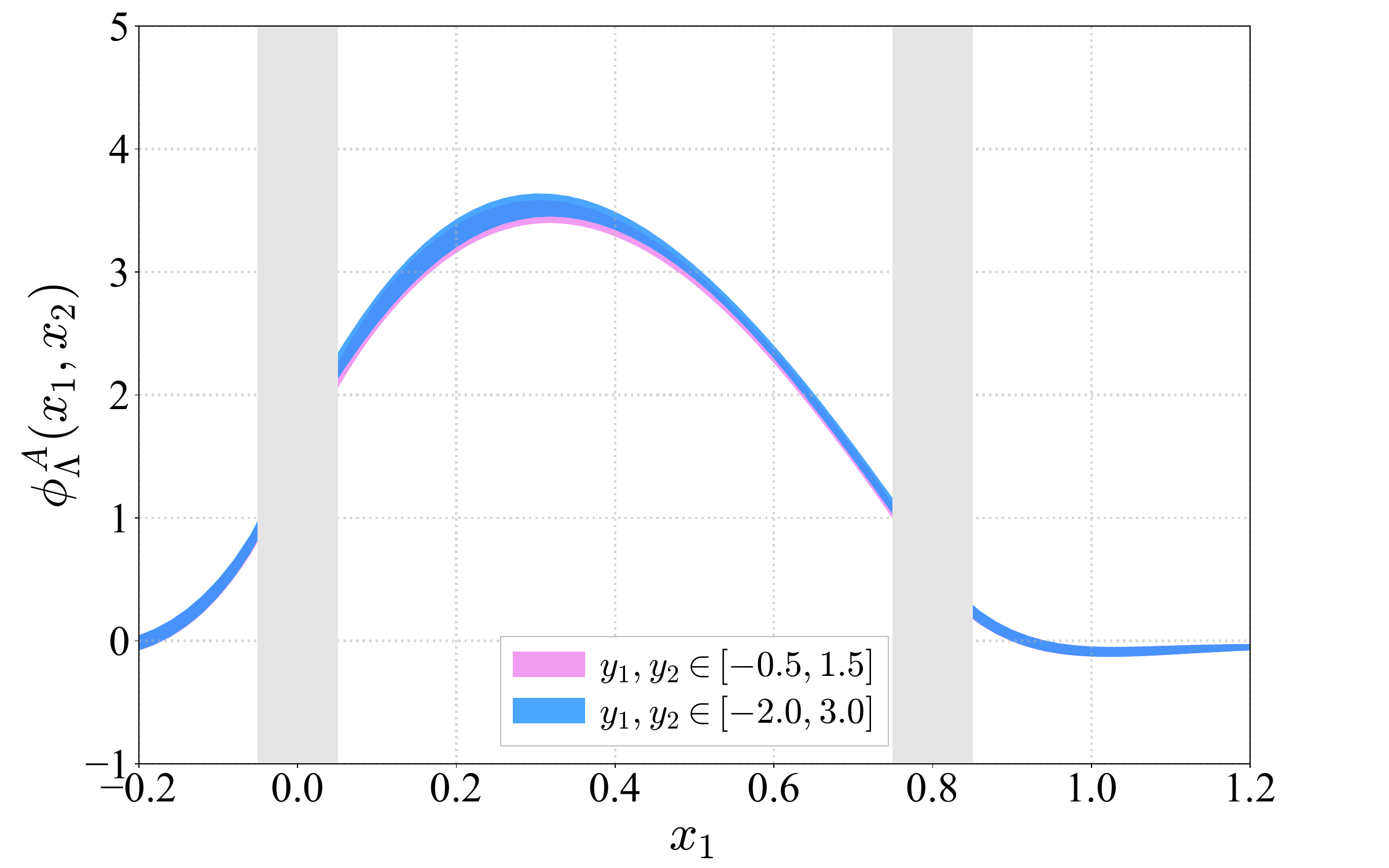}
        }\\
    \subfloat{
        \centering
        \includegraphics[width=\linewidth]{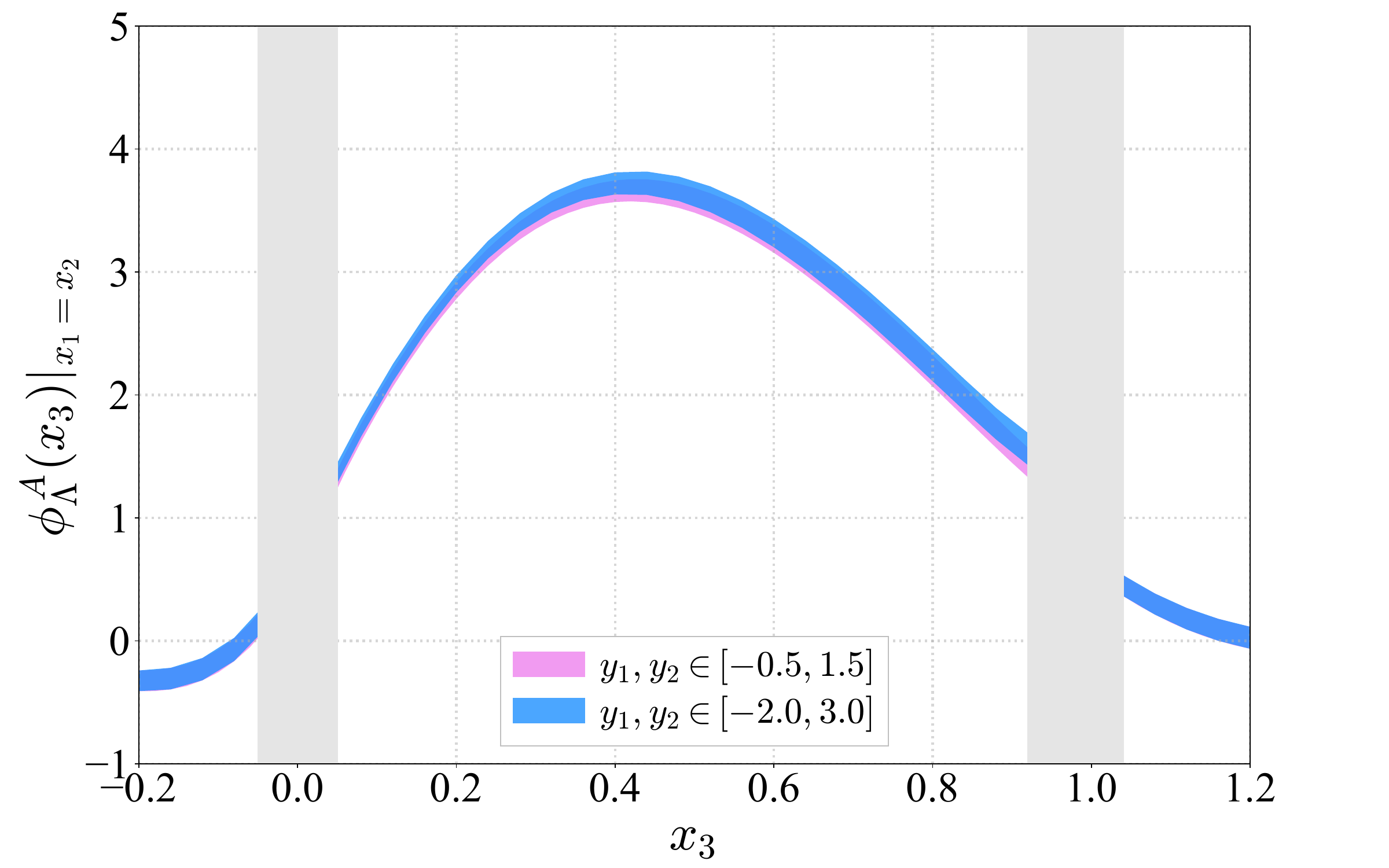}
        }
    \caption{Comparison of different integral limits when matching onto LCDAs, on H48P32 with $P^z\approx2~{\rm GeV}$. Two curves are results obtained by using integral variables $y_1,y_2\in[-0.5,1.5]$ and $y_1,y_2\in[-2.0,3.0]$, respectively.}
    \label{fig:matching_limits}
\end{figure}

\begin{figure*}[htbp]
    \centering
    \subfloat[\ H48P32]{
        \centering
        \includegraphics[width=0.24\textwidth]{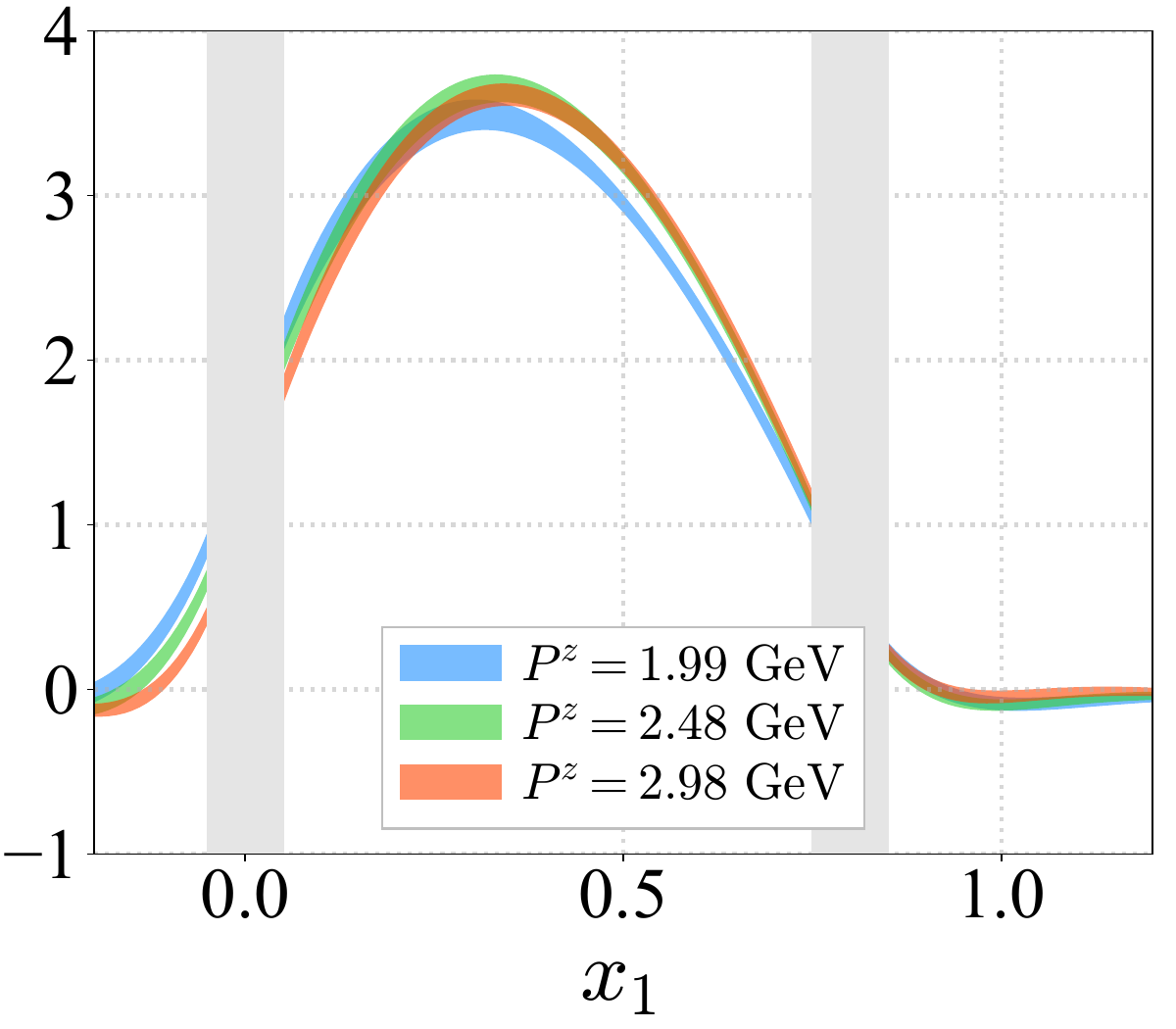}
        }
    \subfloat[\ G36P29]{
        \centering
        \includegraphics[width=0.24\textwidth]{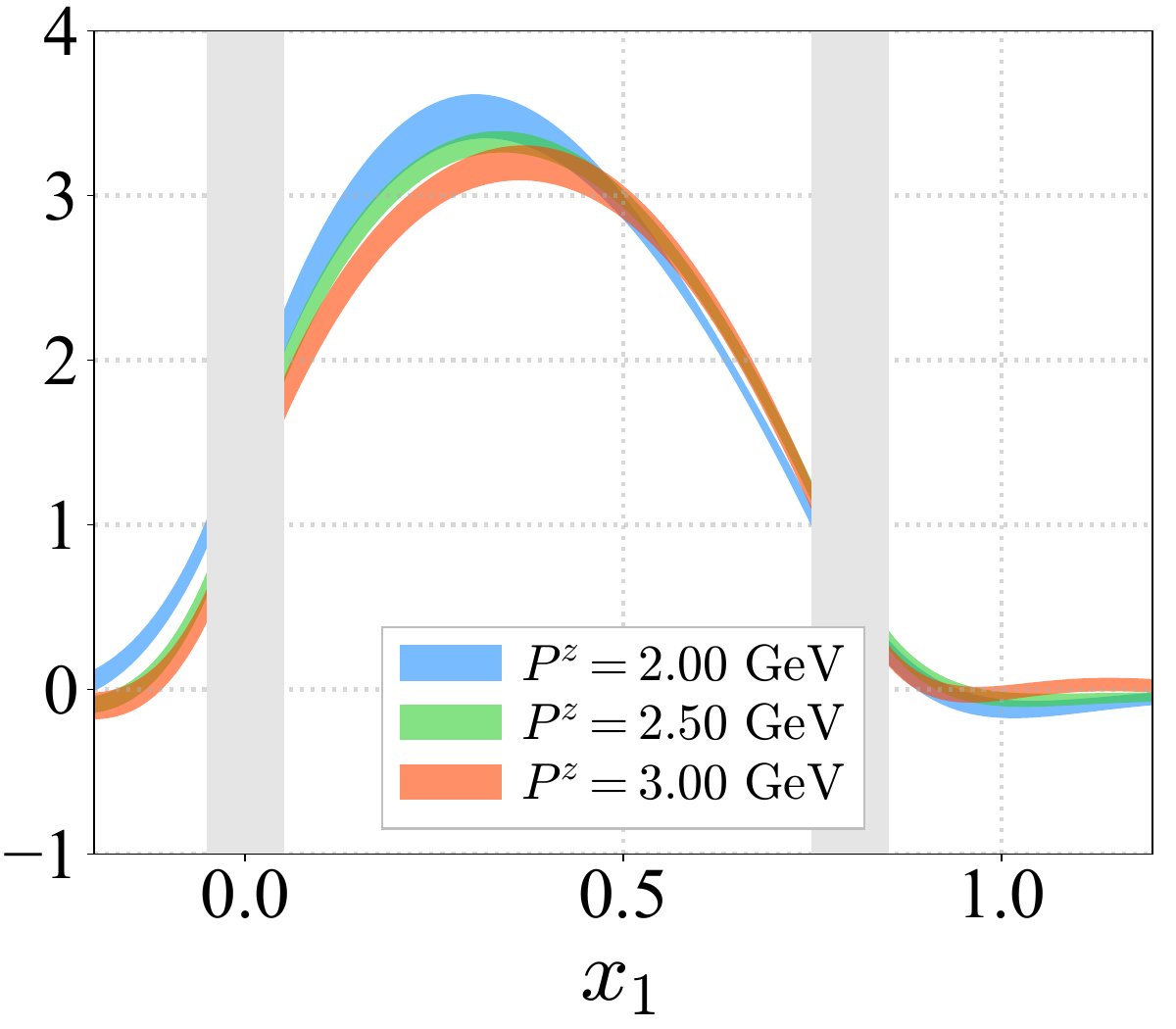}
        }
    \subfloat[\ F32P30]{
        \centering
        \includegraphics[width=0.24\textwidth]{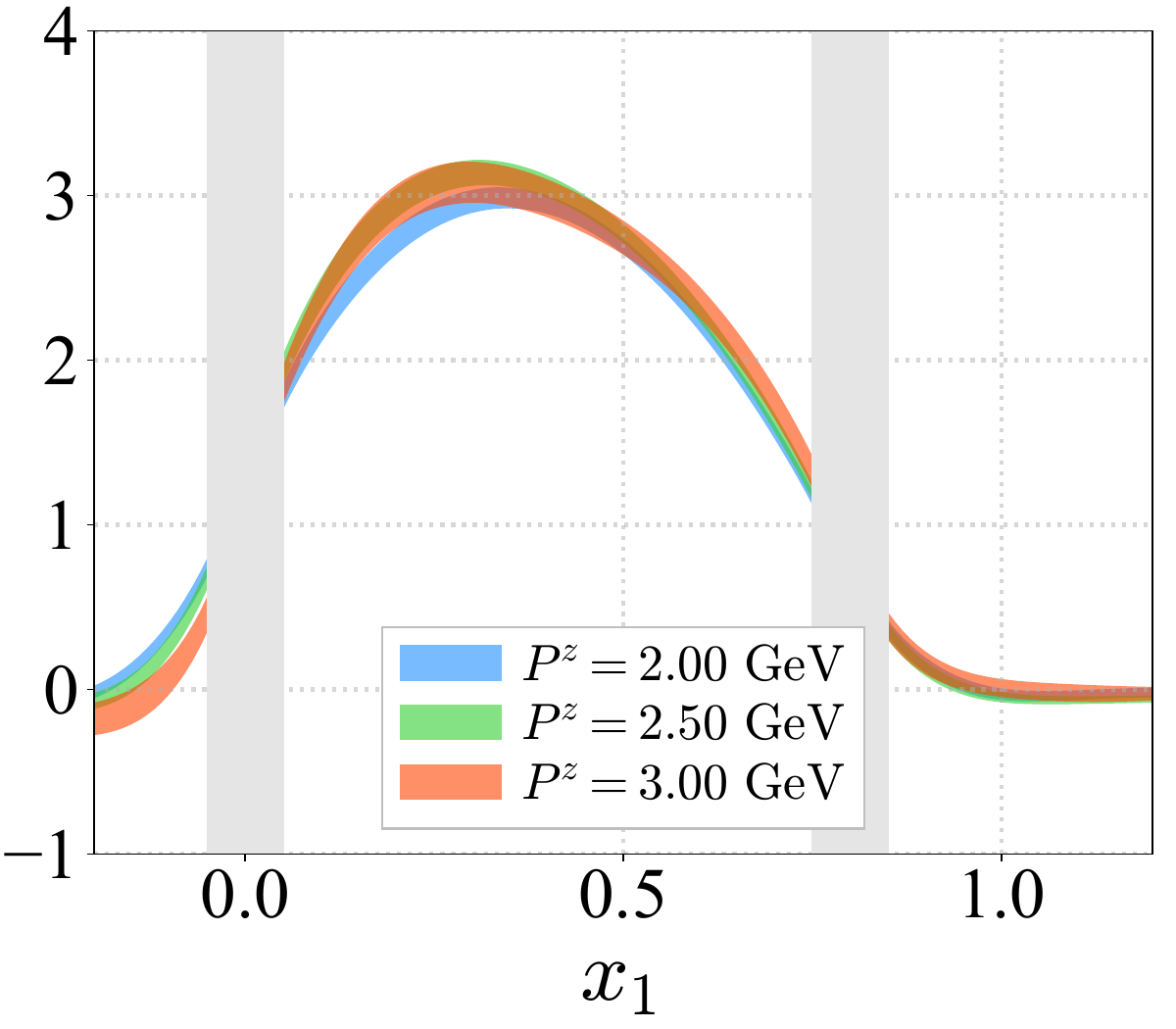}
        }
        \\
    \subfloat[\ F32P21]{
        \centering
        \includegraphics[width=0.24\textwidth]{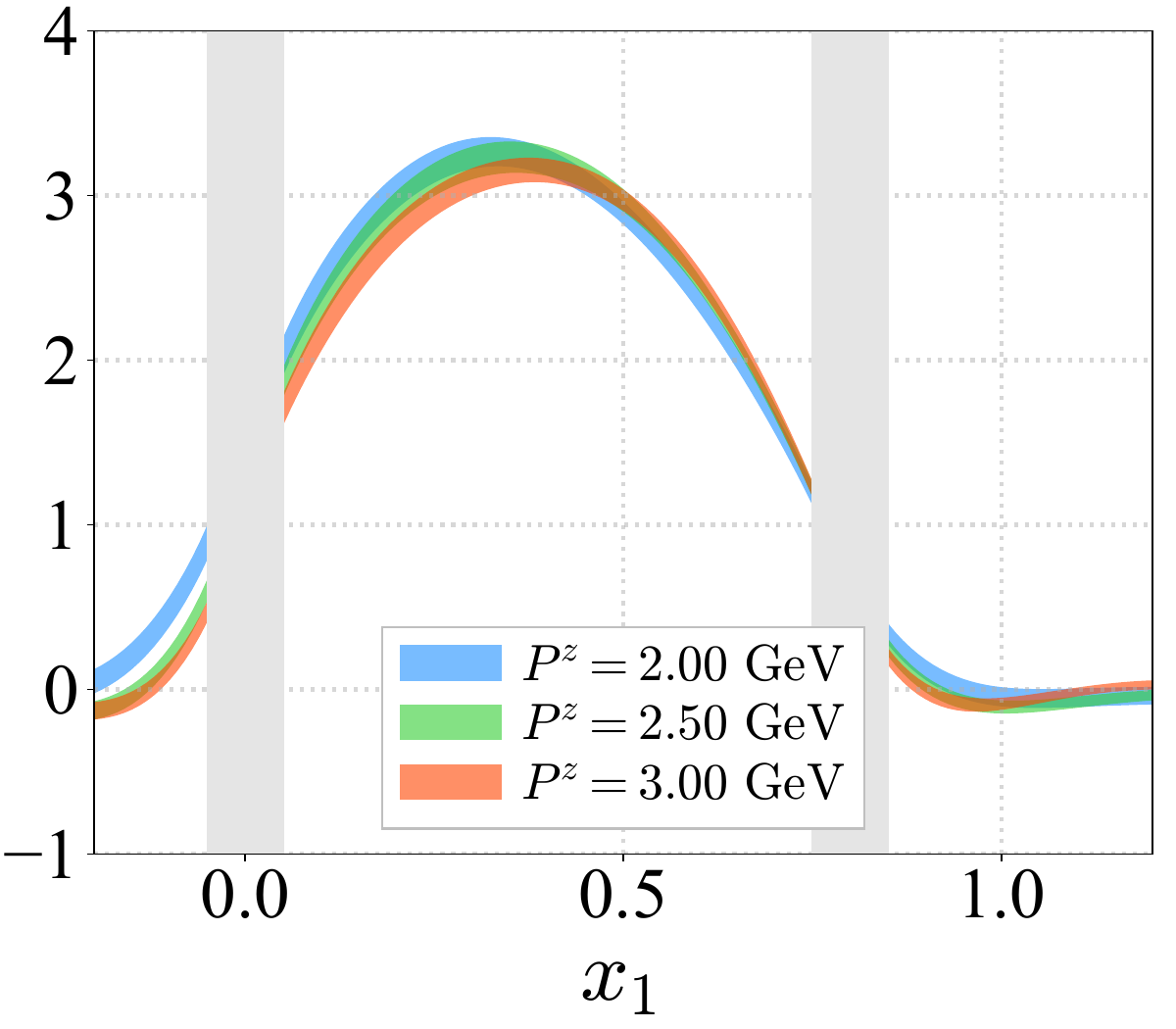}
        }
    \subfloat[\ C48P14]{
        \centering
        \includegraphics[width=0.24\textwidth]{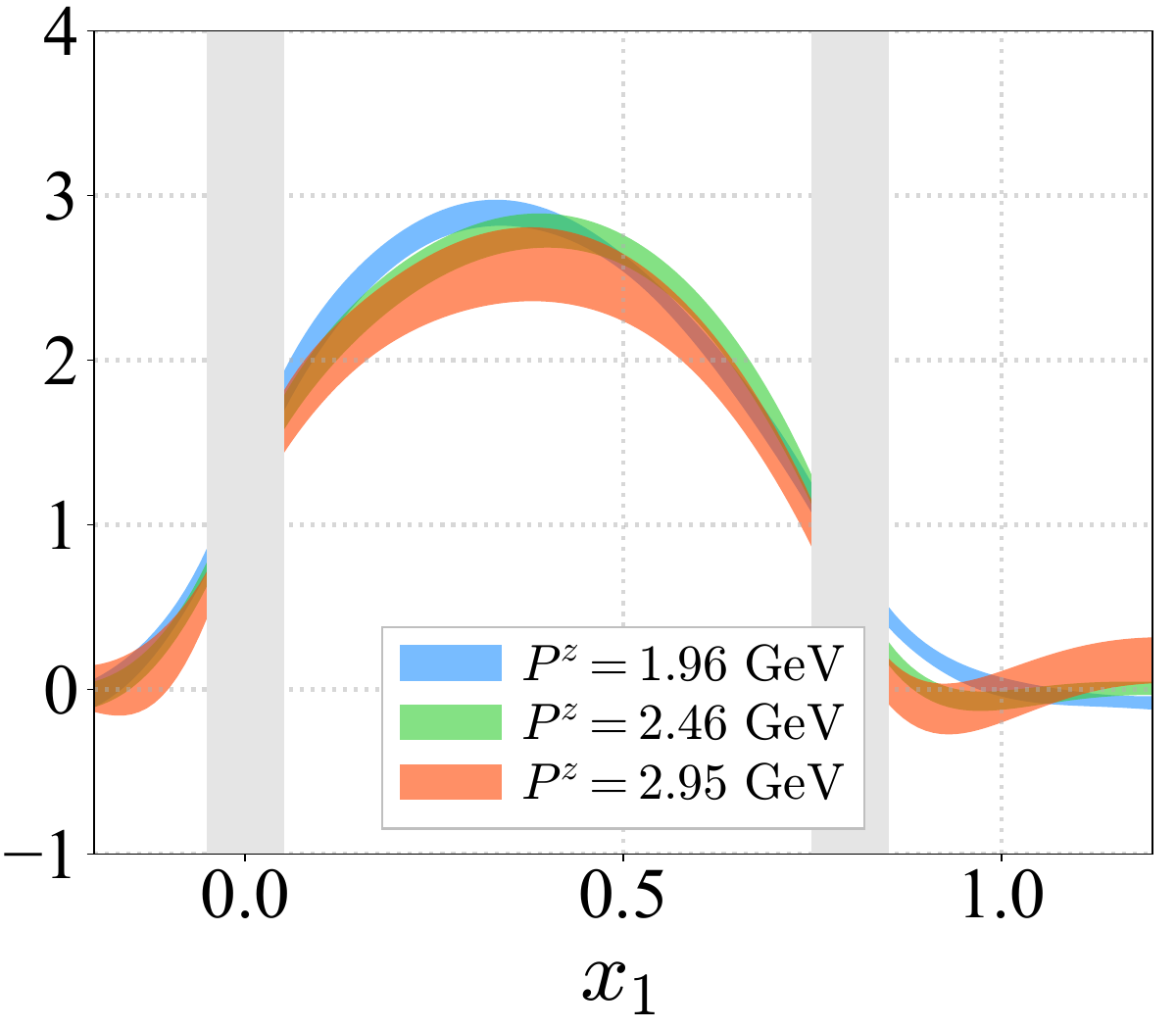}
        }
    \subfloat[\ C32P23]{
        \centering
        \includegraphics[width=0.24\textwidth]{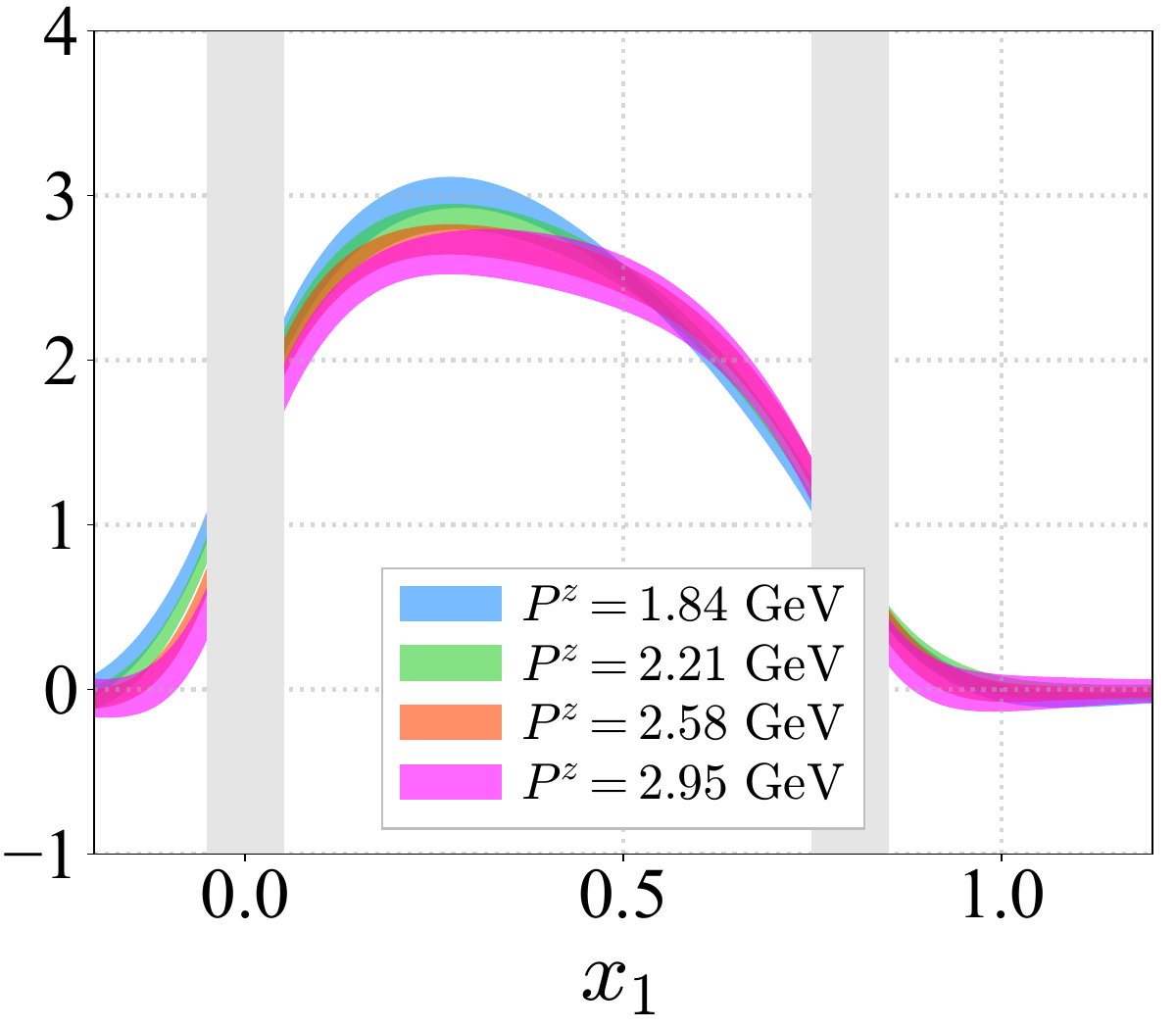}
        }
    \subfloat[\ C24P29]{
        \centering
        \includegraphics[width=0.24\textwidth]{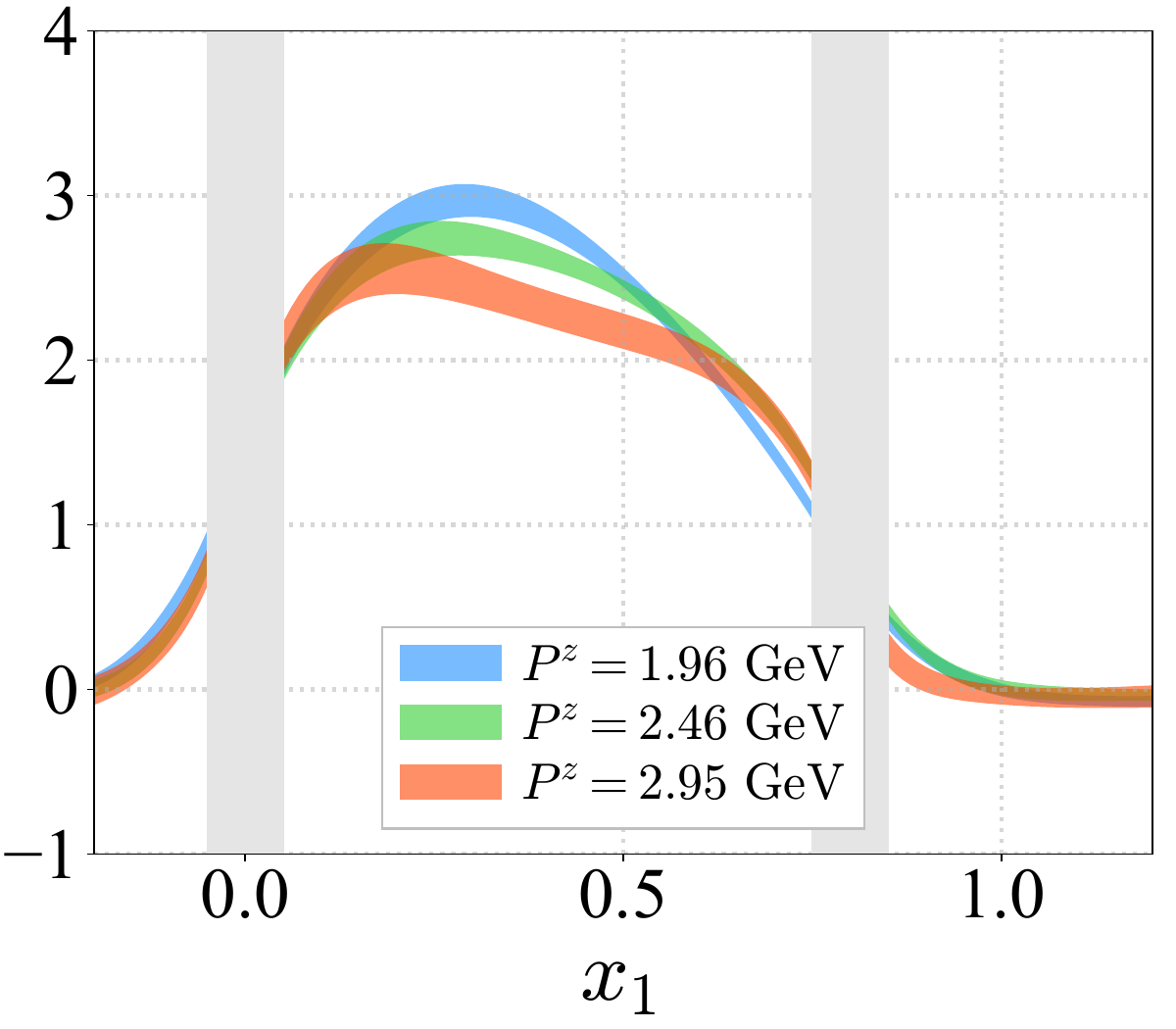}
        }
    \caption{LCDAs obtained from the seven ensembles used in this work. Each panel shows the $x_1$ dependence of the matched LCDA at fixed $x_2=0.2$, obtained at three or four different hadron momenta. The gray bands schematically indicate endpoint regions; together with the unphysical momentum-fraction regions, they mark where LaMET power corrections can be sizable.}
    \label{fig:match_seven_ens}
\end{figure*}

As a stability check of the numerical implementation, we vary the integration range of the two-dimensional matching convolution in Eq.~\eqref{eq:matching_ori} over the quasi momentum fractions $y_1$ and $y_2$. We compare the results obtained with $y_{1,2}\in[-0.5,1.5]$ and with the enlarged range $y_{1,2}\in[-2.0,3.0]$. As shown in Fig.~\ref{fig:matching_limits}, the resulting LCDAs are consistent within the current precision, with no visible dependence on the chosen integration domain. This indicates that the numerical matching convolution is stable and that contributions from the far unphysical momentum-fraction region which are related to high power corrections are well controlled.

Fig.~\ref{fig:match_seven_ens} shows the matched LCDAs obtained from all seven ensembles at the available hadron momenta. Although the ensembles have different lattice spacings and pion masses, the matched distributions exhibit broadly consistent shapes. On each ensemble, the curves become more stable as $P^z$ increases, consistent with the LaMET expectation that finite-momentum corrections are suppressed at larger hadron momentum. This behavior provides a useful check of the matching procedure and motivates the subsequent combined extrapolation in $a$, $m_\pi$, and $P^z$ in Sec.~\ref{sec:Apply_Results}. More illustrations for the numerical impact of LaMET matching on different ensembles and momenta are collected in Appendix~\ref{app:matching_ens}.

\section{Numerical Demonstration and Systematic Analysis for the \texorpdfstring{$\Lambda$}{Lambda} \texorpdfstring{$A$}{A} LCDA} \label{sec:Apply_Results}

After hybrid renormalization, large-$\lambda$ extrapolation, Fourier transform, and perturbative matching, we obtain matched LCDAs for each ensemble and each available hadron momentum. Although these distributions have been converted to the $\overline{\rm MS}$ scheme, they still carry residual dependence on the lattice spacing $a$, the pion mass $m_\pi$, and the finite hadron momentum $P^z$. We therefore perform extrapolations to the continuum, physical-pion-mass, and infinite-momentum limits.

In this section, we use the $\Lambda$-baryon $A$-structure LCDA as a representative amplitude to demonstrate the complete analysis pipeline and to quantify the dominant systematic uncertainties. The combined extrapolation is used as the central analysis, while a sequential extrapolation is used as a cross-check and to estimate the corresponding systematic uncertainty. The same framework applies to the $V$ and $T$ amplitudes, whose complete physical results are presented together with the $A$ amplitude in the companion Letter~\cite{LPC:2026lcj}.

\subsection{Combined Extrapolation to the Physical Limit}\label{sec:extrap_phys}
After perturbative matching, the LCDAs obtained on each ensemble still contains residual effects from the finite lattice spacing, the unphysical-pion-mass, and the finite hadron momentum used in the LaMET calculation. We remove these effects by performing a combined extrapolation in $a$, $m_\pi$, and $P^z$. At each momentum-fraction point $(x_1,x_2)$, the matched LCDAs are fitted with the ansatz:
\begin{equation}\label{eq:joint_extrap}
\begin{aligned}
    \phi(x_1,x_2)& |_{a,m_\pi,P^z} = \phi_{\rm phys}(x_1,x_2) \\
    & + \frac{A(x_1,x_2)}{(P^z)^2} + \left( m_\pi^2 - m_{\pi,\rm phys}^2 \right) B(x_1,x_2)\\
    & + a^2 \left[  D_1(x_1,x_2) + D_2(x_1,x_2) (P^z)^2  \right]\ ,
\end{aligned}
\end{equation}
where $m_{\pi,\rm phys}=139.6~{\rm MeV}$. Here $\phi_{\rm phys}(x_1,x_2)$ is the desired LCDA in continuum and infinite-momentum limits at physical pion mass. The coefficient $A(x_1,x_2)$ parametrizes the leading LaMET power corrections of order $1/(P^z)^2$, while $B(x_1,x_2)$ describes the leading pion-mass dependences. The two terms proportional to $a^2$ describe discretization effects: $D_1(x_1,x_2)$ accounts for the momentum-independent lattice artifacts, and $D_2(x_1,x_2)(P^z)^2$ accounts for the momentum dependent artifacts associated with boosted hadron states and nonlocal operators.

This combined fit treats the continuum, physical-pion-mass, and infinite-momentum extrapolations simultaneously, allowing correlations among these effects to be incorporated in a single fit. In particular, the $a^2(P^z)^2$ term is included to account for momentum-dependent lattice artifacts associated with boosted hadron states and nonlocal operators. Such effects are already indicated by the dispersion-relation analysis in Sec.~\ref{dispersion} and are visible in Fig.~\ref{fig:dispersion_relation_fit}, where mild deviations from the continuum dispersion relation become more pronounced on the coarsest ensembles at large $P^z$. Including this term helps separate these discretization effects from the genuine finite-momentum corrections described by the $1/(P^z)^2$ term.

\begin{figure*}[htbp]
    \centering
    \includegraphics[width=\textwidth]{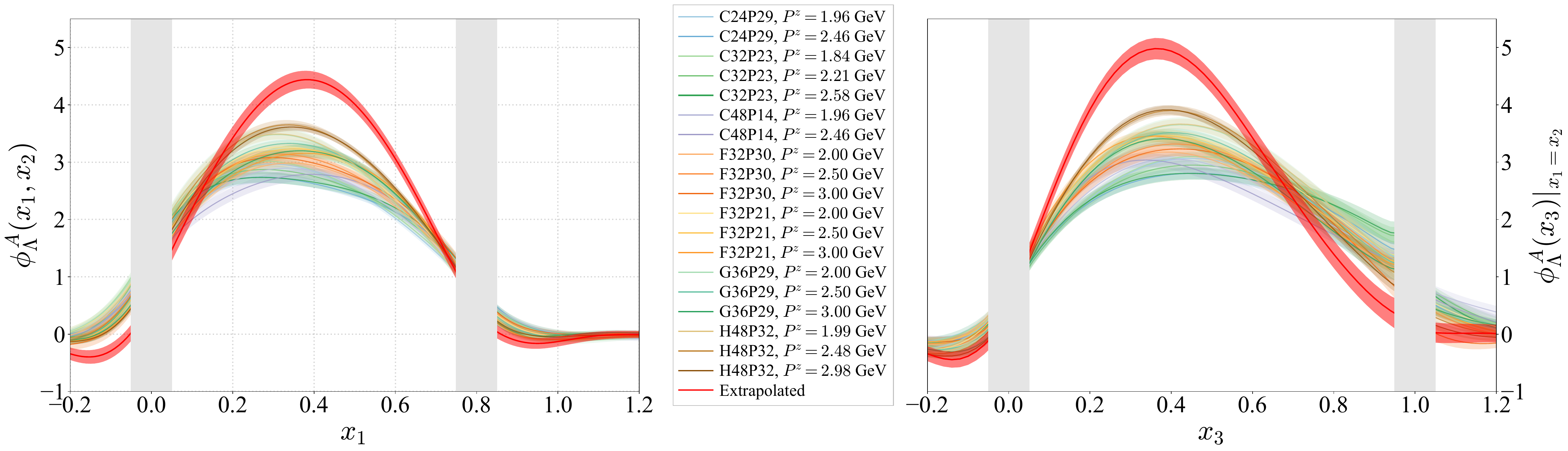}
    \caption{Combined extrapolation for LCDAs on each each ensemble and momentum, with selected one-dimensional slices $x_2=0.2$ and $x_1=x_2$.}
    \label{fig:physical_3fig}
\end{figure*}

The extrapolation is performed independently at each point $(x_1,x_2)$ in the momentum-fraction space. The statistical uncertainty is propagated through the full analysis using the same jackknife samples as in the preceding stages. Representative result for the $\Lambda$-baryon $A$-structure LCDA is shown in  Fig.~\ref{fig:physical_3fig}, where the matched LCDAs from different ensembles and momenta are compared with the extrapolated physical result on selected one-dimensional cuts. 

The extrapolated physical result is separated from the data obtained at finite lattice spacings and finite hadron momenta, with the dominant shift coming from the $P^z$ extrapolation. This behavior is expected for baryon observables, because the relatively large baryon mass $M_B$ limits the achievable Lorentz boost factor of the baryon $\gamma_B=E_B/M_B$ in the present calculation. As a result, residual finite-momentum effects can remain sizable and must be treated carefully. We therefore examine the stability of the $P^z$ extrapolation and estimate the associated systematic uncertainty using an alternative sequential extrapolation strategy in the next subsection.

Special care is required in interpreting the results near the momentum-fraction endpoints, where the LaMET factorization formula receives enhanced power corrections. For a three-quark baryonic system, the endpoint regions include the limits in which any one of the three parton momentum fractions becomes small, namely $x_1\to 0$, $x_2\to 0$, or $x_3=1-x_1-x_2\to 0$. These endpoint regions are indicated by the gray bands in Fig.~\ref{fig:physical_3fig} and other similar figures. At finite $P^z$, the matched distributions can also have residual contributions in the unphysical momentum-fraction regions. These contributions are expected to be suppressed by the subsequent infinite-momentum extrapolation, as observed in our analysis, consistent with the LaMET power-counting expectation. Nevertheless, results from the endpoint regions to the unphysical momentum-fraction regions should be interpreted with caution, since they are more sensitive to finite-$P^z$ power corrections and therefore carry larger systematic uncertainties.


\subsection{Sequential Extrapolation and Stability Check}\label{sec:sequential_extrap}

The combined extrapolation in Eq.~\eqref{eq:joint_extrap} is taken as our central strategy. However, as discussed in the previous subsection, the extrapolated physical result is visibly separated from the data at the finest lattice spacing and the largest available momentum. Therefore, it is important to check the stability of the extrapolation procedure and to estimate the associated systematic uncertainty. For this purpose, we perform an alternative sequential extrapolation: we first take the continuum and physical-pion-mass limits at each fixed $P^z$, and then extrapolate the resulting LCDAs to the infinite-momentum limit using the leading $1/(P^z)^2$ power correction. The difference between the combined and sequential extrapolations is used to estimate the systematic uncertainty associated with the physical-limit extrapolation.

The first step is carried out at fixed hadron momentum $P^z$, where we extrapolate the matched LCDAs to the continuum and physical-pion-mass limits:
\begin{equation}
\begin{aligned}
    \phi(x_1,x_2)&|_{a,m_\pi,P^z} = \phi_0(x_1,x_2)|_{P^z} \\
    &+ \left( m_\pi^2 - m_{\pi,\rm phys}^2 \right) B(x_1,x_2) + a^2 D(x_1,x_2) \ .
\end{aligned}
\end{equation}
Here $\phi_0(x_1,x_2)|_{P^z}$ denotes the LCDA after removing the leading lattice-spacing and pion-mass effects, while still retaining the residual finite-hadron-momentum dependence. This first step is designed to reduce discretization effects before the infinite-$P^z$ extrapolation. In the full combined ansatz, the leading discretization effects contain both a momentum-independent term proportional to $a^2$ and a momentum-dependent term proportional to $a^2(P^z)^2$. At fixed $P^z$, however, these two contributions have the same $a^2$ dependence and can therefore be absorbed into a single effective coefficient $D(x_1,x_2)$. As a result, $\phi_0(x_1,x_2)|_{P^z}$ is independent of the explicit separation between the $a^2$ and $a^2(P^z)^2$ terms. After the continuum and physical-pion-mass extrapolation, the remaining $P^z$ dependence is expected to be dominated by the LaMET finite-momentum power correction rather than by momentum-dependent lattice artifacts.

\begin{figure}[htbp]
    \centering
    \subfloat{
        \centering
        \includegraphics[width=\linewidth]{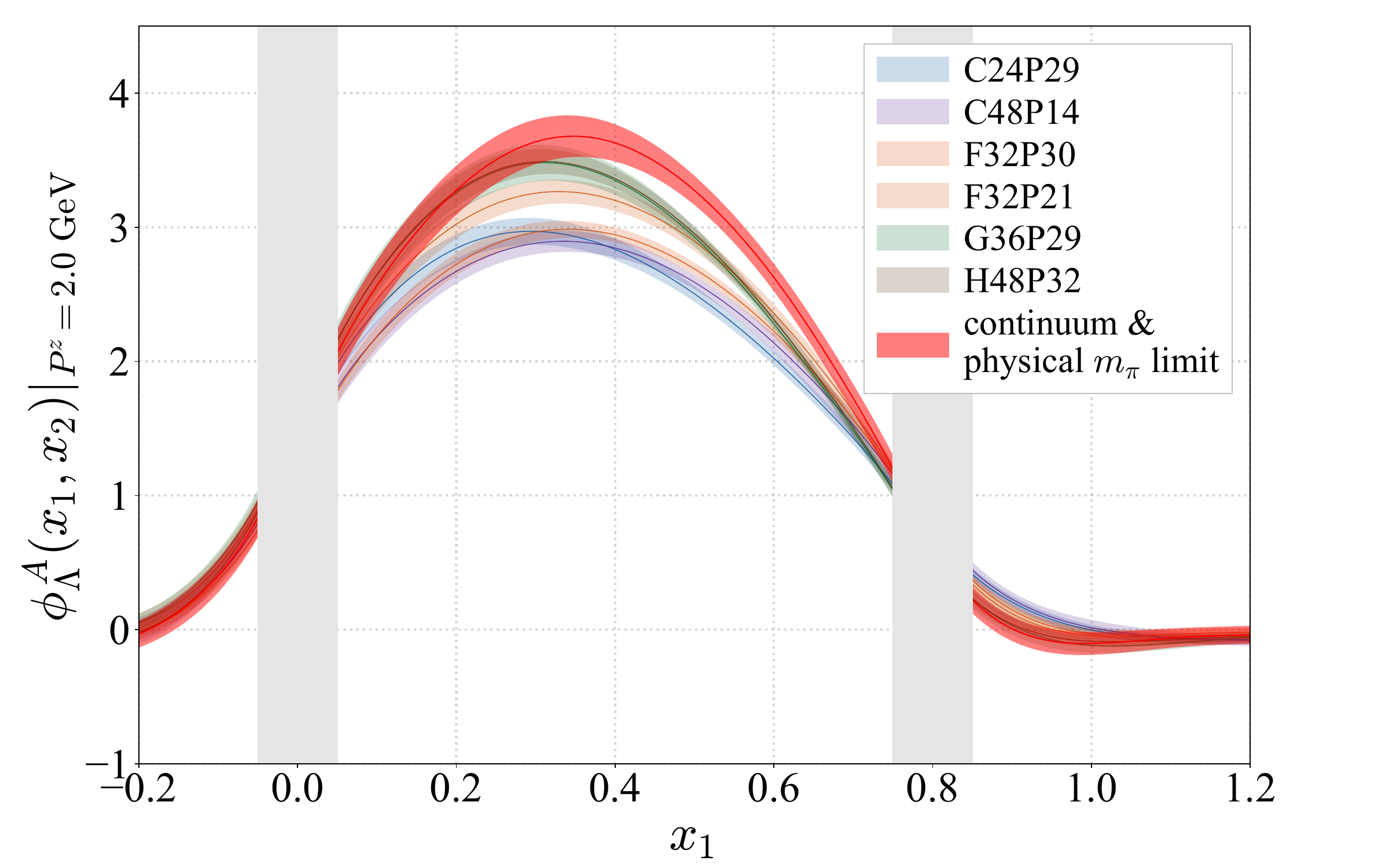}
        }\\
    \subfloat{
        \centering
        \includegraphics[width=\linewidth]{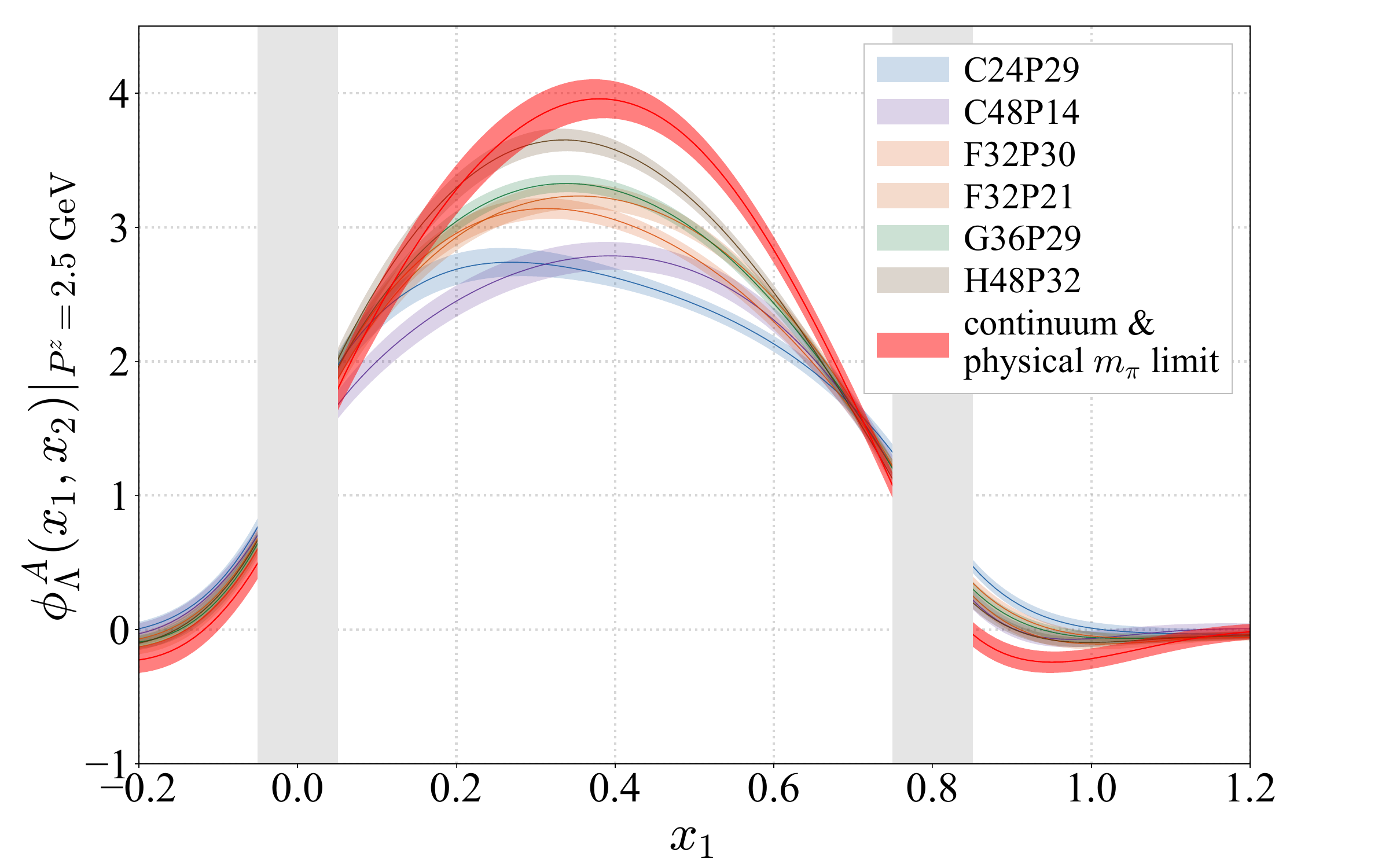}
        }\\
    \subfloat{
        \centering
        \includegraphics[width=\linewidth]{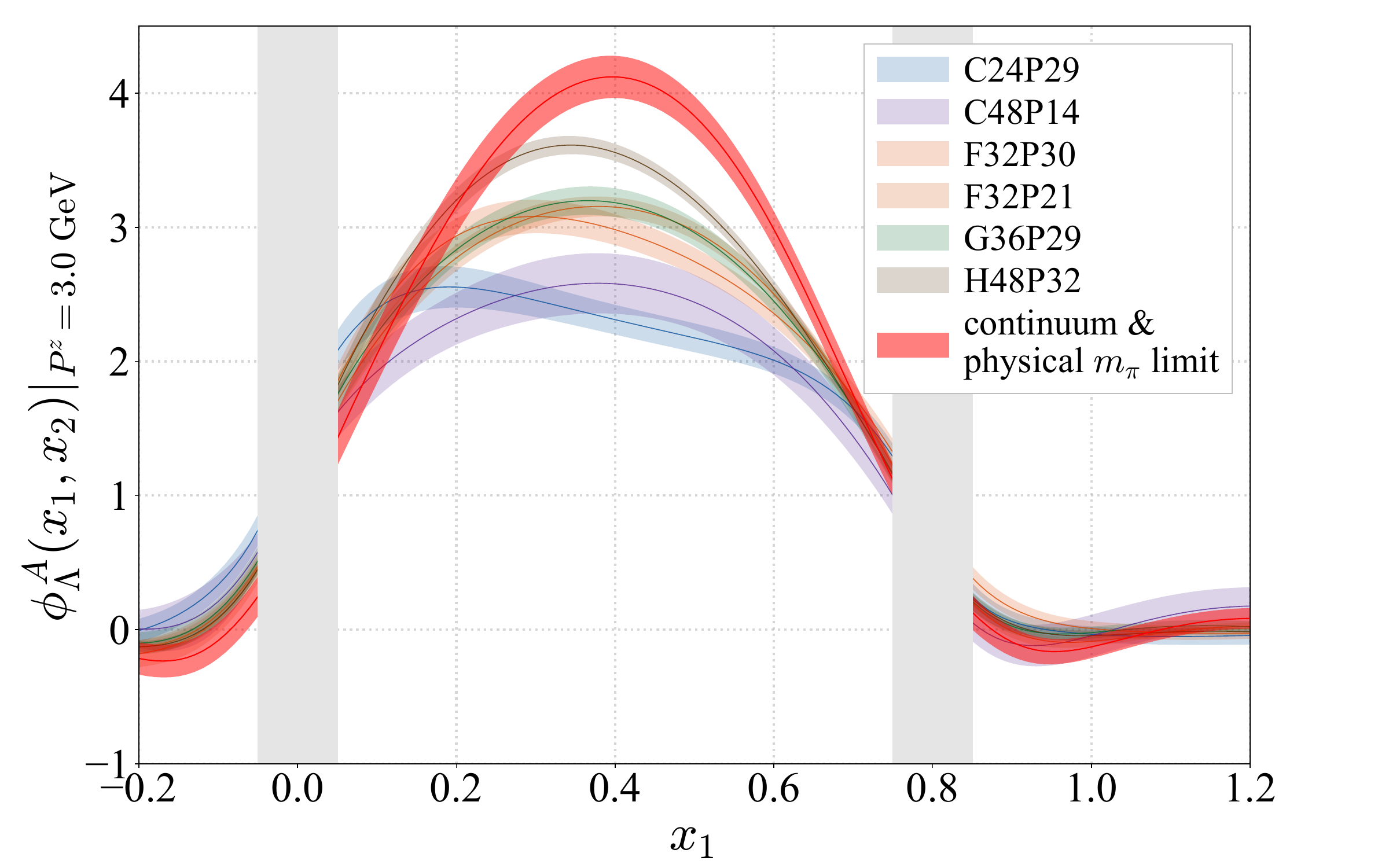}
        }
    \caption{First step of the sequential extrapolation for the $\Lambda$ $A$ LCDA at fixed hadron momenta, with $x_2=0.2$. The three panels correspond to momentum groups $P^z\approx 2.0$, $2.5$, and $3.0~{\rm GeV}$. Six ensembles with aligned momentum values are included, while C32P23 is excluded as its momenta do not align with others.}
    \label{fig:seq_step1}
\end{figure}

The first-step extrapolation is illustrated in Fig.~\ref{fig:seq_step1} for three momentum groups, 
$P^z\approx 2.0$, $2.5$, and $3.0~{\rm GeV}$. 
In this comparison we use the six ensembles whose momenta can be aligned across the lattice spacings, as summarized in Table~\ref{tab:Ensembles}. The C32P23 ensemble is not included in the sequential fit, as its available momenta do not match the common momentum groups used by other ensembles. Fig.~\ref{fig:seq_step1} therefore provides a direct view of the lattice-spacing dependence at fixed $P^z$, after grouping ensembles with approximately the same baryon momentum. The figure also shows that the pion-mass dependence is mild compared with the momentum dependence. This can be seen most clearly by comparing the F32P30 and F32P21 ensembles, which share the same lattice spacing but different pion masses, $m_\pi\approx 300~{\rm MeV}$ and $210~{\rm MeV}$, respectively. Their results are close to each other, especially in the $P^z\approx 2.5~{\rm GeV}$ and $3.0~{\rm GeV}$ groups, indicating that the pion-mass effect is subleading within the current precision for the $\Lambda$ baryon.

In the second step, the remaining finite-momentum dependence of $\phi_0(x_1,x_2)|_{P^z}$ is extrapolated to the infinite-momentum limit using the leading power correction parameterization:
\begin{equation}
\phi_0(x_1,x_2)|_{P^z} = \phi_{\rm phys}^{\rm seq}(x_1,x_2) + \frac{A(x_1,x_2)}{(P^z)^2}\ .
\end{equation}
Since the leading lattice-spacing and pion-mass effects have been removed in the first step, this fit isolates the residual finite-$P^z$ dependence more directly. The resulting $\phi_{\rm phys}^{\rm seq}(x_1,x_2)$ provide a nontrivial cross-check of the physical LCDAs under a different ordering of the extrapolation limits.

\begin{figure}[htbp]
    \centering
    \subfloat{
        \centering
        \includegraphics[width=\linewidth]{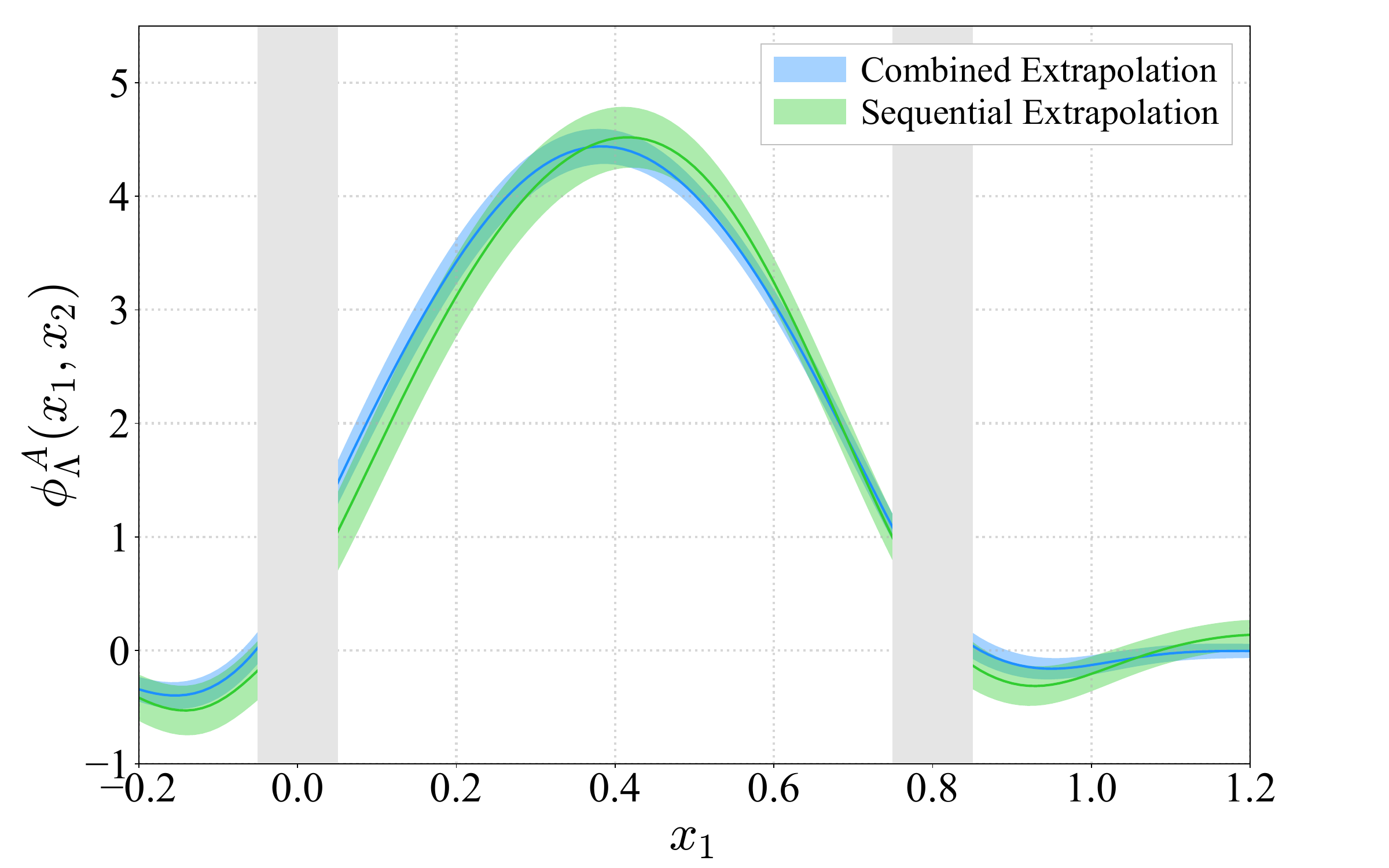}
        }\\
    \subfloat{
        \centering
        \includegraphics[width=\linewidth]{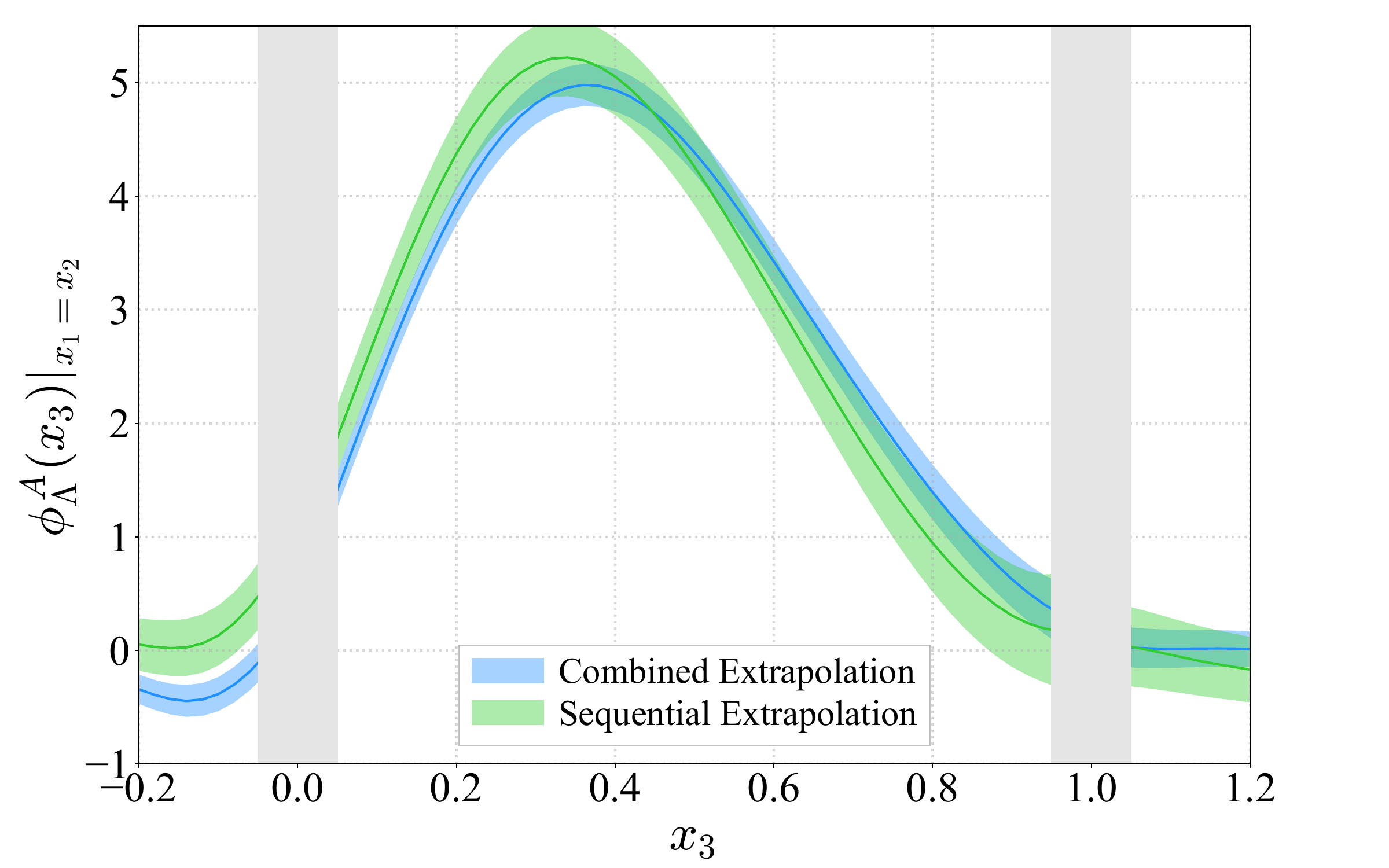}
        }
    \caption{comparison between the results from combined and sequential extrapolation, with  $x_2=0.2$ and $x_1=x_2$. The difference generally remaining within one standard deviation for most of the physical momentum-fraction region.}
    \label{fig:systematic_seq_extra}
\end{figure}

The comparison between the combined and sequential extrapolations is shown in Fig.~\ref{fig:systematic_seq_extra}. The two procedures give very similar results over most of the physical momentum-fraction region, with the difference generally remaining within one statistical standard deviation. This agreement indicates that the final physical LCDAs are stable against the ordering of the continuum, physical-pion-mass, and infinite-momentum extrapolations, and supports the reliability of the combined extrapolation used in the central analysis. The residual difference between the two procedures is taken as the systematic uncertainty associated with the physical-limit extrapolation and is included in the final uncertainty budget discussed in the next subsection.

\subsection{Systematic Uncertainties}
\label{sec:systematics}
In this work, we estimate the systematic uncertainties directly associated with the construction of the physical LCDAs from the lattice quasi-DAs. After hybrid renormalization, large-$\lambda$ extrapolation, Fourier transform, perturbative matching, and extrapolation to the physical limit, the final result can still depend on several analysis choices made along the pipeline. We focus on three dominant sources that can be quantified by repeating the analysis with alternative but reasonable prescriptions: the renormalization and matching scale dependence, the coordinate-space large-$\lambda$ extrapolation, and the extrapolation to the continuum, physical-pion-mass, and infinite-momentum limits. The central result is obtained with the scale $\mu=2~{\rm GeV}$, the $\mathrm{NLA}$-ansatz large-$\lambda$ extrapolation when available, and the combined extrapolation in $a$, $m_\pi$, and $P^z$ in Eq~\eqref{eq:joint_extrap}. The corresponding variations are evaluated point by point in momentum-fraction space and are included as systematic uncertainties.

\subsubsection{Renormalization and Matching Scale Dependence} \label{sec:systematics_mu}

The first source of systematic uncertainty comes from the residual scale dependence associated with the perturbative treatment. In this work, the renormalization scale $\mu_R$ entering the hybrid renormalization procedure through the perturbative expressions $\widehat M_{\rm \overline{MS},pert}^{(1),X}$, and the factorization scale $\mu_F$ entering the one-loop LaMET matching, are chosen to be the same for practical implementation, $\mu=\mu_R=\mu_F$. At finite perturbative order, the final LCDAs retain a residual dependence on this scale. This dependence originates from both the conversion of the hybrid-renormalized matrix elements to the $\overline{\rm MS}$ scheme and the truncation of the LaMET matching kernel at next-to-leading order. It would be cancelled by higher-order contributions in an all-order calculation and is therefore used to estimate the perturbative-treatment-related systematic uncertainty in this work.

We take $\mu_0=2~{\rm GeV}$ as the central scale and repeat the complete analysis at two additional scales, $\mu=\sqrt{2}~{\rm GeV}$ and $\mu=2\sqrt{2}~{\rm GeV}$. At each momentum-fraction point $(x_1,x_2)$, the systematic uncertainty from the scale dependence is defined as:
\begin{equation}
    \delta^\mu_{\rm sys}(x_1,x_2)
    =
    \max_{\mu=\sqrt{2}, 2\sqrt{2}~{\rm GeV}}
    \left|
    \phi(x_1,x_2;\mu)-\phi(x_1,x_2;\mu_0)
    \right|\ ,
\end{equation}
here $\phi(x_1,x_2;\mu)$ denotes the final LCDA obtained with the corresponding value of the renormalization and matching scale, while all other analysis choices are kept the same as in the central analysis.

\begin{figure}[htbp]
    \centering
    \subfloat{
        \centering
        \includegraphics[width=\linewidth]{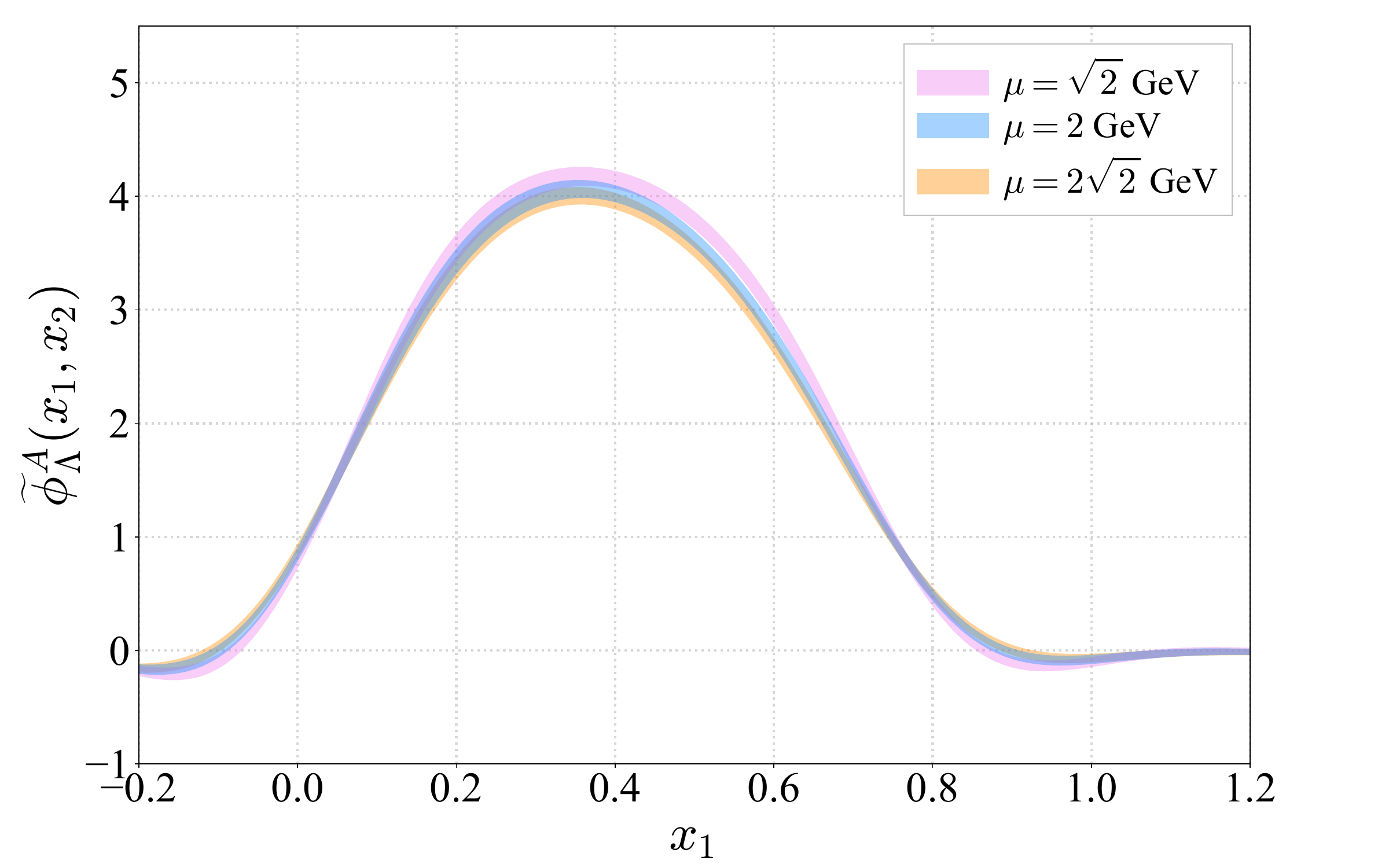}
        }\\
    \subfloat{
        \centering
        \includegraphics[width=\linewidth]{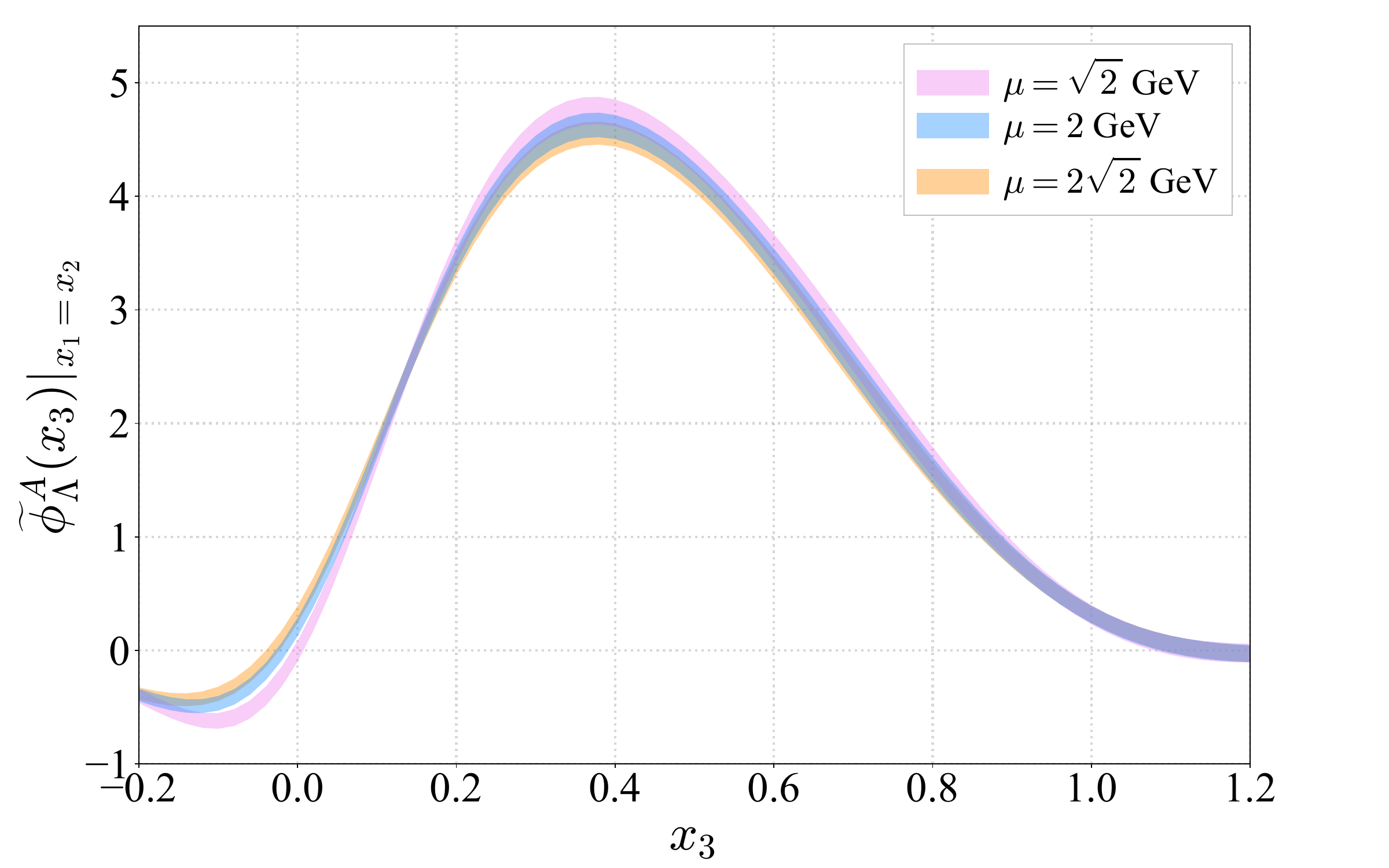}
        }
    \caption{Scale dependence of the $\Lambda$ $A$ quasi-DA on H48P32 with $P^z\approx3~{\rm GeV}$. The three curves correspond to the results obtained with $\mu=\sqrt{2}~{\rm GeV}$, $2~{\rm GeV}$, and $2\sqrt{2}~{\rm GeV}$, while all other analysis choices are kept fixed.}
    \label{fig:system_mu_quasi}
\end{figure}

\begin{figure}[htbp]
    \centering
    \subfloat{
        \centering
        \includegraphics[width=\linewidth]{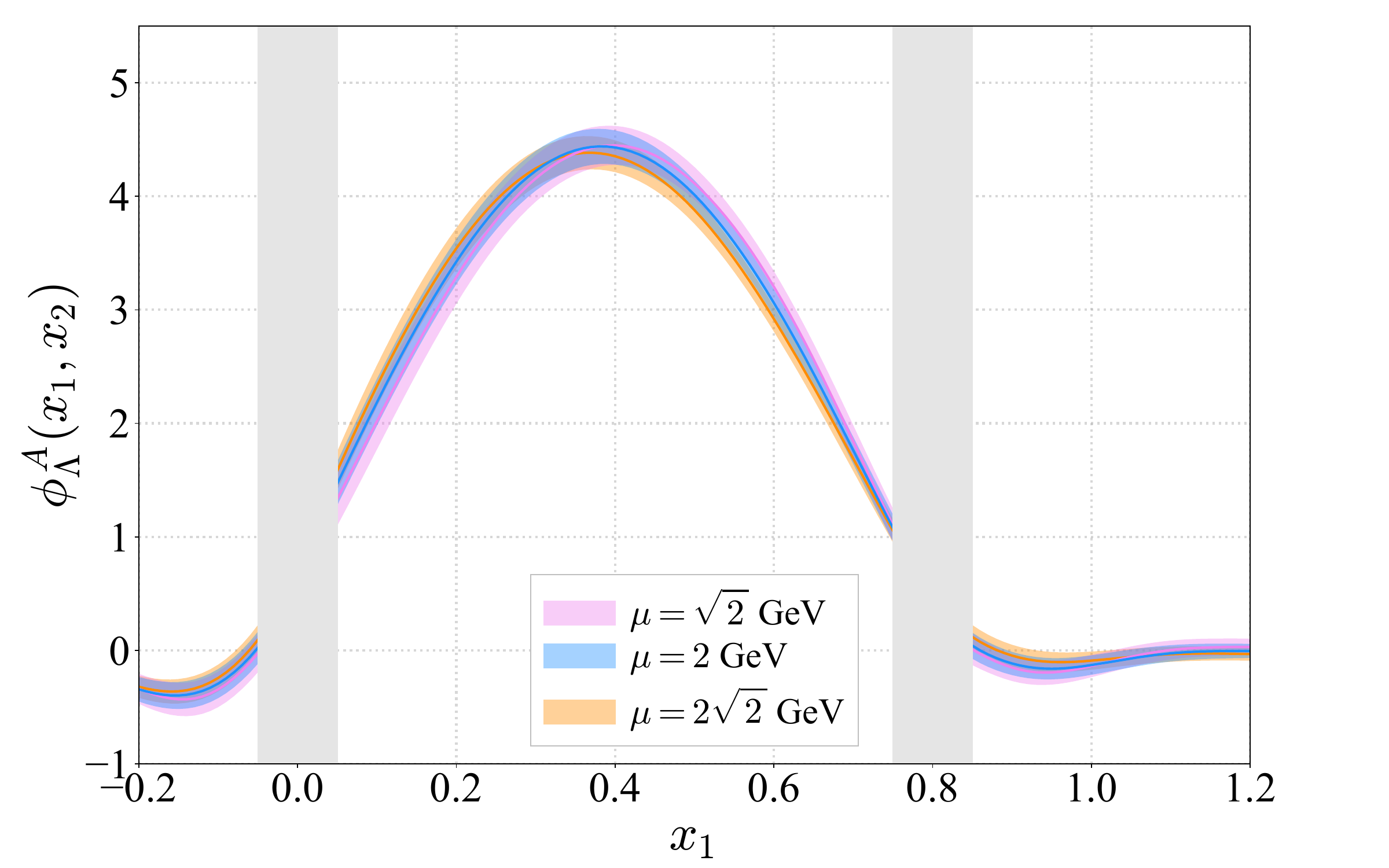}
        }\\
    \subfloat{
        \centering
        \includegraphics[width=\linewidth]{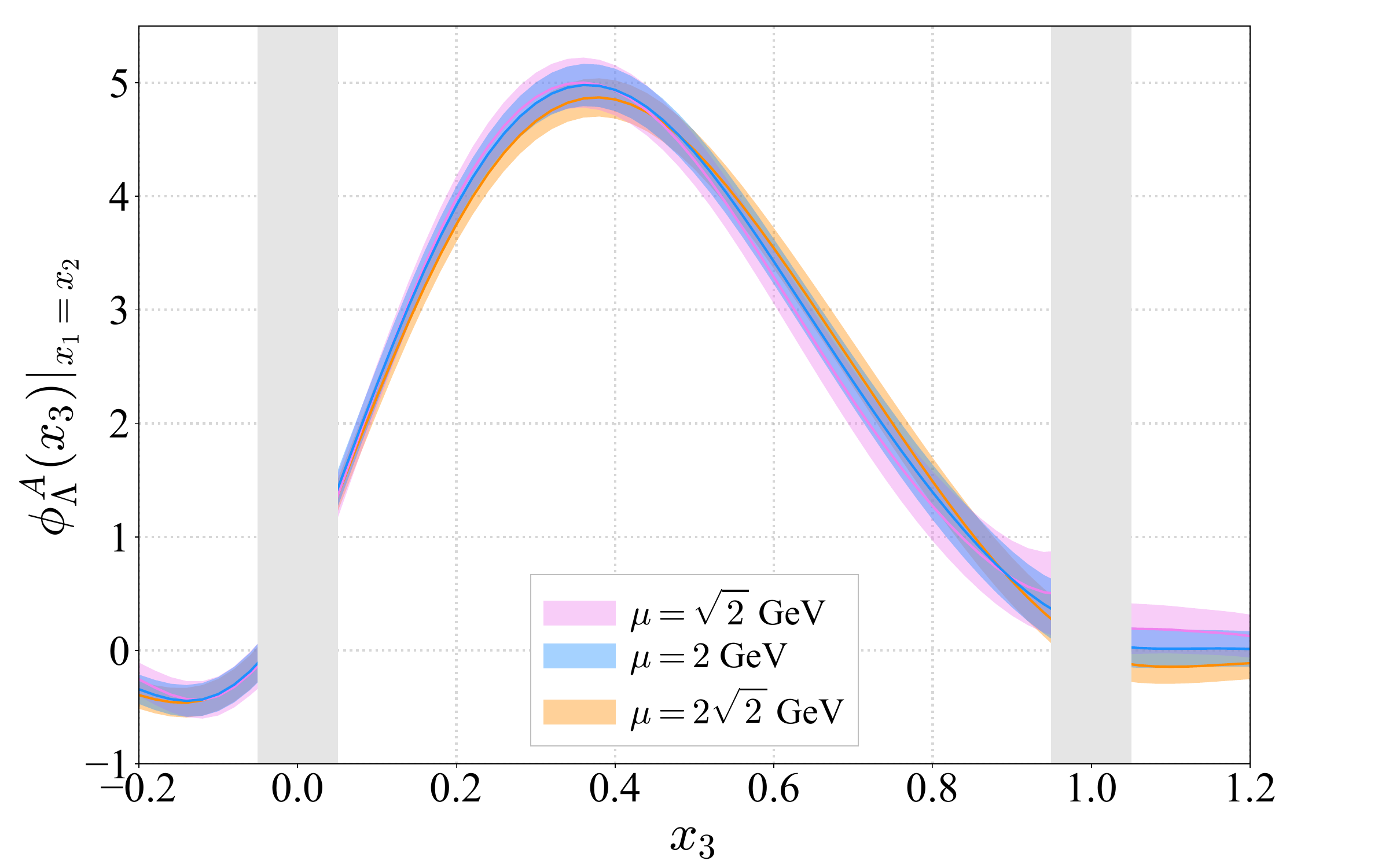}
        }
    \caption{Scale dependence of the final $\Lambda$ $A$ LCDA. The three curves correspond to the results obtained with $\mu=\sqrt{2}~{\rm GeV}$, $2~{\rm GeV}$, and $2\sqrt{2}~{\rm GeV}$, while all other analysis choices are kept fixed. The spread with respect to the central scale $\mu_0=2~{\rm GeV}$ is used to estimate the systematic uncertainty from the renormalization and matching scale dependence.}
    \label{fig:system_mu}
\end{figure}

We first illustrate the scale dependence at an intermediate stage in Fig.~\ref{fig:system_mu_quasi}, where the hybrid-renormalized quasi-DAs on the H48P32 ensemble at $P^z\approx3~{\rm GeV}$ is compared for the three scale choices. This comparison shows the residual scale dependence before the LaMET matching and final physical-limit extrapolation. The scale dependence of the final extrapolated LCDAs is shown in Fig.~\ref{fig:system_mu}. Over most of the physical momentum-fraction region, the variation with $\mu$ is moderate compared with the central distribution. The observed spread among the three scales is included point by point in the final systematic error budget.

\subsubsection{Large-\texorpdfstring{$\lambda$}{lambda} Extrapolation Uncertainty}

The second source of systematic uncertainty is associated with the large-$\lambda$ extrapolation of the coordinate-space quasi-DAs. Since the lattice data are available only in a finite coordinate-space region, the Fourier transform to momentum space requires an assumption about the large-distance behavior of the matrix elements. The corresponding uncertainty is expected to be most pronounced near the momentum-fraction endpoints, but it can also propagate into the interior region through the Fourier reconstruction, normalization, and matching procedure. We therefore retain this uncertainty over the full physical momentum-fraction domain.

To estimate this uncertainty, we follow the prescription proposed in Ref.~\cite{Ji:2026vir} for the large-$\lambda$ extrapolation adopted in this work. For the central analysis of the $\Lambda$ $A$ amplitude, we use the next-to-leading-asymptotic ($\mathrm{NLA}$) ansatz. We then repeat the full analysis using the leading-asymptotic ($\mathrm{LA}$) ansatz, while keeping all other analysis choices fixed. The difference between the two resulting LCDAs is taken as the systematic uncertainty from the large-$\lambda$ extrapolation:
\begin{equation}
    \delta_{\rm sys}^{\lambda}(x_1,x_2) = \left| \phi^{\rm NLA}(x_1,x_2) - \phi^{\rm LA}(x_1,x_2) \right|\ ,
\end{equation}
here $\phi^{\mathrm{NLA}}$ and $\phi^{\mathrm{LA}}$ denote the final LCDAs obtained using the $\mathrm{NLA}$ and $\mathrm{LA}$ coordinate-space extrapolations, respectively.

\begin{figure}[htbp]
    \centering
    \subfloat{
        \centering
        \includegraphics[width=\linewidth]{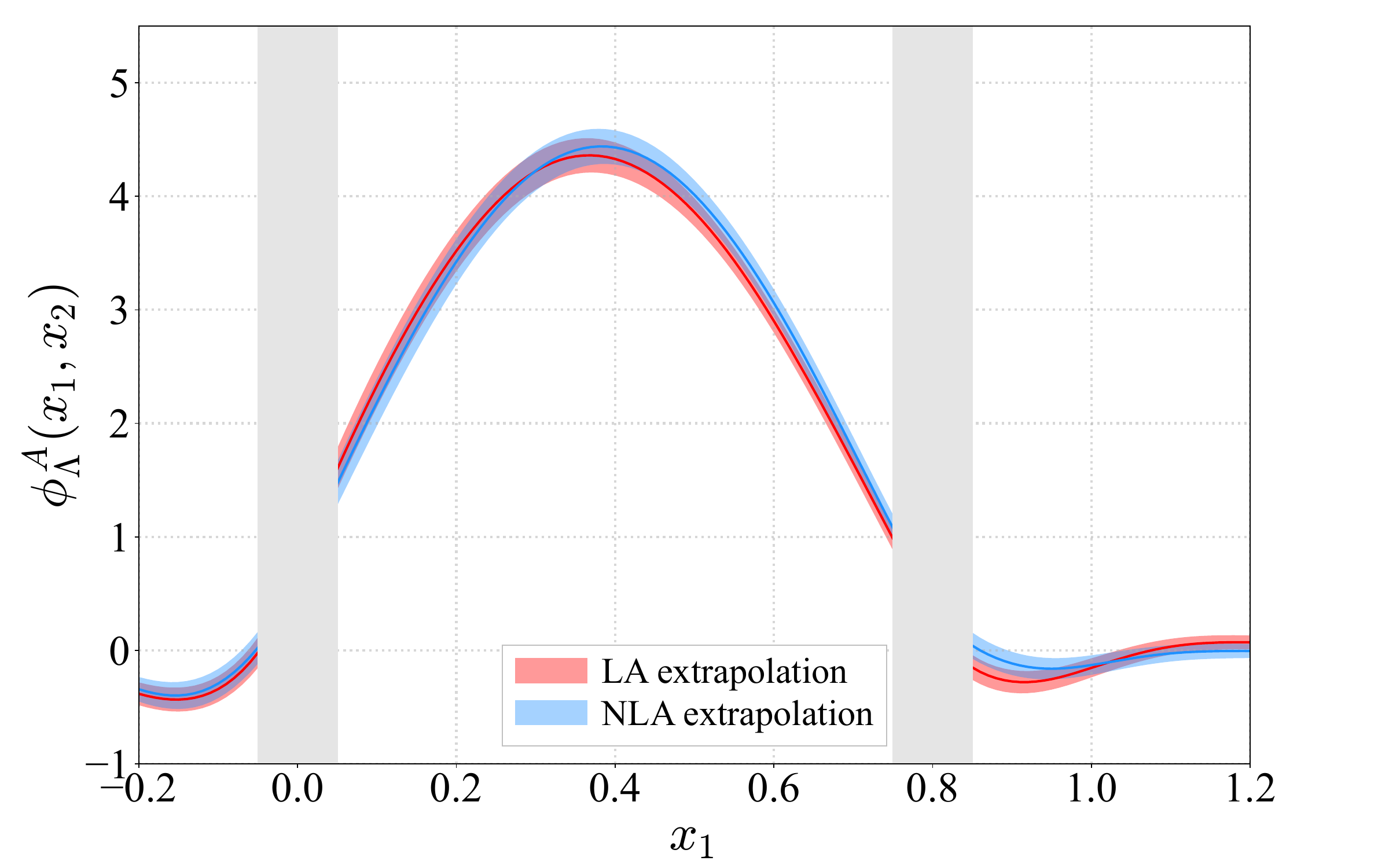}
        }\\
    \subfloat{
        \centering
        \includegraphics[width=\linewidth]{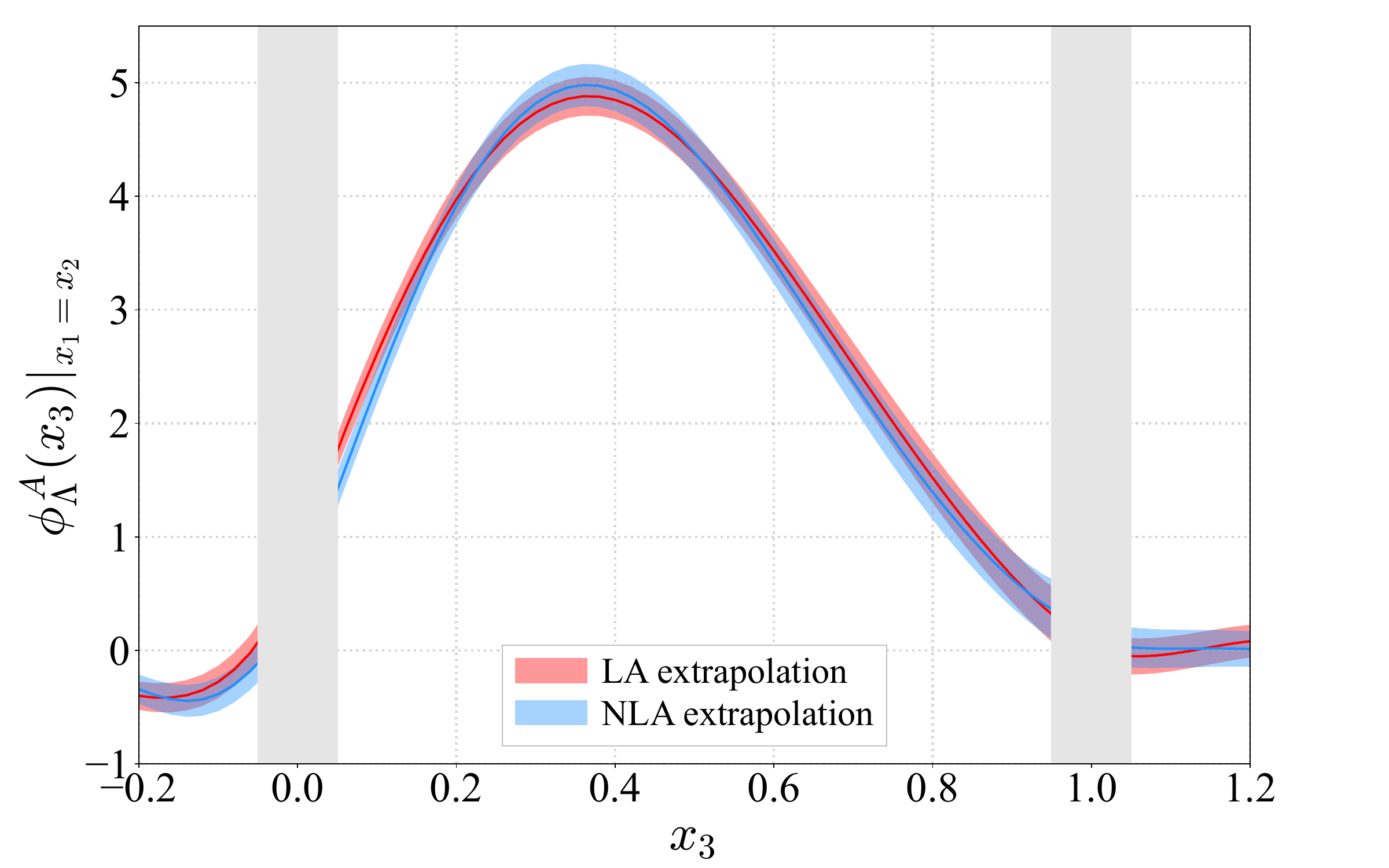}
        }
    \caption{Comparison of the final $\Lambda$ $A$ LCDA obtained using the $\mathrm{LA}$ and $\mathrm{NLA}$ large-$\lambda$ extrapolation forms, with $x_2=0.2$ and $x_1=x_2$.}
    \label{fig:systematic_LA_extrap}
\end{figure}

The comparison between the $\mathrm{LA}$ and $\mathrm{NLA}$ extrapolations is shown in Fig.~\ref{fig:systematic_LA_extrap}. The two results agree well over most of the physical momentum-fraction region, while visible differences remain near the endpoint regions and in regions affected by endpoint reconstruction. This behavior is consistent with the expectation that the coordinate-space tail mainly affects the endpoint behavior after Fourier transform.

In the present analysis, the asymptotic forms used for the large-$\lambda$ extrapolation retain the lowest heavy--light mesonic intermediate states with binding energies $\Lambda^{0^-}$ and $\Lambda^{0^+}$, as discussed in Sec.~\ref{sec:extrapolation}. The general asymptotic expansion can also receive contributions from baryonic intermediate states. A full determination of the corresponding baryonic binding energies would require more precise data at large spatial separations. Given the present statistical precision, including additional baryonic intermediate states would introduce poorly constrained parameters and is therefore not included in the central analysis.

For the $V$ and $T$ amplitudes, the large-distance signals are less precise than that of the $A$ amplitude, and stable $\mathrm{NLA}$ fits cannot be obtained with the current data. We therefore use the $\mathrm{LA}$ form for these two amplitudes and do not assign an additional $\mathrm{LA}$--$\mathrm{NLA}$ variation to them. For the $A$ amplitude, where both $\mathrm{LA}$ and $\mathrm{NLA}$ fits are stable, the $\mathrm{NLA}$ result is taken as the central value, and the difference between the $\mathrm{LA}$ and $\mathrm{NLA}$ results is included as the systematic uncertainty from the large-$\lambda$ extrapolation in the final systematic error budget.

\subsubsection{Physical-limit Extrapolation Uncertainty}

The third source of systematic uncertainty is associated with the extrapolation to the continuum, physical-pion-mass, and infinite-momentum limits. Our central analysis uses the combined extrapolation in Eq.~\eqref{eq:joint_extrap}, where the dependences on $a$, $m_\pi$, and $P^z$ are fitted simultaneously.

To estimate the systematic uncertainty associated with this extrapolation ansatz, we use the sequential extrapolation described in Sec.~\ref{sec:sequential_extrap} as an alternative analysis. In this procedure, the continuum and physical-pion-mass extrapolations are first performed at fixed $P^z$, and the resulting distributions are then extrapolated to the infinite-momentum limit. The difference between the combined and sequential extrapolations is taken as the systematic uncertainty from the physical-limit  extrapolation:
\begin{equation}
    \delta_{\rm sys}^{a,m_\pi,P^z}(x_1,x_2) = \left| \phi^{\rm comb}(x_1,x_2) - \phi^{\rm seq}(x_1,x_2) \right|\ ,
\end{equation}
here $\phi^{\rm comb}$ denotes the result from the combined extrapolation, while $\phi^{\rm seq}$ denotes the result from the sequential extrapolation.

This comparison probes the stability of the final LCDAs against the ordering of the extrapolation limits and the truncation of the extrapolation ansatz. In particular, it estimates the possible impact from higher-order terms not included in Eq.~\eqref{eq:joint_extrap}, such as $a^4$, higher-order pion-mass dependence, higher-power finite-momentum corrections, and additional mixed terms. As shown in Fig.~\ref{fig:systematic_seq_extra}, the two procedures give consistent results over most of the momentum-fraction region. The residual difference is included point by point in the final systematic error budget.

\subsubsection{Total Error Budget}

The central value of the LCDAs is defined by the central analysis setup: the scale $\mu=2~{\rm GeV}$, the $\mathrm{NLA}$ large-$\lambda$ extrapolation when available, and the combined extrapolation in $a$, $m_\pi$, and $P^z$. The total uncertainty is evaluated at each point in momentum-fraction space. It includes the statistical uncertainty from the jackknife analysis and the systematic uncertainties from the scale variation, the large-$\lambda$ extrapolation, and the physical-limit extrapolation procedure.

Since these systematic uncertainties are estimated from different analysis variations and their mutual correlations are not resolved in the present analysis, we combine them in quadrature. The total uncertainty is therefore defined as:
\begin{equation}
\begin{aligned}
    &\ \delta_{\rm total}(x_1,x_2) \\
    =&\ \sqrt{ [\delta_{\rm stat}(x_1,x_2)]^2 + [\delta_{\rm sys}(x_1,x_2)]^2 }\\
    =&\ \sqrt{(\delta_{\rm stat})^2 + (\delta^\mu_{\rm sys})^2 + (\delta^\lambda_{\rm sys})^2 + (\delta^{a,m_\pi,P^z}_{\rm sys})^2 }\ .
\end{aligned}
\end{equation}
Here $\delta_{\rm stat}$ denotes the statistical uncertainty, while $\delta^\mu_{\rm sys}$, $\delta^\lambda_{\rm sys}$, and $\delta^{a,m_\pi,P^z}_{\rm sys}$ denote the systematic uncertainties defined in the previous subsections.

\begin{figure}[htbp]
    \centering
    \includegraphics[width=\linewidth]{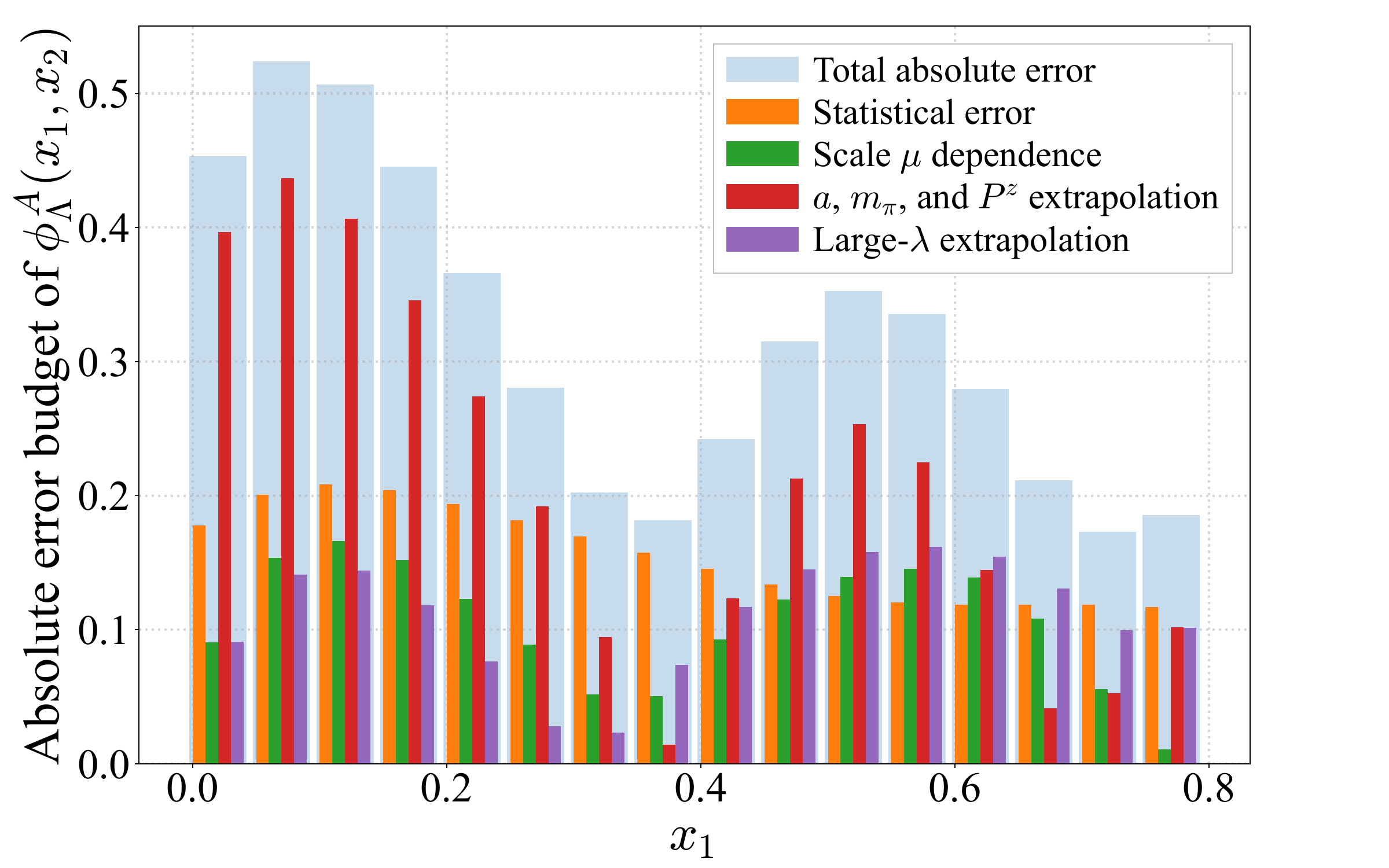}
    \caption{Absolute uncertainty budget for the final $\Lambda$ $A$ LCDA. The total absolute uncertainty is obtained by combining the statistical error, the renormalization and matching scale dependence, the large-$\lambda$ extrapolation uncertainty, and the physical-limit extrapolation uncertainty in quadrature.
    }
    \label{fig:total_error_budget}
\end{figure}

The decomposition of the total absolute uncertainty is shown in Fig.~\ref{fig:total_error_budget}. In the momentum-fraction region displayed here, the statistical uncertainty is not the dominant contribution, indicating that the final precision is mainly limited by systematic effects in the analysis pipeline for the $\Lambda$ $A$ amplitude. Among the systematic sources, the uncertainty from the physical-limit extrapolation in $a$, $m_\pi$, and $P^z$ gives one of the largest contributions over a broad range of momentum fraction $x$. This is consistent with the observation in Sec.~\ref{sec:extrap_phys} and Sec.~\ref{sec:sequential_extrap} that the finite-$P^z$ dependence remains visible even after perturbative matching, as expected for baryon observables where the achievable boost factor is limited by the relatively large baryon masses. The renormalization and matching scale dependence also gives a sizable contribution, reflecting the residual perturbative uncertainty from the one-loop hybrid-scheme conversion and LaMET matching. The large-$\lambda$ extrapolation uncertainty is mainly visible near the endpoint regions. Its effect in the central momentum-fraction region is smaller and more indirect, but it is included in the full error budget for completeness.

\subsection{Summary of the Numerical Demonstration}

We finally collect the representative output of the full analysis pipeline for the $\Lambda$-baryon $A$-structure LCDAs. The result is obtained after hybrid renormalization, large-$\lambda$ extrapolation, Fourier transform, hybrid-scheme perturbative LaMET matching, and the extrapolation to the continuum, physical-pion-mass, and infinite-momentum limits. The central analysis uses $\mu=2~{\rm GeV}$, the $\mathrm{NLA}$ large-$\lambda$ extrapolation, and the combined extrapolation in $a$, $m_\pi$, and $P^z$.

\begin{figure}[htbp]
    \centering
    \subfloat{
        \centering
        \includegraphics[width=\linewidth]{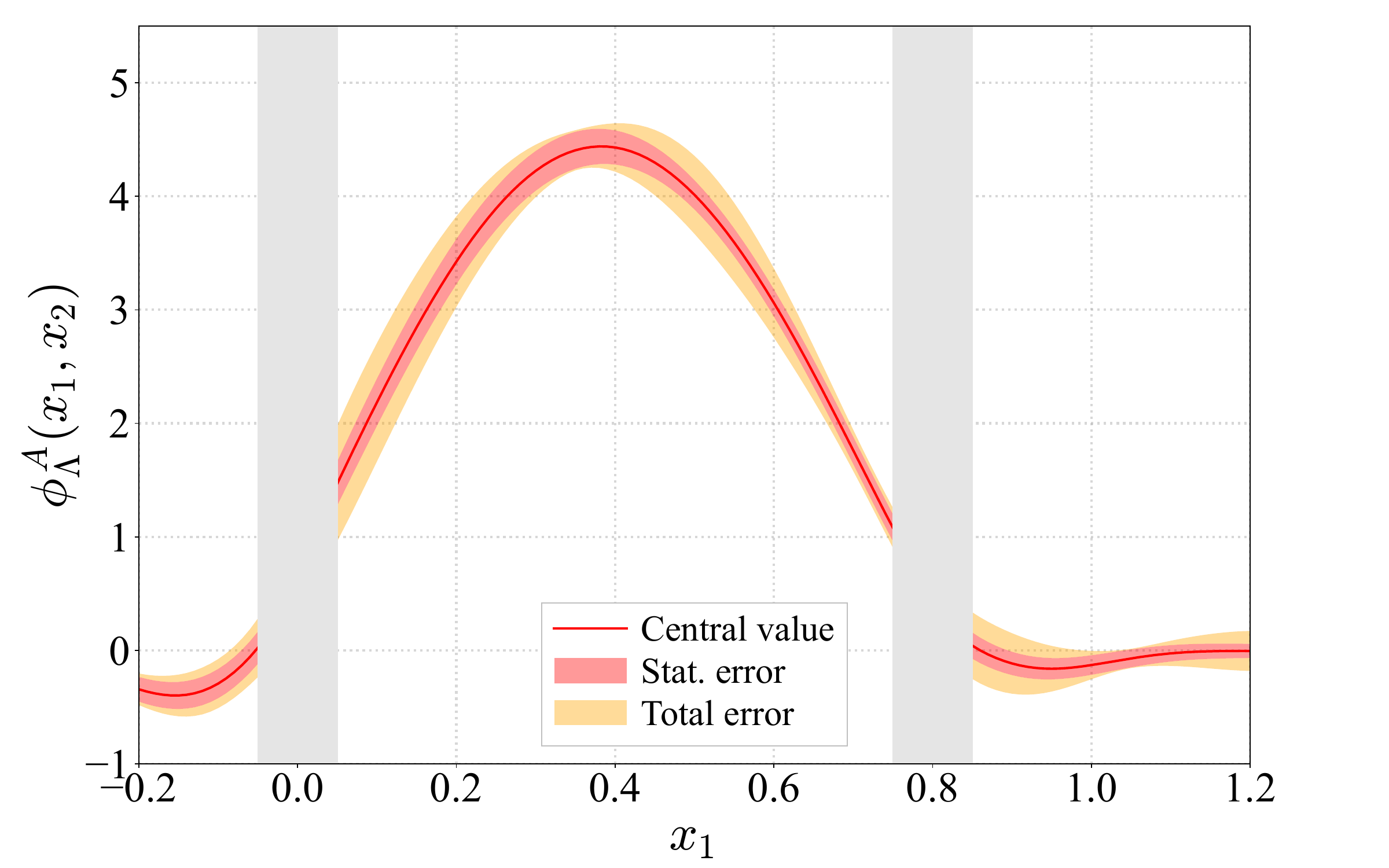}
        }\\
    \subfloat{
        \centering
        \includegraphics[width=\linewidth]{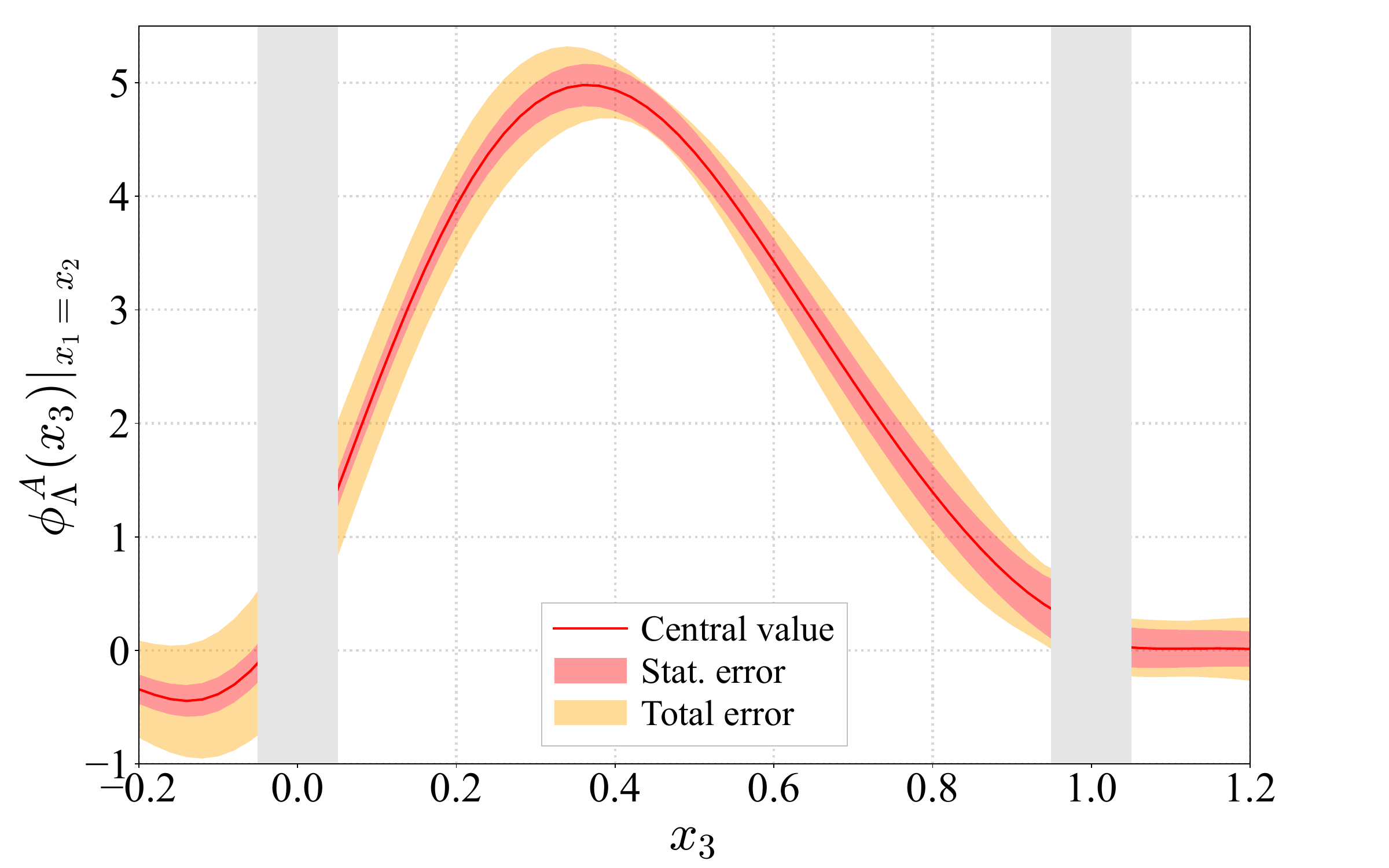}
        }
    \caption{Final central value, statistical uncertainty and systematic uncertainty of $\Lambda$ $A$ LCDA. The upper panel is shown with $x_2 =0.2$ and the lower panel is shown with $x_1=x_2$.}
    \label{fig:final_xcut}
\end{figure}

Fig.~\ref{fig:final_xcut} shows the final result on two representative one-dimensional cuts. The upper panel shows the dependence on $x_1$ at fixed $x_2=0.2$, while the lower panel shows the diagonal direction $x_1=x_2$ as a function of $x_3=1-x_1-x_2$. The gray bands mark the endpoint regions where LaMET power corrections are expected to be enhanced. In the central momentum-fraction region, the $\Lambda$ $A$ LCDA displays a single broad peak, and the total uncertainty is dominated by systematic effects rather than by the statistical error alone.

\begin{figure}[htbp]
    \centering
    \includegraphics[width=\linewidth]{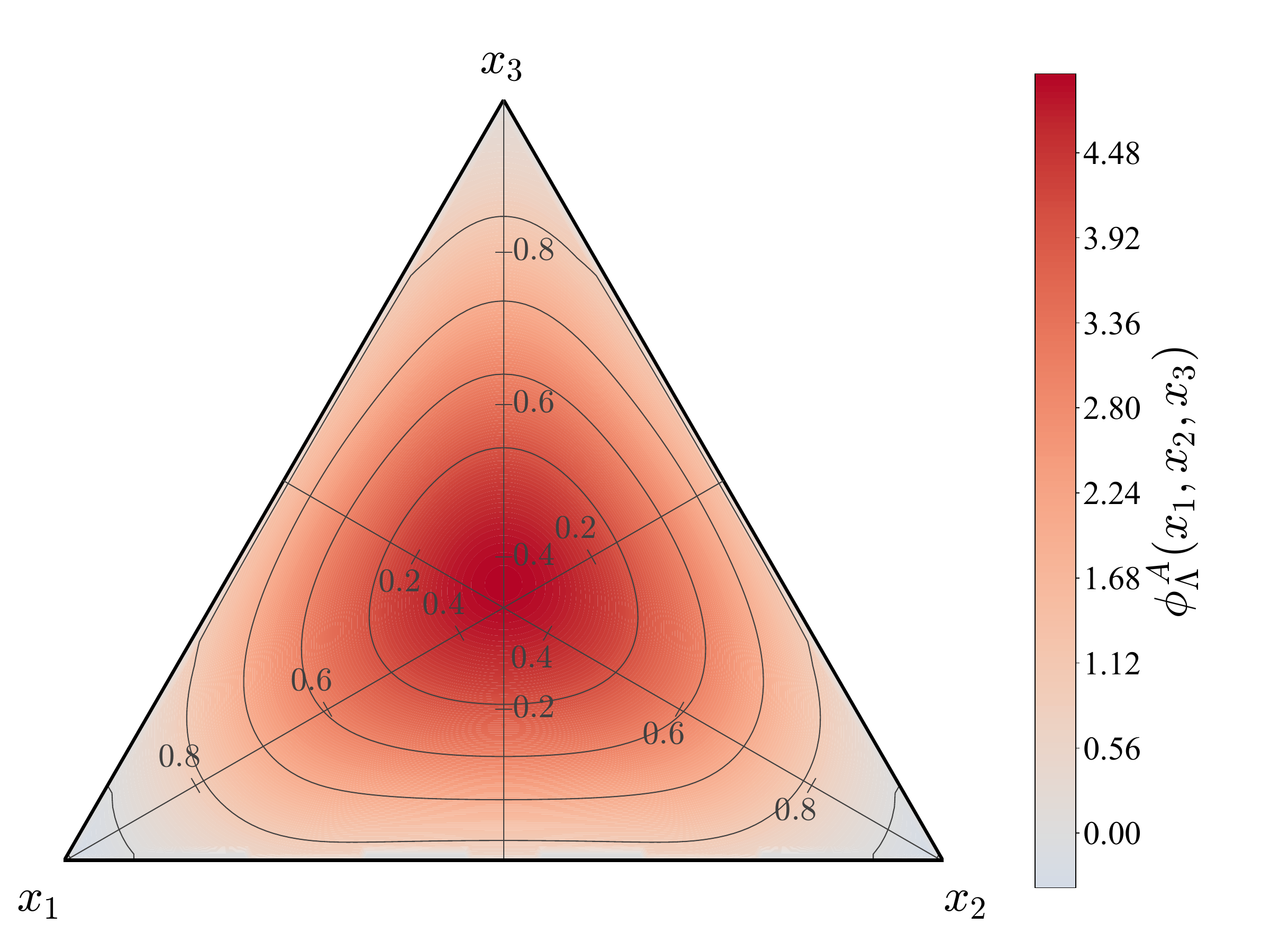}
    \caption{Central value of the extrapolated two-dimensional distribution $\Lambda$ $A$ LCDA in physical momentum-fraction triangular region $0\leq x_1,x_2,x_3\leq 1$.}
    \label{fig:final_heatmap}
\end{figure}

The same result is visualized as a two-dimensional distribution in Fig.~\ref{fig:final_heatmap}, over the physical momentum-fraction triangular region $0\leq x_1,x_2,x_3\leq 1$ with $x_1+x_2+x_3=1$. The heat map provides a compact visualization of the momentum-sharing pattern obtained from the complete pipeline. This numerical demonstration shows that the baryon-LaMET framework developed in this work can produce $x$-dependent baryon LCDAs with controlled uncertainty budget. 
The complete physical results for the $\Lambda$-baryon $V$, $A$, and $T$ LCDAs are presented in the companion Letter~\cite{LPC:2026lcj}, while the polynomial forms and explicit representations used for phenomenological parameterizations are summarized in Appendix~\ref{app:lambda_third_moments}.

\section{Summary}\label{sec:Summary}

In this work, we have developed a baryon-LaMET framework for determining leading-twist baryon light-cone distribution amplitudes from lattice QCD. Compared with meson LCDAs, baryon LCDAs are genuinely two-dimensional functions of the valence-quark momentum fractions and involve several independent structures already at leading twist. These features require a dedicated treatment of the operator construction, coordinate-space symmetries, renormalization, large-distance completion, and perturbative matching. We have formulated the equal-time baryon quasi-DAs for the $V$, $A$, and $T$ structures and analyzed their $z_1 \leftrightarrow z_2$ exchange and Hermiticity properties on the two-dimensional $(z_1,z_2)$ plane. These relations are essential for reducing the independent lattice data, choosing the appropriate normalization for amplitudes with vanishing local limits, and implementing the subsequent renormalization and extrapolation procedures.

A central component of this work is the hybrid renormalization prescription for baryon quasi-DAs. Because the nonlocal three-quark operator depends on the separations $z_1$, $z_2$, and $z_1-z_2$, the renormalization procedure must be defined on the full two-dimensional coordinate plane. We constructed a hybrid renormalization scheme that combines ratio prescription in the short-distance region and self-renormalization in the long-distance region, with boundary continuity prescriptions that connect the different regions smoothly. This construction removes the Wilson-line linear divergences while avoiding the uncontrollable infrared structure at long-distance. We also specified the structure-dependent implementation for the $V$, $A$, and $T$ amplitudes, including the use of appropriate symmetric reference matrix elements for antisymmetric amplitudes whose local limits vanish.

To perform the Fourier transform from coordinate space to momentum-fraction space, we developed a large-$\lambda$ extrapolation strategy for baryon quasi-DAs based on the asymptotic large-distance behavior of Euclidean correlators. The two-dimensional nature of the baryon matrix elements leads to several relevant large-distance variables, $|z_1|$, $|z_2|$, and $|z_1-z_2|$, and therefore requires different asymptotic forms in different coordinate-space regions. Using the asymptotic expansion of Euclidean correlators, we constructed leading- and next-to-leading-asymptotic ans\"atze and implemented them consistently with the symmetry requirements of the $V$, $A$, and $T$ structures. This provides a controlled coordinate-space completion before the Fourier transform and allows the associated uncertainty to be estimated.

We further derived and implemented the LaMET matching in the hybrid scheme. In momentum space, the standard $\overline{\rm MS}$ one-loop kernel contains ultraviolet tails originating from short-distance logarithms in coordinate space. In the hybrid scheme, these logarithms are removed by the short-distance ratio subtraction, which introduces additional perturbative counterterms in the matching kernel. Including these hybrid counterterms makes the two-dimensional double-plus distribution well defined and yields a matching kernel that can be consistently convoluted with the baryon quasi-DAs. We also described the coordinate-space implementation of these counterterms, which avoids introducing an additional two-dimensional convolution in the numerical analysis.

As an end-to-end numerical demonstration, we applied the full pipeline to the $\Lambda$-baryon $A$-structure LCDA using seven $N_f=2+1$ lattice ensembles. The analysis includes signal-improved baryon two-point correlations, hybrid renormalization, large-$\lambda$ extrapolation, Fourier transform, hybrid-scheme LaMET matching, and extrapolation to the continuum, physical-pion-mass, and infinite-momentum limits. We used an alternative sequential extrapolation as a stability check and estimated the dominant systematic uncertainties from the renormalization and matching scale dependence, the large-$\lambda$ extrapolation, and the physical-limit extrapolation. The resulting representative $\Lambda$-baryon $A$-structure LCDA demonstrates that the baryon-LaMET framework developed here can produce $x$-dependent baryon LCDAs with controlled uncertainty budget.

The framework and numerical strategy presented in this work provide the infrastructure for first-principles determinations of baryon LCDAs beyond local moments. The complete physical results for the leading-twist $\Lambda$ $V$, $A$, and $T$ LCDAs, together with their phenomenological implications, are presented in the companion Letter~\cite{LPC:2026lcj}. Future improvements can further reduce the dominant systematic uncertainties by using higher baryon momenta, finer lattice spacings, improved control of the large-distance coordinate-space behavior, and higher-order perturbative matching.

\section*{Acknowledgment} 
We thank Yu-Ming Wang, Jia Yu, Yushan Su for valuable discussions.
We thank the CLQCD collaboration for providing us the gauge configurations with dynamical fermions~\cite{CLQCD:2023sdb}, which are generated on the HPC Cluster of ITP-CAS, the Southern Nuclear Science Computing Center (SNSC), the Siyuan-1 cluster supported by the Center for High Performance Computing at Shanghai Jiao Tong University and the Dongjiang Yuan Intelligent Computing Center.

This work is supported in part by Natural Science Foundation of China under grant No.12575084, 12375069, 12575085, 12222503, 12293060, 12293062, 12435002, 12293065, 12047503, 12125503, 12305103, 12375080, 12275277 and 12435004. M.-H.~C. is supported by the National Science Centre (Poland) grant OPUS No.2021/43/B/ST2/00497. J.~H is supported by Natural Science Foundation of Guangdong Province under Grant No.2025A1515012199. Y.-B.~Y is supported by National Key R\&D Program of China No.2024YFE0109800 and Strategic Priority Research Program of Chinese Academy of Sciences, Grant No.YSBR-101. J.-H.~Z is supported by the Ministry of Science and Technology of China under Grant No.2024YFA1611004, and CUHK-Shenzhen under grant No.UDF01002851.

The computations in this work were run on the Siyuan-1 cluster supported by the Center for High Performance Computing at Shanghai Jiao Tong University, Southern Nuclear Science Computing Center (SNSC), and Advanced Computing East China Sub-center. The LQCD simulations were performed using the PyQUDA software suite~\cite{Jiang:2024lto} and QUDA~\cite{Clark:2009wm,Babich:2011np,Clark:2016rdz} through HIP programming model~\cite{Bi:2020wpt}.

\clearpage

\begin{appendix}
\begin{widetext}

\section{Details on Lattice Calculation of Two-point Correlations}

In this appendix we present the detailed structure of nonlocal two-point correlations used in calculating the baryon DAs, and demonstrate how to obtain their symmetry properties under $z_1 \leftrightarrow z_2$ exchange.

\subsection{Spinor Contraction of Two-point correlations}\label{app:def_intor_contract}

The source-side interpolators we used in calculation are defined as Eq.~\eqref{eq:inter_src}:
\begin{equation}
\begin{aligned}
    \mathcal O^{\rm src,KE}_\Lambda &= \epsilon_{ijk}\ \frac{1}{\sqrt6}\Big( 2 ( u^{i,\rm T}\ C\gamma^5\gamma^t\ d^j )\ s^k + ( u^{i,\rm T}\ C\gamma^5\gamma^t\ s^j )\ d^k + ( s^{i,\rm T}\ C\gamma^5\gamma^t\ d^j )\ u^k \Big)\ ,\\
    \mathcal O^{\rm src,KE}_{\rm p} &= \epsilon_{ijk}\ ( u^{i,\rm T}\ C\gamma^5\gamma^t\ d^j )\ u^k\ .
\end{aligned}
\end{equation}
Using $\overline\psi = \psi^\dagger \gamma^t$ and $\gamma^t (C\gamma^5\gamma^t)^\dagger \gamma^t = - C\gamma^5\gamma^t$, we have:
\begin{equation}\label{eq:inter_src_bar}
\begin{aligned}
    \overline{\mathcal O}^{\rm src,KE}_\Lambda &= -\epsilon_{ijk}\ \frac{1}{\sqrt6}\Big( 2\ \overline s^k\ ( \overline u^i\ C\gamma^5\gamma^t\ \overline d^{j,\rm T} ) + \overline d^k \ ( \overline u^i\ C\gamma^5\gamma^t\ \overline s^{j,\rm T} ) + \overline u^k\ ( \overline s^i\ C\gamma^5\gamma^t\ \overline d^{j,\rm T} ) \Big)\ ,\\
    \overline{\mathcal O}^{\rm src,KE}_{\rm p} &= -\epsilon_{ijk}\ \overline u^k\ ( \overline u^i\ C\gamma^5\gamma^t\ \overline d^{j,\rm T} ) \ .
\end{aligned}
\end{equation}

Contract the sink-side interpolators Eq.~\eqref{eq:inter_sink}, source-side interpolators Eq.~\eqref{eq:inter_src_bar}, and spinor projector $\mathbb T=\slashed{\bar n}=\gamma^t+\gamma^z$ Eq.~\eqref{eq:projector} to obtain the two-point correlations in coordinate space:
\begin{equation}\label{eq:contract_lambda}
\begin{aligned}
    C_{\Lambda,\rm 2pt}(x|x_0) =&\ \frac{\epsilon_{ijm} \epsilon_{k\ell n}}{\sqrt6} (\Gamma^{\rm snk}_1)_{\alpha\beta} \left( C\gamma^5\gamma^t \right)_{\alpha'\beta'} (\mathbb T \Gamma^{\rm snk}_2 )_{\gamma'\gamma} \\
    &\times \Big( 2\ D_u^{-1}(x|x_0)_{\alpha\alpha'}^{ik} D_d^{-1}(x|x_0)_{\beta\beta'}^{j\ell} D_s^{-1}(x|x_0)_{\gamma\gamma'}^{mn}\\
    &\quad + D_u^{-1}(x|x_0)_{\alpha\alpha'}^{ik} D_d^{-1}(x|x_0)_{\beta\gamma'}^{j\ell} D_s^{-1}(x|x_0)_{\gamma\beta'}^{mn}\\
    &\quad + D_u^{-1}(x|x_0)_{\alpha\gamma'}^{ik} D_d^{-1}(x|x_0)_{\beta\beta'}^{j\ell} D_s^{-1}(x|x_0)_{\gamma\alpha'}^{mn} \Big)\ ,
\end{aligned}
\end{equation}
\begin{equation}\label{eq:contract_proton}
\begin{aligned}
    C_{\rm p,2pt}(x|x_0) =&\ \epsilon_{ijm} \epsilon_{k\ell n} (\Gamma^{\rm snk}_1)_{\alpha\beta} \left( C\gamma^5\gamma^t \right)_{\alpha'\beta'} (\mathbb T \Gamma^{\rm snk}_2 )_{\gamma'\gamma} D_d^{-1}(x|x_0)_{\gamma\beta'}^{mn}\\
    &\times \Big(  D_u^{-1}(x|x_0)_{\alpha\alpha'}^{ik} D_u^{-1}(x|x_0)_{\beta\gamma'}^{j\ell} + D_u^{-1}(x|x_0)_{\alpha\gamma'}^{ik} D_u^{-1}(x|x_0)_{\beta\alpha'}^{j\ell} \Big)\ ,
\end{aligned}
\end{equation}
in which the Dirac-$\gamma$ structures $\Gamma^{\rm snk}_1$ and $\Gamma^{\rm snk}_2$ enter the sink-side interpolators, as well as the corresponding symmetry properties are collected in Table~\ref{tab:gamma_2pt}, where ``\rm{Sym}'' represents the symmetric behavior under the $z_1\leftrightarrow z_2$ exchange, while ``\rm{A-Sym}'' represents the antisymmetric behavior.

\begin{table*}[htbp]
    \centering
    \renewcommand{\arraystretch}{1.5}
    \setlength{\tabcolsep}{3mm}
    \begin{tabular}{c c c c}
        \hline\hline
         & $V$ & $A$ & $T$  \\
         \hline
        $\Gamma^{\rm snk}_1$ & $C  \gamma^t$ & $C \gamma^5 \gamma^t$ & $C [\gamma^t, \gamma^{x,y}] /2$   \\

        $\Gamma^{\rm snk}_2$ & $\gamma^5$ & $I$ & $\gamma^5 \gamma_{x,y}$   \\

        \multirow{2}{*}{$z_1\leftrightarrow z_2$}
        
        & \rm{A-Sym} for $\Lambda$ & \rm{Sym} for $\Lambda$ & \rm{A-Sym} for $\Lambda$   \\

         & \rm{Sym} for proton & \rm{A-Sym} for proton & \rm{Sym} for proton   \\
        \hline
    \end{tabular}
    \caption{Dirac-$\gamma$ structures and symmetric properties of nonlocal two-point correlations}  
    \label{tab:gamma_2pt}
\end{table*}

\subsection{Derivation of Symmetry Properties of Baryon DAs}\label{app:operator_exchange_symmetry}

In Sec.~\ref{sec:symmetries}, we stated that the $\Lambda$ and proton quasi-DAs possess definite symmetry or antisymmetry under the exchange $z_1\leftrightarrow z_2$. In this subsection, we show that these properties follow directly from the propagator-level structure of the nonlocal two-point correlation functions defined above, together with the isospin symmetry of $N_f=2+1$ lattice QCD ensembles.

\subsubsection{Nonlocal Two-point Correlation of \texorpdfstring{$\Lambda$}{Lambda}-baryon}

From the contraction Eq.~\eqref{eq:contract_lambda} for the $\Lambda$ baryon in Appendix~\ref{app:def_intor_contract}, the nonlocal two-point correlations have the general structure
\begin{equation}\label{eq:lambda_2pt_sym}
\begin{aligned}
    C_{\Lambda,\rm 2pt}(z_1,z_2) = &\ \frac{\epsilon_{ijm}\epsilon_{k\ell n}}{\sqrt{6}} (\Gamma^{\rm snk}_1)_{\alpha\beta} (C\gamma^5\gamma^t)_{\alpha'\beta'} (\mathbb T \Gamma^{\rm snk}_2)_{\gamma'\gamma} \\
    &\times \Big( 2\ D_l^{-1}(z_1)^{ik}_{\alpha\alpha'} D_l^{-1}(z_2)^{j\ell}_{\beta\beta'} D_s^{-1}(0)^{mn}_{\gamma\gamma'}\\
    &\qquad + D_l^{-1}(z_1)^{ik}_{\alpha\alpha'} D_l^{-1}(z_2)^{jn}_{\beta\gamma'} D_s^{-1}(0)^{m\ell}_{\gamma\beta'}\\
    &\qquad + D_l^{-1}(z_1)^{in}_{\alpha\gamma'} D_l^{-1}(z_2)^{j\ell}_{\beta\beta'} D_s^{-1}(0)^{mk}_{\gamma\alpha'} \Big)\\
    \equiv&\ \frac{1}{\sqrt6}\Big[ 2 \mathscr D_1(z_1,z_2) + \mathscr D_2(z_1,z_2) + \mathscr D_3(z_1,z_2)\Big]\ ,
\end{aligned}
\end{equation}
where we have used the isospin symmetry $D_l^{-1}\equiv D_u^{-1}= D_d^{-1}$ on $N_f=2+1$ ensembles.
The key observation is that in all three contraction terms $\mathscr D_{1,2,3}$, the $z_1$ dependence enters only through the first light-quark propagator $D_l^{-1}(z_1)$, and the $z_2$ dependence only through the second light-quark propagator $D_l^{-1}(z_2)$. The $s$-quark propagator $D_s^{-1}(0)$ and the spinor-projector $\mathbb T$ contributions are independent of both $z_1$ and $z_2$.

Taking the first contraction term $\mathscr D_1(z_1,z_2)$ in Eq.~\eqref{eq:lambda_2pt_sym} as a representative example, after the $z_1\leftrightarrow z_2$ exchange we have:
\begin{equation}
    \mathscr D_1(z_2,z_1) = \epsilon^{ijm}\epsilon^{k\ell n} (\Gamma^{\rm snk}_1)_{\alpha\beta} (C\gamma^5\gamma^t)_{\alpha'\beta'} D_l^{-1}(z_2)^{ik}_{\alpha\alpha'} D_l^{-1}(z_1)^{j\ell}_{\beta\beta'}\times \big[\mathbb T \Gamma^{\rm snk}_2 D_s^{-1}(0) \big]^{mn}\ .
\end{equation}
We now relabel the dummy summation indices: $i\leftrightarrow j$, $k\leftrightarrow \ell$, $\alpha\leftrightarrow\beta$, $\alpha'\leftrightarrow\beta'$.  The two antisymmetric tensors each pick up a sign: $\epsilon^{jim}=-\epsilon^{ijm}$ and $\epsilon^{\ell kn}=-\epsilon^{k\ell n}$, so the overall sign from the two $\epsilon$-tensors is $(-1)^2=+1$. The propagators return to their original forms $D_l^{-1}(z_1)^{ik}_{\alpha\alpha'}$ and $D_l^{-1}(z_2)^{j\ell}_{\beta\beta'}$, and the only change is
\begin{equation}
    (\Gamma^{\rm snk}_1)_{\alpha\beta}\ (C\gamma^5\gamma^t)_{\alpha'\beta'} \ \longrightarrow\  (\Gamma^{\rm snk}_1)_{\beta\alpha}\ (C\gamma^5\gamma^t)_{\beta'\alpha'}\ .
\end{equation}
Defining the transposition sign of a Dirac-$\gamma$ structure $\Gamma$ through $\Gamma_{\beta\alpha}=\eta_\Gamma \Gamma_{\alpha\beta}$, we obtain for the $\mathscr D_1$ contraction term:
\begin{equation}
    \mathscr D_1(z_2,z_1) = \eta_{\Gamma^{\rm snk}_1} \eta_{C\gamma^5\gamma^t} \mathscr D_1(z_1,z_2)\ .
\end{equation}
The same argument applies to the second and third contraction terms $\mathscr D_2$ and $\mathscr D_3$: in each case, after exchanging $z_1\leftrightarrow z_2$ and relabeling $\{i,k,\alpha,\alpha'\}\leftrightarrow\{j,\ell,\beta,\beta'\}$, the only effect is the transposition of $\Gamma^{\rm snk}_1$ and $C\gamma^5\gamma^t$.  Hence, the result holds for the full two-point correlation:
\begin{equation}\label{eq:lambda_sym_final}
    C_{\Lambda,\rm 2pt}(z_2,z_1) = \eta_{\Gamma^{\rm snk}_1} \eta_{C\gamma^5\gamma^t} C_{\Lambda,\rm 2pt}(z_1,z_2)\ .
\end{equation}

The transposition properties of the relevant Dirac-$\gamma$ structures follow from the identity $C\gamma^\mu C^{-1}=-(\gamma^\mu)^T$ and are listed in Table~\ref{tab:transpose_gamma}.

\begin{table}[htbp]
    \centering
    \renewcommand{\arraystretch}{1.5}
    \setlength{\tabcolsep}{4mm}
    \begin{tabular}{c c c}
        \hline\hline
        $\Gamma$ & $\Gamma^T$ & $\eta_\Gamma$  \\
        \hline
        $C\gamma^\mu$ & $+C\gamma^\mu$ & $+1$  \\
        $C\gamma^5\gamma^\mu$ & $-C\gamma^5\gamma^\mu$ & $-1$  \\
        $C\sigma^{\mu\nu}$ & $+C\sigma^{\mu\nu}$ & $+1$  \\
        $C\gamma^5$ & $-C\gamma^5$ & $-1$  \\
        \hline
    \end{tabular}
    \caption{Transposition properties of Dirac-$\gamma$ structures may be used in baryon interpolators.}
    \label{tab:transpose_gamma}
\end{table}

Applying these to the three leading-twist structures defined in Eq.~\eqref{eq:quasi_matrix}, with the chosen kinematically-enhanced source-side structure $C\gamma^5\gamma^t$ ($\eta_{C\gamma^5\gamma^t}=-1$), we could obtain the symmetric properties of the nonlocal two-point correlations og $\Lambda$-baryon under the exchange of $z_1\leftrightarrow z_2$, listed as:

\begin{itemize}
    \item $\Lambda$-baryon $V$ structure: $\Gamma^{\rm snk}_1=C\gamma^t$, $\eta_{\Gamma^{\rm snk}} \eta_{C\gamma^5\gamma^t} = -1$ $\Longrightarrow$ antisymmetric;

    \item $\Lambda$-baryon $A$ structure: $\Gamma^{\rm snk}_1=C\gamma^5\gamma^t$, $\eta_{\Gamma^{\rm snk}} \eta_{C\gamma^5\gamma^t} = +1$ $\Longrightarrow$ symmetric;

    \item $\Lambda$-baryon $T$ structure: $\Gamma^{\rm snk}_1=\tfrac12 C[\gamma^t,\gamma^{x,y}]$, $\eta_{\Gamma^{\rm snk}} \eta_{C\gamma^5\gamma^t} = -1$ $\Longrightarrow$ antisymmetric.
\end{itemize}

These results confirm the exchange symmetries of quasi-DAs in coordinate space, stated in Sec.~\ref{sec:symmetries} Eq.~\eqref{eq:norm_conv_quasi}:
\begin{equation}
    \widetilde\Phi_\Lambda^A(z_2,z_1) = +\widetilde\Phi_\Lambda^A(z_1,z_2)\ ,\qquad
    \widetilde\Phi_\Lambda^{V,T}(z_2,z_1) = -\widetilde\Phi_\Lambda^{V,T}(z_1,z_2)\ .
\end{equation}

We note that the structure of the non-kinematically-enhanced source $C\gamma^5$ is also antisymmetric ($\eta_{C\gamma^5}=-1$), so the same symmetry pattern holds regardless of the source choice.

\subsubsection{Nonlocal Two-point Correlation of proton}

From the contraction Eq.~\eqref{eq:contract_proton} for proton in Appendix~\ref{app:def_intor_contract}, the first two quarks are identical, thus the nonlocal two-point correlation involves two Wick contraction types for the $l$-quark propagators:
\begin{equation}\label{eq:proton_2pt_sym}
\begin{aligned}
    C_{\rm p, \rm 2pt}(z_1,z_2) =&\ \epsilon_{ijm}\epsilon_{k\ell n} (\Gamma^{\rm snk}_1)_{\alpha\beta} (C\gamma^5\gamma^t)_{\alpha'\beta'} \big[\mathbb T \Gamma^{\rm snk}_2 D_l^{-1}(0) \big]_{\gamma'\beta'}^{mn} \\
    &\ \times \Big[ D_l^{-1}(z_1)^{ik}_{\alpha\alpha'} D_l^{-1}(z_2)^{j\ell}_{\beta\gamma'}
    + D_l^{-1}(z_1)^{ik}_{\alpha\gamma'} D_l^{-1}(z_2)^{j\ell}_{\beta\alpha'} \Big]\\
    \equiv&\ \mathscr T_1 + \mathscr T_2\ .
\end{aligned}
\end{equation}
We now perform the exchange $z_1\leftrightarrow z_2$ followed by the relabeling of dummy indices $\{i,k,\alpha\}\leftrightarrow\{j,\ell,\beta\}$. As in the $\Lambda$ case, the two $\epsilon$-tensors contribute $(-1)^2=+1$, and $(\Gamma^{\rm snk}_1)_{\alpha\beta}\to(\Gamma^{\rm snk}_1)_{\beta\alpha}=\eta_{\Gamma^{\rm snk}_1}\ (\Gamma^{\rm snk}_1)_{\alpha\beta}$.

For the propagator products in the square brackets in Eq.~\eqref{eq:proton_2pt_sym}, the exchanging and relabeling give:
\begin{equation}
\begin{aligned}
    D_l^{-1}(z_1)^{ik}_{\alpha\alpha'} D_l^{-1}(z_2)^{j\ell}_{\beta\gamma'} \to&\ D_l^{-1}(z_2)^{j\ell}_{\beta\alpha'} D_l^{-1}(z_1)^{ik}_{\alpha\gamma'} = D_l^{-1}(z_1)^{ik}_{\alpha\gamma'} D_l^{-1}(z_2)^{j\ell}_{\beta\alpha'}\ ,\\
    D_l^{-1}(z_1)^{ik}_{\alpha\gamma'} D_l^{-1}(z_2)^{j\ell}_{\beta\alpha'} \to&\ D_l^{-1}(z_2)^{j\ell}_{\beta\gamma'} D_l^{-1}(z_1)^{ik}_{\alpha\alpha'} = D_l^{-1}(z_1)^{ik}_{\alpha\alpha'} D_l^{-1}(z_2)^{j\ell}_{\beta\gamma'}\ ,
\end{aligned}
\end{equation}
That is, the two products in the square brackets are simply interchanged, while their sum remains invariant. Crucially, no relabeling of the source-side indices $\alpha'$, $\beta'$, or $\gamma'$ is needed, because unlike in the $\Lambda$ case, these indices do not participate symmetrically in the proton contraction.

The only effect is therefore the transposition of $\Gamma^{\rm snk}_1$:
\begin{equation}\label{eq:proton_sym_final}
    C_{\rm p,2pt}(z_2,z_1) = \eta_{\Gamma^{\rm snk}_1} C_{\rm p,2pt}(z_1,z_2)\ .
\end{equation}
Note that, unlike the $\Lambda$ case, the exchange symmetry for the proton is governed solely by the transposition property of the sink-side structure $\Gamma^{\rm snk}_1$, without an additional factor from the source. Applying the transposition signs from Table~\ref{tab:transpose_gamma}:

\begin{itemize}
    \item proton $V$ structure: $\Gamma^{\rm snk}_1=C\gamma^t$, $\eta_{\Gamma^{\rm snk}}=+1$ $\Longrightarrow$ symmetric;
    
    \item proton $A$ structure: $\Gamma^{\rm snk}_1=C\gamma^5\gamma^t$, $\eta_{\Gamma^{\rm snk}}=-1$ $\Longrightarrow$ antisymmetric;
    
    \item proton $T$ structure: $\Gamma^{\rm snk}_1=\tfrac12 C[\gamma^t,\gamma^{x,y}]$, $\eta_{\Gamma^{\rm snk}}=+1$ $\Longrightarrow$ symmetric.
\end{itemize}

This confirms the proton exchange symmetries stated in Sec.~\ref{sec:symmetries} Eq.~\eqref{eq:norm_conv_quasi}:
\begin{equation}
    \widetilde\Phi_{\rm p}^{V,T}(z_2,z_1) = +\widetilde\Phi_{\rm p}^{V,T}(z_1,z_2)\ ,\qquad
    \widetilde\Phi_{\rm p}^A(z_2,z_1) = -\widetilde\Phi_{\rm p}^A(z_1,z_2)\ .
\end{equation}
The pattern is opposite to that of the $\Lambda$ baryon. This difference originates from the additional factor $\eta_{C\gamma^5\gamma^t}=-1$ that appears in the $\Lambda$ case due to the transposition of the source-side structure, whereas for the proton this factor is absent because the two Wick contraction types absorb it through their mutual interchange under $z_1\leftrightarrow z_2$.

\section{Extrapolation Ans\"atze for Baryon LCDAs}
\label{app:extrapolation}

In this appendix, we derive the asymptotic large-distance expansion of baryon quasi-DA matrix elements in coordinate space, by combining the HQET reduction with a dispersive analysis. The resulting expressions provide physics-motivated ans\"atze for the coordinate-space large-$\lambda$ extrapolation used in Sec.~\ref{sec:extrapolation}, with the large-distance behavior determined directly from Euclidean equal-time correlators.

\subsection{Heavy-quark Representation of Wilson Lines}
\label{app:heavy_quark_wilson_line}

Following Ref.~\cite{Ji:2026vir}, the first step in deriving the large-distance behavior of Euclidean correlators is to replace the Wilson lines in the nonlocal baryon operators by the propagators of auxiliary heavy quarks. This trick makes it possible to use a standard spectral decomposition in terms of physical color-singlet states.

Consider a heavy quark field $Q$ with mass $m_Q$. In the heavy-quark limit $m_Q\to\infty$, the coordinate-space propagator of $Q$ reduces to a straight Wilson line multiplied by a known scalar factor:
\begin{equation}
    Q(x)\bar Q(0)
    \ \longrightarrow\ 
    U(x,0)\ 
    D_Q(x^2,m_Q^2)\ 
    \frac{\rmi\slashed{x}+\sqrt{-x^2}}{2\sqrt{-x^2}}
    \left[
    1+\mathcal{O}\left(
    \frac{1}{m_Q\sqrt{-x^2}},
    \frac{\Lambda_{\rm QCD}}{m_Q}
    \right)
    \right]\ ,
    \label{eq:HQET_reduction}
\end{equation}
where $U(x,0)$ is the straight Wilson line connecting from $0$ to $x$. The scalar function $D_Q$ contains the universal heavy-quark propagation factor. For a equal-time separation $x=(0,0,0,z)$, it takes the asymptotic form:
\begin{equation}
    D_Q(z^2,m_Q^2)
    =
    \frac{m_Q^3}{2\sqrt{2}\pi^{3/2}}\ 
    \frac{\rme^{-m_Q |z|}}{(m_Q |z|)^{3/2}}\ .
    \label{eq:DQ_factor}
\end{equation}
Thus, after dividing out the known factor $D_Q$, the heavy-quark correlator reproduces the desired Wilson-line matrix element up to corrections suppressed by $1/(m_Q |z|)$ and $\Lambda_{\rm QCD}/m_Q$.

For the baryon quasi-DAs considered in this work, the nonlocal operator contains two Wilson lines extending to the quark fields located at $z_1$ and $z_2$. We therefore introduce two auxiliary heavy-quark fields, denoted by $Q$ and $G$, to reproduce the two Wilson lines. The original coordinate-space baryon matrix element:
\begin{equation}
    \epsilon^{ijk}
    \langle 0|
    q^{i'}_\alpha(z_1) {U_{i'}}^{i}(z_1,0)  
    g^{j'}_\beta(z_2) {U_{j'}}^{j}(z_2,0)  
    h^k_\gamma(0)
    |B(P)\rangle\ ,
    \label{eq:baryon_quasiDA_original}
\end{equation}
can then be related to the heavy-quark form:
\begin{equation}
\begin{aligned}
    &\ \epsilon_{ijk}\langle 0|
    \bar Q_{\alpha',i'}(z_1) q^{i'}_\alpha(z_1)\ 
    Q^i_{\alpha''}(0) h^k_\gamma(0) G^j_{\beta''}(0)\ 
    \bar G_{\beta',j'}(z_2) g^{j'}_\beta(z_2)
    |B(P)\rangle \\[4pt]
    =&\  \bigg(\frac{\rmi \slashed z_1-|z_1|}{2|z_1|}\bigg)_{\alpha''\alpha'}
    \bigg(\frac{\rmi \slashed z_2-|z_2|}{2|z_2|}\bigg)_{\beta''\beta'}
    \times H_Q H_G H_{QGh}\ 
    D_Q(z_1^2,m_Q^2)\ 
    D_G(z_2^2,m_G^2) \\
    &\ \qquad\qquad \times
    \epsilon_{ijk}
    \langle 0|
    q^{i'}_\alpha(z_1) {U_{i'}}^{i}(z_1,0)  
    g^{j'}_\beta(z_2) {U_{j'}}^{j}(z_2,0)  
    h^k_\gamma(0)
    |B(P)\rangle\ ,
\end{aligned}
\label{eq:baryon_HQET_reduction}
\end{equation}
where $H_Q$, $H_G$, and $H_{QGh}$ denote short-distance matching coefficients, and $q,g,h$ are the light quark fields following the convention in Table~\ref{tab:valancequark}. To project onto the $V$, $A$, and $T$ structures, we inserts the appropriate Dirac matrices
$(\Gamma_1)_{\alpha'\alpha}$, $(\Gamma_2)_{\alpha''\beta''}$, and $(\Gamma_3)_{\beta'\beta}$.

\subsection{Dispersive Relation Analysis}
\label{app:dispersive}

To analyze the asymptotic behavior of the nonlocal matrix elements, one inserts two complete sets of intermediate states into the matrix elements:
\begin{equation}
\begin{aligned}
    &\epsilon_{ijk}\langle 0 |\bar{Q}_{\alpha',i'}(z_1)q_\alpha^{i'}(z_1)\ 
    Q_{\alpha''}^i(0) h_\gamma^k(0) G^j_{\beta''}(0)\ 
    \bar{G}_{\beta',j'}(z_2) g_\beta^{j'}(z_2)|B(P)\rangle \\
    =& \sum_{X_1,X_2}
    \int \rmd\Gamma_{X_1}(k_1)\ \rmd\Gamma_{X_2}(k_2)
    \ \langle 0 |\bar{Q}_{\alpha'}q_\alpha(z_1)|X_1(k_1)\rangle
    \ \langle X_1(k_1)|Q_{\alpha''}h_\gamma G_{\beta''}(0)|X_2(k_2)\rangle
    \ \langle X_2(k_2)|\bar G_{\beta'}g_\beta(z_2)|B(P)\rangle\ ,
\end{aligned}
\label{eq:dispersion}
\end{equation}
where
\begin{equation}
    \rmd\Gamma_X(k):=\frac{1}{S_X}
    \prod_{\ell\in X}\int \frac{\rmd^3k_\ell}{(2\pi)^3}\frac{1}{2E_\ell}
\end{equation}
denotes the Lorentz-invariant phase-space (LIPS) measure, and $\epsilon^{ijk}$ is omitted for simplicity.

The asymptotic behavior for $z\to\infty$ is obtained by the saddle-point approximation of the phase-space integral. When $z\to\infty$, one finds:
\begin{equation}
\begin{aligned}
    \int\frac{\rmd^3k}{(2\pi)^3}
    \frac{1}{2\sqrt{k^2+m^2}}
    \rme^{-\rmi zk^z}|X(k)\rangle\langle X(k)|
    \ \to\ 
    P(m,z)
    \equiv\frac{m^{1/2}\rme^{-m|z|}}{4\sqrt{2}\pi^{3/2}}
    \frac{1}{|z|^{3/2}}
    |\Lambda_{\mathcal C}(m,\vec{0})\rangle
    \langle \Lambda_{\mathcal C}(m,\vec0)| + \cdots\ .
\end{aligned}
\label{eq:saddle_point}
\end{equation}
Eq.~\eqref{eq:saddle_point} is the fundamental building block in this analysis: each intermediate state $|\Lambda_{\mathcal C}(m,\vec{0})\rangle$ contributes $\rme^{-m|z|}/|z|^{3/2}$ to the asymptotic expansion.

\subsection{Classification of Connected Channels and Asymptotic Behavior}
\label{app:channel_classification}

The asymptotic behavior of the baryon quasi-DA matrix elements can be analyzed by inserting complete sets of intermediate states between the heavy-quark operators in Eq.~\eqref{eq:baryon_HQET_reduction}. Depending on how the external baryon state $|B(P) \rangle$ is connected to the quark operators, the contributions are classified into three distinct channels. This classification allows a controlled large-distance expansion in each case.

In this section and the next, we take the $z_1<0$ and $z_2>0$ sector as an illustrative example. The generalization to the other sectors in the $(z_1,z_2)$ plane proceeds analogously, and the full results are collected in Appendix~\ref{app:full_asym_express}. The remaining sign choices are obtained by relabeling which quark field is treated as locating at $0$. The corresponding Dirac-$\gamma$ matrices then follow straightforwardly from Fierz transformations of the leading-twist baryon DA structures~\cite{Braun:2000kw}.

\subsubsection{Case 1: \texorpdfstring{$|B(P) \rangle$}{|B(P)>} connected to \texorpdfstring{$\bar{G} \Gamma_3 g$}{Gbar Gamma3 g}}\label{channel_case1}

\begin{figure}[htbp]
    \centering
    \includegraphics[width=0.7\textwidth]{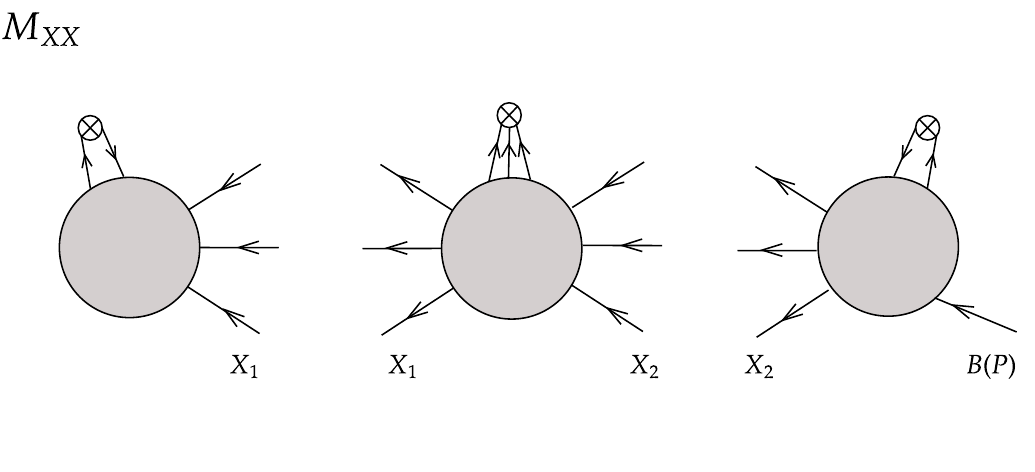}
    \caption{Schematic representation of the first dispersive channel in the large-distance expansion of the baryon quasi-DA matrix element. In this channel, the external baryon state $|B(P)\rangle$ is attached to the heavy--light operator $\bar G\Gamma_3 g$, with two complete sets of intermediate states denoted by $X_1(k_1)$ and $X_2(k_2)$. From left to right, the three matrix elements are $\langle 0|\bar Q\Gamma_1 q(0)|X_1(k_1)\rangle$, $\langle X_1(k_1)|Q\Gamma_2 G h(0)|X_2(k_2)\rangle$, and $\langle X_2(k_2)|\bar G\Gamma_3 g(0)|B(P)\rangle$.}
    \label{topxx}
\end{figure}

In this channel, the external baryon is attached to the heavy--light operator $\bar{G} \Gamma_3 g$. The intermediate state adjacent to this operator is denoted by $X_2$, and the other inserted state is denoted as $X_1$. After translating all operators to the $z=0$ and keeping the resulting longitudinal phases, the corresponding contribution can be written as:
\begin{equation}
\begin{aligned}
    M_{XX}=&\ \sum_{X_1,X_2}\int \rmd\Gamma_{X_1}(k_1) \rmd\Gamma_{X_2}(k_2)
    \ \rme^{\rmi z_1 k_1^z} \rme^{\rmi z_2(P^z-k_2^z)}\\
    & \times \langle 0 |\bar{Q} \Gamma_1 q(0)|X_1(k_1)\rangle
    \ \langle X_1(k_1)|Q \Gamma_2 G  h (0)|X_2(k_2)\rangle
    \ \langle X_2(k_2)|\bar G \Gamma_3 g(0)|B(P)\rangle\ .
\end{aligned}
\label{eq:M_XX}
\end{equation}

When the first separation is taken to the asymptotic region, $z_1\to -\infty$, the $X_1$ phase-space integral is saturated by the lowest states selected by the operator $\bar{Q}\Gamma_1q$. The saddle-point expansion then gives:
\begin{equation}
\begin{aligned}
    M_{XX}=&\ \rme^{\rmi z_2P^z}\sum_{X_1,X_2}\frac{1}{S_{X_1}}
    \prod_{\ell\in X_1}P(m_\ell,z_1)
    \int \rmd\Gamma_{X_2}(k_2) \  \rme^{-\rmi z_2k^z_2}
    \ \langle X_2(k_2)|\bar G\Gamma_3 g(0)|B(P) \rangle\\
    & \times \langle 0 |\bar{Q}\Gamma_1 q(0)| \prod_{\ell\in X_1}n_\ell(\rmi\hat{z}_1 m_\ell)\rangle
    \ \langle \prod_{\ell\in X_1}n_\ell(\rmi\hat{z}_1 m_\ell)| Q \Gamma_2 G h (0)|X_2(k_2)\rangle\ .
\end{aligned}
\label{eq:M_XX_z1inf}
\end{equation}
Similarly, in the limit $z_2\to \infty$, the asymptotic behavior is governed by the states in the $X_2$ channel, leading to:
\begin{equation}
\begin{aligned}
    M_{XX}=&\ \rme^{\rmi z_2P^z}\sum_{X_1,X_2}\frac{1}{S_{X_2}}
    \prod_{\ell\in X_2}P(m_\ell,z_2)
    \int \rmd\Gamma_{X_1}(k_1) \  \rme^{\rmi z_1k^z_1}
    \ \langle \prod_{\ell\in X_2}n_\ell(-\rmi\hat{z}_2 m_\ell)|\bar G \Gamma_3 g(0)|B(P) \rangle\\
    & \times \langle 0 |\bar{Q}\Gamma_1 q(0)|X_1(k_1)\rangle
    \ \langle X_1(k_1)|Q \Gamma_2 G  h (0)|\prod_{\ell\in X_2}n_\ell(-\rmi\hat{z}_2 m_\ell)\rangle\ .
\end{aligned}
\label{eq:M_XX_z2inf}
\end{equation}
In the joint asymptotic limit $z_1\to -\infty$ and $z_2\to +\infty$, the two saddle-point reductions can be applied simultaneously. One obtains
\begin{equation}
\begin{aligned}
    M_{XX}=&\ \rme^{\rmi z_2P^z}\sum_{X_1,X_2}\frac{1}{S_{X_1}}\prod_{\ell\in X_1}P(m_\ell,z_1)
    \frac{1}{S_{X_2}}\prod_{{\ell'}\in X_2}P(m_{\ell'},z_2)
    \ \langle \prod_{{\ell'}\in X_2}n_{\ell'}(-\rmi\hat{z}_2 m_{\ell'})|\bar G\Gamma_3 g(0)|B(P) \rangle\\
    & \times \langle 0 |\bar{Q}\Gamma_1 q(0)|\prod_{\ell\in X_1}n_\ell(\rmi\hat{z}_1 m_\ell)\rangle
    \ \langle\prod_{\ell\in X_1}n_\ell(\rmi\hat{z}_1 m_\ell)|Q \Gamma_2 G  h (0)|\prod_{{\ell'}\in X_2}n_{\ell'}(-\rmi\hat{z}_2 m_{\ell'})\rangle\ .
\end{aligned}
\label{eq:M_XX_double}
\end{equation}

\subsubsection{Case 2: \texorpdfstring{$|B(P) \rangle$}{|B(P)>} connected to \texorpdfstring{$Q \Gamma_2 G h$}{Q Gamma2 G h}}\label{channel_case2}

\begin{figure}[htbp]
    \centering
    \includegraphics[width=0.7\textwidth]{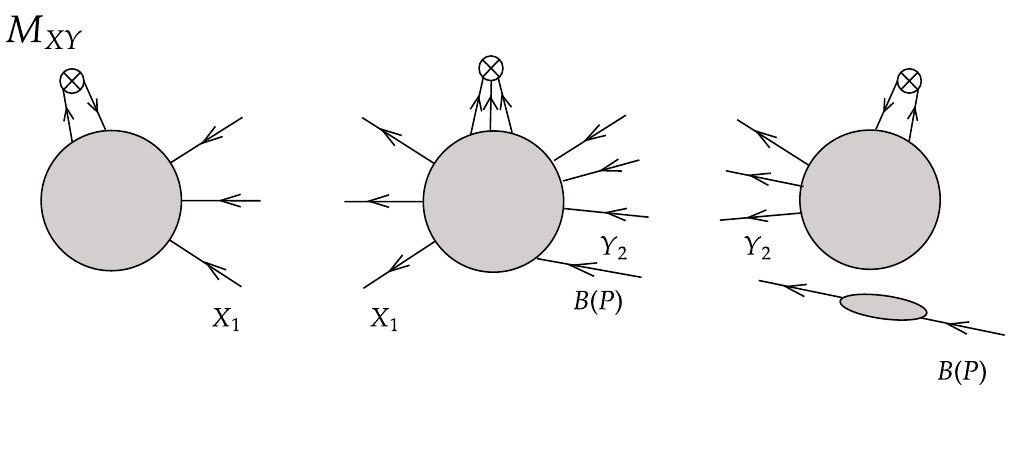}
    \caption{Schematic representation of the second connected channel used in the dispersive analysis of the baryon quasi-DA matrix element. In this channel, the external baryon state $|B(P)\rangle$ is attached to the central operator $Q\Gamma_2 G h$, and the remaining intermediate states are denoted by $X_1(k_1)$ and $Y_2(k_2)$. From left to right, the three matrix elements are $\langle 0|\bar Q\Gamma_1 q(0)|X_1(k_1)\rangle$, $\langle X_1(k_1)|Q\Gamma_2 G h(0)|Y_2(k_2),B(P)\rangle$, and $\langle Y_2(k_2)|\bar G\Gamma_3 g(0)|0\rangle$.}
    \label{topxy}
\end{figure}

In the second channel, the external baryon is connected to the central three-quark operator $Q \Gamma_2 G  h$. The intermediate state to the right of this operator is denoted by $Y_2$, while the left-side intermediate state is described by the same $X_1$ as in Appendix~\ref{channel_case1}. The corresponding spectral representation is:
\begin{equation}
\begin{aligned}
    M_{XY}=&\ \sum_{X_1,Y_2}\int \rmd\Gamma_{X_1}(k_1)\rmd\Gamma_{Y_2}(k_2)
    \ \rme^{\rmi z_1 k_1^z}\rme^{-\rmi z_2 k_2^z}\\
    & \times \langle 0 |\bar{Q}\Gamma_1 q(0)|X_1(k_1)\rangle
    \ \langle X_1(k_1)|Q \Gamma_2 G  h (0)|Y_2(k_2), B(P) \rangle
    \ \langle Y_2(k_2)|\bar G\Gamma_3 g(0)| 0 \rangle\ .
\end{aligned}
\label{eq:M_XY}
\end{equation}

For $z_1\to -\infty$, only the left intermediate state $X_1$ is forced into its large-distance saddle-point form. This gives:
\begin{equation}
\begin{aligned}
    M_{XY}=&\ \sum_{X_1,Y_2}\frac{1}{S_{X_1}}
    \prod_{\ell\in X_1}P(m_\ell,z_1)
    \int \rmd\Gamma_{Y_2}(k_2) \  \rme^{-\rmi z_2k^z_2}
    \ \langle \prod_{\ell\in X_1}n_\ell(\rmi\hat{z}_1 m_\ell)|Q \Gamma_2 G  h (0)|Y_2(k_2), B(P) \rangle\\
    & \times \langle 0 |\bar{Q}\Gamma_1 q(0)|\prod_{\ell\in X_1}n_\ell(\rmi\hat{z}_1 m_\ell)\rangle
    \ \langle Y_2(k_2)|\bar G\Gamma_3 g(0)| 0 \rangle\ .
\end{aligned}
\label{eq:M_XY_z1inf}
\end{equation}
For $z_2\to \infty$, the right intermediate state $Y_2$ controls the asymptotic falloff, and the contribution becomes:
\begin{equation}
\begin{aligned}
    M_{XY}=&\ \sum_{X_1,Y_2}\frac{1}{S_{Y_2}}
    \prod_{\ell\in Y_2}P(m_\ell,z_2)
    \int \rmd\Gamma_{X_1}(k_1) \  \rme^{\rmi z_1k^z_1}
    \ \langle X_1(k_1)|Q \Gamma_2 G  h (0)|\prod_{\ell\in Y_2}n_\ell(-\rmi\hat{z}_2 m_\ell), B(P) \rangle\\
    & \times \langle 0 |\bar{Q}\Gamma_1 q(0)|X_1(k_1)\rangle
    \ \langle \prod_{\ell\in Y_2}n_\ell(-\rmi\hat{z}_2 m_\ell)|\bar G\Gamma_3 g(0)| 0 \rangle\ .
\end{aligned}
\label{eq:M_XY_z2inf}
\end{equation}
In the joint limit $z_1\to -\infty$ and $z_2\to +\infty$,
\begin{equation}
\begin{aligned}
    M_{XY}=&\ \sum_{X_1,Y_2}\frac{1}{S_{X_1}}\prod_{\ell\in X_1}P(m_\ell,z_1)
    \frac{1}{S_{Y_2}}\prod_{{\ell'}\in Y_2}P(m_{\ell'},z_2)
    \ \langle\prod_{\ell\in X_1}n_\ell(\rmi\hat{z}_1 m_\ell)|Q \Gamma_2 G  h (0)|\prod_{{\ell'}\in Y_2}n_{\ell'}(-\rmi\hat{z}_2 m_{\ell'}), B(P) \rangle\\
    & \times \langle 0 |\bar{Q}\Gamma_1 q(0)|\prod_{\ell\in X_1}n_\ell(\rmi\hat{z}_1 m_\ell)\rangle
    \ \langle \prod_{{\ell'}\in Y_2}n_{\ell'}(-\rmi\hat{z}_2 m_{\ell'})|\bar G\Gamma_3 g(0)| 0 \rangle\ .
\end{aligned}
\label{eq:M_XY_double}
\end{equation}

\subsubsection{Case 3: \texorpdfstring{$|B(P) \rangle$}{|B(P)>} connected to \texorpdfstring{$\bar{Q} \Gamma_1 q$}{Qbar Gamma1 q}}\label{channel_case3}

\begin{figure}[htbp]
    \centering
    \includegraphics[width=0.7\textwidth]{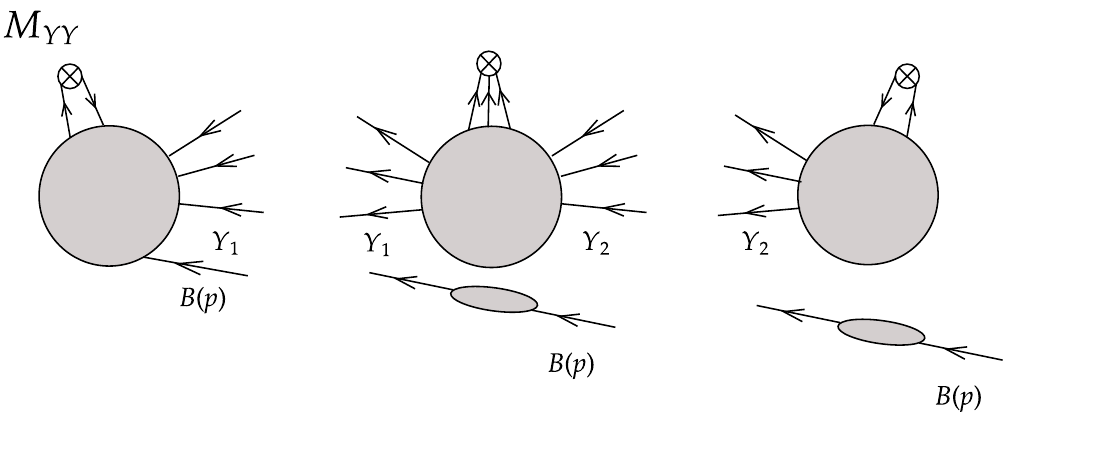}
    \caption{Schematic representation of the third connected channel used in the dispersive analysis of the baryon quasi-DA matrix element. In this channel, the external baryon state $|B(P)\rangle$ is attached to the left operator $\bar Q\Gamma_1 q$, and the remaining intermediate states are denoted by $Y_1(k_1)$ and $Y_2(k_2)$. From left to right, the three matrix elements are $\langle 0|\bar Q\Gamma_1 q(0)|Y_1(k_1),B(P)\rangle$, $\langle Y_1(k_1)|Q\Gamma_2 G h(0)|Y_2(k_2)\rangle$, and $\langle Y_2(k_2)|\bar G\Gamma_3 g(0)|0\rangle$.}
    \label{topyy}
\end{figure}

In the third channel, the external baryon is attached to the heavy--light operator $\bar{Q} \Gamma_1 q$. The state adjacent to this operator is labeled as $Y_1$, while the farther one is described by $Y_2$. The spectral decomposition then takes the form:
\begin{equation}
\begin{aligned}
    M_{YY}=&\ \sum_{Y_1,Y_2}\int \rmd\Gamma_{Y_1}(k_1)\rmd\Gamma_{Y_2}(k_2)
    \ \rme^{\rmi z_1 (k_1^z+P^z)}\rme^{-\rmi z_2 k_2^z}\\
    & \times \langle 0 |\bar{Q}\Gamma_1 q(0)|Y_1(k_1), B(P) \rangle
    \ \langle Y_1(k_1)|Q \Gamma_2 G  h (0)|Y_2(k_2)\rangle
    \ \langle Y_2(k_2)|\bar G\Gamma_3 g(0)| 0 \rangle\ .
\end{aligned}
\label{eq:M_YY}
\end{equation}

In the limit $z_1\to -\infty$, the large-distance behavior is governed by the $Y_1$ intermediate state. Applying the saddle-point approximation to this channel yields:
\begin{equation}
\begin{aligned}
    M_{YY}=&\ \rme^{\rmi z_1P^z}\sum_{Y_1,Y_2}\frac{1}{S_{Y_1}}
    \prod_{\ell\in Y_1}P(m_\ell,z_1)
    \int \rmd\Gamma_{Y_2}(k_2) \  \rme^{-\rmi z_2k^z_2}
    \ \langle 0 |\bar{Q}\Gamma_1u_\alpha(0)|
    \prod_{\ell\in Y_1}n_\ell(\rmi\hat{z}_1 m_\ell), B(P) \rangle\\
    & \times \langle \prod_{\ell\in Y_1}n_\ell(\rmi\hat{z}_1 m_\ell)|Q \Gamma_2 G  h (0)|Y_2(k_2)\rangle
    \ \langle Y_2(k_2)|\bar G\Gamma_3 g(0)| 0 \rangle\ .
\end{aligned}
\label{eq:M_YY_z1inf}
\end{equation}
For $z_2\to \infty$, the right-side intermediate state $Y_2$ is instead projected onto its asymptotic form, resulting in:
\begin{equation}
\begin{aligned}
    M_{YY}=&\ \rme^{\rmi z_1P^z}\sum_{Y_1,Y_2}\frac{1}{S_{Y_2}}
    \prod_{\ell\in Y_2}P(m_\ell,z_2)
    \int \rmd\Gamma_{Y_1}(k_1) \  \rme^{\rmi z_1k^z_1}
    \ \langle 0 |\bar{Q}\Gamma_1 q(0)|Y_1(k_1), B(P) \rangle\\
    & \times \langle Y_1(k_1)|Q \Gamma_2 G  h (0)|\prod_{\ell\in Y_2}n_\ell(-\rmi\hat{z}_2 m_\ell)\rangle
    \ \langle \prod_{\ell\in Y_2}n_\ell(-\rmi\hat{z}_2 m_\ell)|\bar G\Gamma_3 g(0)| 0 \rangle\ .
\end{aligned}
\label{eq:M_YY_z2inf}
\end{equation}
In the joint limit $z_1\to -\infty$ and $z_2\to +\infty$,
\begin{equation}
\begin{aligned}
    M_{YY}=&\ \rme^{\rmi z_1P^z}\sum_{Y_1,Y_2}\frac{1}{S_{Y_1}}
    \prod_{\ell\in Y_1}P(m_\ell,z_1)
    \frac{1}{S_{Y_2}}\prod_{{\ell'}\in Y_2}P(m_{\ell'},z_2)
    \ \langle 0 |\bar{Q}\Gamma_1 q(0)|
    \prod_{\ell\in Y_1}n_\ell(\rmi\hat{z}_1 m_\ell), B(P) \rangle\\
    & \times \langle\prod_{\ell\in Y_1}n_\ell(\rmi\hat{z}_1 m_\ell)|Q \Gamma_2 G h(0)|\prod_{{\ell'}\in Y_2}n_{\ell'}(-\rmi\hat{z}_2 m_{\ell'})\rangle
    \ \langle \prod_{{\ell'}\in Y_2}n_{\ell'}(-\rmi\hat{z}_2 m_{\ell'})|\bar G\Gamma_3 g(0)| 0 \rangle\ .
\end{aligned}
\label{eq:M_YY_double}
\end{equation}

For all the channels discussed above, we insert six complete sets of intermediate states in total. However, these six insertions fall into only four distinct quantum-number classes, which we denote by $X_1$, $X_2$, $Y_1$, and $Y_2$. The corresponding states satisfy the quantum-number relations:
\begin{equation}
\begin{aligned}
    X_1 &\sim \bar q (\Gamma_1)^{\rm T} Q | 0 \rangle\ , &
    X_2 &\sim \bar G \Gamma_3 g |B(P) \rangle\ ,\\[5pt]
    Y_1 &\sim \bar q (\Gamma_1)^{\rm T} Q |\bar{B}(P) \rangle\ , &
    Y_2 &\sim \bar G \Gamma_3 g | 0 \rangle\ .
\end{aligned}
\label{eq:quantum_numbers}
\end{equation}

\subsection{General Structure of the Extrapolation}
\label{app:general_structure}

In this work, the exponentials in the asymptotic forms of quasi-DAs contain the binding energies selected by the channel quantum numbers. Here we analyze the general spin structures and binding energies enter the asymptotic forms.

\subsubsection{Spin Structures of different amplitudes}
\label{app:spin_structure}

The spin structure connecting the $q$ and $g$ quarks in the baryon matrix element can be simplified using heavy-quark spin symmetry. After contracting all Dirac indices in Eq.~\eqref{eq:baryon_HQET_reduction}, one finds:
\begin{equation}
\begin{aligned}
    {(\Gamma_1)}_{\alpha'\alpha}
    &\bigg(\frac{\rmi \slashed z_1-|z_1|}{2|z_1|}\bigg)_{\alpha''\alpha'}
    {(\Gamma_2)}_{\alpha''\beta''}
    \bigg(\frac{\rmi \slashed z_2-|z_2|}{2|z_2|}\bigg)_{\beta''\beta'}
    {(\Gamma_3)}_{\beta'\beta} \\
    &= \frac{1}{4} {(\Gamma_1)}^{\rm T} (1-i\gamma^3)^{\rm T} \Gamma_2 (1+i\gamma^3)\Gamma_3 \\
    &= \frac{1}{4} {(\Gamma_1)}^{\rm T} C(1+i\gamma^3) C^{-1} \Gamma_2 (1+i\gamma^3) \Gamma_3 \\
    &= \pm\frac{1}{4} C {\Gamma_1} (1+i\gamma^3) C^{-1} \Gamma_2 (1+i\gamma^3) \Gamma_3\ .
\end{aligned}
\label{eq:spin_reduction}
\end{equation}
The choice of $\Gamma_1$, $\Gamma_2$, and $\Gamma_3$ is fixed by the desired Dirac-$\gamma$ structure of metrix element in Eq.~\eqref{eq:baryon_quasiDA_original}. We use the standard heavy--light projectors that select static hadrons at the saddle point. In heavy-quark effective theory, the spin--parity quantum numbers $J^P$ of a static heavy--light meson are related to the Dirac-$\gamma$ structure of the interpolating operator as summarized in Table~\ref{tab:dirac_structures}.

\begin{table}[t]
    \centering
    \renewcommand{\arraystretch}{1.15}
    \setlength{\tabcolsep}{8mm}
    \begin{tabular}{c c}
        \hline\hline
        $J^P$ & $\Gamma$ \\
        \hline
        $0^-$ & $\gamma^5,\ \gamma^0\gamma^5$ \\
        $0^+$ & $\mathbb{I},\ \gamma^0$ \\
        $1^-$ & $\gamma^i,\ \gamma^0\gamma^i$ \\
        $1^+$ & $\gamma^i\gamma^5,\ \gamma^i\gamma^j$ \\
        \hline\hline
    \end{tabular}
    \caption{Dirac-$\gamma$ structures corresponding to different spin--parity quantum numbers $J^P$ of a static heavy--light meson (summarized from Table~6.1 of Ref.~\cite{Gattringer:2010zz}).}
    \label{tab:dirac_structures}
\end{table}

For the $A$ amplitude (take $C\gamma^0\gamma^5$ as example),
the allowed $(\Gamma_1,\Gamma_2,\Gamma_3)$ triples are:
\begin{equation}
\begin{cases}
    -(\gamma^5,\  C\gamma^0\gamma^5,\ 
    \gamma^5) - (\gamma^i\gamma^0,\ 
    C\gamma^0\gamma^5,\ 
    \gamma^0\gamma^i)\ ,
    \\[4pt]
    (\mathbb{I},\  C\gamma^0\gamma^5,\ 
    \mathbb{I}) - (\gamma^5,\ 
    C\gamma^0\gamma^5,\ 
    \gamma^5)\ ,
    \\[4pt]
    (\mathbb{I},\  C\gamma^0\gamma^5,\ 
    \mathbb{I}) + (\gamma^i,\ 
    C\gamma^0\gamma^5,\ 
    \gamma^i )\ .
\end{cases}
\label{eq:A_projectors}
\end{equation}

For the $V$ amplitude (take $C\gamma^0$ as example):
\begin{equation}
    (\mathbb{I},\  C\gamma^0\gamma^5,\ 
    \gamma^5) - (\gamma^5,\ 
    C\gamma^0\gamma^5,\ 
    \mathbb{I})\  .
\label{eq:V_projectors}
\end{equation}

For the $T$ amplitude (take $\frac{1}{2}C[\gamma^0,\gamma^{1}]$ as example):
\begin{equation}
\begin{cases}
    (\gamma^5,\  C\gamma^0\gamma^1,\ 
    \gamma^5) + (\mathbb{I},\ 
    C\gamma^0\gamma^1,\ 
    \mathbb{I})\ ,
    \\[4pt]
    (\gamma^5,\  C\gamma^0\gamma^1,\ 
    \gamma^5) + (\gamma^3,\ 
    C\gamma^0\gamma^1,\ 
    \gamma^3)\ ,
    \\[4pt]
    (\mathbb{I},\  C\gamma^0\gamma^5,\ 
    \gamma^1\gamma^5) + (\gamma^1\gamma^5,\ 
    C\gamma^0\gamma^1,\ 
    \mathbb{I})\ .
\end{cases}
\label{eq:T_projectors}
\end{equation}

\subsubsection{Determining the Binding Energies}

The key consequence of the dispersive analysis is that the binding energies appearing in the exponential suppression are fixed by the quantum numbers of the intermediate states. According to Eq.~\eqref{eq:quantum_numbers}, each channel projects onto states with specific quantum numbers determined by $\bar{Q}\Gamma_1 q$, $Q\Gamma_2 G h$, and $\bar{G}\Gamma_3 g$, which are precisely those of heavy--light hadrons.

Heavy-quark spin symmetry~\cite{Manohar:2000dt} dictates that the total spin of a heavy--light hadron can be divided into
\begin{equation}
	\vec{J}=\vec{s}_{\rm h}+\vec{s}_{\rm l}\ ,
\end{equation}
where $\vec{s}_{\rm h}$ is the spin of the heavy quark and $\vec{s}_{\rm l}$ is the spin of the total light degrees of freedom, which can be further decomposed into 
\begin{equation}
	\vec{s}_{\rm l}=\vec{s}_{q}+\vec{l}\ ,
\end{equation}
which factorize out the orbital angular momentum $\vec{l}$. Thus the heavy--light meson spectrum can be organized into doublets labeled by the spin--parity $(s^P)_{\rm l}$ of the light degrees of freedom:
\begin{itemize}
  \item Ground state doublet ($l=0$): parity $P=(-1)^{l+1}=-1$, $(s^P)_{\rm l} = 1/2^-$, $J=0,1$.
  This case contains the pseudo-scalar ($0^-$, e.g.\ $D^0$, $B^0$) and vector
  ($1^-$, e.g.\ $D^{*0}$, $B^{*0}$) mesons, which share the same binding energy $\Lambda^{0^-} = \Lambda^{1^-}$ up to
  $\mathcal{O}(\Lambda_{\rm QCD}/m_Q)$ corrections.

  \item First excited doublet ($l=1$): parity $P=(-1)^{l+1}=+1$, $(s^P)_{\rm l} = 1/2^+$, $J=0,1$.
  This case contains the scalar ($0^+$, e.g.\ $D_0^*$, $B_0^*$) and axial-vector
  ($1^+$, e.g.\ $D_1$, $B_1$) mesons, which share the same binding energy $\Lambda^{0^+} = \Lambda^{1^+}$ up to
  $\mathcal{O}(\Lambda_{\rm QCD}/m_Q)$ corrections.
\end{itemize}
In addition, the intermediate states can also include heavy--light baryons with
$J^P = 1/2^-$ and $J^P = 1/2^+$, whose binding energies we denote by
$\Lambda^{1/2^-}$ and $\Lambda^{1/2^+}$, respectively.

For each channel, the specific set of binding energies that appear is determined by which spin--parity states can be interpolated by the operators $\bar{Q}\Gamma_1 q$, $Q\Gamma_2 G h$, and $\bar{G}\Gamma_3 g$. However, one should note that the intermediate states are not physical particles, and a complex Lorentz boost must be applied to the $\Gamma_i$ structures to obtain the correct spin quantum numbers~\cite{Ji:2026vir}.

In summary, the exponential suppression factors in the extrapolation ansatz are not free parameters --- they are fixed by the heavy--light hadron spectrum through heavy-quark spin symmetry. This is what makes the extrapolation ans\"atze physics-based rather than phenomenological models.

\subsection{Full Asymptotic Expressions}\label{app:full_asym_express}

The $(z_1,z_2)$ plane is divided into several regions according to which coordinate separations become large, as discussed in Sec.~\ref{sec:asym-ext}, following the same convention and region partition shown in Fig.~\ref{fig:asymptotic_region_division}. In the derivation above, we used the case $z_1z_2<0$ as an illustrative example. In the full baryon quasi-DA matrix element, however, two sign sectors have to be distinguished:
\begin{itemize}
    \item For $z_1 z_2<0$, the two quark fields are located on opposite sides of the origin. In this case, the two Wilson-line segments are oppositely oriented, thus the separations $z_1$ and $z_2$ are the relevant long-distance scales that connect the nonlocal quark fields and govern the asymptotic behavior of correlators.
    \item For $z_1 z_2>0$, the two quark fields lie on the same side of the origin. The corresponding Wilson-line segments are oriented in the same direction. In this sector, therefore, the correlator behavior depends on three relative sizes of $z_1$, $z_2$, and $z_1-z_2$: for example, $|z_1|$ and $|z_2|$ can become large while $|z_1-z_2|$ remains short, or all three separations can become large simultaneously.
\end{itemize}
For each region, the asymptotic form is written as a linear combination of exponentials multiplied by smooth coefficient functions. The phase factors $\exp(\pm \rmi z_i P^z)$ are determined by the momentum-flow pattern of the corresponding dispersive channel, as derived in Appendix~\ref{app:channel_classification}.

For completeness, here we provide the general large-distance expansion forms of baryon quasi-DAs, including both mesonic ($\Lambda^{0^\pm}$) and baryonic ($\Lambda^{1/2^\pm}$) binding energies. Dropping the baryonic and higher-binding-energy channels and retaining only the leading-asymptotic $(\mathrm{LA})$ or next-to-leading-asymptotic $(\mathrm{NLA})$ terms in $1/|z|$ yields the simplified forms used in the main text analysis.

The explicit region-by-region expressions for the symmetric $A$ amplitude of the $\Lambda$ baryon are:
\begin{itemize}
    \item \textbf{Region G}: $z_1z_2<0$, $|z_1|\to\infty$, $|z_2|\to\infty$ and $|z_1-z_2|\to\infty$,
    \begin{equation}
    \begin{aligned}
        &\ \widetilde A_\Lambda(z_1,z_2;P^z) \\
        =&\ \rme^{-\Lambda^{0^-}|z_1|} \rme^{-\Lambda^{0^-}|z_2|}
        \left[ \mathcal G_{A,1} + \frac{\mathcal G_{A,1}^{(1)}}{|z_1|} + \frac{\mathcal G_{A,1}^{(2)}}{|z_2|} +\cdots \right] \\
        & + \rme^{\rmi z_2 P^z} \rme^{-\Lambda^{0^-}|z_1|} \rme^{-\Lambda^{1/2^-}|z_2|}
        \left[ \mathcal G_{A,2} + \frac{\mathcal G_{A,2}^{(1)}}{|z_1|} + \frac{\mathcal G_{A,2}^{(2)}}{|z_2|} +\cdots \right] \\
        & + \rme^{\rmi z_1 P^z} \rme^{-\Lambda^{0^-}|z_2|} \rme^{-\Lambda^{1/2^-}|z_1|}
        \left[ \mathcal G_{A,2}^* + \frac{\mathcal G_{A,2}^{*(1)}}{|z_1|} + \frac{\mathcal G_{A,2}^{*(2)}}{|z_2|} +\cdots \right]+\cdots \ .
    \end{aligned}
    \end{equation}

    \item \textbf{Region P}: $z_1z_2<0$, $|z_1|\to\infty$, $|z_1-z_2|\to\infty$ while $|z_2|$ remains finite,
    \begin{equation}
    \begin{aligned}
        &\ \widetilde A_\Lambda(z_1,z_2;P^z) \\
        = &\ \rme^{\rmi z_1P^z} \rme^{-\Lambda^{0^-}|z_2|}
        \left[ \mathcal P_{A,1} + \frac{\mathcal P_{A,1}^{(1)}}{|z_1|} +\cdots \right] \\
        & + \rme^{-\Lambda^{0^-}|z_2|}
        \left[ \mathcal P_{A,2} + \frac{\mathcal P_{A,2}^{(1)}}{|z_1|} +\cdots \right] \\
        & + \rme^{\rmi z_2P^z} \rme^{-\Lambda^{1/2^-}|z_2|}
        \left[ \mathcal P_{A,3} + \frac{\mathcal P_{A,3}^{(1)}}{|z_1|} +\cdots \right]+\cdots \ .
    \end{aligned}
    \end{equation}

    \item \textbf{Region B}: $z_1z_2>0$, $|z_1|\to\infty$, $|z_1-z_2|\to\infty$ while $|z_2|$ remains finite,
    \begin{equation}
    \begin{aligned}
        &\ \widetilde A_\Lambda(z_1,z_2;P^z) \\
        = &\ \rme^{-\Lambda^{0^-}|z_2-z_1|}
        \left[ \mathcal B_{A,1} + \frac{\mathcal B_{A,1}^{(1)}}{|z_1|} +\cdots \right] \\
        & + \rme^{\rmi z_1P^z} \rme^{-\Lambda^{0^-}|z_2-z_1|}
        \left[ \mathcal B_{A,2} + \frac{\mathcal B_{A,2}^{(1)}}{|z_1|} +\cdots \right] \\
        & + \rme^{\rmi z_2P^z} \rme^{-\Lambda^{1/2^-}|z_2-z_1|}
        \left[ \mathcal B_{A,3} + \frac{\mathcal B_{A,3}^{(1)}}{|z_1|} +\cdots \right]+\cdots \ .
    \end{aligned}
    \end{equation}

    \item \textbf{Region O}: $z_1z_2>0$, $|z_1|\to\infty$, $|z_2|\to\infty$ and $|z_1-z_2|\to\infty$,
    \begin{equation}
    \begin{aligned}
        &\ \widetilde A_\Lambda(z_1,z_2;P^z) \\
        =&\ \rme^{\rmi z_1P^z} \rme^{-\Lambda^{0^-}|z_1|} \rme^{-\Lambda^{0^-}|z_2-z_1|}
        \left[ \mathcal O_{A,1} + \frac{\mathcal O_{A,1}^{(1)}}{|z_1|} + \frac{\mathcal O_{A,1}^{(12)}}{|z_1-z_2|} +\cdots \right] \\
        &+ \rme^{\rmi z_2P^z} \rme^{-\Lambda^{0^-}|z_1|} \rme^{-\Lambda^{1/2^-}|z_2-z_1|}
        \left[ \mathcal O_{A,2} + \frac{\mathcal O_{A,2}^{(1)}}{|z_1|} + \frac{\mathcal O_{A,2}^{(12)}}{|z_1-z_2|} +\cdots \right] \\
        &+ \rme^{-\Lambda^{0^-}|z_2-z_1|} \rme^{-\Lambda^{1/2^-}|z_1|}
        \left[ \mathcal O_{A,3} + \frac{\mathcal O_{A,3}^{(1)}}{|z_1|} + \frac{\mathcal O_{A,3}^{(12)}}{|z_1-z_2|} +\cdots \right]+\cdots \ .
    \end{aligned}
    \end{equation}

    \item \textbf{Region Y}: $z_1z_2>0$, $|z_1|\to\infty$, $|z_2|\to\infty$ while $|z_1-z_2|$ remains finite,
    \begin{equation}
    \begin{aligned}
        &\ \widetilde A_\Lambda(z_1,z_2;P^z) \\
        = &\ \rme^{\rmi z_2P^z} \rme^{-\Lambda^{0^-}|z_1|}
        \left[ \mathcal Y_{A,1} + \frac{\mathcal Y_{A,1}^{(2)}}{|z_2|} +\cdots \right] \\
        & + \rme^{\rmi z_1P^z} \rme^{-\Lambda^{0^-}|z_1|}
        \left[ \mathcal Y_{A,2} + \frac{\mathcal Y_{A,2}^{(2)}}{|z_2|} +\cdots \right] \\
        & + \rme^{-\Lambda^{1/2^-}|z_1|}
        \left[ \mathcal Y_{A,3} + \frac{\mathcal Y_{A,3}^{(2)}}{|z_2|} +\cdots \right]+\cdots \ .
    \end{aligned}
    \end{equation}
\end{itemize}

For the antisymmetric $V$ and $T$ amplitudes of the $\Lambda$ baryon, the explicit expressions are:

\begin{itemize}
    \item \textbf{Region G}: $z_1z_2<0$, $|z_1|\to\infty$, $|z_2|\to\infty$ and $|z_1-z_2|\to\infty$,
    \begin{equation}
    \begin{aligned}
        &\ \widetilde V_\Lambda/\widetilde T_\Lambda(z_1,z_2;P^z) \\
        =&\ \rme^{-\Lambda^{0^-}|z_1|} \rme^{-\Lambda^{0^+}|z_2|}
        \left[ \mathcal G_{V/T,1} + \frac{\mathcal G_{V/T,1}^{(1)}}{|z_1|} + \frac{\mathcal G_{V/T,1}^{(2)}}{|z_2|} +\cdots \right] \\
        &- \rme^{-\Lambda^{0^+}|z_1|} \rme^{-\Lambda^{0^-}|z_2|}
        \left[ \mathcal G_{V/T,1}^* + \frac{\mathcal G_{V/T,1}^{*(1)}}{|z_1|} + \frac{\mathcal G_{V/T,1}^{*(2)}}{|z_2|} +\cdots \right] \\
        &+ \rme^{\rmi z_2 P^z} \rme^{-\Lambda^{0^-}|z_1|} \rme^{-\Lambda^{1/2^+}|z_2|}
        \left[ \mathcal G_{V/T,2} + \frac{\mathcal G_{V/T,2}^{(1)}}{|z_1|} + \frac{\mathcal G_{V/T,2}^{(2)}}{|z_2|} +\cdots \right] \\
        &- \rme^{\rmi z_2 P^z} \rme^{-\Lambda^{0^+}|z_1|} \rme^{-\Lambda^{1/2^-}|z_2|}
        \left[ \mathcal G_{V/T,3}^* + \frac{\mathcal G_{V/T,3}^{*(1)}}{|z_1|} + \frac{\mathcal G_{V/T,3}^{*(2)}}{|z_2|} +\cdots \right] \\
        &+ \rme^{\rmi z_1 P^z} \rme^{-\Lambda^{0^+}|z_2|} \rme^{-\Lambda^{1/2^-}|z_1|}
        \left[ \mathcal G_{V/T,3} + \frac{\mathcal G_{V/T,3}^{(1)}}{|z_1|} + \frac{\mathcal G_{V/T,3}^{(2)}}{|z_2|} +\cdots \right] \\
        &- \rme^{\rmi z_1 P^z} \rme^{-\Lambda^{0^-}|z_2|} \rme^{-\Lambda^{1/2^+}|z_1|}
        \left[ \mathcal G_{V/T,2}^* + \frac{\mathcal G_{V/T,2}^{*(1)}}{|z_1|} + \frac{\mathcal G_{V/T,2}^{*(2)}}{|z_2|} +\cdots \right]+\cdots \ .
    \end{aligned}
    \end{equation}

    \item \textbf{Region P}: $z_1z_2<0$, $|z_1|\to\infty$, $|z_1-z_2|\to\infty$ while $|z_2|$ remains finite,
    \begin{equation}
    \begin{aligned}
        &\ \widetilde V_\Lambda/\widetilde T_\Lambda(z_1,z_2;P^z) \\
        = &\ \rme^{\rmi z_1P^z} \rme^{-\Lambda^{0^-}|z_2|}
        \left[ \mathcal P_{V/T,1} + \frac{\mathcal P_{V/T,1}^{(1)}}{|z_1|} +\cdots \right] \\
        & + \rme^{-\Lambda^{0^-}|z_2|}
        \left[ \mathcal P_{V/T,2} + \frac{\mathcal P_{V/T,2}^{(1)}}{|z_1|} +\cdots \right] \\
        & + \rme^{\rmi z_2P^z} \rme^{-\Lambda^{1/2^+}|z_2|}
        \left[ \mathcal P_{V/T,3} + \frac{\mathcal P_{V/T,3}^{(1)}}{|z_1|} +\cdots \right]+\cdots \ .
    \end{aligned}
    \end{equation}

    \item \textbf{Region B}: $z_1z_2>0$, $|z_1|\to\infty$, $|z_1-z_2|\to\infty$ while $|z_2|$ remains finite,
    \begin{equation}
    \begin{aligned}
        &\ \widetilde V_\Lambda/\widetilde T_\Lambda(z_1,z_2;P^z) \\
        = &\ \rme^{-\Lambda^{0^-}|z_2-z_1|}
        \left[ \mathcal B_{V/T,1} + \frac{\mathcal B_{V/T,1}^{(1)}}{|z_1|} +\cdots \right] \\
        & + \rme^{\rmi z_1P^z} \rme^{-\Lambda^{0^-}|z_2-z_1|}
        \left[ \mathcal B_{V/T,2} + \frac{\mathcal B_{V/T,2}^{(1)}}{|z_1|} +\cdots \right] \\
        & + \rme^{\rmi z_2P^z} \rme^{-\Lambda^{1/2^+}|z_2-z_1|}
        \left[ \mathcal B_{V/T,3} + \frac{\mathcal B_{V/T,3}^{(1)}}{|z_1|} +\cdots \right]+\cdots \ .
    \end{aligned}
    \end{equation}

    \item \textbf{Region O}: $z_1z_2>0$, $|z_1|\to\infty$, $|z_2|\to\infty$ and $|z_1-z_2|\to\infty$,
    \begin{equation}
    \begin{aligned}
        &\ \widetilde V_\Lambda/\widetilde T_\Lambda(z_1,z_2;P^z) \\
        =&\ \rme^{\rmi z_1P^z} \rme^{-\Lambda^{0^-}|z_1|} \rme^{-\Lambda^{0^-}|z_2-z_1|}
        \left[ \mathcal O_{V/T,1} + \frac{\mathcal O_{V/T,1}^{(1)}}{|z_1|} + \frac{\mathcal O_{V/T,1}^{(12)}}{|z_1-z_2|} +\cdots \right] \\
        &+ \rme^{\rmi z_2P^z} \rme^{-\Lambda^{0^-}|z_1|} \rme^{-\Lambda^{1/2^+}|z_2-z_1|}
        \left[ \mathcal O_{V/T,2} + \frac{\mathcal O_{V/T,2}^{(1)}}{|z_1|} + \frac{\mathcal O_{V/T,2}^{(12)}}{|z_1-z_2|} +\cdots \right] \\
        &+ \rme^{-\Lambda^{0^-}|z_2-z_1|} \rme^{-\Lambda^{1/2^+}|z_1|}
        \left[ \mathcal O_{V/T,3} + \frac{\mathcal O_{V/T,3}^{(1)}}{|z_1|} + \frac{\mathcal O_{V/T,3}^{(12)}}{|z_1-z_2|} +\cdots \right]+\cdots \ .
    \end{aligned}
    \end{equation}

    \item \textbf{Region Y}: $z_1z_2>0$, $|z_1|\to\infty$, $|z_2|\to\infty$ while $|z_1-z_2|$ remains finite,
    \begin{equation}
    \begin{aligned}
        &\ \widetilde V_\Lambda/\widetilde T_\Lambda(z_1,z_2;P^z) \\
        = &\ \rme^{\rmi z_2P^z} \rme^{-\Lambda^{0^-}|z_1|}
        \left[ \mathcal Y_{V/T,1} + \frac{\mathcal Y_{V/T,1}^{(2)}}{|z_2|} +\cdots \right] \\
        & + \rme^{\rmi z_1P^z} \rme^{-\Lambda^{0^-}|z_1|}
        \left[ \mathcal Y_{V/T,2} + \frac{\mathcal Y_{V/T,2}^{(2)}}{|z_2|} +\cdots \right] \\
        & + \rme^{-\Lambda^{1/2^+}|z_1|}
        \left[ \mathcal Y_{V/T,3} + \frac{\mathcal Y_{V/T,3}^{(2)}}{|z_2|} +\cdots \right]+\cdots \ .
    \end{aligned}
    \end{equation}
\end{itemize}

At sufficiently large distances, the asymptotic behavior is governed by the terms with the weakest exponential suppression. We therefore retain only the contribution with the smallest binding energies $\Lambda^{0^-}$ and $\Lambda^{0^+}$, together with the leading and, when needed, next-to-leading inverse-power prefactors. In this work, these contributions are referred to as the leading-asymptotic (LA) and next-to-leading-asymptotic (NLA) terms, respectively.

\section{Explicit Definitions of the Matching Kernel}

This appendix collects the explicit expressions for the matching kernels used in the hybrid-scheme LaMET matching of this work, as introduced in Secs.~\ref{sec:MSbar_kernel} and~\ref{sec:hybrid_kernel}.

\subsection{Expressions of the \texorpdfstring{$\mathcal C_n$}{Cn} Functions}\label{app:kernel_MSbar}

The functions $\mathcal C_n$ appearing in one-loop matching kernel in the $\overline{\rm MS}$ scheme Eqs.~\eqref{eq:MSbar_kernel_VA}--\eqref{eq:MSbar_kernel_T} are explicitly defined as

\begin{equation}\label{eq:C235}
\begin{aligned}
    &\mathcal C_2(x_1, x_2; y_1, y_2; P^z,\mu)= \\[1em]
    &\begin{cases}
        \displaystyle
            \frac{(x_1+y_1) (x_3+y_3)}{y_1 y_3 (y_1-x_1)} \ln \frac{y_1-x_1}{-x_1}
            -\frac{x_3 (x_1+y_1+2 y_3) }{y_3 (y_1-x_1) (y_1+y_3)} \ln \frac{x_3}{-x_1}
        & x_1<0\\[1em]
        \displaystyle
            \frac{(x_1-3 y_1-2 y_3) x_1}{y_1 (x_3-y_3) (y_1+y_3)}
            -\frac{\left[(x_3-y_3)^2-2 x_3 y_1\right]}{(x_3-y_3) y_1 y_3} \ln \frac{x_3-y_3}{x_3}
            +\frac{2 x_1 \ln \frac{4 x_1  (x_3-y_3) (P^z)^2}{\mu^2}}{(x_3-y_3)y_1 }
            +\frac{x_1 \ln\frac{4 x_1 x_3 (P^z)^2}{\mu^2} }{y_1 (y_1+y_3)}
        & 0<x_1<y_1\\[1em]
        \displaystyle
            \frac{(x_3-2 y_1-3 y_3) x_3}{y_3 (x_1-y_1) (y_1+y_3)}
            -\frac{\left[(x_1-y_1)^2-2 x_1 y_3\right]}{(x_1-y_1) y_1 y_3} \ln \frac{x_1-y_1}{x_1}
            + \frac{2 x_3 \ln \frac{4 x_3 (x_1-y_1) (P^z)^2}{\mu^2}}{(x_1-y_1) y_3}
            +\frac{x_3 \ln \frac{4 x_1 x_3 (P^z)^2}{\mu^2}}{y_3 (y_1+y_3)}
        & y_1<x_1<y_1+y_3\\[1em]
        \displaystyle
            \frac{(x_1+y_1) (x_3+y_3) }{ y_1 y_3 (y_3-x_3)} \ln \frac{y_3-x_3}{-x_3}
            -\frac{x_1 (x_3+2 y_1+y_3) }{ y_1 (y_3-x_3) (y_1+y_3)} \ln \frac{x_1}{-x_3}
        & x_1>y_1+y_3
    \end{cases}\ .\\
    \\
    &\mathcal C_3(x_1, x_2; y_1, y_2; P^z,\mu)= \\[1em]
    &\begin{cases}
        \displaystyle
            \frac{x_1 x_2+y_1 y_2}{y_1 (x_2-y_2) y_2}\ln \frac{x_2-y_2}{x_2} 
            -\frac{x_1 (x_2+y_1) }{y_1 (x_2-y_2) (y_1+y_2)}\ln \frac{-x_1}{x_2}
        & x_1<0\\[1em]
        \displaystyle
            \frac{1}{x_1-y_1}+\frac{2 x_1+x_2}{y_1 \left(y_1+y_2\right)}
            +\frac{\left[ \left(x_1+y_2\right)y_1-x_1{}^2\right] }{\left(x_2-y_2\right) y_1 y_2} \ln \frac{x_2-y_2}{x_2}
            +\frac{x_1 \ln  \frac{4 x_1 \left(x_2-y_2\right) (P^z)^2}{\mu^2} }{(x_2-y_2)y_1 }
            +\frac{x_1 \ln \frac{4 x_1 x_2 (P^z)^2}{\mu^2} }{y_1 \left(y_1+y_2\right)}
        & 0<x_1<y_1\\[1em]
        \displaystyle
            \frac{1}{x_2-y_2}+\frac{  x_1+2 x_2 }{y_2 \left(y_1+y_2\right)}
            +\frac{ \left[\left(x_2+y_1\right) y_2-x_2{}^2\right]}{\left(x_1-y_1\right) y_1 y_2} \ln \frac{x_1-y_1}{x_1}
            +\frac{ x_2 \ln \frac{4 x_2 \left(x_1-y_1 \right)(P^z)^2}{\mu^2}}{\left(x_1-y_1\right) y_2}
            +\frac{ x_2 \ln \frac{4 x_1 x_2 (P^z)^2}{\mu^2}}{y_2 \left(y_1+y_2\right)}
        & y_1<x_1<y_1+y_2\\[1em]
        \displaystyle
            \frac{x_1 x_2+y_1 y_2}{y_1 (x_1-y_1) y_2} \ln \frac{x_1-y_1}{x_1}
            -\frac{x_2 (x_1+y_2)}{y_2 (x_1-y_1) \left(y_1+y_2\right)} \ln \frac{-x_2}{x_1}
        & x_1>y_1+y_2
    \end{cases}\ .\\
    \\
    &\mathcal C_5(x_1, x_2; y_1, y_2; P^z,\mu)=\\[1em]
    &\begin{cases}
        \displaystyle
            \frac{x_1}{y_1\left(y_1+y_2\right)} \ln \frac{x_2}{-x_1}
            +\frac{y_2-x_2}{y_1 y_2} \ln \frac{x_2-y_2}{x_2}
        & x_1<0\\[1em]
        \displaystyle
            \frac{x_1\left(\ln\frac{4 x_2 (P^z)^2}{\mu^2} +1\right)}{y_1( y_1 +y_2)}
            +\frac{x_2}{y_2( y_1 +y_2)}\ln x_2
            -\frac{x_2 -y_2}{y_1 y_2}\ln( x_2 -y_2)
        & 0<x_1<y_1\\[1em]
        \displaystyle
            \frac{x_2\left(\ln\frac{4 x_1 (P^z)^2}{\mu^2} +1\right)}{y_2( y_1 +y_2)}
            +\frac{x_1}{y_1( y_1 +y_2)}\ln x_1
            -\frac{ x_1 -y_1}{y_1 y_2}\ln( x_1 -y_1)
        & y_1<x_1<y_1+y_2\\[1em]
        \displaystyle
            \frac{x_1}{y_1\left(y_1+y_2\right)} \ln \frac{x_1}{-x_2}
            +\frac{y_2-x_2}{y_1 y_2} \ln \frac{-x_2}{y_2-x_2}
        & x_1>y_1+y_2
    \end{cases}\ .
\end{aligned}
\end{equation}

\subsection{Explicit Forms of the Hybrid Counterterms}\label{app:hybrid_count}

In hybrid renormalization scheme, to computing the hard kernels and express the complicated hybrid counterterms, it is convenient to define a set of master integrals as follows:

\begin{equation}
\begin{aligned}
    \mathrm{I1} \big( \{L_2,L_1 \},p \big) 
    \equiv &\ \int_{L_1}^{L_2} \frac{\rmd z}{2\pi} \rme^{-\rmi p z} \ln(z^2) \\
    =&\ +\frac{\rmi \Big( \gamma_E +\log (-\rmi p L_1) + \big(-1+\rme^{\rmi p L_1} \big) \log(L_1) + \Gamma (0,-\rmi p L_1 ) \Big)}{2 \pi  p}\\
    &\ -\frac{\rmi \Big( \gamma_E +\log(-\rmi p L_2) + \big(-1+\rme^{\rmi p L_2} \big) \log(L_2) + \Gamma (0,-\rmi p L_2 ) \Big)}{2 \pi  p}\ ,
\end{aligned}
\end{equation}

\begin{equation}
\begin{aligned}
    \mathrm{I1t} \big( \{L,-L \},p \big) 
    \equiv &\  \rmi \Big( \mathrm{I1} ( \{L,0 \},p ) -\mathrm{I1} ( \{0,-L \},p ) \Big)\\
    =&\  \frac{2 \Big(-\mathrm{Ci}(p L)+\log (L) \cos (p L)+\log (p)+\gamma_E \Big)}{\pi  p}\ ,
\end{aligned}
\end{equation}

\begin{equation}
\begin{aligned}
    \mathrm{I0} \big( \{L_2,L_1 \},p \big) 
    \equiv \int_{L_1}^{L_2} \frac{\rmd z}{2\pi} \rme^{\rmi p z}
    =-\frac{\rmi \big(-1+\rme^{\rmi p L_2} \big)}{2 \pi  p}+\frac{\rmi \big(-1+\rme^{\rmi p L_1} \big)}{2 \pi  p}\ ,
\end{aligned}
\end{equation}

\begin{equation}
\begin{aligned}
    \mathrm{I0} \big( \{\infty,L \},p \big) 
    = \frac{\delta (p)}{2} + \frac{\rmi \big(-1+\rme^{\rmi p L} \big)}{2 \pi  p} + \frac{\rmi}{2 \pi  p}\ ,
\end{aligned}
\end{equation}

\begin{equation}
\begin{aligned}
    \mathrm{I0} \big( \{-L,-\infty \},p \big) 
    = \frac{\delta (p)}{2} - \frac{\rmi \big(-1+\rme^{-\rmi p L} \big)}{2 \pi  p} - \frac{\rmi}{2 \pi  p}\ ,
\end{aligned}
\end{equation}

\begin{equation}
\begin{aligned}
    \mathrm{I0}\left(\left\{\infty,-\infty\right\},p\right) 
    =\delta (p)\ .
\end{aligned}
\end{equation}

The master integrals listed above provide the elementary building blocks for the hybrid counterterms in the hybrid renormalization scheme, allowing us to express and organize the analytic forms of the matching coefficients in a systematic manner. It should be emphasized that the logarithmic, trigonometric, and exponential functions appearing in these expressions are to be understood with dimensionless arguments. This is achieved by using a consistent convention for the units of coordinate-space separations $z$ and momenta $p$, or equivalently by working in natural units. With these conventions, all integrals entering the following formulas can be evaluated explicitly. In what follows, we take $z_s>0$.

\subsubsection{\texorpdfstring{$V$}{V} and \texorpdfstring{$A$}{A} counterterms}

For both the baryon octet and decuplet, the $V$ and $A$ amplitudes exhibit the same short-distance structures. As discussed in Sec.~\ref{sec:hybrid_2D}, in constructing the hard kernels, the zero-momentum matrix-element contribution is decomposed into different sub-regions, which allows each contribution to be computed separately. The functions entering Eq.~\eqref{eq:hyb_count_VA} are defined as:

\begin{equation}
\begin{aligned}
    &\ I^{V/A}_{\rm H} (p_1,p_2) \\
    \equiv&\ \int \frac{\rmd z_1}{2\pi} \frac{\rmd z_2}{2\pi} \rme^{-\rmi p_1 z_1 - \rmi p_2 z_2}  \Bigg[ \frac{7}{8} \ln(z_1^2) + \frac{7}{8} \ln(z_2^2) + \frac{3}{4} \ln((z_1-z_2)^2) \Bigg] \\
    & \qquad \qquad \times \Big(\theta(2z_s-|z_1|)\theta(z_s-|z_2|)+\theta(z_s-|z_1|)\theta(|z_2|-z_s)\theta(2z_s-|z_2|) \Big) \\
    =& \frac{7}{8} \left[
    \begin{aligned}
        +&\mathrm{I0}(\{2 z_s,-2 z_s\},p_2) \mathrm{I1}(\{z_s,-z_s\},p_1) +\mathrm{I1}(\{z_s,-z_s\},p_2) \Big(\mathrm{I0}(\{2 z_s,-2 z_s\},p_1)-\mathrm{I0}(\{z_s,-z_s\},p_1) \Big)\\
        +& \mathrm{I0}(\{z_s,-z_s\},p_1) \mathrm{I1}(\{2 z_s,-2 z_s\},p_2) + \mathrm{I0}(\{z_s,-z_s\},p_2) \Big(\mathrm{I1}(\{2 z_s,-2 z_s\},p_1)-\mathrm{I1}(\{z_s,-z_s\},p_1) \Big) 
    \end{aligned} \right]\\
    & +\frac{3}{8 \pi  (p_1+p_2)} \left[
    \begin{aligned}
        +& \sin ((p_1+p_2) z_s) \Big( +\mathrm{I1}(\{z_s,-z_s\},p_1)+\mathrm{I1}(\{z_s,-z_s\},-p_2) -\mathrm{I1}(\{2 z_s,-2 z_s\},p_1)\\
        & \qquad \qquad \qquad -\mathrm{I1}(\{2 z_s,-2 z_s\},-p_2) +\mathrm{I1}(\{3 z_s,-3 z_s\},p_1)+\mathrm{I1}(\{3 z_s,-3 z_s\},-p_2) \Big) \\
        +& \sin (2 (p_1+p_2) z_s) \Big( -\mathrm{I1}(\{z_s,-z_s\},p_1)-\mathrm{I1}(\{z_s,-z_s\},-p_2) \\
        & \qquad \qquad \qquad \qquad +\mathrm{I1}(\{3 z_s,-3 z_s\},p_1)+\mathrm{I1}(\{3 z_s,-3 z_s\},-p_2)\Big) \\
        +& \cos (2 (p_1+p_2) z_s) \Big( -\mathrm{I1t}(\{z_s,-z_s\},p_1)+\mathrm{I1t}(\{z_s,-z_s\},-p_2) \\
        & \qquad \qquad \qquad \qquad +\mathrm{I1t}(\{3 z_s,-3 z_s\},p_1)-\mathrm{I1t}(\{3 z_s,-3 z_s\},-p_2)\Big) \\
        +& \cos ((p_1+p_2) z_s) \Big(-\mathrm{I1t}(\{z_s,-z_s\},p_1)+\mathrm{I1t}(\{z_s,-z_s\},-p_2) -\mathrm{I1t}(\{2 z_s,-2 z_s\},p_1) \\
        & \qquad \qquad \qquad +\mathrm{I1t}(\{2 z_s,-2 z_s\},-p_2) +\mathrm{I1t}(\{3 z_s,-3 z_s\},p_1)-\mathrm{I1t}(\{3 z_s,-3 z_s\},-p_2)\Big)
    \end{aligned} \right]\ ,
\end{aligned}
\end{equation}

\begin{equation}
\begin{aligned}
    &\ I^{V/A}_{\rm HSI} (p_1,p_2) \\
    \equiv&\ \int \frac{\rmd z_1}{2\pi} \frac{\rmd z_2}{2\pi} \rme^{-\rmi p_1 z_1 - \rmi p_2 z_2} \theta(z_s-|z_1|)\theta(|z_2|-2z_s) 
   \times\Bigg[ \frac{7}{8} \ln(z_1^2) + \frac{7}{8} \ln((2 z_s)^2) + \frac{3}{4} \ln((z_1-2 z_s {\rm sign}[z_2])^2)\Bigg] \\
    =&\ \frac{1}{8} \left[
    \begin{aligned}
        +& 6 \rme^{2 \rmi p_1 z_s} \mathrm{I1}(\{-z_s,-3 z_s\},p_1) \mathrm{I0}(\{\infty ,2 z_s\},p_2) 
         +6 \rme^{-2 \rmi p_1 z_s} \mathrm{I1}(\{3 z_s,z_s\},p_1) \mathrm{I0}(\{-2 z_s,-\infty \},p_2)\\
        +& 7 \Big(\mathrm{I0}(\{\infty ,-\infty\},p_2)-\mathrm{I0}(\{2 z_s,-2 z_s\},p_2) \Big) \Big(\log (4 z_s^2) \mathrm{I0}(\{z_s,-z_s\},p_1)+\mathrm{I1}(\{z_s,-z_s\},p_1) \Big)
    \end{aligned} \right]\ ,
\end{aligned}
\end{equation}

\begin{equation}
\begin{aligned}
    &\ I^{V/A}_{\rm HSII} (p_1,p_2) \\
    \equiv&\ \int \frac{\rmd z_1}{2\pi} \frac{\rmd z_2}{2\pi} \rme^{-\rmi p_1 z_1 - \rmi p_2 z_2} \theta(|z_1|-2z_s)\theta(z_s-|z_2|) \times  \Bigg[ \frac{7}{8} \ln((2 z_s)^2) + \frac{7}{8} \ln(z_2^2) + \frac{3}{4} \ln(({\rm sign}[z_1]2 z_s- z_2)^2) \Bigg] \\
    =&\ \frac{1}{8} \left[
    \begin{aligned}
        +& 6 \rme^{2 \rmi p_2 z_s} \mathrm{I1}(\{-z_s,-3 z_s\},p_2) \mathrm{I0}(\{\infty ,2 z_s\},p_1) +6 \rme^{-2 \rmi p_2 z_s} \mathrm{I1}(\{3 z_s,z_s\},p_2) \mathrm{I0}(\{-2 z_s,-\infty \},p_1) \\
        +& 7 \Big(\mathrm{I0}(\{\infty ,-\infty \},p_1)-\mathrm{I0}(\{2 z_s,-2 z_s\},p_1) \Big) \Big(\log (4 z_s^2) \mathrm{I0}(\{z_s,-z_s\},p_2)+\mathrm{I1}(\{z_s,-z_s\},p_2) \Big)
    \end{aligned} \right]\ ,
\end{aligned}
\end{equation}

\begin{equation}
\begin{aligned}
    &\ I^{V/A}_{\rm HSIII} (p_1,p_2) \\
    \equiv&\ \int \frac{\rmd z_1}{2\pi} \frac{\rmd z_2}{2\pi} \rme^{-\rmi p_1 z_1 - \rmi p_2 z_2} \theta(|z_1|-z_s)\theta(|z_2|-z_s)\theta(z_s-|z_1-z_2|) \\
    &\ \times \Bigg[ \frac{7}{8} \ln ((z_s+(z_1-z_2) \theta (z_1-z_2)){}^2)+\frac{7}{8} \ln ((z_s+(z_2-z_1) \theta (z_2-z_1)){}^2) +\frac{3}{4} \ln ((z_1-z_2){}^2) \Bigg] \\
    =& \frac{\mathrm{I0}(\{\infty ,-\infty \},p_1+p_2)}{8} \left[
    \begin{aligned}
        +& 7\log (z_s^2) \mathrm{I0}(\{z_s,-z_s\},\frac{1}{2}(p_1-p_2)) +7\rme^{-\frac{1}{2} \rmi (p_1-p_2) z_s} \mathrm{I1}(\{2 z_s,z_s\},\frac{1}{2} (p_1-p_2))\\
        +& 6 \mathrm{I1}(\{z_s,-z_s\},\frac{1}{2} (p_1-p_2)) +7\rme^{\frac{1}{2} \rmi (p_1-p_2) z_s} \mathrm{I1}(\{-z_s,-2 z_s\},\frac{1}{2} (p_1-p_2))
    \end{aligned} \right] \\
    & + \frac{\rmi \rme^{-\rmi (p_1+p_2) z_s}}{16 \pi  (p_1+p_2)} \left[
    \begin{aligned}
        &-7\log (z_s^2) \mathrm{I0}(\{0,-z_s\},p_1) +7\rme^{2 \rmi (p_1+p_2) z_s} \log (z_s^2) \mathrm{I0}(\{0,-z_s\},-p_2) -6 \mathrm{I1}(\{0,-z_s\},p_1) \\
        &-7\log (z_s^2) \mathrm{I0}(\{z_s,0\},-p_2) +7\rme^{2 \rmi (p_1+p_2) z_s} \log (z_s^2) \mathrm{I0}(\{z_s,0\},p_1) -7\rme^{\rmi p_2 z_s} \mathrm{I1}(\{2 z_s,z_s\},-p_2)\\
        &+6 \rme^{2 \rmi (p_1+p_2) z_s} \mathrm{I1}(\{0,-z_s\},-p_2) +6 \rme^{2 \rmi(p_1+p_2) z_s} \mathrm{I1}(\{z_s,0\},p_1) -7\rme^{\rmi p_1 z_s} \mathrm{I1}(\{-z_s,-2 z_s\},p_1)\\
        &+7\rme^{\rmi (2 p_1+p_2) z_s} \mathrm{I1}(\{-z_s,-2 z_s\},-p_2) +7\rme^{\rmi (p_1+2 p_2) z_s} \mathrm{I1}(\{2 z_s,z_s\},p_1) -6 \mathrm{I1}(\{z_s,0\},-p_2)
    \end{aligned} \right] \ ,
\end{aligned}
\end{equation}

\begin{equation}
\begin{aligned}
    &\ I^{V/A}_{\rm HSIV} (p_1,p_2) \\
    \equiv&\ \int \frac{\rmd z_1}{2\pi} \frac{\rmd z_2}{2\pi} \rme^{-\rmi p_1 z_1 - \rmi p_2 z_2} \theta(|z_1|-z_s)\theta(|z_2|-z_s)\theta(z_s-|z_1+z_2|) \\
    &\ \times \Bigg[ \frac{7}{8} \ln ((z_s+(z_1+z_2) \theta (z_1+z_2)){}^2)+\frac{7}{8} \ln ((-z_s+(z_2+z_1) \theta (-z_2-z_1)){}^2) + \frac{3}{4} \ln ((2 z_s + |z_1+z_2|){}^2) \Bigg] \\
    =& \frac{\mathrm{I0}(\{\infty ,-\infty \},\frac{1}{2} (p_1-p_2))}{16} \left[
    \begin{aligned}
            +& 7\log (z_s^2) \mathrm{I0}(\{z_s,-z_s\},\frac{1}{2} (p_1+p_2)) \\
            +& 6 \rme^{\rmi (p_1+p_2) z_s} \mathrm{I1}(\{-2 z_s,-3z_s\},\frac{1}{2} (p_1+p_2)) + 7\rme^{-\frac{1}{2} \rmi (p_1+p_2) z_s} \mathrm{I1}(\{2 z_s,z_s\},\frac{1}{2} (p_1+p_2))\\
            +& 7\rme^{\frac{1}{2} \rmi (p_1+p_2) z_s} \mathrm{I1}(\{-z_s,-2 z_s\},\frac{1}{2} (p_1+p_2)) + 6 \rme^{-\rmi (p_1+p_2) z_s} \mathrm{I1}(\{3 z_s,2 z_s\},\frac{1}{2} (p_1+p_2))\\
    \end{aligned} \right] \\
    & +\frac{\rmi \rme^{-\rmi (p_1+p_2) z_s}}{16 \pi(p_1-p_2)} \left[
    \begin{aligned}
        &-7\rme^{2 \rmi p_2 z_s} \log (z_s^2) \mathrm{I0}(\{0,-z_s\},p_1) -7\rme^{\rmi (p_1+2 p_2) z_s} \mathrm{I1}(\{-z_s,-2z_s\},p_1) +6 \mathrm{I1}(\{3 z_s,2 z_s\},p_1)\\
        &+7\rme^{2 \rmi p_1 z_s} \log (z_s^2) \mathrm{I0}(\{0,-z_s\},p_2) +7\rme^{\rmi (2 p_1+p_2) z_s} \mathrm{I1}(\{-z_s,-2 z_s\},p_2) -6 \mathrm{I1}(\{3 z_s,2 z_s\},p_2)\\
        &+7\rme^{2 \rmi p_1 z_s} \log (z_s^2) \mathrm{I0}(\{z_s,0\},p_1) -6 \rme^{2 \rmi(p_1+p_2) z_s} \mathrm{I1}(\{-2 z_s,-3 z_s\},p_1) +7\rme^{\rmi p_1 z_s} \mathrm{I1}(\{2 z_s,z_s\},p_1)\\
        &-7\rme^{2 \rmi p_2 z_s} \log (z_s^2) \mathrm{I0}(\{z_s,0\},p_2) +6 \rme^{2 \rmi (p_1+p_2) z_s} \mathrm{I1}(\{-2 z_s,-3z_s\},p_2) -7\rme^{\rmi p_2 z_s} \mathrm{I1}(\{2 z_s,z_s\},p_2)\\
    \end{aligned} \right] \ ,
\end{aligned}
\end{equation}

\begin{equation}
\begin{aligned}
    &\ I^{V/A}_{\rm S} (p_1,p_2) \\
    \equiv&\ \int \frac{\rmd z_1}{2\pi} \frac{\rmd z_2}{2\pi} \rme^{-\rmi p_1 z_1 - \rmi p_2 z_2}  \theta(|z_1|-z_s)\theta(|z_2|-z_s)\theta(|z_1-z_2|-z_s)\theta(|z_1+z_2|-z_s)\\
    &\ \times\Bigg[ \frac{7}{8} \ln(z_s^2) + \frac{7}{8} \ln((2z_s)^2) + \frac{3}{4} \ln(({\rm sign}[z_1]z_s-{\rm sign}[z_2]2z_s)^2)\Bigg] \\
    =&\ \delta (p_1-p_2) \delta (p_1+p_2) (\frac{7}{4} \log (4 z_s^4)+\frac{3}{4} \log (9 z_s^4)) \\
    &\ -\frac{\delta (p_1+p_2) (6 \log (z_s^2)+7 \log (4 z_s^4)) \sin (\frac{1}{2} (p_1-p_2) z_s)}{4 \pi  (p_1-p_2)} -\frac{\delta (p_2) (20 \log (z_s)+\log (3456)) \sin (p_1 z_s)}{4 \pi  p_1}\\
    &\ -\frac{\delta (p_1-p_2) (6 \log (9 z_s^2)+7 \log (4 z_s^4)) \sin (\frac{1}{2} (p_1+p_2) z_s)}{4 \pi  (p_1+p_2)} -\frac{\delta (p_1) (20 \log (z_s)+\log (3456)) \sin (p_2 z_s)}{4 \pi  p_2} \\
    &\ -\frac{(20 \log (z_s)+\log (128)) (p_1 \cos ((p_1+2 p_2) z_s)+p_2 \cos ((2 p_1+p_2) z_s))}{8 \pi ^2 p_1 p_2 (p_1+p_2)} \\
    &\ +\frac{(20 \log (z_s)+\log (93312)) (p_1 \cos ((p_1-2 p_2) z_s)-p_2 \cos (2 p_1 z_s-p_2 z_s))}{8 \pi ^2 p_1 (p_1-p_2) p_2}\ .
\end{aligned}
\end{equation}

\subsubsection{\texorpdfstring{$T$}{T} counterterms}

The $T$ amplitude has different spin structure from $V$ and $A$ amplitudes, certain perturbative loop corrections vanish. Therefore, the $T$ amplitude exhibits the different behavior in hybrid counterterms. The functions appearing in Eqs.~\eqref{eq:hyb_count_T} are explicitly given by:

\begin{equation}
\begin{aligned}
    &\ I^{T}_{\rm H} (p_1,p_2) \\
    \equiv&\ \int \frac{\rmd z_1}{2\pi} \frac{\rmd z_2}{2\pi} \rme^{-\rmi p_1 z_1 - \rmi p_2 z_2} \Bigg[ \frac{7}{8} \ln(z_1^2) + \frac{7}{8} \ln(z_2^2) + \frac{1}{2} \ln((z_1-z_2)^2) \Bigg] \\
    & \qquad \qquad \times \Big(\theta(2z_s-|z_1|)\theta(z_s-|z_2|)+\theta(z_s-|z_1|)\theta(|z_2|-z_s)\theta(2z_s-|z_2|) \Big) \\
    =& \frac{7}{8} \left[
    \begin{aligned}
        +&\mathrm{I0}(\{2 z_s,-2 z_s\},p_2) \mathrm{I1}(\{z_s,-z_s\},p_1) + \mathrm{I1}(\{z_s,-z_s\},p_2) \Big(\mathrm{I0}(\{2 z_s,-2 z_s\},p_1)-\mathrm{I0}(\{z_s,-z_s\},p_1)\Big)\\
        +& \mathrm{I0}(\{z_s,-z_s\},p_1) \mathrm{I1}(\{2 z_s,-2 z_s\},p_2) + \mathrm{I0}(\{z_s,-z_s\},p_2) \Big(\mathrm{I1}(\{2 z_s,-2 z_s\},p_1)-\mathrm{I1}(\{z_s,-z_s\},p_1)\Big)
    \end{aligned} \right]\\
    & +\frac{1}{4 \pi  (p_1+p_2)} \left[
    \begin{aligned}
        +&\sin ((p_1+p_2) z_s) \Big( +\mathrm{I1}(\{z_s,-z_s\},p_1)+\mathrm{I1}(\{z_s,-z_s\},-p_2) -\mathrm{I1}(\{2 z_s,-2 z_s\},p_1)\\
        & \qquad \qquad \qquad -\mathrm{I1}(\{2 z_s,-2 z_s\},-p_2) +\mathrm{I1}(\{3 z_s,-3 z_s\},p_1)+\mathrm{I1}(\{3 z_s,-3 z_s\},-p_2) \Big) \\
        +& \sin (2 (p_1+p_2) z_s) \Big(-\mathrm{I1}(\{z_s,-z_s\},p_1)-\mathrm{I1}(\{z_s,-z_s\},-p_2) \\
        & \qquad \qquad \qquad \qquad +\mathrm{I1}(\{3 z_s,-3 z_s\},p_1)+\mathrm{I1}(\{3 z_s,-3 z_s\},-p_2)\Big) \\
        +& \cos (2 (p_1+p_2) z_s) \Big(-\mathrm{I1t}(\{z_s,-z_s\},p_1)+\mathrm{I1t}(\{z_s,-z_s\},-p_2) \\
        & \qquad \qquad \qquad \qquad +\mathrm{I1t}(\{3 z_s,-3 z_s\},p_1)-\mathrm{I1t}(\{3 z_s,-3 z_s\},-p_2) \Big) \\
        +& \cos ((p_1+p_2) z_s) \Big( -\mathrm{I1t}(\{z_s,-z_s\},p_1)+\mathrm{I1t}(\{z_s,-z_s\},-p_2) -\mathrm{I1t}(\{2 z_s,-2 z_s\},p_1) \\
        & \qquad \qquad \qquad +\mathrm{I1t}(\{2 z_s,-2 z_s\},-p_2) +\mathrm{I1t}(\{3 z_s,-3 z_s\},p_1)-\mathrm{I1t}(\{3 z_s,-3 z_s\},-p_2) \Big)
    \end{aligned} \right]\ ,
\end{aligned}
\end{equation}

\begin{equation}
\begin{aligned}
    &\ I^{T}_{\rm HSI} (p_1,p_2) \\
    \equiv&\ \int \frac{\rmd z_1}{2\pi} \frac{\rmd z_2}{2\pi} \rme^{-\rmi p_1 z_1 - \rmi p_2 z_2} \theta(z_s-|z_1|)\theta(|z_2|-2z_s) \times \Bigg[ \frac{7}{8} \ln(z_1^2) + \frac{7}{8} \ln((2 z_s)^2) + \frac{1}{2} \ln((z_1-2 z_s {\rm sign}[z_2])^2) \Bigg] \\
    =&\ \frac{1}{8} \left[
    \begin{aligned}
        +& 4 \rme^{2 \rmi p_1 z_s} \mathrm{I1}(\{-z_s,-3 z_s\},p_1) \mathrm{I0}(\{\infty ,2 z_s\},p_2) +4 \rme^{-2 \rmi p_1 z_s} \mathrm{I1}(\{3 z_s,z_s\},p_1) \mathrm{I0}(\{-2 z_s,-\infty \},p_2)\\
        +& 7 \Big(\mathrm{I0}(\{\infty ,-\infty \},p_2)-\mathrm{I0}(\{2 z_s,-2 z_s\},p_2) \Big) \Big(\log (4 z_s^2) \mathrm{I0}(\{z_s,-z_s\},p_1)+\mathrm{I1}(\{z_s,-z_s\},p_1) \Big)
    \end{aligned} \right]\ ,
\end{aligned}
\end{equation}

\begin{equation}
\begin{aligned}
    &\ I^{T}_{\rm HSII} (p_1,p_2) \\
    \equiv&\ \int \frac{\rmd z_1}{2\pi} \frac{\rmd z_2}{2\pi} \rme^{-\rmi p_1 z_1 - \rmi p_2 z_2} \theta(|z_1|-2z_s)\theta(z_s-|z_2|) \times  \Bigg[ \frac{7}{8} \ln((2 z_s)^2) + \frac{7}{8} \ln(z_2^2) + \frac{1}{2} \ln(({\rm sign}[z_1]2 z_s- z_2)^2)\Bigg] \\
    =&\ \frac{1}{8} \left[
    \begin{aligned}
        +& 4 \rme^{2 \rmi p_2 z_s} \mathrm{I1}(\{-z_s,-3 z_s\},p_2) \mathrm{I0}(\{\infty ,2 z_s\},p_1) + 4 \rme^{-2 \rmi p_2 z_s} \mathrm{I1}(\{3 z_s,z_s\},p_2) \mathrm{I0}(\{-2 z_s,-\infty \},p_1) \\
        +& 7 \Big(\mathrm{I0}(\{\infty ,-\infty \},p_1)-\mathrm{I0}(\{2 z_s,-2 z_s\},p_1) \Big) \Big(\log (4 z_s^2) \mathrm{I0}(\{z_s,-z_s\},p_2)+\mathrm{I1}(\{z_s,-z_s\},p_2) \Big)
    \end{aligned} \right]\ ,
\end{aligned}
\end{equation}

\begin{equation}
\begin{aligned}
    &\ I^{T}_{\rm HSIII} (p_1,p_2) \\
    \equiv&\ \int \frac{\rmd z_1}{2\pi} \frac{\rmd z_2}{2\pi} \rme^{-\rmi p_1 z_1 - \rmi p_2 z_2} \theta(|z_1|-z_s)\theta(|z_2|-z_s)\theta(z_s-|z_1-z_2|) \\
    &\ \times \Bigg[ \frac{7}{8} \ln ((z_s+(z_1-z_2) \theta (z_1-z_2)){}^2)+\frac{7}{8} \ln ((z_s+(z_2-z_1) \theta (z_2-z_1)){}^2) +\frac{1}{2} \ln ((z_1-z_2){}^2) \Bigg] \\
    =&\ \frac{\mathrm{I0}(\{\infty ,-\infty \},p_1+p_2)}{8} \left[
    \begin{aligned}
        & +7\log (z_s^2) \mathrm{I0}(\{z_s,-z_s\},\frac{1}{2} (p_1-p_2)) +7\rme^{-\frac{1}{2} \rmi (p_1-p_2) z_s} \mathrm{I1}(\{2 z_s,z_s\},\frac{1}{2} (p_1-p_2)) \\
        & +4 \mathrm{I1}(\{z_s,-z_s\},\frac{1}{2} (p_1-p_2)) +7\rme^{\frac{1}{2} \rmi (p_1-p_2) z_s} \mathrm{I1}(\{-z_s,-2 z_s\},\frac{1}{2} (p_1-p_2))
    \end{aligned} \right]\\
    & +\frac{\rmi \rme^{-\rmi (p_1+p_2) z_s}}{16 \pi  (p_1+p_2)} \left[
    \begin{aligned}
        & -7\log (z_s^2) \mathrm{I0}(\{0,-z_s\},p_1) +7\rme^{2 \rmi (p_1+p_2) z_s} \log (z_s^2) \mathrm{I0}(\{0,-z_s\},-p_2) -4 \mathrm{I1}(\{0,-z_s\},p_1) \\
        & -7\log (z_s^2) \mathrm{I0}(\{z_s,0\},-p_2) +7\rme^{2 \rmi (p_1+p_2) z_s} \log (z_s^2) \mathrm{I0}(\{z_s,0\},p_1) -7\rme^{\rmi p_2 z_s} \mathrm{I1}(\{2 z_s,z_s\},-p_2)\\
        & +4 \rme^{2 \rmi (p_1+p_2) z_s} \mathrm{I1}(\{0,-z_s\},-p_2) +4 \rme^{2 \rmi(p_1+p_2) z_s} \mathrm{I1}(\{z_s,0\},p_1) -7\rme^{\rmi p_1 z_s} \mathrm{I1}(\{-z_s,-2 z_s\},p_1)\\
        & +7\rme^{\rmi (2 p_1+p_2) z_s} \mathrm{I1}(\{-z_s,-2 z_s\},-p_2) +7\rme^{\rmi (p_1+2 p_2) z_s} \mathrm{I1}(\{2 z_s,z_s\},p_1) -4 \mathrm{I1}(\{z_s,0\},-p_2)
    \end{aligned} \right]\ ,
\end{aligned}
\end{equation}

\begin{equation}
\begin{aligned}
    &\ I^{T}_{\rm HSIV} (p_1,p_2) \\
    \equiv&\ \int \frac{\rmd z_1}{2\pi} \frac{\rmd z_2}{2\pi} \rme^{-\rmi p_1 z_1 - \rmi p_2 z_2} \theta(|z_1|-z_s)\theta(|z_2|-z_s)\theta(z_s-|z_1+z_2|) \\
    &\ \times \Bigg[ \frac{7}{8} \ln ((z_s+(z_1+z_2) \theta (z_1+z_2)){}^2)+\frac{7}{8} \ln ((-z_s+(z_2+z_1) \theta (-z_2-z_1)){}^2) +\frac{1}{2} \ln ((2 z_s + |z_1+z_2|){}^2) \Bigg] \\
    =&\ \frac{\mathrm{I0}(\{\infty ,-\infty \},\frac{1}{2} (p_1-p_2))}{16} \left[
    \begin{aligned}
        & +7\log (z_s^2) \mathrm{I0}(\{z_s,-z_s\},\frac{1}{2} (p_1+p_2)) \\
        & +4 \rme^{\rmi (p_1+p_2) z_s} \mathrm{I1}(\{-2 z_s,-3z_s\},\frac{1}{2} (p_1+p_2)) +4 \rme^{-\rmi (p_1+p_2) z_s} \mathrm{I1}(\{3 z_s,2 z_s\},\frac{1}{2} (p_1+p_2))\\
        & +7\rme^{-\frac{1}{2} \rmi (p_1+p_2) z_s} \mathrm{I1}(\{2 z_s,z_s\},\frac{1}{2} (p_1+p_2)) +7\rme^{\frac{1}{2} \rmi (p_1+p_2) z_s} \mathrm{I1}(\{-z_s,-2 z_s\},\frac{1}{2} (p_1+p_2))
    \end{aligned} \right]\\
    &\ +\frac{\rmi \rme^{-\rmi (p_1+p_2) z_s}}{16 \pi (p_1-p_2)} \left[
    \begin{aligned}
        & -7\rme^{2 \rmi p_2 z_s} \log (z_s^2) \mathrm{I0}(\{0,-z_s\},p_1) -7\rme^{\rmi (p_1+2 p_2) z_s} \mathrm{I1}(\{-z_s,-2 z_s\},p_1) +4 \mathrm{I1}(\{3 z_s,2 z_s\},p_1)\\
        & +7\rme^{2 \rmi p_1 z_s} \log (z_s^2) \mathrm{I0}(\{0,-z_s\},p_2) +7\rme^{\rmi (2 p_1+p_2) z_s} \mathrm{I1}(\{-z_s,-2 z_s\},p_2) -4 \mathrm{I1}(\{3 z_s,2 z_s\},p_2)\\
        & +7\rme^{2 \rmi p_1 z_s} \log (z_s^2) \mathrm{I0}(\{z_s,0\},p_1) -4 \rme^{2 \rmi(p_1+p_2) z_s} \mathrm{I1}(\{-2 z_s,-3 z_s\},p_1) +7\rme^{\rmi p_1 z_s} \mathrm{I1}(\{2 z_s,z_s\},p_1)\\
        & -7\rme^{2 \rmi p_2 z_s} \log (z_s^2) \mathrm{I0}(\{z_s,0\},p_2) +4 \rme^{2 \rmi (p_1+p_2) z_s} \mathrm{I1}(\{-2 z_s,-3z_s\},p_2) -7\rme^{\rmi p_2 z_s} \mathrm{I1}(\{2 z_s,z_s\},p_2)
    \end{aligned} \right]\ ,
\end{aligned}
\end{equation}

\begin{equation}
\begin{aligned}
    &\ I^{T}_{\rm S} (p_1,p_2) \\
    \equiv&\ \int \frac{\rmd z_1}{2\pi} \frac{\rmd z_2}{2\pi} \rme^{-\rmi p_1 z_1 - \rmi p_2 z_2}  \theta(|z_1|-z_s)\theta(|z_2|-z_s)\theta(|z_1-z_2|-z_s)\theta(|z_1+z_2|-z_s) \\
    &\ \times \Bigg[ \frac{7}{8} \ln(z_s^2) + \frac{7}{8} \ln((2z_s)^2) + \frac{1}{2} \ln(({\rm sign}[z_1]z_s-{\rm sign}[z_2]2z_s)^2)\Bigg]\\
    =&\ \delta (p_1-p_2) \delta (p_1+p_2) (\frac{7}{4} \log (4 z_s^4)+\frac{1}{2} \log (9 z_s^4)) \\
    &\ -\frac{\delta (p_1+p_2) (4 \log (z_s^2)+7 \log (4 z_s^4)) \sin (\frac{1}{2} (p_1-p_2) z_s)}{4 \pi  (p_1-p_2)} -\frac{\delta (p_2) (18 \log (z_s)+\log (1152)) \sin (p_1 z_s)}{4 \pi  p_1}\\
    &\ -\frac{\delta (p_1-p_2) (4 \log (9 z_s^2)+7 \log (4 z_s^4)) \sin (\frac{1}{2} (p_1+p_2) z_s)}{4 \pi  (p_1+p_2)} -\frac{\delta (p_1) (18 \log (z_s)+\log (1152)) \sin (p_2 z_s)}{4 \pi  p_2}\\
    &\ -\frac{(18 \log (z_s)+\log (128)) (p_1 \cos ((p_1+2 p_2) z_s)+p_2 \cos ((2 p_1+p_2) z_s))}{8 \pi ^2 p_1 p_2 (p_1+p_2)} \\
    &\ +\frac{(18 \log (z_s)+\log (10368)) (p_1 \cos ((p_1-2 p_2) z_s)-p_2 \cos (2 p_1 z_s-p_2 z_s))}{8 \pi ^2 p_1 (p_1-p_2) p_2}\ .
\end{aligned}
\end{equation}

\section{More Results for LaMET Matching}\label{app:matching_ens}

We present more details of the LCDAs obtained after LaMET matching in Figs.~\ref{fig:match_sub_C24P29}--\ref{fig:match_sub_H48P32}, on the different ensembles covering lattice spacings from $a\approx 0.105~{\rm fm}$ down to $0.052~{\rm fm}$, and pion masses from $m_\pi\approx 317~{\rm MeV}$ to $136~{\rm MeV}$. The calculations are performed at three or four large hadron momenta on each ensemble.

In each figure, the left panel shows the $x_1$-dependence at fixed $x_2=0.2$, the middle panel shows the $x_3$-dependence along the diagonal trajectory $x_1=x_2$, and the right panel shows the corresponding two-dimensional distribution in the physical momentum-fraction triangular region, $0\le x_1,x_2,x_3\le 1$ with $x_1+x_2+x_3=1$.

\begin{figure}[htbp]
    \centering
    \subfloat{
        \centering
        \includegraphics[width=0.28\textwidth]{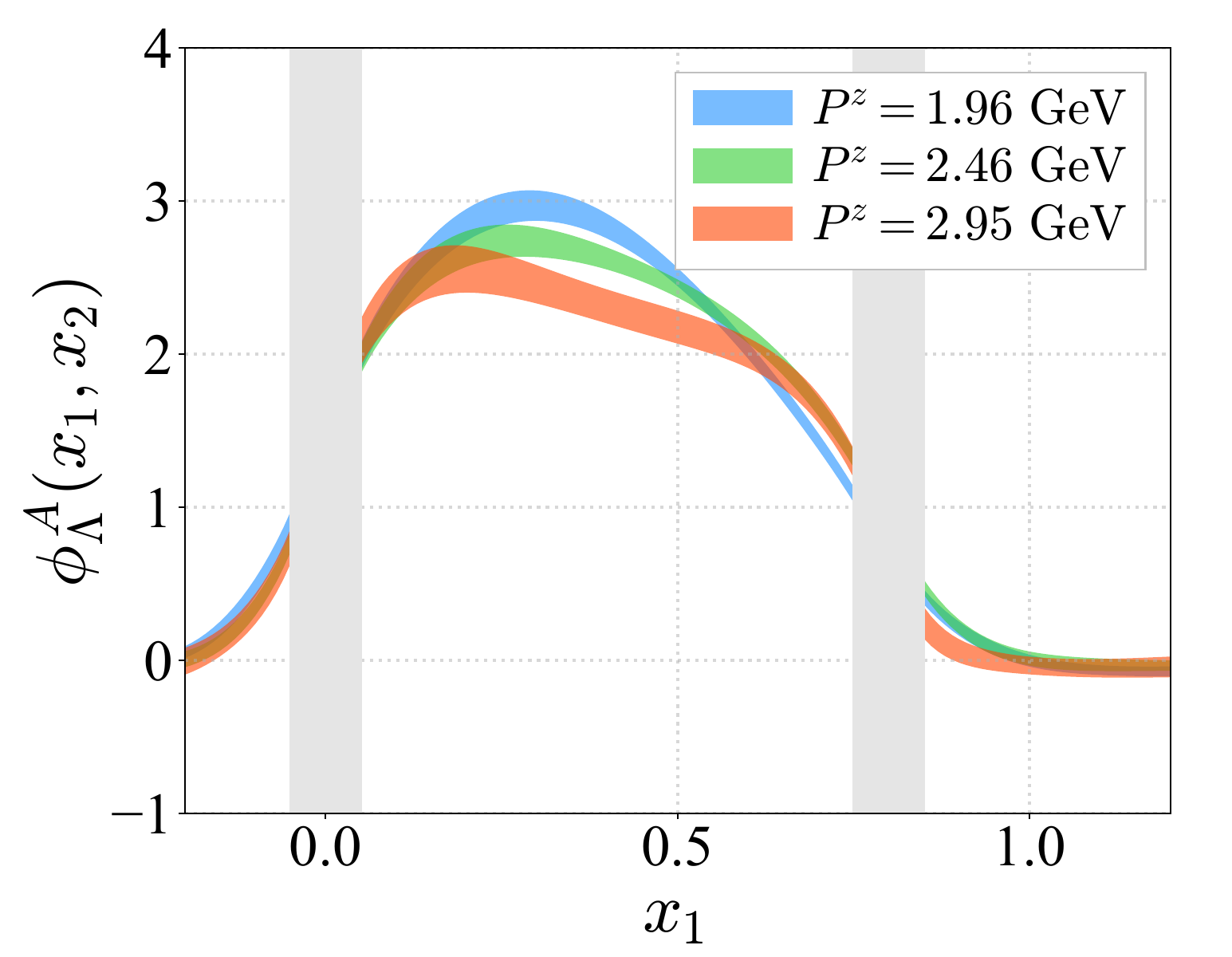}
        }\hspace{0.02\textwidth}
    \subfloat{
        \centering
        \includegraphics[width=0.28\textwidth]{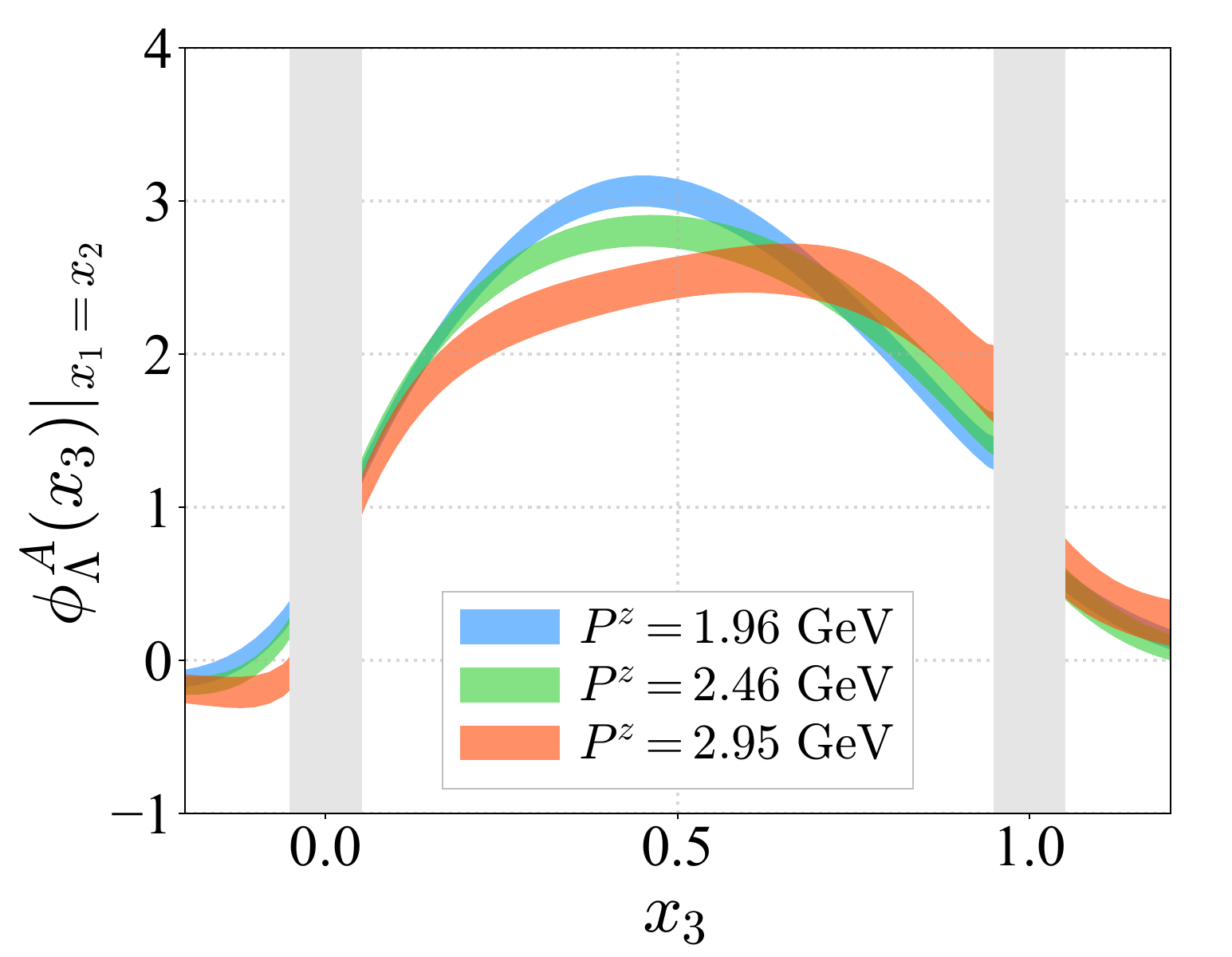}
        }\hspace{0.02\textwidth}
    \subfloat{
        \centering
        \includegraphics[width=0.28\textwidth]{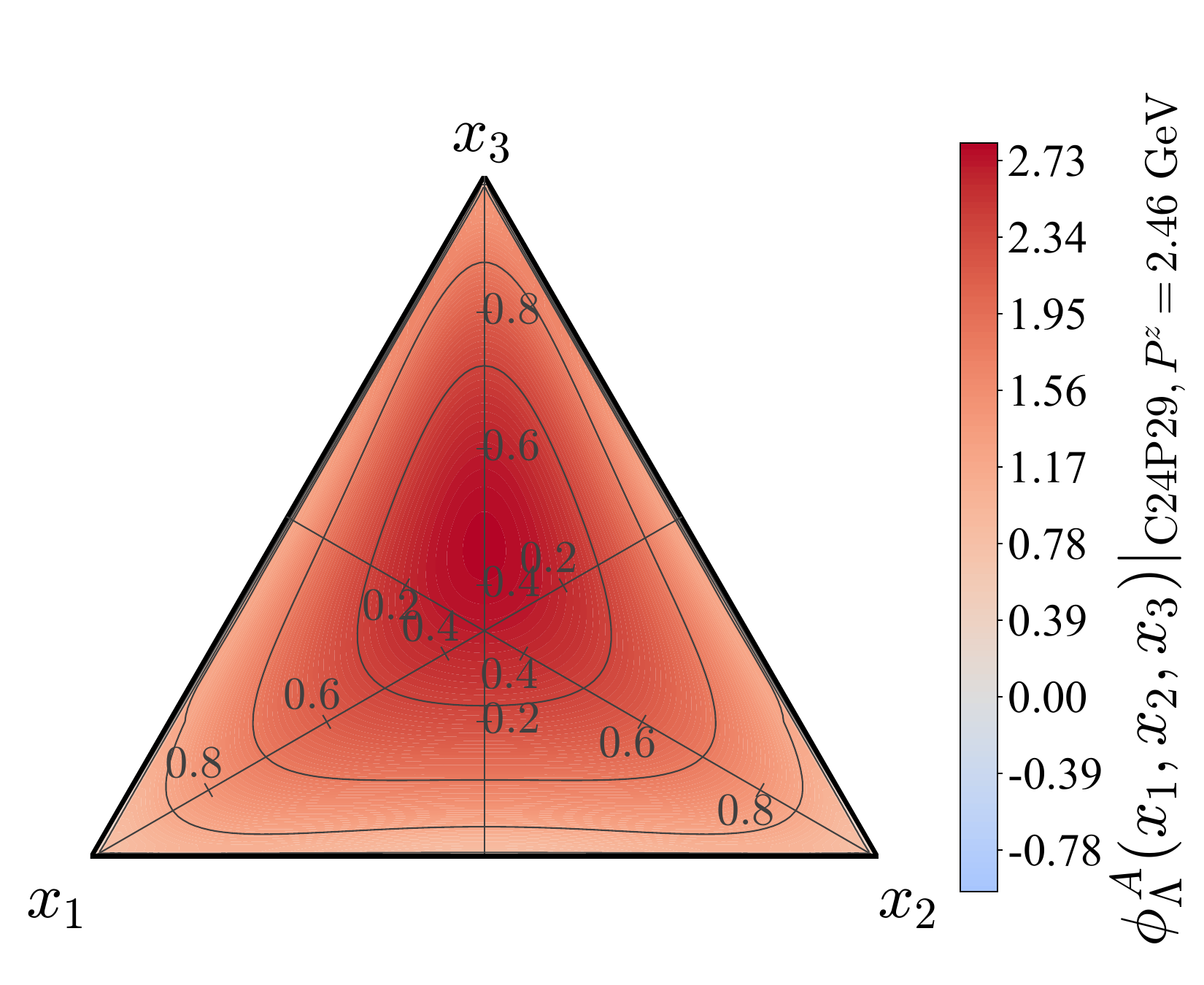}}
    \caption{Matched LCDAs for $\Lambda$ $A$ on C24P29 ($a \approx 0.1052~{\rm fm}$, $m_\pi\approx 292.3~{\rm MeV}$). Left: $x_1$-dependence at fixed $x_2=0.2$. Middle: $x_3$-dependence along $x_1=x_2$. Right: the heat map in the two-dimensional physical region.}\label{fig:match_sub_C24P29}
\end{figure}

\begin{figure}[htbp]
    \centering
    \subfloat{
        \centering
        \includegraphics[width=0.28\textwidth]{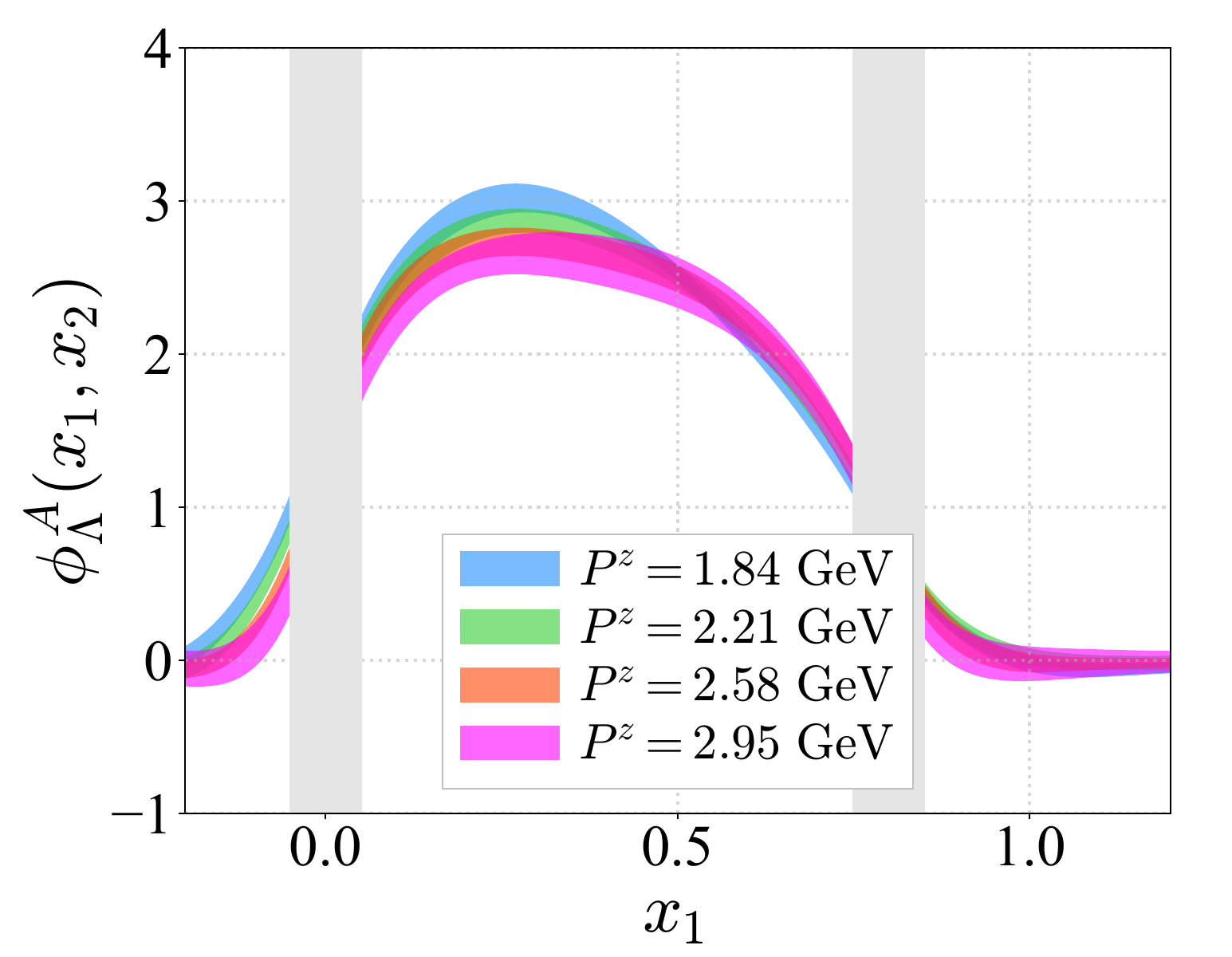}
        }\hspace{0.02\textwidth}
    \subfloat{
        \centering
        \includegraphics[width=0.28\textwidth]{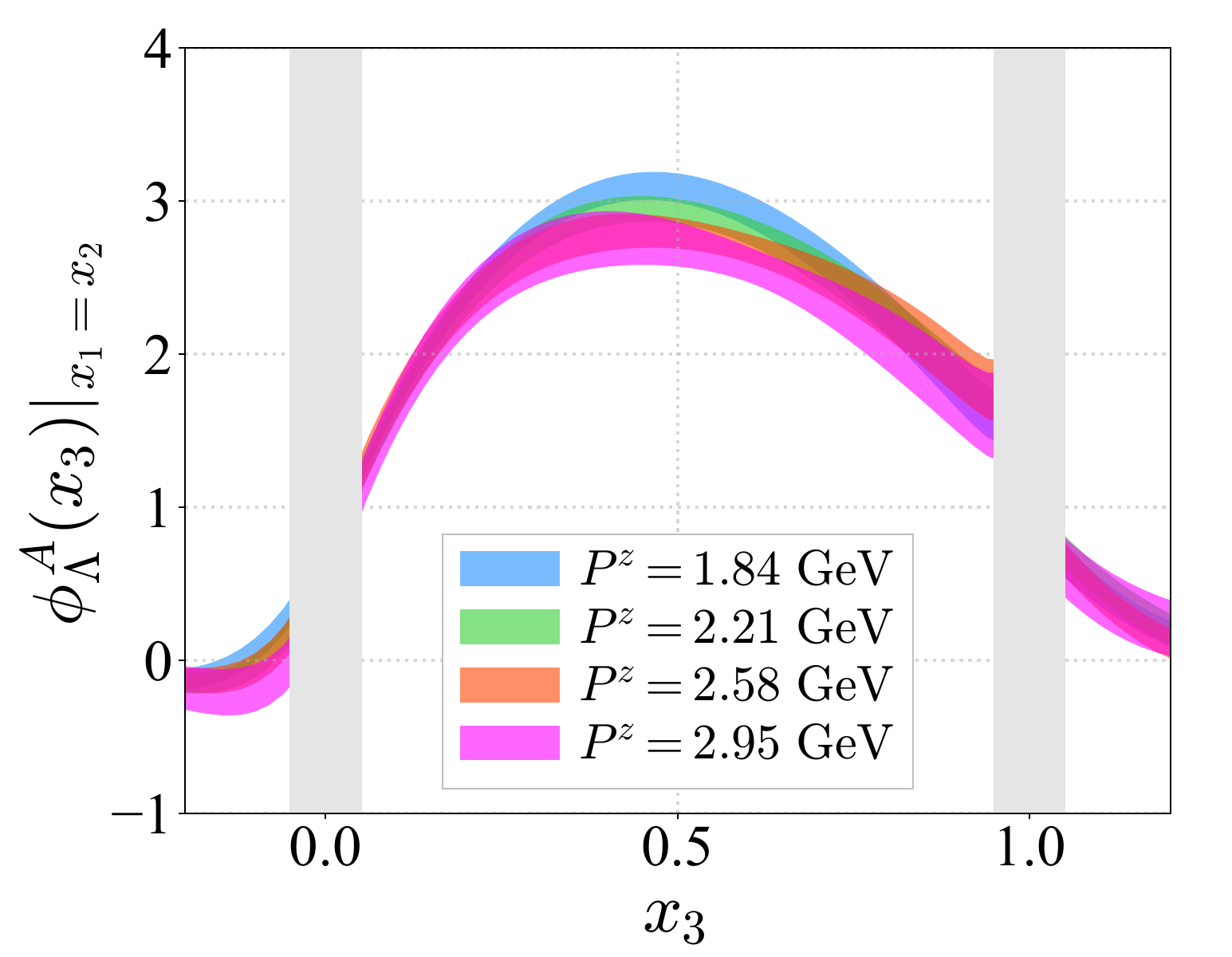}
        }\hspace{0.02\textwidth}
    \subfloat{
        \centering
        \includegraphics[width=0.28\textwidth]{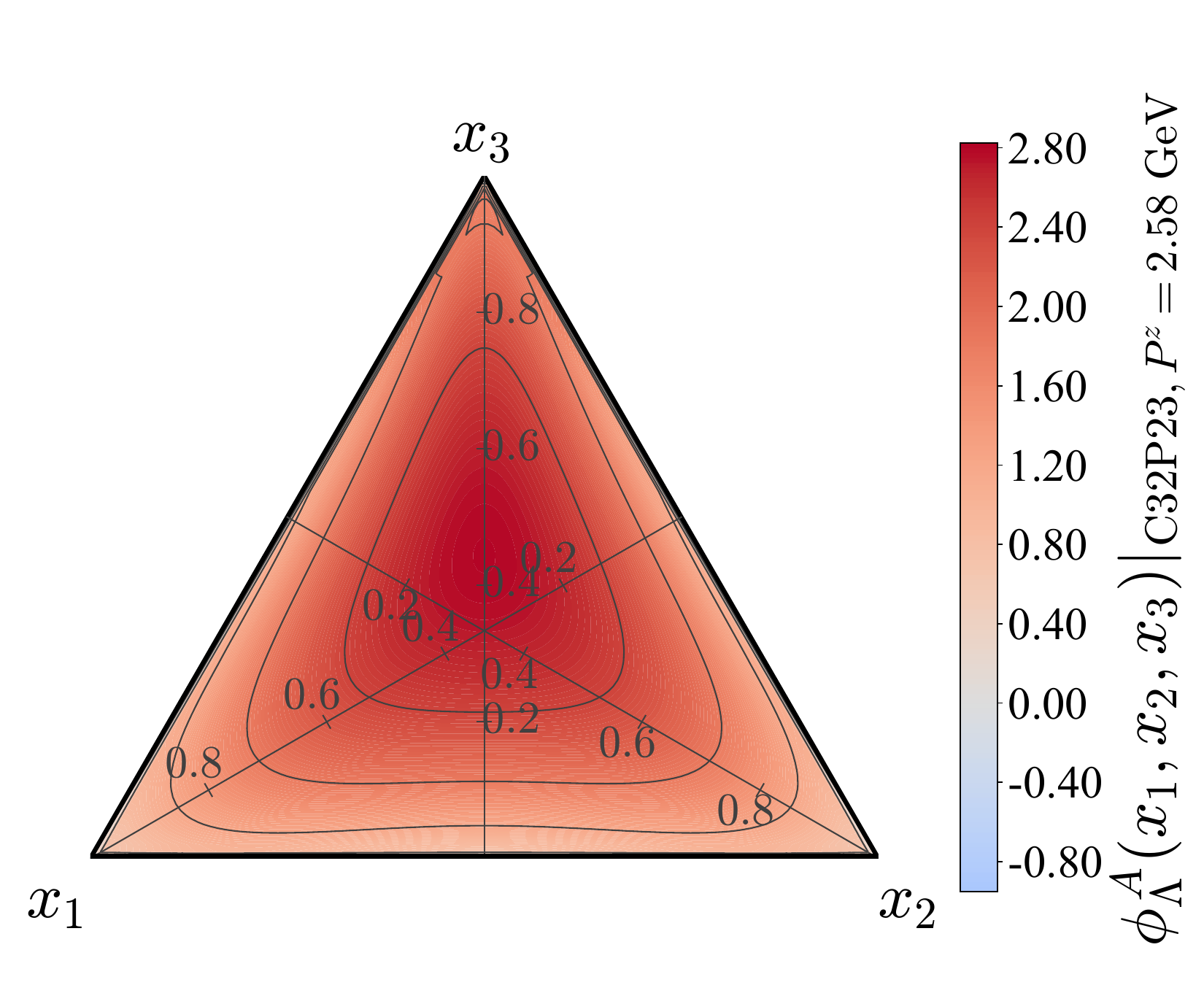}}
    \caption{Matched LCDAs for $\Lambda$ $A$ on C32P23 ($a \approx 0.1052~{\rm fm}$, $m_\pi\approx 227.9~{\rm MeV}$). Left: $x_1$-dependence at fixed $x_2=0.2$. Middle: $x_3$-dependence along $x_1=x_2$. Right: the heat map in the two-dimensional physical region.}\label{fig:match_sub_C32P23}
\end{figure}

\begin{figure}[htbp]
    \centering
    \subfloat{
        \centering
        \includegraphics[width=0.28\textwidth]{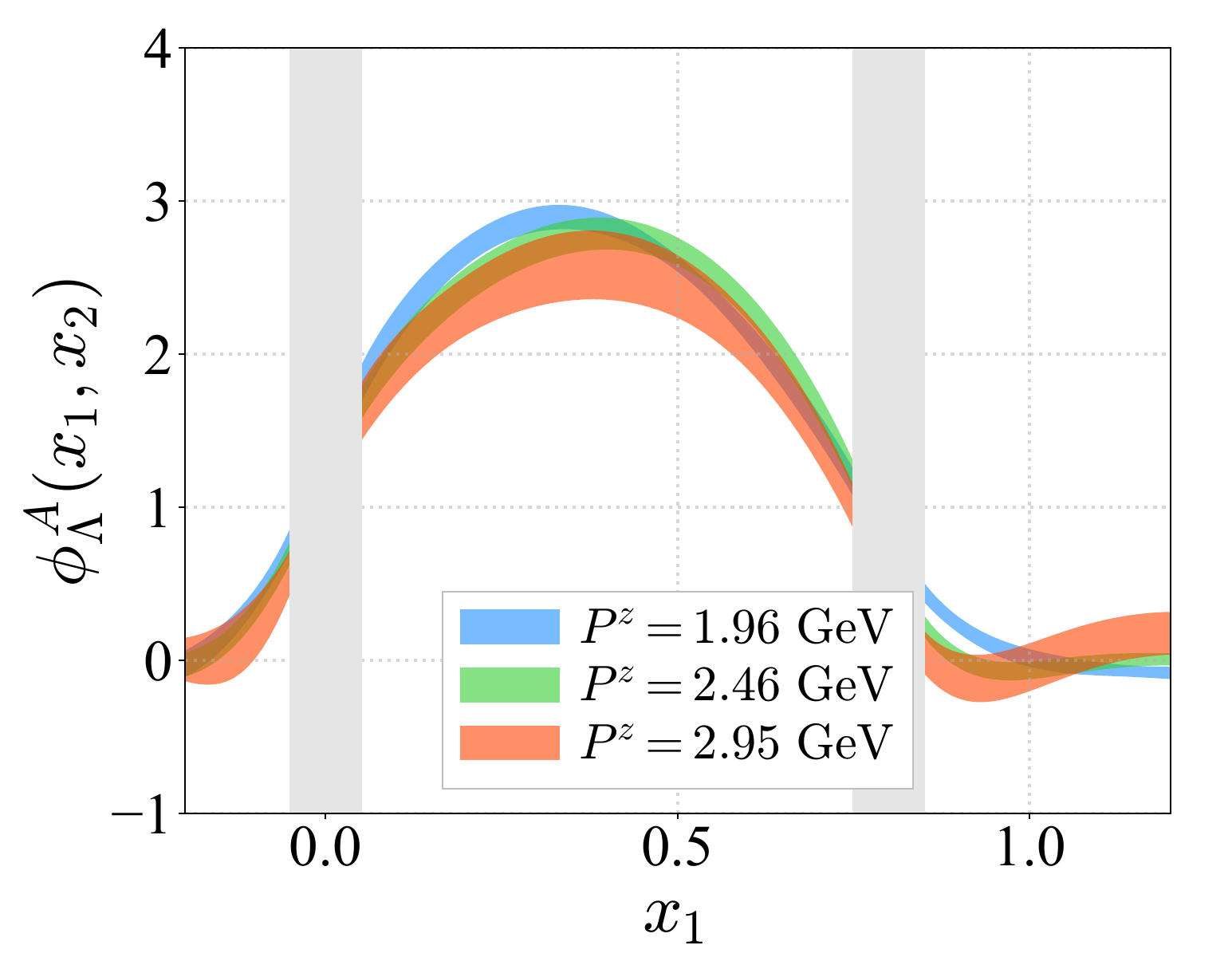}
        }\hspace{0.02\textwidth}
    \subfloat{
        \centering
        \includegraphics[width=0.28\textwidth]{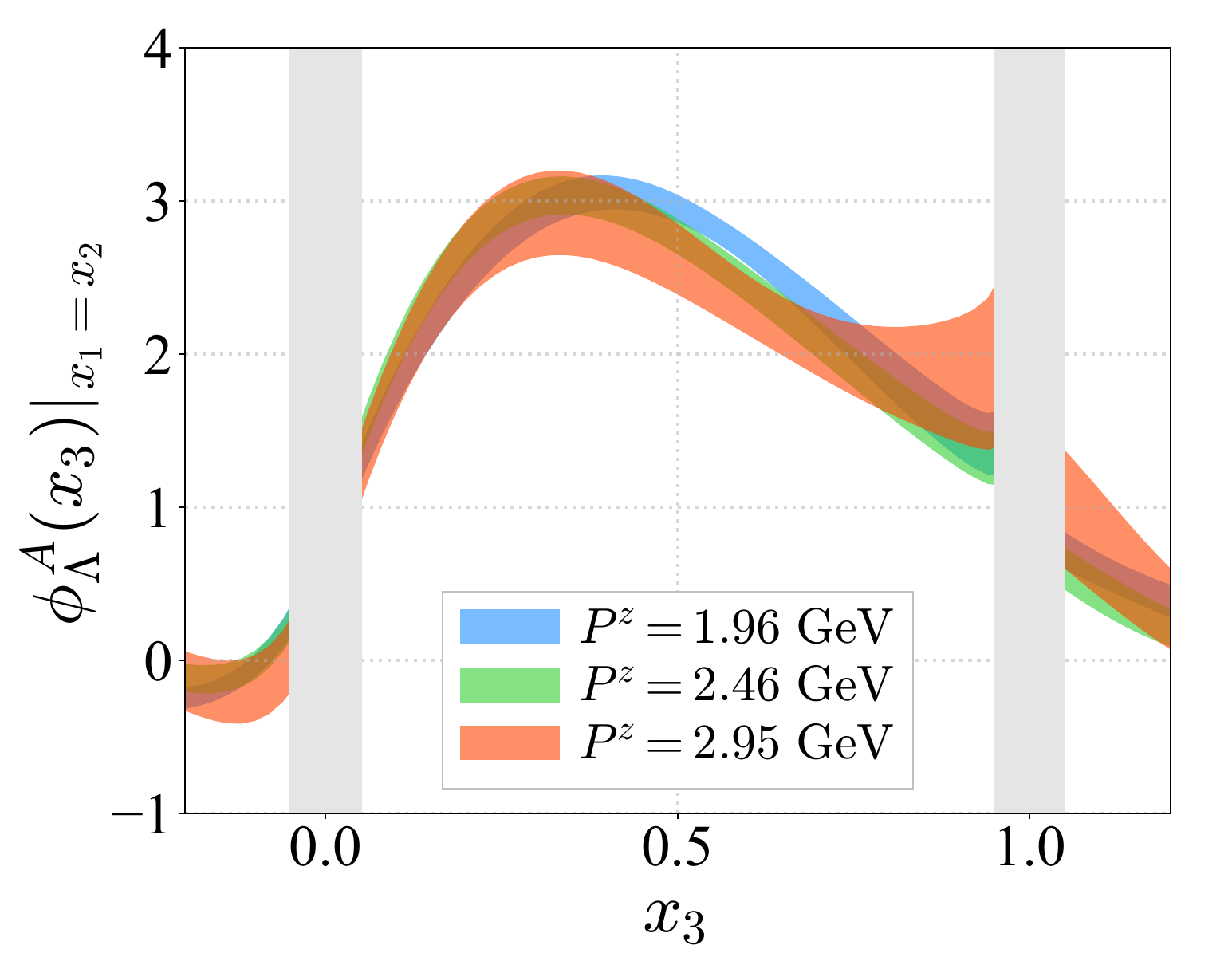}
        }\hspace{0.02\textwidth}
    \subfloat{
        \centering
        \includegraphics[width=0.28\textwidth]{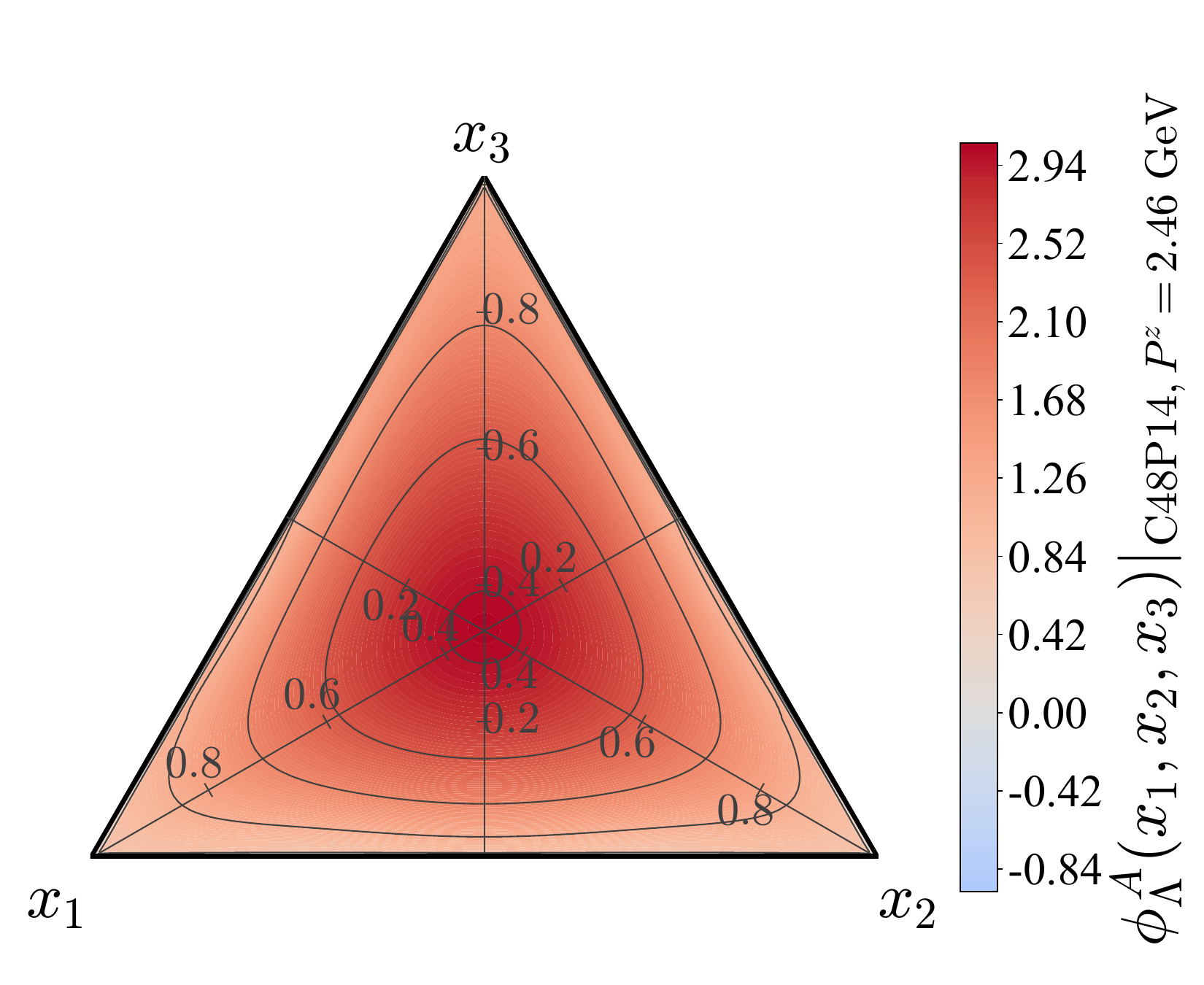}}
    \caption{Matched LCDAs for $\Lambda$ $A$ on C48P14 ($a \approx 0.1052~{\rm fm}$, $m_\pi\approx 136.4~{\rm MeV}$). Left: $x_1$-dependence at fixed $x_2=0.2$. Middle: $x_3$-dependence along $x_1=x_2$. Right: the heat map in the two-dimensional physical region.}\label{fig:match_sub_C48P14}
\end{figure}

\begin{figure}[htbp]
    \centering
    \subfloat{
        \centering
        \includegraphics[width=0.28\textwidth]{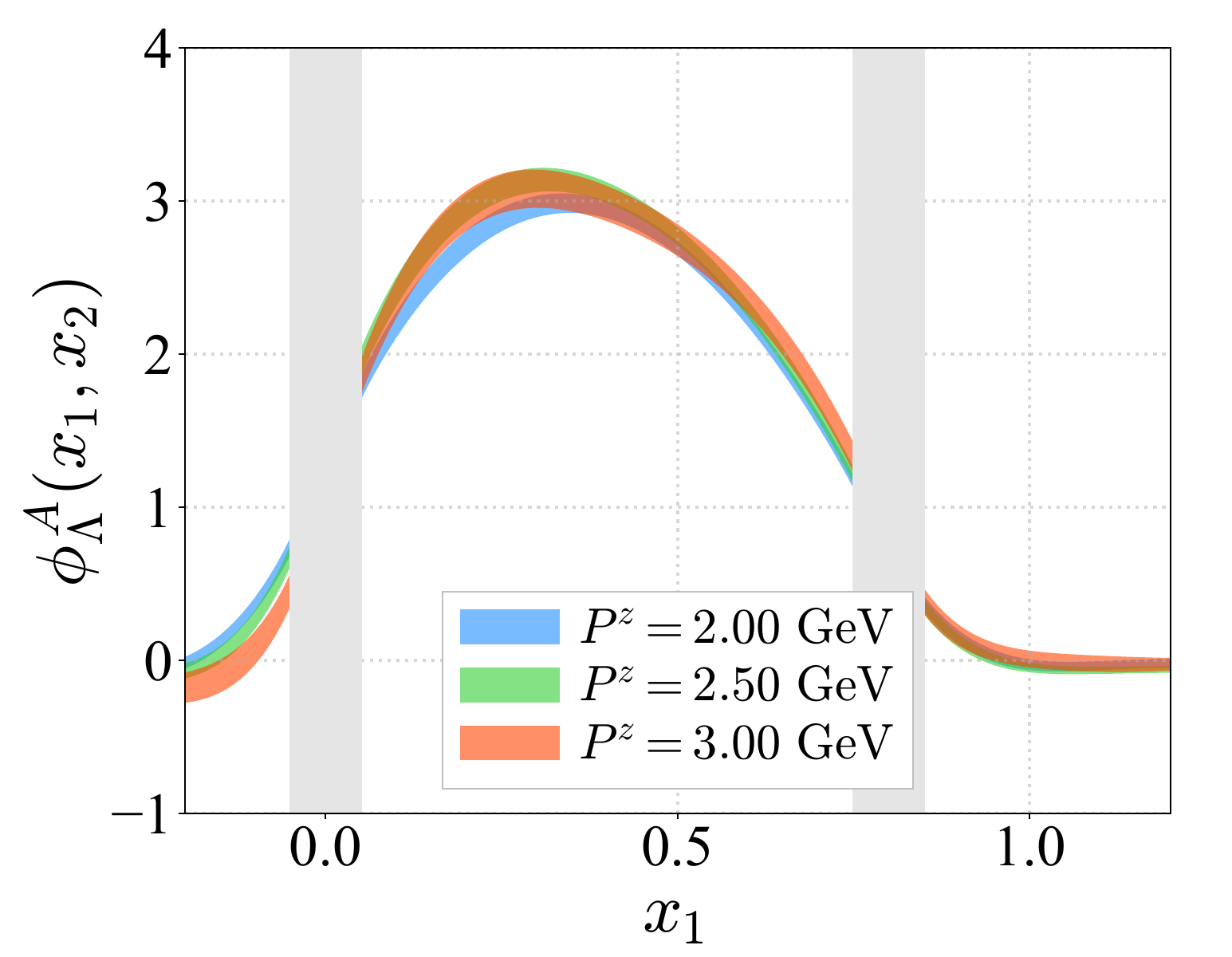}
        }\hspace{0.02\textwidth}
    \subfloat{
        \centering
        \includegraphics[width=0.28\textwidth]{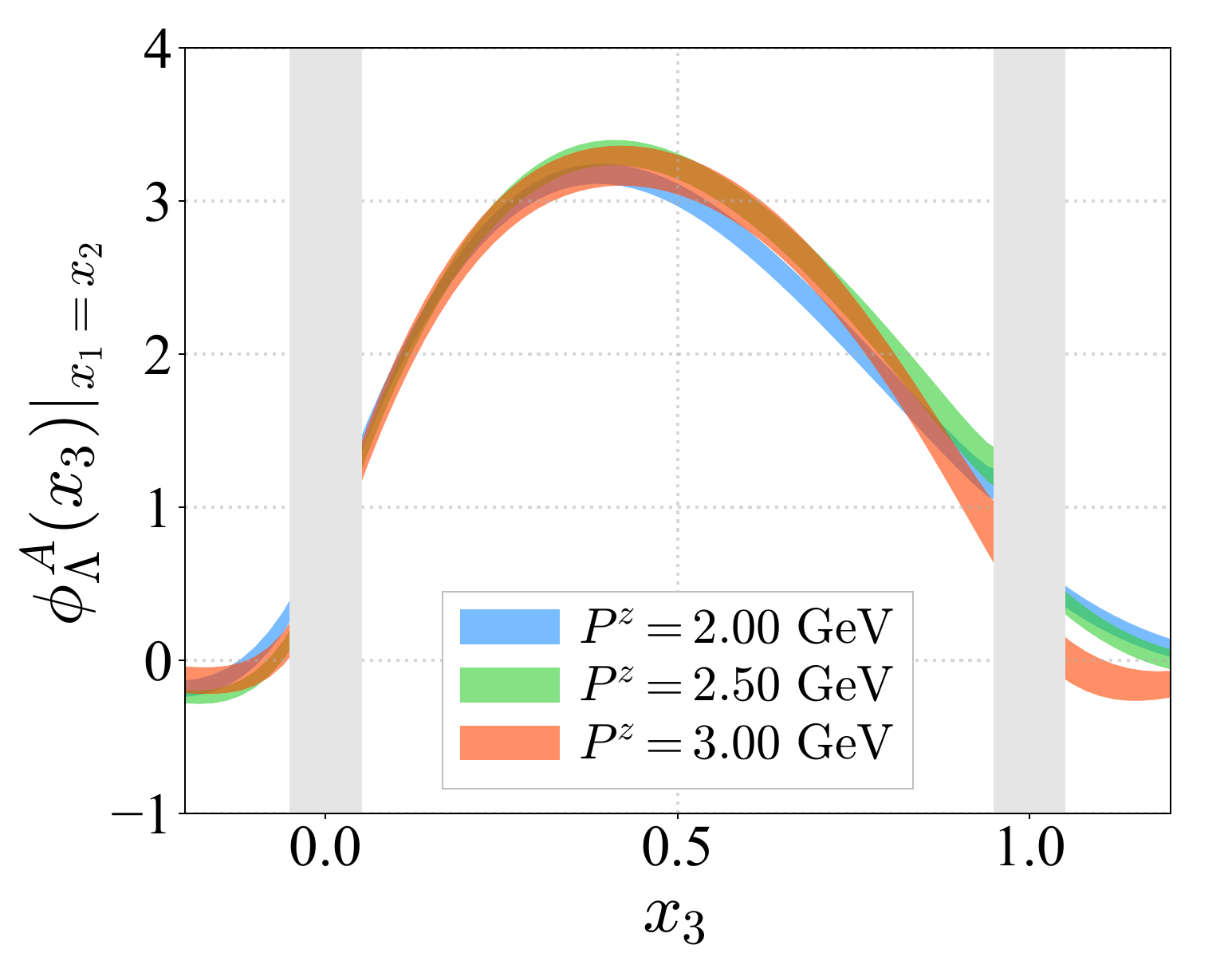}
        }\hspace{0.02\textwidth}
    \subfloat{
        \centering
        \includegraphics[width=0.28\textwidth]{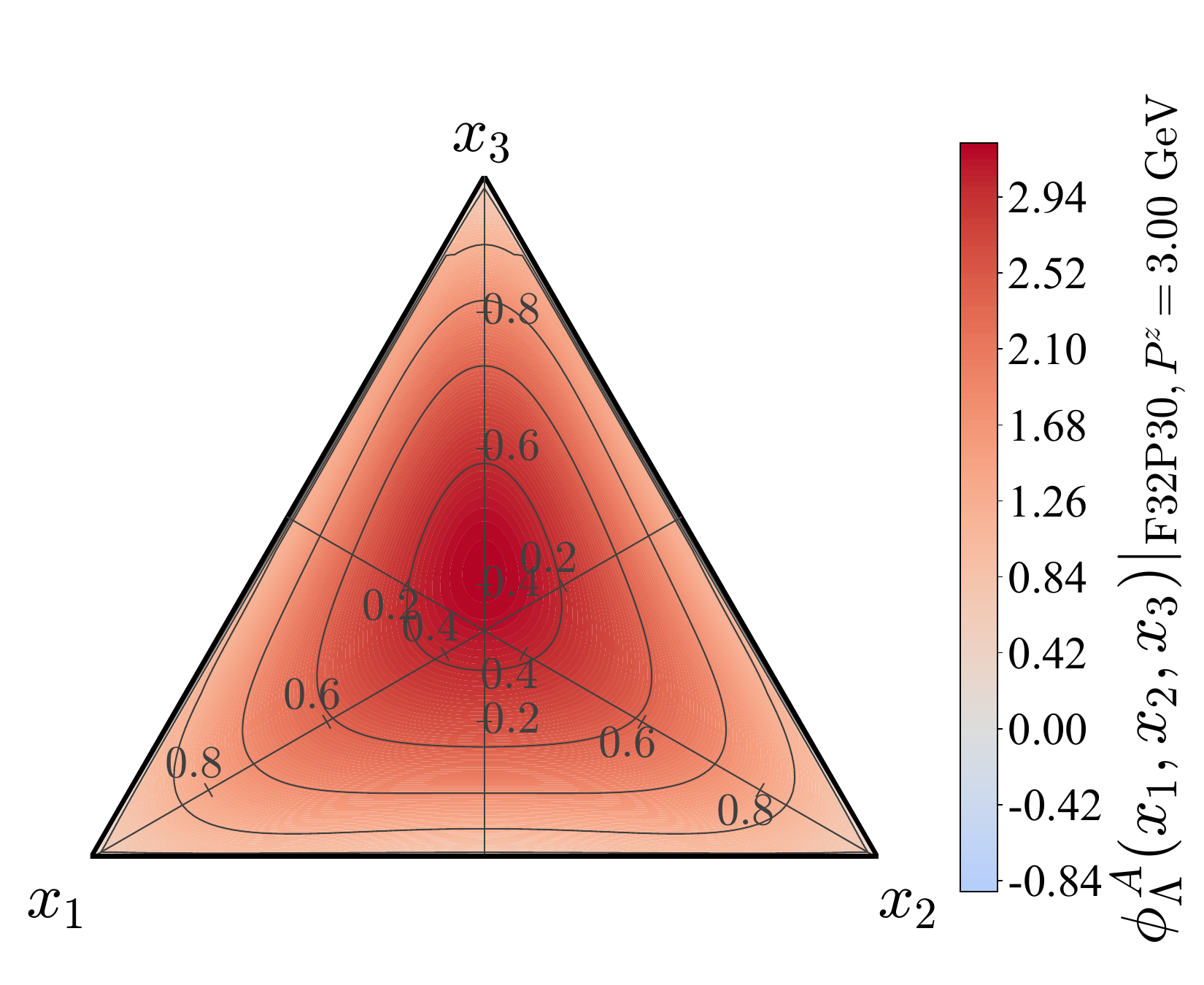}}
    \caption{Matched LCDAs for $\Lambda$ $A$ on F32P30 ($a \approx 0.0775~{\rm fm}$, $m_\pi\approx 300.4~{\rm MeV}$). Left: $x_1$-dependence at fixed $x_2=0.2$. Middle: $x_3$-dependence along $x_1=x_2$. Right: the heat map in the two-dimensional physical region.}\label{fig:match_sub_F32P30}
\end{figure}

\begin{figure}[htbp]
    \centering
    \subfloat{
        \centering
        \includegraphics[width=0.28\textwidth]{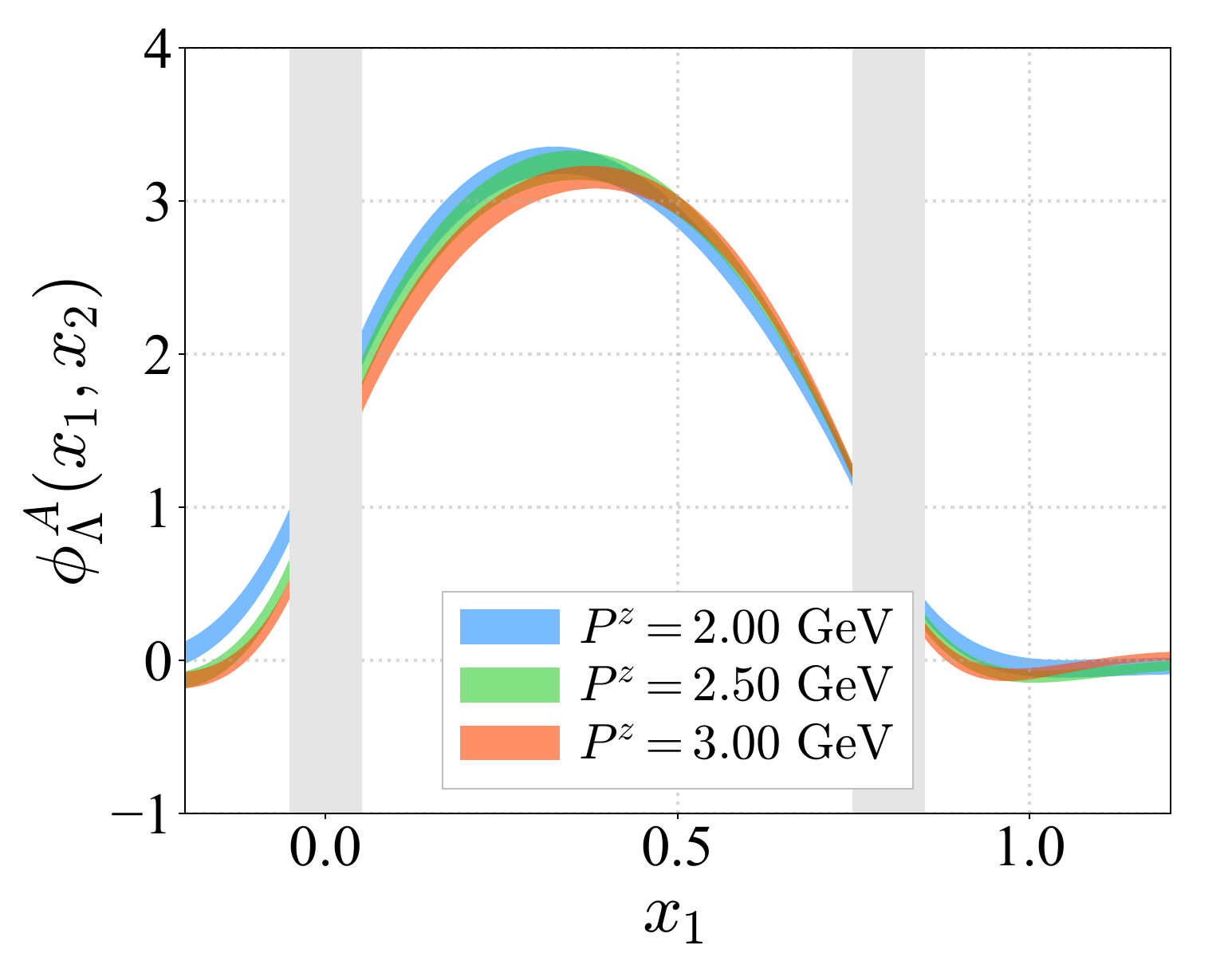}
        }\hspace{0.02\textwidth}
    \subfloat{
        \centering
        \includegraphics[width=0.28\textwidth]{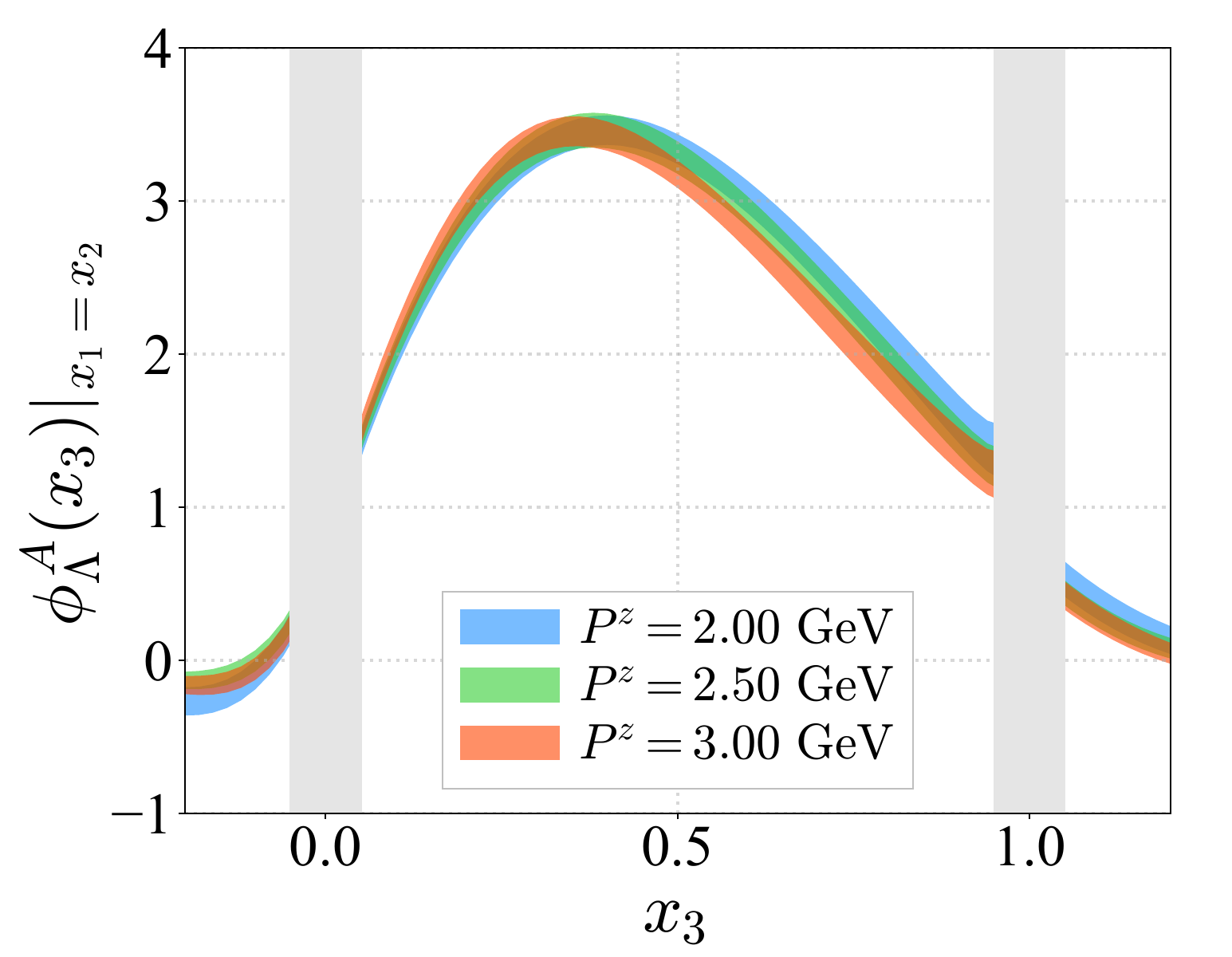}
        }\hspace{0.02\textwidth}
    \subfloat{
        \centering
        \includegraphics[width=0.28\textwidth]{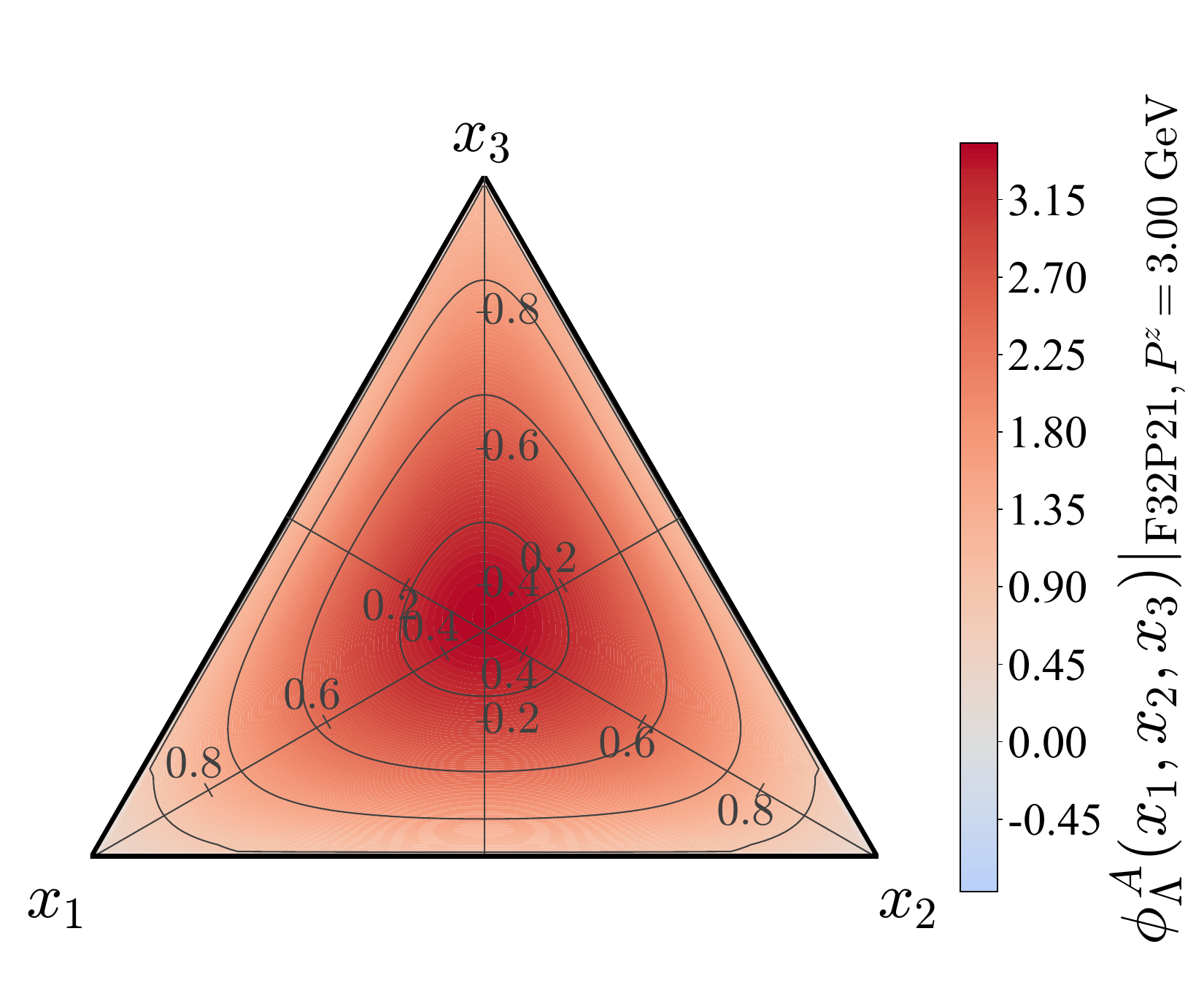}}
    \caption{Matched LCDAs for $\Lambda$ $A$ on F32P21 ($a \approx 0.0775~{\rm fm}$, $m_\pi\approx 210.3~{\rm MeV}$). Left: $x_1$-dependence at fixed $x_2=0.2$. Middle: $x_3$-dependence along $x_1=x_2$. Right: the heat map in the two-dimensional physical region.}\label{fig:match_sub_F32P21}
\end{figure}

\begin{figure}[htbp]
    \centering
    \subfloat{
        \centering
        \includegraphics[width=0.28\textwidth]{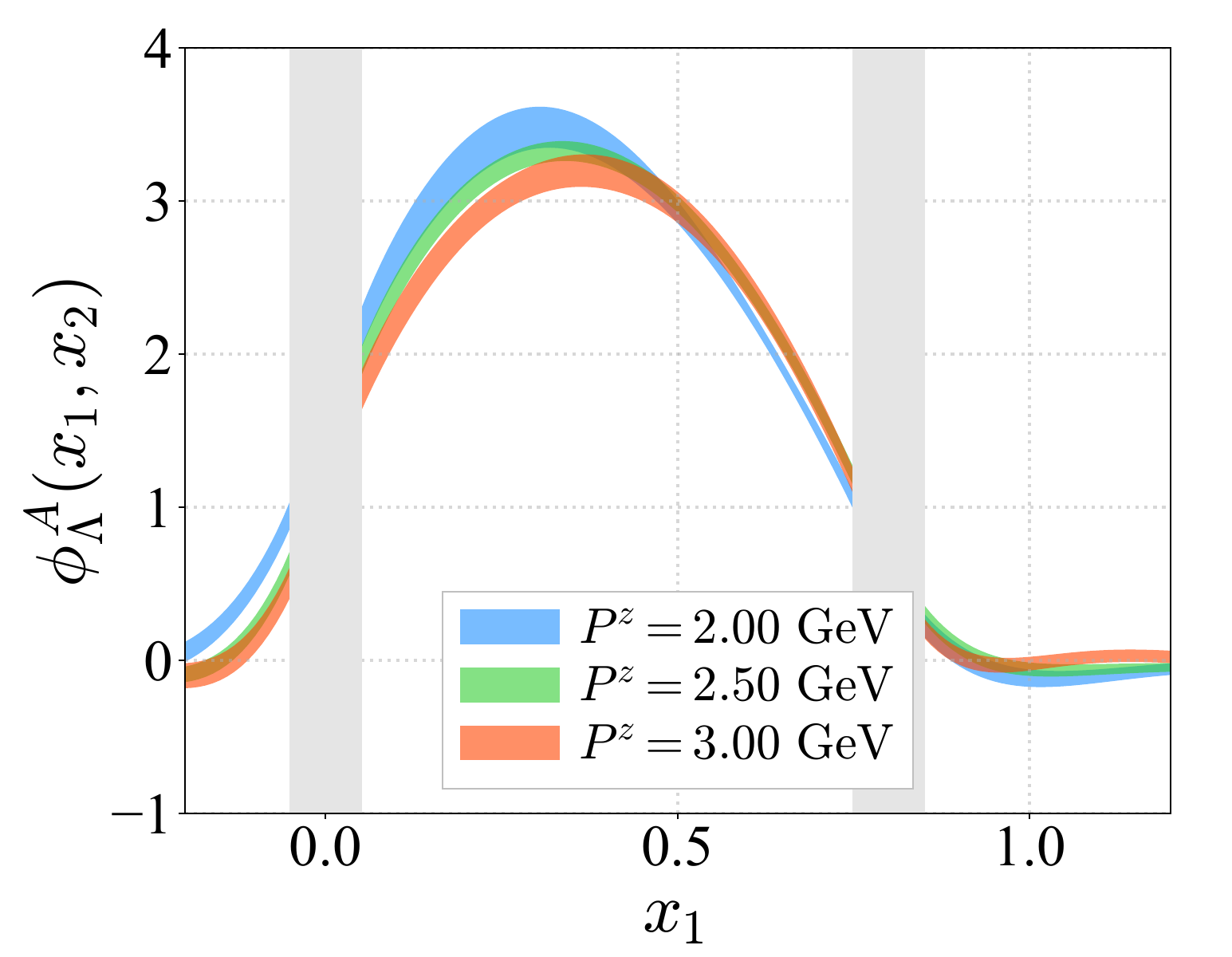}
        }\hspace{0.02\textwidth}
    \subfloat{
        \centering
        \includegraphics[width=0.28\textwidth]{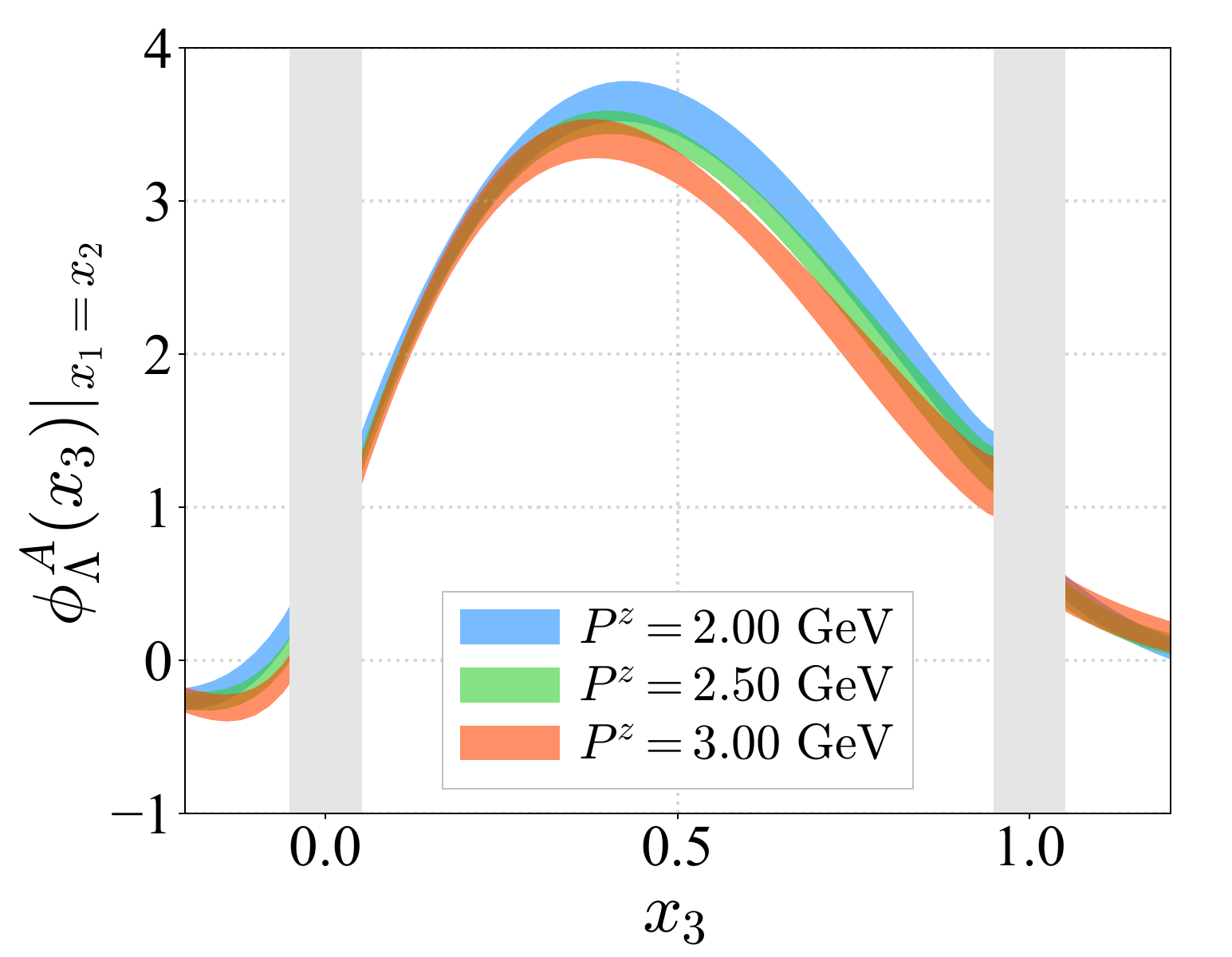}
        }\hspace{0.02\textwidth}
    \subfloat{
        \centering
        \includegraphics[width=0.28\textwidth]{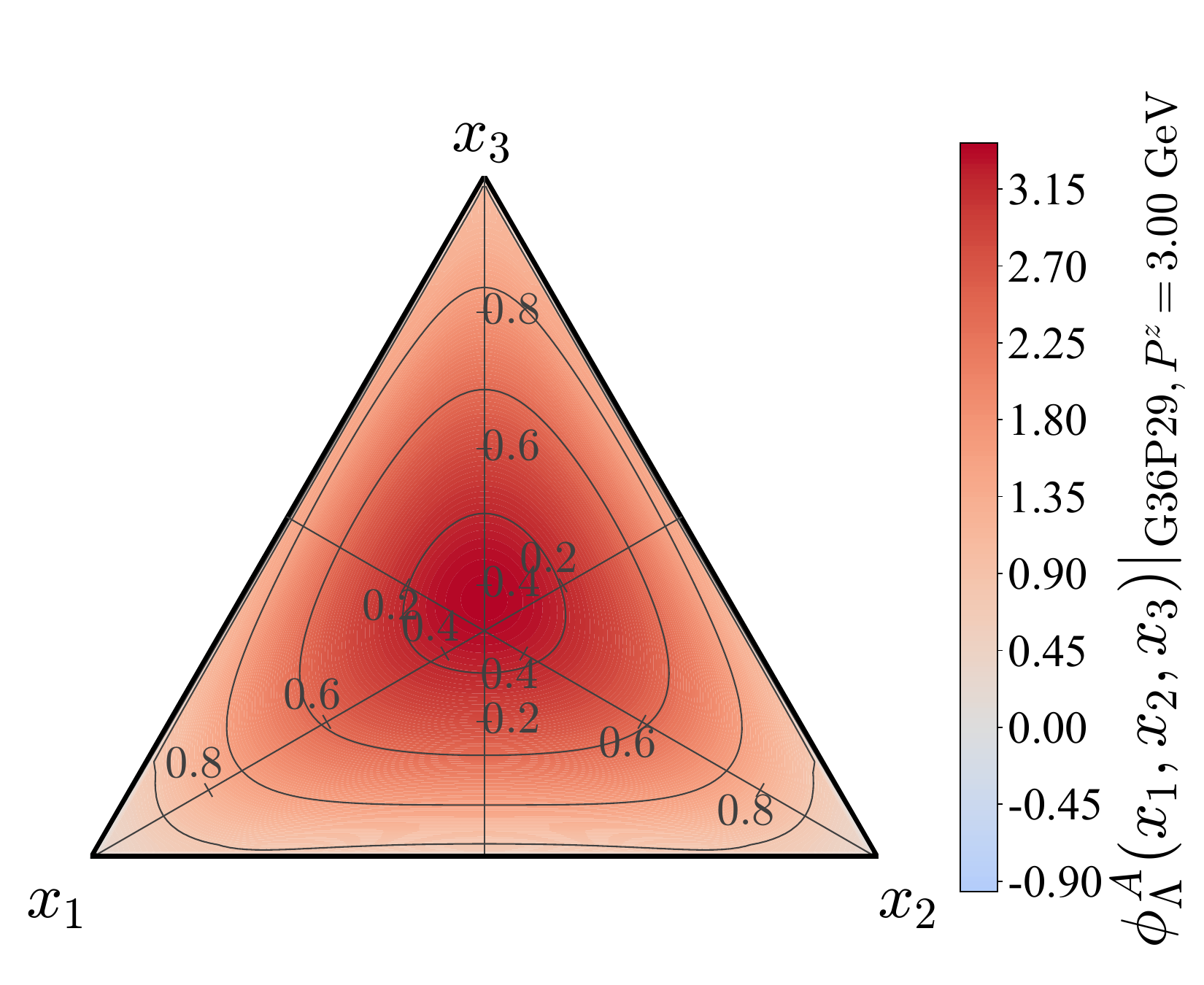}}
    \caption{Matched LCDAs for $\Lambda$ $A$ on G36P29 ($a \approx 0.0689~{\rm fm}$, $m_\pi\approx 297.2~{\rm MeV}$). Left: $x_1$-dependence at fixed $x_2=0.2$. Middle: $x_3$-dependence along $x_1=x_2$. Right: the heat map in the two-dimensional physical region.}\label{fig:match_sub_G36P29}
\end{figure}

\begin{figure}[htbp]
    \centering
    \subfloat{
        \centering
        \includegraphics[width=0.28\textwidth]{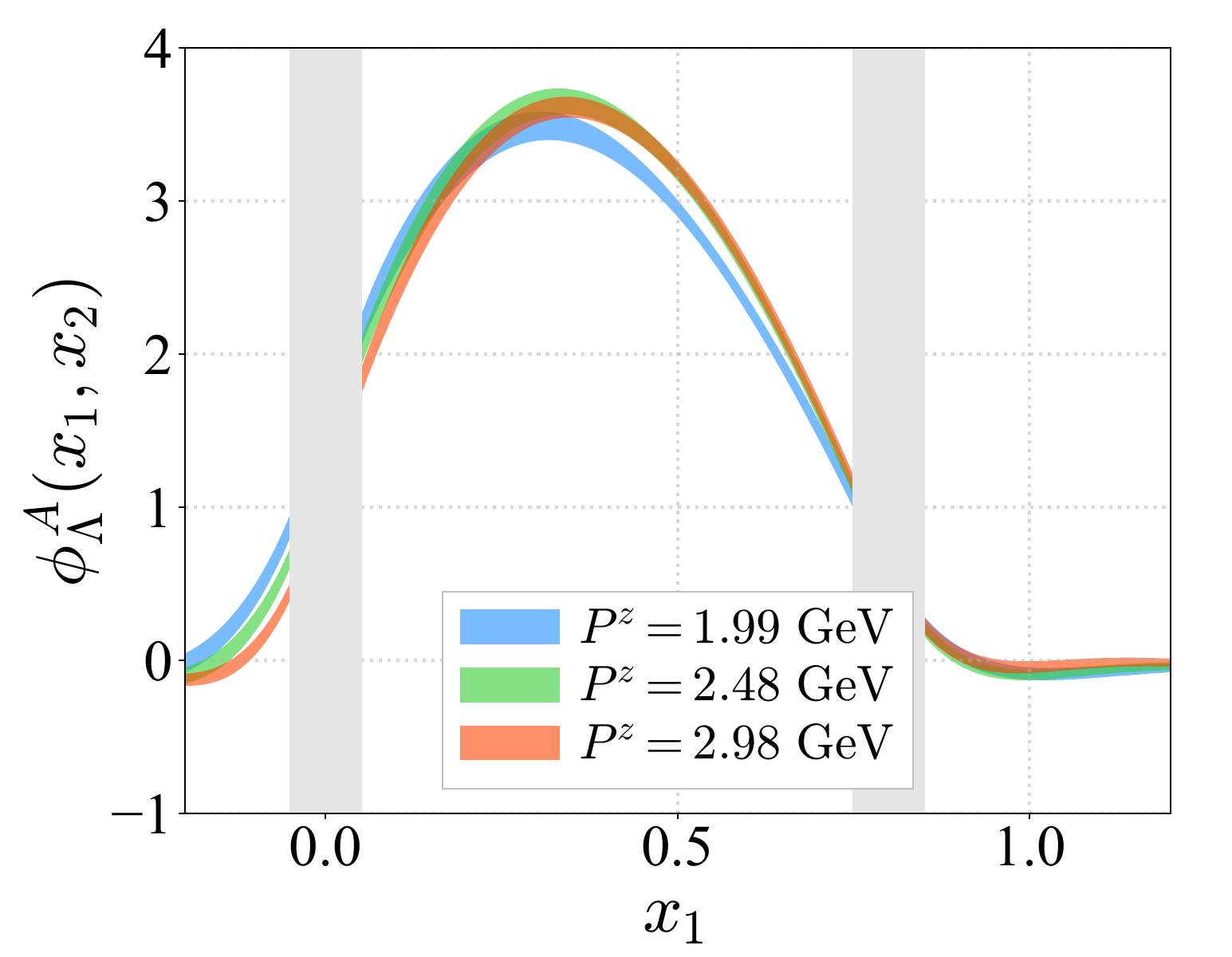}
        }\hspace{0.02\textwidth}
    \subfloat{
        \centering
        \includegraphics[width=0.28\textwidth]{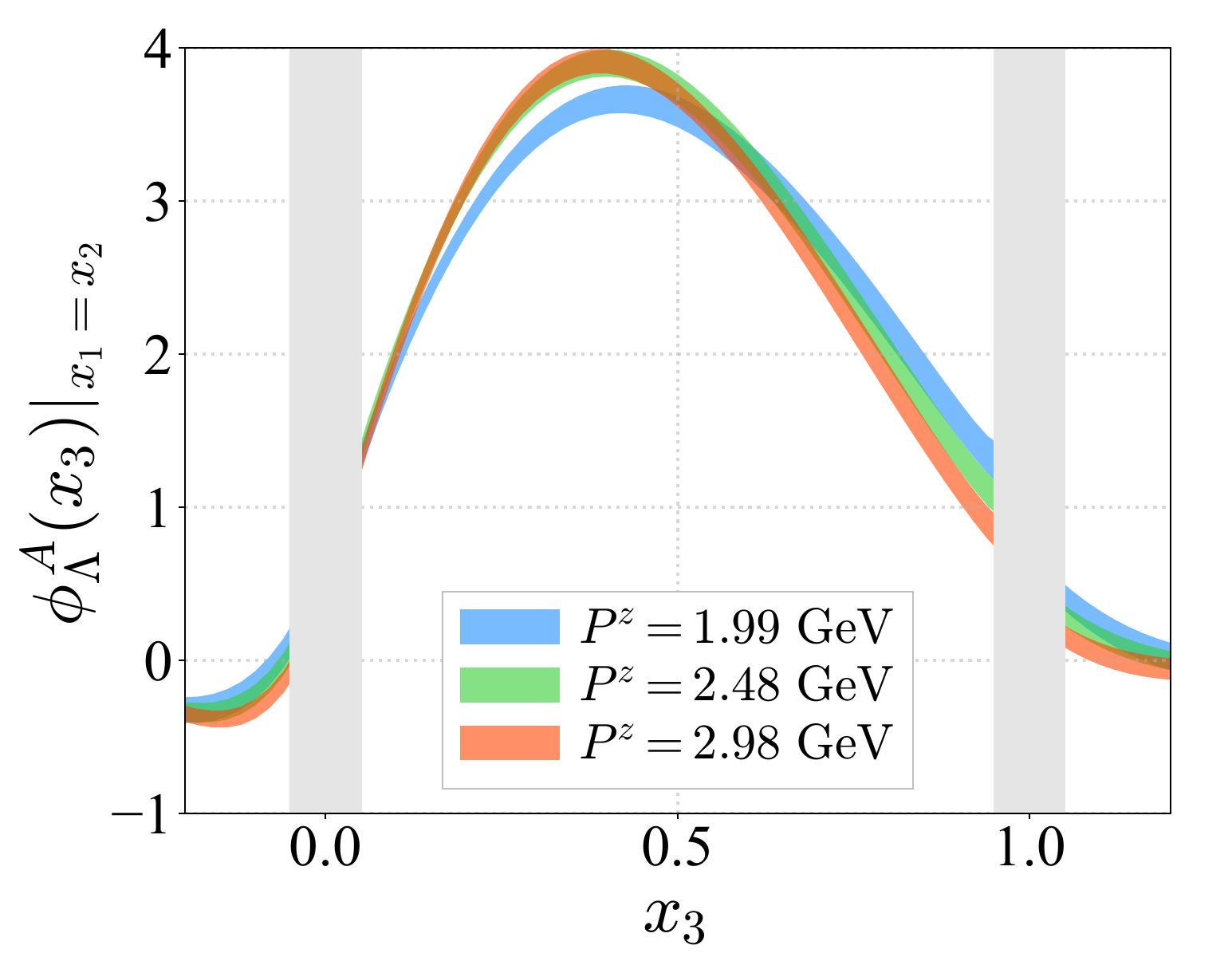}
        }\hspace{0.02\textwidth}
    \subfloat{
        \centering
        \includegraphics[width=0.28\textwidth]{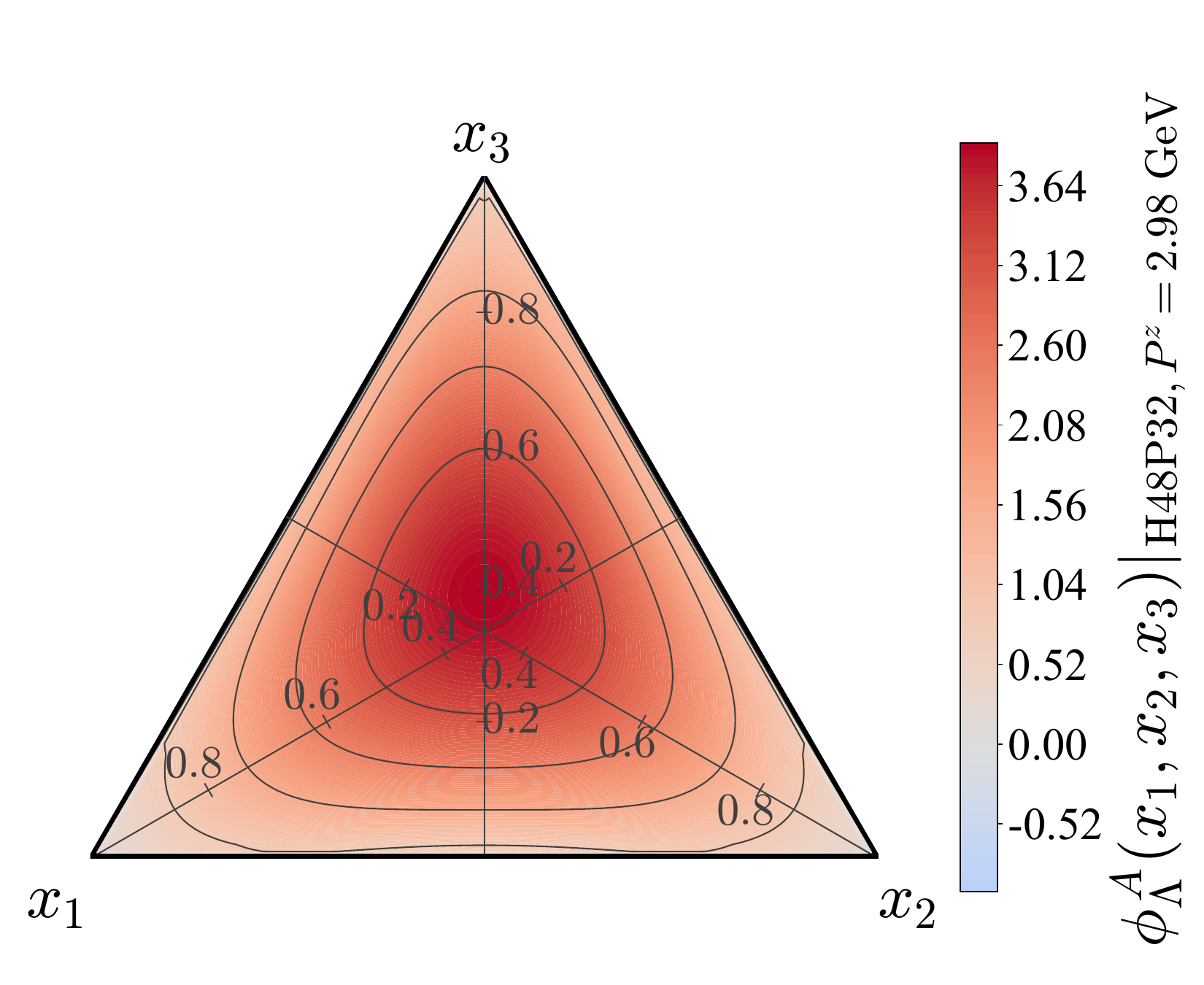}}
    \caption{Matched LCDAs for $\Lambda$ $A$ on H48P32 ($a \approx 0.0520~{\rm fm}$, $m_\pi\approx 316.6~{\rm MeV}$). Left: $x_1$-dependence at fixed $x_2=0.2$. Middle: $x_3$-dependence along $x_1=x_2$. Right: the heat map in the two-dimensional physical region.}\label{fig:match_sub_H48P32}
\end{figure}

\section{Moment parameterization for the \texorpdfstring{$x$}{x}-dependent \texorpdfstring{$\Lambda$}{Lambda} LCDAs}\label{app:lambda_third_moments}

To facilitate phenomenological applications, we provide a compact Gegenbauer-polynomial parameterization of the lattice-determined leading-twist $\Lambda$ LCDAs. The numerical fit results for the corresponding shape parameters are summarized in the companion Letter~\cite{LPC:2026lcj}, while this appendix records the explicit polynomial parameterizations $V_\Lambda(x_{123})$, $A_\Lambda(x_{123})$, and $T_\Lambda(x_{123})$, for the $V$, $A$, and $T$ amplitudes of $\Lambda$-baryon calculated in this work, respectively.

We adopt the standard conformal-polynomial basis for baryon LCDAs and keep the notation for the shape parameters consistent with Ref~\cite{Chernyak:1984bm,Braun:2008ia}.  The general Jacobi/Gegenbauer construction of the conformal basis is discussed in Refs.~\cite{Braun:2008ia,Huang:2025bbk}. The dimensionless parameterization forms quoted below are normalized such that the leading contribution to $A_\Lambda(x_{123})$ is proportional to $120x_1x_2x_3$.

\subsection{Orthogonal polynomials up to third moments}\label{app:ortho_polynomial}

We use the shorthand $x_{ijk}\equiv (x_i,x_j,x_k)$, $x_i+x_j+x_k=1$ in following discussions. For a generic leading-twist amplitude, the truncated conformal parameterization up to third moments takes the form:
\begin{equation}
    F_\Lambda(x_{123})
    =
    120x_1x_2x_3
    \left[
        \eta_{\Lambda,0}
        +\sum_{n=1}^{3}\sum_{m=0}^{n}
        \eta_{\Lambda,nm}
        \mathscr P_{nm}(x_{123})
        +\cdots
    \right]\ ,
\end{equation}
where the coefficients $\eta_{\Lambda,nm}$ denote dimensionless coefficients related to moments, and $\mathscr P_{nm}(x_{123})$ for orthogonal conformal polynomials. 

The polynomial basis $\mathscr P_{nm}$ adopted here can be written explicitly as functions of $x_{ijk}$ as follows.  The first two orders $n=1,2$ follow the standard octet-baryon LCDA convention used in Refs.~\cite{Bali:2015ykx,RQCD:2019hps,Bali:2024oxg}, while the higher-order polynomials are obtained from the same Jacobi/Gegenbauer construction of conformal three-quark operators discussed in Refs.~\cite{Braun:2008ia,Huang:2025bbk}.

The polynomials listed below define the fitting forms and determine the fitted coefficients in this appendix.  When using the results, each coefficient $\eta_{\Lambda,nm}$ should be combined directly with the polynomial $\mathscr P_{nm}$ carrying the same $(n,m)$ label.

The polynomials $\mathscr P_{nm}(x_{123})$ up to third order can be writen as:

\begin{equation}
\begin{aligned}
    \mathscr P_{00}(x_{123})
    &=
    1\ ,
    \\[0.5em]
    \mathscr P_{10}(x_{123})
    &=
    21x_1 - 21x_3\ ,
    \\[0.5em]
    \mathscr P_{11}(x_{123})
    &=
    7x_1 - 14x_2 + 7x_3\ ,
    \\[0.5em]
    \mathscr P_{20}(x_{123})
    &=
    \frac{
    189 x_{1}^{2}
    -189 x_{1}x_{2}
    -378 x_{1}x_{3}
    +126 x_{2}^{2}
    -189 x_{2}x_{3}
    +189 x_{3}^{2}
    }{10}\ ,
    \\[0.5em]
    \mathscr P_{21}(x_{123})
    &=
    \frac{
    63 x_{1}^{2}
    -189 x_{1}x_{2}
    +189 x_{2}x_{3}
    -63 x_{3}^{2}
    }{2}\ ,
    \\[0.5em]
    \mathscr P_{22}(x_{123})
    &=
    \frac{
    9 x_{1}^{2}
    +81 x_{1}x_{2}
    -108 x_{1}x_{3}
    -54 x_{2}^{2}
    +81 x_{2}x_{3}
    +9 x_{3}^{2}
    }{5}\ ,
    \\[0.5em]
    \mathscr P_{30}(x_{123})
    &=
    \frac{1}{38408}
    \left(
    \begin{aligned}
        & + 247 \sqrt{4801}\  x_{1}^{3}
          + 33607 x_{1}^{3}
          - 201642 x_{1}^{2}x_{3}
          - 1482 \sqrt{4801}\  x_{1}^{2}x_{3}
        \\
        & + 3960 \sqrt{4801}\  x_{1}x_{2}^{2}
          - 2376 \sqrt{4801}\  x_{1}x_{2}
          + 1482 \sqrt{4801}\  x_{1}x_{3}^{2}
        \\
        & + 201642 x_{1}x_{3}^{2}
          + 264 \sqrt{4801}\  x_{1}
          - 3960 \sqrt{4801}\  x_{2}^{2}x_{3}
        \\
        & + 2376 \sqrt{4801}\  x_{2}x_{3}
          - 33607 x_{3}^{3}
          - 247 \sqrt{4801}\  x_{3}^{3}
          - 264 \sqrt{4801}\  x_{3}
    \end{aligned}
    \right)\ ,
    \\[0.5em]
    \mathscr P_{31}(x_{123})
    &=
    \frac{1}{15520}
    \left(
    \begin{aligned}
        & - 4235 \sqrt{97}\  x_{1}^{2}x_{2}
          + 186725 x_{1}^{2}x_{2}
          - 37345 x_{1}^{2}
          + 847 \sqrt{97}\  x_{1}^{2}
        \\
        & - 560175 x_{1}x_{2}x_{3}
          + 12705 \sqrt{97}\  x_{1}x_{2}x_{3}
          - 2541 \sqrt{97}\  x_{1}x_{3}
        \\
        & + 112035 x_{1}x_{3}
          + 36960 \sqrt{97}\  x_{2}^{3}
          - 44352 \sqrt{97}\  x_{2}^{2}
        \\
        & - 4235 \sqrt{97}\  x_{2}x_{3}^{2}
          + 186725 x_{2}x_{3}^{2}
          + 14784 \sqrt{97}\  x_{2}
        \\
        & - 37345 x_{3}^{2}
          + 847 \sqrt{97}\  x_{3}^{2}
          - 1232 \sqrt{97}
    \end{aligned}
    \right)\ ,
    \\[0.5em]
    \mathscr P_{32}(x_{123})
    &=
    \frac{1}{15520}
    \left(
    \begin{aligned}
        & + 4235 \sqrt{97}\  x_{1}^{2}x_{2}
          + 186725 x_{1}^{2}x_{2}
          - 37345 x_{1}^{2}
          - 847 \sqrt{97}\  x_{1}^{2}
        \\
        & - 560175 x_{1}x_{2}x_{3}
          - 12705 \sqrt{97}\  x_{1}x_{2}x_{3}
          + 2541 \sqrt{97}\  x_{1}x_{3}
        \\
        & + 112035 x_{1}x_{3}
          - 36960 \sqrt{97}\  x_{2}^{3}
          + 44352 \sqrt{97}\  x_{2}^{2}
        \\
        & + 4235 \sqrt{97}\  x_{2}x_{3}^{2}
          + 186725 x_{2}x_{3}^{2}
          - 14784 \sqrt{97}\  x_{2}
        \\
        & - 37345 x_{3}^{2}
          - 847 \sqrt{97}\  x_{3}^{2}
          + 1232 \sqrt{97}
    \end{aligned}
    \right)\ ,
    \\[0.5em]
    \mathscr P_{33}(x_{123})
    &=
    \frac{1}{38408}
    \left(
    \begin{aligned}
        & -247 \sqrt{4801}\  x_{1}^{3}
          + 33607 x_{1}^{3}
          - 201642 x_{1}^{2}x_{3}
          + 1482 \sqrt{4801}\  x_{1}^{2}x_{3}
        \\
        & - 3960 \sqrt{4801}\  x_{1}x_{2}^{2}
          + 2376 \sqrt{4801}\  x_{1}x_{2}
          - 1482 \sqrt{4801}\  x_{1}x_{3}^{2}
        \\
        & + 201642 x_{1}x_{3}^{2}
          - 264 \sqrt{4801}\  x_{1}
          + 3960 \sqrt{4801}\  x_{2}^{2}x_{3}
        \\
        & - 2376 \sqrt{4801}\  x_{2}x_{3}
          - 33607 x_{3}^{3}
          + 247 \sqrt{4801}\  x_{3}^{3}
          + 264 \sqrt{4801}\  x_{3}
    \end{aligned}
    \right)\ .
\end{aligned}
\end{equation}

\subsection{Fully expanded LCDAs in the normalization convention of this work}

We now give the explicit $V_\Lambda(x_{123})$, $A_\Lambda(x_{123})$, and $T_\Lambda(x_{123})$ parameterizations used in our numerical fit.  These expressions are obtained by imposing the exchange symmetries of the $\Lambda$-baryon leading-twist amplitudes,
\begin{equation}
\begin{aligned}
    V_\Lambda(x_{123})=-V_\Lambda(x_{213})\ , \qquad
    A_\Lambda(x_{123})=+A_\Lambda(x_{213})\ , \qquad 
    T_\Lambda(x_{123})=-T_\Lambda(x_{213})\ .
\end{aligned}
\end{equation}
The resulting expressions are dimensionless functions. Since the $V$ and $A$ amplitudes share the same decay constant, we denote their dimensionless coefficients by $\phi_{\Lambda,nm}$. The corresponding coefficients for the $T$ amplitude are denoted by $\pi_{\Lambda,nm}$.

These parameterizations follow the same normalization prescription as in the main lattice analysis: in coordinate space, the nonlocal matrix elements for the $V$, $A$, and $T$ structures are all normalized by the same local, non-vanishing $A$-structure two-point correlation, as discussed in Sec.~\ref{sec:quasi}.

This normalization convention introduces an additional factor for the antisymmetric amplitude $T$. In particular, the quantity obtained from this normalization is in fact $\frac{ f_{\Lambda}^{T}}{f_{\Lambda}}\ \phi_{\Lambda}^{T}(x_1,x_2)$, see Eq.~\eqref{eq:norm_T} and Eq.~\eqref{eq:ratio_T}; see also the companion Letter~\cite{LPC:2026lcj}. The same convention is reflected in the corresponding normalized coefficients of $T_\Lambda(x_{123})$, which are defined as:
\begin{equation}
    \widetilde{\pi}_{\Lambda,nm}=\frac{ f_{\Lambda}^{T}}{f_{\Lambda}}\ 
    \pi_{\Lambda,nm}\ .
\end{equation}
Here, $f_{\Lambda}^{T}$ is the genuine decay constant associated with the $T$ amplitude, while $f_{\Lambda}$ for $V$ and $A$ amplitudes.

Since the antisymmetric $T$ structure has a vanishing local limit, $f_\Lambda^T$ cannot be determined directly from the corresponding local matrix element. This does not affect phenomenological applications, because the LCDA enters the relevant convolution together with its associated decay constant. Therefore, using normalized $\widetilde{\pi}_{\Lambda,nm}$ and multiplying it by $f_{\Lambda}$ still provides a well-defined phenomenological input.

\subsubsection{Parameterization of \texorpdfstring{$\Lambda$}{Lambda}-baryon \texorpdfstring{$A$}{A} LCDA}
The following expressions then represent the direct input forms used to fit the final $x$-dependent LCDAs. The parameterization for $\Lambda$-baryon $A$ amplitude $\phi_{\Lambda}^{A}(x_1,x_2)$:
\begin{equation}
\begin{aligned}
    & A_{\Lambda}\left(x_{123}\right)=120 x_1 x_2 x_3\Bigg\{\phi_{\Lambda, 0}[1] \\
    & +\phi_{\Lambda, 10}\left[\frac{-7 x_1-7 x_2+14 x_3}{2}\right] \\
    & +\phi_{\Lambda, 11}\left[\frac{-7 x_1-7 x_2+14 x_3}{2}\right] \\
    & +\phi_{\Lambda, 20}\left[\frac{315 x_1^2-378 x_1 x_2-567 x_1 x_3+315 x_2^2-567 x_2 x_3+378 x_3^2}{20}\right] \\
    & +\phi_{\Lambda, 21}\left[\frac{-21 x_1^2+126 x_1 x_2-63 x_1 x_3-21 x_2^2-63 x_2 x_3+42 x_3^2}{4}\right] \\
    & +\phi_{\Lambda, 22}\left[\frac{-45 x_1^2+162 x_1 x_2-27 x_1 x_3-45 x_2^2-27 x_2 x_3+18 x_3^2}{10}\right] \\
    & +\frac{\phi_{\Lambda, 30}}{230448}\left[\begin{array}{l}
    -33607 x_1^3-247 \sqrt{4801} x_1^3 \\
    -3960 \sqrt{4801} x_1^2 x_2+201642 x_1^2 x_3 \\
    +5442 \sqrt{4801} x_1^2 x_3-3960 \sqrt{4801} x_1 x_2^2 \\
    +4752 \sqrt{4801} x_1 x_2-201642 x_1 x_3^2 \\
    -1482 \sqrt{4801} x_1 x_3^2-2376 \sqrt{4801} x_1 x_3 \\
    -264 \sqrt{4801} x_1-33607 x_2^3-247 \sqrt{4801} x_2^3 \\
    +201642 x_2^2 x_3+5442 \sqrt{4801} x_2^2 x_3 \\
    -201642 x_2 x_3^2-1482 \sqrt{4801} x_2 x_3^2 \\
    -2376 \sqrt{4801} x_2 x_3-264 \sqrt{4801} x_2 \\
    +494 \sqrt{4801} x_3^3+67214 x_3^3+528 \sqrt{4801} x_3
    \end{array}\right]
    +\frac{\phi_{\Lambda, 31}}{31040}\left[\begin{array}{l}
    36960 \sqrt{97} x_1^3-4235 \sqrt{97} x_1^2 \\
    +186725 x_1^2 x_2-43505 \sqrt{97} x_1^2-37345 x_1^2 \\
    -4235 \sqrt{97} x_1 x_2^2+186725 x_1 x_2^2 \\
    -1120350 x_1 x_2 x_3+25410 \sqrt{97} x_1 x_2 x_3 \\
    -4235 \sqrt{97} x_1 x_3^2+186725 x_1 x_3^2 \\
    -2541 \sqrt{97} x_1 x_3+112035 x_1 x_3+14784 \sqrt{97} x_1 \\
    +36960 \sqrt{97} x_2^3-43505 \sqrt{97} x_2^2-37345 x_2^2 \\
    -4235 \sqrt{97} x_2 x_3^2+186725 x_2 x_3^2 \\
    -2541 \sqrt{97} x_2 x_3+112035 x_2 x_3+14784 \sqrt{97} x_2 \\
    -74690 x_3^2+1694 \sqrt{97} x_3^2-2464 \sqrt{97}
    \end{array}\right] \\
    & \left.+\frac{\phi_{\Lambda, 32}}{31040}\left[\begin{array}{l}
    -36960 \sqrt{97} x_1^3+4235 \sqrt{97} x_1^2 x_2 \\
    +186725 x_1^2 x_2-37345 x_1^2+43505 \sqrt{97} x_1^2 \\
    +4235 \sqrt{97} x_1 x_2^2+186725 x_1 x_2^2 \\
    -1120350 x_1 x_2 x_3-25410 \sqrt{97} x_1 x_2 x_3 \\
    +4235 \sqrt{97} x_1 x_3^2+186725 x_1 x_3^2 \\
    +2541 \sqrt{97} x_1 x_3+112035 x_1 x_3-14784 \sqrt{97} x_1 \\
    -36960 \sqrt{97} x_2^3-37345 x_2^2+43505 \sqrt{97} x_2^2 \\
    +4235 \sqrt{97} x_2 x_3^2+186725 x_2 x_3^2 \\
    +2541 \sqrt{97} x_2 x_3+112035 x_2 x_3-14784 \sqrt{97} x_2 \\
    -74690 x_3^2-1694 \sqrt{97} x_3^2+2464 \sqrt{97}
    \end{array}\right]
    +\frac{\phi_{\Lambda, 33}}{230448}\left[\begin{array}{l}
    -33607 x_1^3+247 \sqrt{4801} x_1^3 \\
    +3960 \sqrt{4801} x_1^2 x_2-5442 \sqrt{4801} x_1^2 x_3 \\
    +201642 x_1^2 x_3+3960 \sqrt{4801} x_1 x_2^2 \\
    -4752 \sqrt{4801} x_1 x_2-201642 x_1 x_3^2 \\
    +1482 \sqrt{4801} x_1 x_3^2+2376 \sqrt{4801} x_1 x_3 \\
    +264 \sqrt{4801} x_1-33607 x_2^3+247 \sqrt{4801} x_2^3 \\
    -5442 \sqrt{4801} x_2^2 x_3+201642 x_2^2 x_3 \\
    -201642 x_2 x_3^2+1482 \sqrt{4801} x_2 x_3^2 \\
    +2376 \sqrt{4801} x_2 x_3+264 \sqrt{4801} x_2 \\
    -494 \sqrt{4801} x_3^3+67214 x_3^3-528 \sqrt{4801} x_3
    \end{array}\right]\right\}\ .
\end{aligned}
\end{equation}

\subsubsection{Parameterization of \texorpdfstring{$\Lambda$}{Lambda}-baryon \texorpdfstring{$V$}{V} LCDA}

The $V$ amplitude is fitted together with $A$ amplitude and shares the same dimensionless coefficients $\phi_{\Lambda,nm}$.  It is antisymmetric under $x_1\leftrightarrow x_2$, and the parameterization for $\Lambda$-baryon $V$ amplitude $\phi_{\Lambda}^{V}(x_1,x_2)$:
\begin{equation}
\begin{aligned}
    & V_{\Lambda}\left(x_{123}\right)=120 x_1 x_2 x_3\Bigg\{\phi_{\Lambda, 10}\left[\frac{7 x_1-7 x_2}{2}\right] \\
    & +\phi_{\Lambda, 11}\left[\frac{-21 x_1+21 x_2}{2}\right] \\
    & +\phi_{\Lambda, 20}\left[\frac{-63 x_1^2+189 x_1 x_3+63 x_2^2-189 x_2 x_3}{20}\right] \\
    & +\phi_{\Lambda, 21}\left[\frac{21 x_1^2-63 x_1 x_3-21 x_2^2+63 x_2 x_3}{4}\right] \\
    & +\phi_{\Lambda, 22}\left[\frac{-63 x_1^2+189 x_1 x_3+63 x_2^2-189 x_2 x_3}{10}\right] \\
    & +\frac{\phi_{\Lambda, 30}}{230448}\left[\begin{array}{l}
    247 \sqrt{4801} x_1^3+33607 x_1^3-3960 \sqrt{4801} x_1^2 x_2 \\
    -201642 x_1^2 x_3+2478 \sqrt{4801} x_1^2 x_3 \\
    +3960 \sqrt{4801} x_1 x_2^2+1482 \sqrt{4801} x_1 x_3^2 \\
    +201642 x_1 x_3^2-2376 \sqrt{4801} x_1 x_3 \\
    +264 \sqrt{4801} x_1-33607 x_2^3-247 \sqrt{4801} x_2^3 \\
    -2478 \sqrt{4801} x_2^2 x_3+201642 x_2^2 x_3 \\
    -201642 x_2 x_3^2-1482 \sqrt{4801} x_2 x_3^2 \\
    +2376 \sqrt{4801} x_2 x_3-264 \sqrt{4801} x_2
    \end{array}\right]
    +\frac{\phi_{\Lambda, 31}}{31040}\left[\begin{array}{l}
    36960 \sqrt{97} x_1^3-186725 x_1^2 x_2 \\
    +4235 \sqrt{97} x_1^2 x_2-45199 \sqrt{97} x_1^2+37345 x_1^2 \\
    -4235 \sqrt{97} x_1 x_2^2+186725 x_1 x_2^2 \\
    -4235 \sqrt{97} x_1 x_3^2+186725 x_1 x_3^2-112035 x_1 x_3 \\
    +2541 \sqrt{97} x_1 x_3+14784 \sqrt{97} x_1 \\
    -36960 \sqrt{97} x_2^3-37345 x_2^2+45199 \sqrt{97} x_2^2 \\
    -186725 x_2 x_3^2+4235 \sqrt{97} x_2 x_3^2 \\
    -2541 \sqrt{97} x_2 x_3+112035 x_2 x_3-14784 \sqrt{97} x_2
    \end{array}\right] \\
    & \left.+\frac{\phi_{\Lambda, 32}}{31040}\left[\begin{array}{l}
    -36960 \sqrt{97} x_1^3-186725 x_1^2 x_2 \\
    -4235 \sqrt{97} x_1^2 x_2+37345 x_1^2+45199 \sqrt{97} x_1^2 \\
    +4235 \sqrt{97} x_1 x_2^2+186725 x_1 x_2^2 \\
    +4235 \sqrt{97} x_1 x_3^2+186725 x_1 x_3^2-112035 x_1 x_3 \\
    -2541 \sqrt{97} x_1 x_3-14784 \sqrt{97} x_1 \\
    +36960 \sqrt{97} x_2^3-45199 \sqrt{97} x_2^2-37345 x_2^2 \\
    -186725 x_2 x_3^2-4235 \sqrt{97} x_2 x_3^2 \\
    +2541 \sqrt{97} x_2 x_3+112035 x_2 x_3+14784 \sqrt{97} x_2
    \end{array}\right]
    +\frac{\phi_{\Lambda, 33}}{230448}\left[\begin{array}{l}
    -247 \sqrt{4801} x_1^3+33607 x_1^3+264 \sqrt{4801} x_2 \\
    +3960 \sqrt{4801} x_1^2 x_2-201642 x_1^2 x_3 \\
    -2478 \sqrt{4801} x_1^2 x_3-3960 \sqrt{4801} x_1 x_2^2 \\
    -1482 \sqrt{4801} x_1 x_3^2+201642 x_1 x_3^2 \\
    +2376 \sqrt{4801} x_1 x_3-264 \sqrt{4801} x_1 \\
    +247 \sqrt{4801} x_2^3+2478 \sqrt{4801} x_2^2 x_3 \\
    +201642 x_2^2 x_3-201642 x_2 x_3^2-33607 x_2^3 \\
    +1482 \sqrt{4801} x_2 x_3^2-2376 \sqrt{4801} x_2 x_3 
    \end{array}\right]\right\}\ .
\end{aligned}
\end{equation}

\subsubsection{Parameterization of \texorpdfstring{$\Lambda$}{Lambda}-baryon \texorpdfstring{$T$}{T} LCDA}

The $T$ amplitude is fitted separately from $V$ and $A$ amplitudes, and is written in terms of its own coefficients $\widetilde\pi_{\Lambda,nm}$ defined above. The parameterization for $\Lambda$-baryon $T$ amplitude $\phi_{\Lambda}^{T}(x_1,x_2)$:
\begin{equation}
\begin{aligned}
    &T_{\Lambda}\left(x_{123}\right)=  120 x_1 x_2 x_3\Bigg\{\widetilde{\pi}_{\Lambda, 10}\left[-7 x_1+7 x_2\right] \\
    & +\widetilde{\pi}_{\Lambda, 21}\left[\frac{-21 x_1^2+63 x_1 x_3+21 x_2^2-63 x_2 x_3}{2}\right] \\
    & +\left.\frac{\widetilde{\pi}_{\Lambda, 30}}{115224}\left[\begin{array}{l}
    -33607 x_1^3-247 \sqrt{4801} x_1^3 \\
    +1482 \sqrt{4801} x_1^2 x_2+201642 x_1^2 x_2 \\
    -201642 x_1 x_2^2-1482 \sqrt{4801} x_1 x_2^2 \\
    -3960 \sqrt{4801} x_1 x_3^2+2376 \sqrt{4801} x_1 x_3 \\
    -264 \sqrt{4801} x_1+247 \sqrt{4801} x_2^3+33607 x_2^3 \\
    +3960 \sqrt{4801} x_2 x_3^2-2376 \sqrt{4801} x_2 x_3 \\
    +264 \sqrt{4801} x_2
    \end{array}\right]
    +\frac{\widetilde{\pi}_{\Lambda, 33}}{115224}\left[\begin{array}{l}
    -33607 x_1^3+247 \sqrt{4801} x_1^3 \\
    -1482 \sqrt{4801} x_1^2 x_2+201642 x_1^2 x_2 \\
    -201642 x_1 x_2^2+1482 \sqrt{4801} x_1 x_2^2 \\
    +3960 \sqrt{4801} x_1 x_3^2-2376 \sqrt{4801} x_1 x_3 \\
    +264 \sqrt{4801} x_1-247 \sqrt{4801} x_2^3+33607 x_2^3 \\
    -3960 \sqrt{4801} x_2 x_3^2+2376 \sqrt{4801} x_2 x_3 \\
    -264 \sqrt{4801} x_2
    \end{array}\right]\right\}\ .
\end{aligned}
\end{equation}

\subsection{Convention difference with RQCD moments}\label{app:conventionRCD}

We now relate the above $V_\Lambda(x_{123})$, $A_\Lambda(x_{123})$, and $T_\Lambda(x_{123})$ parameterizations to the $\Phi/\Pi$ amplitudes used in RQCD studies~\cite{Bali:2015ykx,RQCD:2019hps,Bali:2024oxg}. These $\Phi/\Pi$ amplitudes are linear combinations of the same leading-twist LCDAs $V$, $A$, and $T$, rather than independent non-perturbative functions. This identification follows, for example, from the RQCD definition of $\Phi_{\Lambda,+}$ in terms of the $[V-A]_\Lambda$ combination.

For this comparison, we first define dimensionless quantities by factoring out the RQCD decay constant $f_\Lambda^{\rm RQCD}$:
\begin{equation}
\begin{aligned}
    \widehat\Phi_{\Lambda,+}(x_{123})
    &\equiv
    \frac{\Phi_{\Lambda,+}^{\rm RQCD}(x_{123})}
         {f_\Lambda^{\rm RQCD}}\ ,\\
    \widehat\Phi_{\Lambda,-}(x_{123})
    &\equiv
    \frac{\Phi_{\Lambda,-}^{\rm RQCD}(x_{123})}
         {f_\Lambda^{\rm RQCD}}\ ,\\
    \widehat\Pi_{\Lambda}(x_{123})
    &\equiv
    \frac{\Pi_{\Lambda}^{\rm RQCD}(x_{123})}
         {f_\Lambda^{\rm RQCD}}\ .
\end{aligned}
\end{equation}
The exchange parity classification refers to the transformation $x_1 \leftrightarrow x_3$ in the RQCD representation:
\begin{itemize}
    \item Even ($+1$):
    $\mathscr P_{00}$, $\mathscr P_{11}$, $\mathscr P_{20}$,
    $\mathscr P_{22}$, $\mathscr P_{31}$, $\mathscr P_{32}$, $\cdots$;

    \item Odd ($-1$):
    $\mathscr P_{10}$, $\mathscr P_{21}$, $\mathscr P_{30}$,
    $\mathscr P_{33}$, $\cdots$.
\end{itemize}
With these conventions, the dimensionless $\Phi/\Pi$ combinations are given by:
\begin{equation}
\begin{aligned}
    \widehat\Phi_{\Lambda,+}(x_{123})
    &=
    120x_1x_2x_3
    \Bigl[
          \phi_{\Lambda,0}\mathscr P_{00}
        + \phi_{\Lambda,11}\mathscr P_{11}
        + \phi_{\Lambda,20}\mathscr P_{20}
        + \phi_{\Lambda,22}\mathscr P_{22}
        + \phi_{\Lambda,31}\mathscr P_{31}
        + \phi_{\Lambda,32}\mathscr P_{32}
        + \cdots
    \Bigr]\ ,
    \\
    \widehat\Phi_{\Lambda,-}(x_{123})
    &=
    120x_1x_2x_3
    \Bigl[
          \phi_{\Lambda,10}\mathscr P_{10}
        + \phi_{\Lambda,21}\mathscr P_{21}
        + \phi_{\Lambda,30}\mathscr P_{30}
        + \phi_{\Lambda,33}\mathscr P_{33}
        + \cdots
    \Bigr]\ ,
    \\
    \widehat\Pi_{\Lambda}(x_{123})
    &=
    120x_1x_2x_3
    \Bigl[
          \widetilde\pi_{\Lambda,10}\mathscr P_{10}
        + \widetilde\pi_{\Lambda,21}\mathscr P_{21}
        + \widetilde\pi_{\Lambda,30}\mathscr P_{30}
        + \widetilde\pi_{\Lambda,33}\mathscr P_{33}
        + \cdots
    \Bigr]\ .
\end{aligned}
\end{equation}
The relations used to obtain the $V_\Lambda$, $A_\Lambda$, and $T_\Lambda$ expressions from the RQCD $\Phi/\Pi$ combinations are:
\begin{equation}
\begin{aligned}
    A_\Lambda(x_{123})
    &=
    +\frac{1}{2}
    \left[
        \widehat\Phi_{\Lambda,+}(x_{123})
        +
        \widehat\Phi_{\Lambda,+}(x_{213})
    \right]
    -\frac{1}{6}
    \left[
        \widehat\Phi_{\Lambda,-}(x_{123})
        +
        \widehat\Phi_{\Lambda,-}(x_{213})
    \right]\ ,
    \\
    V_\Lambda(x_{123})
    &=
    -\frac{1}{2}
    \left[
        \widehat\Phi_{\Lambda,+}(x_{123})
        -
        \widehat\Phi_{\Lambda,+}(x_{213})
    \right]
    +\frac{1}{6}
    \left[
        \widehat\Phi_{\Lambda,-}(x_{123})
        -
        \widehat\Phi_{\Lambda,-}(x_{213})
    \right]\ ,
    \\
    T_\Lambda(x_{123})
    &=
    -\frac{1}{3}\ \widehat\Pi_\Lambda(x_{132})\ .
\end{aligned}
\end{equation}

Finally, we need to clarify the normalization conventions. In the RQCD convention, the leading normalization of the $\Lambda$-baryon LCDA is carried by
$\Phi_{\Lambda,+}$:
\begin{equation}
    \Phi_{\Lambda,+}^{(0),{\rm RQCD}}(x_{123})
    =
    120x_1x_2x_3\  f_\Lambda^{\rm RQCD}\ ,
\end{equation}
where the central value $|f_\Lambda^{\rm RQCD}| = 4.75 \times 10^{-3}~{\rm GeV}^2$ is taken from Table~IV of Ref.~\cite{Bali:2024oxg}. In contrast, in the present work the leading contribution to the dimensionless $A_\Lambda(x_{123})$ is
normalized as:
\begin{equation}
    A_\Lambda^{(0)}(x_{123})
    =
    120x_1x_2x_3\ .
\end{equation}
A point to note is that the decay-constant convention used in this appendix differ from that adopted in the RQCD works. Assuming the same baryon-state phase convention, the magnitude of the decay constant in the present convention is related to the RQCD one by:
\begin{equation}\label{eq:decay_const_relation}
    |f_\Lambda|
    =
    \sqrt{\frac{3}{2}}\ 
    |f_\Lambda^{\rm RQCD}|\ .
\end{equation}
Therefore, when the dimensionless parameterizations given in this appendix are used as a phenomenological input, the corresponding decay constant may be taken as $|f_\Lambda| \approx 5.82\times10^{-3}~{\rm GeV}^2$, obtained from the RQCD value after applying the factor $\sqrt{3/2}$.

\end{widetext}
\end{appendix}

\bibliography{PRDref}

\end{document}